\documentclass[11pt]{article}
\usepackage{latexsym,amssymb,amsmath}
\usepackage[dvips]{graphics,graphicx,epsfig,color}
\usepackage[lofdepth,lotdepth]{subfig}
\begin{document}
%\linenumbers
%
\thispagestyle{empty}
\title{\bf  On the constant constitutive parameter (e.g., mass density) assumption in integral equation  approaches to (acoustic) wave scattering}
\author{\bf Armand Wirgin \thanks {
LMA, CNRS, UMR 7031, Aix-Marseille Univ, Centrale Marseille, F-13453 Marseille Cedex 13, France.}}
\date{\today}
\maketitle
\begin{abstract}
In 2D acoustic and elastodynamic  problems the spatial variability of a constitutive parameter such as the mass density makes it difficult to employ boundary integral and domain integral techniques to solve the forward and inverse wave scattering problems. The oft-employed method for avoiding this problem is to assume  this constitutive parameter (which is chosen herein to be the mass density) to be spatially-invariant throughout all space. The reliability of this assumption is evaluated both theoretically and numerically and it is shown, in the example of a canonical-shaped scattering obstacle, that the scattered field can be obtained in the form of a series of powers of the mass density contrast (the latter vanishing for constant mass density). The first term of this series is the solution for the scattered field corresponding to the constant density assumption and it is shown that taking into account only two more terms in the series enables to correct for practically all the errors incurred by the constant mass density assumption for a wide range of the other constitutive parameters and frequencies. It is shown how to apply this result for obstacles of non-canonical shape. 
\end{abstract}
Keywords: 2D acoustics, 2D elastodynamics, scattering, domain integral equations, boundary integral equations, constant density
\newline
\newline
Abbreviated title: Constant mass density assumption in integral equation methods of wave scattering
\newline
\newline
Corresponding author: Armand Wirgin, \\ e-mail: wirgin@lma.cnrs-mrs.fr
%%%%%%%%%%%%%%%%%%%%%%%%%%%%%%%%%%%%%%%%%%%%%%%%%%%%%%%%%%%%%%%%%%%%%%%%%%%%%%%%%%%%
 \newpage
%%%%%%%%%%%%%%%%%%%%%%%%%%%%%%%%%%%%%%%%%%%%%%%%%%%%%%%%%%%%%%%%%%%%%%%%%%%%%%%%%%%%
\tableofcontents
%%%%%%%%%%%%%%%%%%%%%%%%%%%%%%%%%%%%%%%%%%%%%%%%%%%%%%%%%%%%%%%%%%%%%%%%%%%%%%%%%%%%
\newpage
%%%%%%%%%%%%%%%%%%%%%%%%%%%%%%%%%%%%%%%%%%%%%%%%%%%%%%%%%%%%%%%%%%%%%%%%%%%%%%%%%%%%
%%%%%%%%%%%%%%%%%%%%%%%%%%%%%%%%%%%%%%%%%%%%%%%%%%%%%%%%%%%%%%%%%%%%%%%%%%%%%%%%%%%%%%%%%%%%%%%%%%%%%%%%%%%%%%%%%%%%%%%%%%%%%%%%%%%%%%%%%%%%%%%%%
\section{Introduction}\label{intro}
When an oscillating or impulsive source  is present in an unbounded, homogeneous, isotropic medium, it produces radiation (i.e., a radiated acoustic, elastic or electromagnetic wave). When the medium is locally-inhomogeneous (as for instance, due to the presence of a spatially-bounded or unbounded in one or two directions (e.g., half-space) region in which the constitutive properties are different from those of the previously-mentioned homogeneous medium), the wave produced by the source is said to be scattered. Interest in scattering is primarily of practical nature (see \cite{bg04,ke16,wi16} for many references), but also of theoretical (mathematical) nature.

Scattering theory initially concerned forward (scattering) problems that deal with the prediction of the spatial and spectral distributions of the scattered wave, assuming known everything (boundaries and  constitutive functions (in terms of space, time/frequency variables) of the  source and scatterer (or scatterers which can be grouped into a single entity termed 'the obstacle') as well as the partial differential equations (PDE's) and boundary conditions (BC's, the radiation condition at infinity being a sort of BC on the 'boundary' at infinity) that describe the way the radiated field acts on the obstacle. At present, scattering theory concentrates mostly on inverse (scattering) problems \cite{ro04} that are concerned with the recovery (from knowledge of the radiated and/or scattered fields) of something or everything (i.e., the boundaries and constitutive functions) about the source and obstacle, assuming once again that the PDE's and BC's are known.

Fundamentally, the PDE's of the forward and inverse problems are the same \cite{mo87}, but in the forward case the unknown is a field (acoustic, elastic, electromagnetic) and usually linear in terms of this field, whereas in the inverse case the unknown(s) is(are) the coefficient(s) or driving term of the PDE (associated with the aforementioned source which we now qualify as 'applied' because it is assumed to not depend on the scattered field) and are nonlinear in terms of these entities. Another difference between the two cases is that forward scattering problems require a single resolution of the PDE per (scattering) configuration and frequency, whereas inverse scattering problems usually require several (if not many) resolutions of the PDE per configuration and frequency \cite{kd91}. For this reason, it has become increasingly-important to either modify (in the sense of simplification and enabling ease of resolution) the PDE and/or obtain the most appropriate method of resolution (notably to reduce the computational burden required for inversion).

The important point to remember is that both forward and inverse scattering problems require the resolution(s) of a  PDE (or PDE's) for a function of space and time (or space and frequency) that represents the wavefield of interest. This function can be scalar (representing pressure, as in acoustics) or vectorial (representing particle displacement, as in elastodynamics, or the electric or magnetic field, as in electrodynamics). In the vectorial case, the various field components are usually coupled so that the previously-mentioned PDE is, in fact, a set of coupled PDE's for the scalar components of the vectorial field. In both cases, we refer to the set of PDE's as 'the PDE'. Similarly, we shall refer to the multitude of boundary conditions of each problem as 'the boundary condition'.

The PDE can be solved in brute-force manner by some sort of discretization scheme (e.g., finite difference (FD; \cite{mo87,ro04,ve12}), finite element (FE; \cite{bj00,gt06}), spectral element (SE; \cite{kv98}), etc.). This is the approach in many of the so-called nonlinear full-wave inversion (FWI) schemes \cite{bk97,wb12} that are increasingly adopted \cite{ro04} in the geophysics community. We shall not be further concerned with this purely-numerical approach other than to underline the fact that, at present, little or no efforts are made, nor required in this approach (assuming, as is often the case, that one has considerable computer power at one's disposal) to simplify the governing PDE, notably by eliminating a term that one could estimate a priori to be small compared to the others. This, of course, was not the situation several decades ago, especially in the marine acoustics and geophysical (i.e., seismology in relation to the solid earth) contexts for large-scale scattering problems, when the tendency was rather to first reduce the set of coupled PDE's to  a single PDE for a scalar function (i.e., the so-called acoustic approximation) \cite{ta84,lv92,ns09,ve12,cc15} and then perhaps assume a coefficient  of the PDE to be either small or constant with respect to position (these coefficients were (and still are often) assumed to be independent of time when the time domain PDE was (is) solved, but this limitation was (is) artificially-overcome in the frequency domain PDE by making the velocity complex).

 Of course, less numerically-oriented methods also exist to solve directly (i.e., without resorting to integral transformations) the PDE. The most widely-known is a high-frequency procedure \cite{sr11} wherefrom the solution of the second-order scalar-wave PDE emerges iteratively via a series of simpler, first-order PDE's, the first one of which is none other than the eikonal equation, enabling the establishment of the notions of wavefronts and rays. Kline \cite{kl62} extends this method to the vectorial PDE of electrodynamics and shows that the successive solutions of the first-order equations combine to form an asymptotic (high-frequency) representation of the solution of the second-order equation. A less well-known procedure \cite{st53,da99} is low-frequency in nature, and consists in expanding the solution of the governing PDE in a series of powers of the wavenumber whereby the solution emerges via a set of simpler PDE's, the first of which is the static PDE. This series of powers was first thought to be convergent in certain regions outside the obstacle, but it is more generally an asymptotic series. Last but not least, we mention the parabolic approximation procedure \cite{co89,st93,wb12} which is particularly suited to problems in which the waves in the inhomogeneous medium travel preferentially in one (horizontally in the sea) direction. Using this fact, it is possible, using some approximations valid in the far-field zone, to reduce the hyberbolic (second-order) wave equation to a parabolic (first-order) equation which lends itself to the non-iterative so-called 'marching method' (also related to the 'splitting method' mentioned further on) for its solution. An interesting feature of this parabolic equation is that it resembles the Schr\"odinger equation of quantum mechanics and since the solution of the latter can be represented in path-integral form, it has been suggested that this feature might provide new insights on how to efficiently solve scalar wave scattering problems such as those which arise in acoustics \cite{fm84,fw86,fm87,sc99}.

Perhaps the oldest method for solving the PDE is by separation of variables (SOV). This is usually preceded by domain decomposition (DD) (e.g., to separate the regions inside and outside the obstacle and/or various homogeneous subregions within the obstacle) in order to be confronted with a set of PDE's having (spatially-) constant coefficients. SOV then enables (provided each decomposed domain of the obstacle is homogeneous and its boundary has certain shapes) the PDE(s) to be reduced to several ordinary differential equations (ODE), each of which can be solved (this term is inexact because the ODE's actually only give rise to {\it representations} of the wavefield, and the actual solution is obtained by employing these representations in the boundary conditions) in terms of special functions (exponentials, ordinary and spherical Bessel functions, etc.). The final solution of each PDE is then a sum or integral of the products of these various functions \cite{ke16}. This method (outlined herein in sect. \ref{ddsov} is very powerful (not just because computer libraries enable the computation of these special functions very efficiently), but is restricted to very special scattering problems (notably as concerns the shape of the boundary (planar, circular \cite{tr71,sw14,wi16}, spherical \cite{hh81,ma02,ke16}, etc.) of the scatterer).  The reasons for paying attention to the DD-SOV method are that: (a) it provides solutions of so-called 'canonical scattering problems' (with reference to the shape of the boundary) which are exact (to within numerical errors) and therefore useful to test the validity \cite{gt06} of other methods of resolution of the forward scattering problem, (b) it leads to a functional (i.e., mathematical instead of purely-numerical) expression of the solution which can be further manipulated by perturbation or asymptotic schemes so as to furnish functionally-approximate approximations (of the solution), which can be very useful in the inverse problem context \cite{ta76,wi95} when one or several parameters of the obstacle and/or of the source is/are small \cite{da99,sw04} or large \cite{wi95}, or when the obstacle has non-canonical shape but its response to the radiated  wave can be approximated locally by that of obstacles with canonical shape (i.e., the ICBA \cite{lw71,sw95,wi99,wi02,bg04,fm07}), and (c) the canonical solutions can be incorporated in homogenization schemes {\cite{ra92,si02,ke16,qk11,wi16} for composite or multi-body obstacles.

A useful variant of the DD-SOV method is its 'extension', by Rayleigh \cite{ra07,bw74,wi99}, to obstacles of other-than-canonical shape. Another interesting variant of the DD-SOV method is the so-called null-field or T-matrix method, attributed to Waterman \cite{wa65,wa69,bw74,wi99}. Both the Rayleigh and Waterman methods will not be discussed any more in the present study although they too provide grounds for further approximations (via perturbations \cite{ra07,er68,wi99} for a size parameter of the obstacle that is small compared to the wavelength, and \cite{ra45} for a size parameter that is large compared to the wavelength,  as well as other asymptotic procedures \cite{wi99}), and are, in any case, useful candidates for treating inverse scattering problems \cite{sw95,gr98}. A final line of attack, not discussed further on in our study, is that of the so-called wave-splitting method, which has proved to be useful too in inverse problems for continuously- or discontinuously-varying (in terms of position) constitutive parameters. The wave splitting technique is exact and easily-applicable only  for layered one-dimensional slab-or layer-like obstacles \cite{dr07,dr08}, but approximate and not easily-applicable for 2D and 3D obstacles \cite{gus00}.

This brings us to the very object of our study which concerns integral equation methods for solving forward, as well as inverse (for the aforementioned reasons) scattering problems.

In the preceding paragraphs, we followed the modern practice of discussing the governing equations  of elastodynamic and electrodynamic phenomena in their (partial) differential forms (PDE's). Actually, the procedure of going from PDE's to integral equations is contrary to the  way the said dynamic phenomena were first (physically-) formulated in integral terms and then reformulated in differential terms. For instance, Faraday's law, linking the electric field $\mathbf{E}$ to temporal ($t$) changes of the magnetic field $\mathbf{B}$, is (spatially) in integral form, and from this, one can, with the help of Stoke's theorem, derive  the PDE known as (one of the) Maxwell equation(s)  $\nabla\times\mathbf{E}=-\frac{\partial\mathbf{B}}{\partial t}$. Our procedure is perhaps justified by the fact that we are searching for ways, other than those mentioned previously, of actually solving the governing PDE, in the hope that these ways, based on integral representations of the wavefield, may offer useful alternatives to FD, FE, SE and DD-SOV methods, particularly in the inverse problem context.

Re-consider the radiation problem in a homogeneous, isotropic medium of infinite extent in all directions. In electrostatics, a radiated field reduces to a static electric field $\mathbf{E}$ and Coulomb's law is expressed by a domain (volume for a 3D source, area for a 2D source) integral  whose integrand is the product of the source density with the Green's function of the medium (called the free-space Green's function (FPGF) of this medium), the latter being the field radiated by a point or line source (and known by the law of action at a distance). By employing the divergence theorem, one can obtain a PDE relating the divergence of $\mathbf{E}$ to the source (i.e., charge) density divided by a (spatially-constant) constitutive parameter of the medium, and from this one can show \cite{pp56} that $\nabla\times\mathbf{E}=\mathbf{0}$ which is another PDE. This illustrates once again that a physically-established (Coulomb's) law, expressed in integral form, is, in fact, the starting point of the means by which the usual object of interest for mathematical theory (i.e., a PDE) is obtained. Of course, this does not exclude the reverse procedure, which is to obtain the integral representation from the PDE.

Consequently, we now address the issue  of how, starting with the PDE of our scattering problem, to represent, by a boundary or domain integral, the scattered field. Henceforth,  we shall be concerned mostly with elastic wave, and especially with, acoustic wave problems (because the latter are scalar in nature; note that for this reason, many publications dealing with geophysical elastic wave problems, or acoustic problems in which the obstacles are manifestly solid surrounded by a fluid host, make the so-called acoustic approximation which partially-enables the description of vectorial field phenomena by scalar field phenomena) \cite{cc15,ch86,gv86,ll07,lv92}. Moreover, since the time domain expressions of the governing equations relative to the wavefield are related to their frequency domain expressions by a simple Fourier transform, we will be concerned (unless stated otherwise) with frequency domain expressions of the wavefield.

The PDE of interest is of the form $\mathcal{L}(\mathbf{\mathfrak{c}})u=-s$, where $\mathcal{L}$ is a linear partial differential operator, $u$ the field and $s$ the source density. $\mathcal{L}$, $u$ and $s$ are functions of the space and  frequency) and $\mathcal{L}$ is second-order in terms of the space variables. $\mathcal{L}(\mathbf{\mathfrak{c}})$ contains certain  coefficients (grouped into the vector $\mathbf{\mathfrak{c}}$) which define  the constitutive properties of the medium filling the domain $\Omega$ in which the field $u$ is searched.

In the aforementioned {\it radiation} problem, $\Omega$ is all space, $s$ is usually of finite support in $\Omega$ and the coefficient vector $\mathbf{\mathfrak{c}}$ is constant (denoted by $\mathbf{\mathfrak{c}}^{[0]}$) with respect to the space, and also usually the frequency, variables. The free-space Green's function (FPGF), denoted by $g$, is the solution of the radiation problem when $s=-\delta$, wherein $\delta$ is the Dirac delta distribution which is a  function of the space variables of both the 'receiver' (i.e., where the field is 'sensed') and the 'emitter' (i.e., the point or line {\it applied} source). Since $\Omega$ (for this problem) is unbounded, the radiation condition tells us that the radiated wave must be outgoing at points far-removed from the location of the source. Furthermore, the field is, by the very nature of $\mathcal{L}$,  bounded everywhere except at the location of the source. It turns out that the field $g$ of this PDE can be represented by a sum of plane waves, and that the amplitudes of these waves can be fairly-easily obtained by projection with the help of the radiation condition \cite{mf53}. Thus, the solution $g$ of $\mathcal{L}(\mathbf{\mathfrak{c}}^{[0]})g=-\delta$ can be considered to be known (a similar technique is employed in \cite{ef56} to obtain the free-space Green's function for the elastodynamic wave equation).

When a part of the previous $\Omega$ is occupied by the obstacle (assumed to be isotropic), we say that the latter is included in the domain $\Omega_{1}$ whereas the remaining portion of $\Omega$ is $\Omega_{0}$, both being separated by the boundary $\Gamma$. We already know that the medium in $\Omega_{0}$ is characterized by the constant vector $\mathbf{\mathfrak{c}}^{[0]}$. Generally speaking, the medium in $\Omega_{1}$ is inhomogeneous, which means that $\mathfrak{c}$ therein, denoted by $\mathbf{\mathfrak{c}}^{[1]}$, is not constant with respect to the space variables. This fact makes the problem of the representation of the field $u$ more complex.

The first trick is to write the governing PDE $\mathcal{L}(\mathbf{\mathfrak{c}})u=-s$ in $\Omega$ as $\mathcal{L}(\mathbf{\mathfrak{c}}^{[0]})u=-s-[\mathcal{L}(\mathbf{\mathfrak{c}})u-\mathcal{L}(\mathbf{\mathfrak{c}}^{[0]})u]$ and to consider the factor $[~]$ on the  right-hand side of this equation as a new source density which is termed the {\it induced} source density because it is what induces (i.e., gives rise to) the scattered (=diffracted) field $u^{d}$, much as the the {\it applied} source density $s$ gives rise to the 'incident' field $u^{i}$, their sum being none other than the total field $u=u^{i}+u^{d}$. The second trick is to perform the multiplications:
$g\mathcal{L}(\mathbf{\mathfrak{c}}^{[0]})u=-gs-g[\mathcal{L}(\mathbf{\mathfrak{c}})u-\mathcal{L}(\mathbf{\mathfrak{c}}^{[0]})u]$ and $u\mathcal{L}(\mathbf{\mathfrak{c}}^{[0]})g=-u\delta$, subtract these two equations, and then integrate the result over $\Omega$. The third, and last, trick is to employ Green's second identity, the sifting property of the Dirac delta, and the radiation condition satisfied ~~by the ~~field ~~on ~~the 'boundary at ~infinity',~~ to~~ obtain ~~the~~ domain~~integral~ (DI) ~{\it representation} ~ $u=u^{i}+\int_{\Omega}g\left[\mathcal{L}(\mathbf{\mathfrak{c}})u-\mathcal{L}(\mathbf{\mathfrak{c}}^{[0]})u\right]d\Omega$ of $u$ in $\Omega$, wherein $u^{i}=\int_{\Omega}gsd\Omega$. We insist on the word 'representation' because this DI relation does not solve the problem since what we are looking for, i.e., $u$, appears not only on the left-hand side of the relation, but also  within the induced source density term $[~]$ on the right hand side. Solving the forward problem, i.e., actually determining $u$ throughout $\Omega$ by means of this DI representation, given $s$, $\mathbf{\mathfrak{c}}^{[0]}$, $\mathbf{\mathfrak{c}}^{[1]}$, and $\Gamma$, is another story that will be schematized further on in this introduction and treated in more detail in sect. \ref{DI}.

The  PDE for a complete (incident plus scattered) acoustic wavefield in a fluid-like medium with spatially-varying density and compressibility is generally-attributed to Bergman \cite{be46}. We call this the Bergman PDE (BPDE).  A more detailed development of the way to obtain the BPDE  from first mechanical continua principles can be found in \cite{bg04} from which is borrowed the material  in sect.  \ref{bgwx} of the present study. A DI representation of the wavefield is also associated with  Bergman's name (see also \cite{ma03}) and is named the Bergman domain integral representation (BDIR) hereafter. Actually there exist  several versions of the BDIR \cite{be46,ta76,ma02,lo09,co15} that are shown, in sect. \ref{recon} herein to be equivalent under certain conditions. In the same section, we also show that the 2D version of the elastodynamic domain integral representation of the antiplane component of the scattered displacement field in \cite{mk67} is equivalent to the BDIR provided certain associations are made between the fields and constitutive parameters of the two problems. The 3D DI representation of the elastodynamic wavefield was  proposed in \cite{mk67,lm95,le08,pa03}, but, since the stress herein will be on scalar fields, we shall not be interested in this DI representation any longer. Finally, an integral (actually, integrodifferential) representation similar to the BDIR, but for the electrodynamic wavefield, was first obtained by Shifrin \cite{sh51}. Further discussions of Shifrin's representation and how to employ it to solve electrodynamic scattering problems can be found in \cite{ac80,bi89,bl81,ta76} and we shall not discuss it any further for the reasons mentioned previously.

  Now, let us focus on the scattered-wave component of the BDIR. The domain $\Omega$ of this representation is actually all space (i.e., $\mathbb{R}^{n}$ in one ($n=1$), two or three-dimensional problems). As we saw previously, the integrand of the DI is actually the product of the FSGF (i.e., g) with  the induced source density, the latter involving: (1) the field within the obstacle and its gradient, and (2) the velocities and densities of the inner and outer media, or (in each of the two media) any two of the acoustic constitutive parameters (which may, or may not, depend on the spatial variables): mass density, wavespeed, bulk modulus and compressibility (we choose mass density and wavespeed hereafter). Since (1) depends on (2) and multiplies (2) in the integrand, the scattered field is a nonlinear function of the $\mathbf{\mathfrak{c}}$, this being the principal source of difficulty in the inverse problem (of determining one, several, or all of $s$, $\mathbf{\mathfrak{c}}^{[1]}$ and $\Gamma$ from the knowledge of $u$ in some subdomain of $\Omega$ and for a given set of frequencies). However, in the forward problem context we are not confronted with such a difficulty since we search for $u$ which appears only linearly in the integrand of the DI (although it is in a multiplicative relation with $g$ which is another field function, but the latter is known).

  At present, we address the question of how to employ the BDIR to solve for $u$ in $\Omega$ (i.e., the forward problem). The purely-numerical method is schematically to first divide $\Omega$ into two subdomains $\Omega_{ext}$ and $\Omega_{int}$, with the properties of $\Omega_{ext}$ being the spatially-invariant $\mathbf{\mathfrak{c}}^{[0]}$. It then follows that the portion of the DI concerning $\Omega_{ext}$ vanishes so that we obtain a sort of integral equation of the second kind $u=u^{i}+\int_{\Omega_{int}}g[~]d\Omega~;~\in \Omega_{int}$ which could be the means for obtaining $u\in \Omega_{int}$ if $[~]$ were only a function of $u$. Unfortunately, this is not the case when the mass density depends on position within $\Omega_{int}$ because then $[~]$ depends both on $u$ and $\nabla u$ (within $\Omega_{int}$).

  Two strategies have been proposed to cope with this problem. The first \cite{cc84,lo09,pa19}, which we call the {\it p-q transformation} in sect. \ref{pq}, enables the elimination of the field gradient in the induced source density term, but this is obtained at the expense of rendering more difficult (especially when there are discontinuites of density) the  way the density component of $\mathbf{\mathfrak{c}}$ is taken into account. The second strategy is to suppose that the mass density does not depend on position within all of  $\Omega$ \cite{nr06,ns09} and the majority of what is offered in the present investigation will deal with a critical examination of this supposition.

  The constant-density supposition enables the DI to be reduced to what is often called the Lippmann-Schwinger DI (LSDI) and the aforementioned integral equation of the second kind is then called the Lippmann-Schwinger integral equation (LSIE) \cite{ls50,mob87,ri65,wi02,zb02} for the sole unknown function $u\in\Omega_{int}$. Note that this unknown is of finite support when, as is often the case, the obstacle occupies a domain of finite extension.

  The preference for the DI rather then for the PDE methods, in relation to solving scattering problems in brute-force numerical manner, is due to the fact that the support of the unknown function of the PDE is all space (i.e., $\Omega$) and must be fenced-in by artificial (usually of the absorption variety \cite{gt06,ve12}) boundary conditions whereas that of the LSDI is only a subdomain (i.e., $\Omega_{int}$) of $\Omega$. Thus, the LSIE lends itself quite easily to brute-force numerical resolution \cite{ri65,gu00,gl01,wi99}, but also, and most often (in the scientific literature), to iterative resolution, starting with the so-called Born approximation \cite{rg86,gl01}, which amounts to assuming $u=u^{i}~;~\in\Omega_{int}$. The last step in the LSIE scheme is to obtain $u$ in $\Omega_{ext}$ by inserting the previously-obtained $u~;~in\Omega_{int}$ into the LS field representation.

  So much for the use of the LS method in conjunction with the {\it forward} scattering problem. The question now is whether there is any advantage in using such a DI method in the {\it inverse} scattering problem. For many years (actually, until powerful desktop computers became available), the scientific community responded almost unanimously in the positive manner for several reasons. The first has to do with what was written in the last lines of the preceding paragraph. But, quite early, it was realized that, even with the LS representation (which is based on the constant-density assumption, so that implicitly one renounces at retrieving the mass density in $\Omega_{int}$), the remaining to-be-determined parameter (more often a position-dependent function), i.e., the wavespeed in $\Omega_{int}$, intervenes in  nonlinear manner in the induced source density so that the LSIE is nonlinear in terms of this parameter (or function). Recall that it was linear in terms of $u$. Thus, retrieving  the wavespeed from the LSIE turns out to be a  difficult task.

   Let us examine the ingredients  of this task. The problem, for a known $s$, $\Gamma$, $\mathbf{\mathfrak{c}}^{[ext]}$ and measured field (in $\Omega_{mes}\subset\Omega_{ext}$), is to determine the wavenumber contrast function $\chi(\mathbf{\mathfrak{c}}^{[ext]},\mathbf{\mathfrak{c}}^{[int]})$ via the LSIE $u=u^{i}+\int_{\Omega_{int}}g\chi u d\Omega~;~\in \Omega_{mes}$. Since the $u$ in the integrand is unknown (i.e., not measured), there does not appear to exist any obvious way to 'invert' this IE so as to retrieve $\chi$ \cite{hw10}.

   This led to the idea of approximating $u$ within the induced source density term (the latter giving rise to what is termed the first-order Born approximation of the scattered field \cite{rg86,gl01,hs11}) so that the LSIE becomes a linear function of the to-be-retrieved parameter. Above all, $u$ in the integrand is now a known function (just like $g$) so that the only unknown (in the integrand) is $\chi$. This, of course, makes it much easier to solve the LSIE for the wavespeed. To do this, various approximations (asymptotic far-field, high frequency, etc.) were made of the Green's function \cite{to99} whereupon it was realized that $\chi$ emerges as a sort of (explicit) Fourier transform of $\chi$, which is the basis of what has become to be known as Diffraction Tomography ((DT)\cite{ra81,dl85,dt85,le85,rg86,wv86,mc93,dl00,gl01}, widely used in geophysical and bioacoustical applications \cite{hs11}. Other approximations of the Green's function \cite{mob87} enabled the explicit computation of the Fr\'echet derivative \cite{mo87} in the full-wave inversion (FWI) methods of inversion \cite{gv07}, the latter based on retrieval by optimizing (i.e., seeking for the minimum) of a functional involving the difference between $u$ in $\Omega_{mes}$ and $u^{i}+\int_{\Omega_{int}}g\chi u^{i} d\Omega~;~\in \Omega_{mes}$. Naturally, there exist refinements of the FWI method \cite{wi99,zb02,wb12,sw14,wi16} which make use of the DI representation of the scattered field, but since our study is not focused on the inverse problem, we shall not delve further into this issue.

   The main issue under our scrutiny is that of whether (or not) the constant mass-density assumption is justified, this, of course,  being of some importance in the forward problem, and perhaps of greater importance in the inverse problem. Surprisingly, there exist practically no studies (perhaps \cite{ta86,be10,hs11}) that treat this issue in depth. In the bioacoustics community, which is concerned with imaging organs within the body \cite{wb12}, it is customary, and probably justified for in situ studies, to either consider the organ to be surrounded by a medium whose mass density is quite close to  that of the organ \cite{wb12}, or in laboratory studies to surround the obstacle with a fluid medium (usually water) for the same purpose \cite{gl01}. In the geophysics community, this assumption appears a priori to be less justified, but is nonetheless frequently made for reasons of 'simplicity' or 'mathematical convenience' \cite{rg86}. In the porous media acoustics context (notably for applications dealing with the absorption of airborne sound)  it has been found \cite{wi18} that the effective (bulk) density  of typical air-saturated porous materials can be complex, dispersive, and quite different (real part) from the density of air. Sometimes,  theoretical studies of scattering are undertaken in a quite general setting (i.e., without assuming constant mass density), but the mass density is taken to be either continuously-varying \cite{to10} or  constant in the final, numerical stage \cite{gv86}. The reason for the latter assumption is (as was suggested in the preceding lines) that it is not easy to carry out the numerical calculations via the DI formulation when the mass density is not constant in $\Omega$.

   This is the reason why a less-drastic assumption has been invoked in connection with the integral formulation. What surely comes to mind is to assume that the mass density of the obstacle be different from that of the host medium, but constant within $\Omega_{int}$. At first, one is not required to make such an assumption regarding the wavespeed, so that one can show, quite easily (this is done  in sect. \ref{constrhoone} of the present study and appeals to domain decomposition) that the scattered field can be represented by a sum of a DI and a boundary integral (BI), involving $u$ in $\Omega_{int}$ and $u$ on the boundary $\Gamma_{ext-int}$ separating $\Omega_{ext}$ from $\Omega_{int}$. In the forward problem, these two functions are unknown, and can be shown to be obtained via a system of two coupled equations involving both DI's and BI's \cite{gu00,fu02,fb04}, this being called the DI-BI scheme. This system lends itself to an iterative resolution scheme similar to that of the previously-evoked sole domain integral equation, or else, it can be solved in brute-force numerical manner. As regards the brute-force manner, the amount of required computations is less than in the DI method (before invocation of the constant mass assumption, which is more-properly called the DI-DI method because it involves two domain integrals, one involving variations of the mass density and the other variations of the wavespeed) due to the fact that one domain integral (involving a large amount of discretized unknowns) is replaced by a boundary integral (involving a much lesser amount of discretized unknowns). However, it should be recalled that this advantage is obtained at the expense of making a strong assumption regarding the mass density.

   With the same assumptions on the mass density and wavespeed as in the preceding paragraph, it can be shown (as is done in sect. \ref{bidi}  herein), starting with domain decomposition, that the scattered field can be represented by a sum of a DI and a boundary integral (BI), involving $u$ in $\Omega_{int}$ and $\nabla u$ on the boundary $\Gamma_{ext-int}$. In the forward problem, these two functions are unknown, and are obtained via a system of two coupled equations involving both BI's and DI's, this being called the BI-DI scheme. This system again lends itself to an iterative resolution scheme similar to that of the previously-evoked sole domain integral equation, or else, it can be solved in brute-force numerical manner. For the same reasons as previously, the BI-DI brute-force scheme involves less computations than the DI-DI scheme. However, it should be recalled that this advantage is obtained at the expense of making a strong assumption regarding the mass density.

   With more drastic assumptions: (a) mass densities spatially-constant, but different from each other, in the host and obstacle and (b) wavespeeds spatially-constant, but different from each other, in the host and obstacle, one can show (as is done in sect. \ref{constrhoonecone} herein), starting with domain decomposition, that the scattered field in $\Omega_{ext}$ is represented by a BI involving $u$ and $\nabla u$ on    $\Gamma_{ext-int}$ and the scattered field in $\Omega_{int}$ is represented by another BI involving the same two field functions on the boundary. The latter can then be shown to constitute the unknowns of a coupled system of boundary integral equations (this method is thus termed the BI-BI method \cite{bo95,wi99}) which again lends itself to an interative type of resolution scheme or else can be solved in brute-force manner. Since two domain integrals in the DI-DI scheme are replaced by two boundary integrals in the BI-BI scheme \cite{sc85}, the latter requires much less computations than the DI-DI scheme when brute force is employed. However, this enormous advantage is obtained at the expense of making the drastic assumptions regarding both the wavespeed and the density (see also sect.\ref{mix}).

   At this point, one may wonder whether the DI-DI, DI-BI, BI-DI and BI-BI methods give rise to the same solution (i.e., the exact solution, as they should) when the assumptions of the BI-BI regarding mass density and wavespeed are made. We have been unable to make this demonstration in the general case of an obstacle of arbitrary shape, because, among other reasons, we do not dispose of an exact,  reference solution for obstacles of arbitrary shape. However, we do dispose of such a solution for an obstacle of canonical shape, and since we are here interested essentially in 2D problems, this obstacle was chosen to be a circular cylinder. As mentioned before, the scattering problem for this homogeneous obstacle can be found by the DD-SOV method. Thus, we shall employ this  solution (which is in explicit, functional form) as the reference (a similar procedure, using the DD-SOV solution for elastic waves impinging on a spherical obstacle as a reference, was followed in \cite{hh81})  for determining whether the other four methods lead to the same (or different) solutions. As will be shown, it turns out that all these methods lead to the same solution for the homogeneous circular cylinder canonical obstacle even though the way of obtaining the solution is relatively simple only via the BI-BI method. Futhermore, and more importantly, only the DI-DI method leads to an intermediate form of the solution which readily separates the contributions of the mass density contrast from the wavespeed contrast. This enables us to generate the solution via an iterative scheme that is equivalent to its expansion as a geometrical series in terms of the mass density contrast parameter. Taking only the first term of this series is equivalent to approximating the solution by the one obtained from the constant mass density assumption. Finally, we show that retaining only three terms in this series is generally sufficient to    obtain a quite accurate approximation of the solution for realistic mass density contrasts.
%%%%%%%%%%%%%%%%%%%%%%%%%%%%%%%%%%%%%%%%%%%%%%%%%%%%%%%%%%%%%%%%%%%%%%%%%%%%%%%%%%%%%%%%%%%%%%%%%%%%%%%%%%%%%%%%%%%%%%%%%%%%%%%%%%%%%%%%%%%%%%%%%
%%%%%%%%%%%%%%%%%%%%%%%%%%%%%%%%%%%%%%%%%%%%%%%%%%%%%%%%%%%%%%%%%%%%%%%%
\section{Governing equations}\label{bgwx}
The material presented in this section is taken mostly from
\cite{bg04}. A solid or fluid (the medium of interest in acoustics) is here considered to be a linear, isotropic elastic continuum.
%%%%%%%%%%%%%%%%%%%%%%%%%%%%%%%%%%%%%%%%%%%%%%%%%%%%%%%
\subsection{Conservation of momentum}
The conservation of momentum relation in a mechanical material
continuum is expressed by
\begin{equation}\label{bgwx1}
  \sigma_{kl,k}+\rho(f_{l}-\dot{v}_{l})=0~~;~~l,k=1,2,3~,
\end{equation}
wherein: $\sigma_{kl}$ are the (cartesian) components of the
stress tensor, $\rho$ the mass density, $f_{l}$ the components of
the applied body force, and $\mathcal{F}_{,k}$ designates the
partial derivative of $\mathcal{F}$ with respect to $x_{k}$ (with
$x_{k}$ the $k$-th cartesian coordinate). By extension, $\mathcal{F}_{,t}$ designates the
partial derivative of $\mathcal{F}$ with respect to $t$, with $t$ the time. Furthermore, $v_{l}$ are
the components of particle velocity, $\dot{v}_{l}$ the components
of particle acceleration (the dot above a variable meaning
"material derivative", i.e.,
$\dot{\mathcal{F}}_{k}:=\mathcal{F}_{,kt}+
\mathcal{F}_{k,l}\mathcal{F}_{l}$).
%%%%%%%%%%%%%%%%%%%%%%%%%%%%%%%%%%%%%%%%%%%%%%%%%%%%%%%%
\subsection{Conservation of momentum for small deformations}
For small deformations, $\dot{v}_{l}\simeq u_{l,tt}$, wherein
$u_{l}$ are the cartesian components of the particle displacement
vector, and $\mathcal{F}_{,tt}$ designates the  second partial
derivative of $\mathcal{F}$ with respect to $t$, so that
\begin{equation}\label{bgwx2}
  \sigma_{kl,k}+\rho(f_{l}-{u}_{l,tt})=0~~;~~l,k=1,2,3~,
\end{equation}
%%
%%%%%%%%%%%%%%%%%%%%%%%%%%%%%%%%%%%%%%%%%%%%%%%%%%%%%%%%%
\subsection{Constitutive relations in a linear material continuum}
In a {\it linear} mechanical material continuum the stress tensor
is related to the strain tensor (whose components are
$\varepsilon_{mn}$) by
\begin{equation}\label{bgwx3}
  \sigma_{kl}=c_{klmn}\varepsilon_{mn}~~;~~k,l,m,n=1,2,3~,
\end{equation}
with
\begin{equation}\label{bgwx4}
  \varepsilon_{kl}=\frac{1}{2}(u_{k,l}+u_{l,k})~~;~~k,l=1,2,3~,
\end{equation}
and $c_{klmn}$  the fourth-order material coefficient tensor.
%%%%%%%%%%%%%%%%%%%%%%%%%%%%%%%%%%%%%%%%%%%%%%%%%%%%%%%%%%%
\subsection{Constitutive relations in a linear, isotropic,  elastic continuum}
In a linear, {\it elastic} continuum, (\ref{bgwx3}) takes the form
given by the Cauchy-Hooke relation
\begin{equation}\label{bgwx5}
  \sigma_{kl}=\lambda\delta_{kl}\varepsilon_{mm}+2\mu\varepsilon_{kl}=
  \sigma_{lk}~~;~~k,l,m=1,2,3~,
\end{equation}
wherein $\lambda,~\mu$ are the Lam\'e coefficients, $\delta_{kl}$
the Kronecker symbol, and repeated indices meaning summation over
this index (Einstein convention), i.e.,
$\varepsilon_{mm}:=\varepsilon_{11}+\varepsilon_{22}+\varepsilon_{33}$.

Consequently,
\begin{multline}\label{bgwx6}
  \sigma_{kl,k}=\lambda\delta_{kl}\varepsilon_{mm,k}+
  \lambda_{,k}\delta_{kl}\varepsilon_{mm}+2\mu\varepsilon_{kl,k}+2\mu_{,k}\varepsilon_{kl}=
\\
\lambda\varepsilon_{mm,l}+
  \lambda_{,l}\varepsilon_{mm}+2\mu\varepsilon_{kl,k}+2\mu_{,k}\varepsilon_{kl}  ~~;~~k,l,m=1,2,3~.
\end{multline}
%%
%%%%%%%%%%%%%%%%%%%%%%%%%%%%%%%%%%%%%%%%%%%%%%%%%%%%%%%%%%%%%
\subsection{Wave equation in a linear, isotropic, elastic
medium (Navier's equation)}
Introducing (\ref{bgwx4}) into (\ref{bgwx6}) gives
\begin{multline}\label{bgwx7}
\sigma_{kl,k}=\lambda u_{m,ml}+
  \lambda_{,l}u_{m,m}+\mu(u_{k,lk}+u_{l,kk})+
  \mu_{,k}(u_{k,l}+u_{l,k})=
  \\
  \lambda u_{m,lm}+
  \lambda_{,l}u_{m,m}+\mu(u_{k,lk}+u_{l,kk})+
  \mu_{,k}(u_{k,l}+u_{l,k})=
  \\
  \lambda u_{k,lk}+
  \lambda_{,l}u_{m,m}+\mu(u_{k,lk}+u_{l,kk})+
  \mu_{,k}(u_{k,l}+u_{l,k})~~;~~k,l,m=1,2,3~,
\end{multline}
or, collecting terms
\begin{equation}\label{bgwx8}
\sigma_{kl,k}=(\lambda+\mu) u_{k,kl}+\mu u_{l,kk}+
  \lambda_{,l}u_{m,m}+
  \mu_{,k}(u_{k,l}+u_{l,k})
 ~~;~~k,l,m=1,2,3~,
\end{equation}
so that the conservation of momentum relation (\ref{bgwx2})
becomes
\begin{equation}\label{bgwx9}
(\lambda+\mu) u_{k,kl}+\mu u_{l,kk}+
  \lambda_{,l}u_{m,m}+
  \mu_{,k}(u_{k,l}+u_{l,k})+\rho(f_{l}-\dot{v}_{l})=0
 ~~;~~k,l,m=1,2,3~,
\end{equation}
or, in its linearized version
\begin{equation}\label{bgwx10}
(\lambda+\mu) u_{k,kl}+\mu u_{l,kk}+
  \lambda_{,l}u_{m,m}+
  \mu_{,k}(u_{k,l}+u_{l,k})+\rho(f_{l}-u_{l,tt})=0
 ~;~k,l,m=1,2,3~.
\end{equation}
%%
%%%%%%%%%%%%%%%%%%%%%%%%%%%%%%%%%%%%%%%%%%%%%%%%%%%%%%%%%%%%%%%%%
\subsection{Governing equations in an inviscid fluid}
When the medium is a (heterogeneous) ideal (i.e., inviscid; see e.g., \cite{gt06} for a method of handling the viscous fluid case) fluid,
$\mu=0$, so that
\begin{equation}\label{bgwx12}
  \sigma_{kl}=\lambda\delta_{kl}\varepsilon_{mm}=\lambda u_{m,m}\delta_{kl}
  \sigma_{lk}~~;~~k,l,m=1,2,3~.
\end{equation}
With the definition
\begin{equation}\label{bgwx13}
  p:=-\lambda\varepsilon_{mm}=-\lambda u_{m,m}~~;~~
  m=1,2,3~,
\end{equation}
wherein $p$ denotes the {\it pressure} in the ideal fluid, it
follows that
\begin{equation}\label{bgwx14}
  \sigma_{kl}=-p\delta_{kl}=
  \sigma_{lk}~~;~~k,l=1,2,3~.
\end{equation}
which shows that the stress tensor is diagonal in an ideal fluid.

Then
\begin{equation}\label{bgwx15}
  \sigma_{kl,k}=-p_{,k}\delta_{kl}=-p_{,l}
  ~~;~~k,l=1,2,3~.
\end{equation}
so that the conservation of momentum relation in an ideal fluid
becomes
\begin{equation}\label{bgwx16}
-p_{,l}+\rho(f_{l}-\dot{v}_{l})=0
 ~~;~~k,l,m=1,2,3~,
\end{equation}
or, in its linearized form
\begin{equation}\label{bgwx17}
-p_{,l}+\rho(f_{l}-u_{l,tt})=0
 ~~;~~k,l,m=1,2,3~.
\end{equation}
%%
%%%%%%%%%%%%%%%%%%%%%%%%%%%%%%%%%%%%%%%%%%%%%%%%%%%%%%%%%%%%%%%%%
\subsection{Time domain wave equation in a fluid}
Eq. (\ref{bgwx13}) gives rise to
\begin{equation}\label{bgwx18}
  p_{,l}=-\lambda u_{m,ml}-\lambda_{,l} u_{m,m}~~;~~
  l,m=1,2,3~,
\end{equation}
so that (\ref{bgwx17}) becomes
\begin{equation}\label{bgwx19}
\lambda u_{m,ml}+\lambda_{,l} u_{m,m}+\rho(f_{l}-u_{l,tt})=0
 ~~;~~l,m=1,2,3~.
\end{equation}
which is the wave equation in an ideal fluid in terms of particle
displacement.

An alternative (i.e., a partial differential equation exclusively
in terms of pressure) can be found as follows. Taking derivatives
of (\ref{bgwx17}) gives
\begin{equation}\label{bgwx20}
-p_{,ll}+\rho_{,l}(f_{l}-u_{l,tt})+\rho(f_{l,l}-u_{l,ttl})=0
 ~~;~~l=1,2,3~.
\end{equation}
However, (\ref{bgwx13}) gives rise, under the assumption that
$\lambda$ does not depend on $t$, to
\begin{equation}\label{bgwx21}
  p_{,tt}:=-\lambda u_{m,mtt}~~\Rightarrow~~
  u_{l,ltt}=u_{l,ttl}=-\frac{p_{,tt}}{\lambda}
 ~.
\end{equation}
Another consequence of (\ref{bgwx17}) is
\begin{equation}\label{bgwx22}
 u_{l,tt}=-\frac{p_{l}}{\rho}+f_{l}~~;~~l=1,2,3
 ~,
\end{equation}
so that inserting (\ref{bgwx21}) and (\ref{bgwx22}) into
(\ref{bgwx20}) results in
\begin{equation}\label{bgwx23}
-p_{,ll}+\rho_{,l}(f_{l}+\frac{p_{l}}{\rho}-f_{l})+\rho(f_{l,l}+\frac{p_{,tt}}{\lambda})=0
 ~~;~~l=1,2,3~.
\end{equation}
or
\begin{equation}\label{bgwx24}
-p_{,ll}+\frac{\rho}{\lambda}p_{,tt}+\frac{\rho_{,l}}{\rho}p_{,l}+\rho
f_{l,l}=0
 ~~;~~l=1,2,3~.
\end{equation}
which is the wave equation, in terms of pressure, in an ideal
non-homogeneous fluid.

Letting
\begin{equation}\label{bgwx24a}
\frac{1}{c^{2}}:=\frac{\rho}{\lambda}=\frac{\rho}{K}=\rho\kappa
,
\end{equation}
with $c$, $K$ and $\kappa$ the phase velocity, isentropic bulk modulus and isentropic compressibility within/of the fluid medium, we obtain
\begin{equation}\label{bgwx24b}
-p_{,ll}+\frac{1}{c^{2}}p_{,tt}+\frac{\rho_{,l}}{\rho}p_{,l}+\rho
f_{l,l}=0
 ~~;~~l=1,2,3~.
\end{equation}
or, in vectorial notation (recalling that $\rho$ and $c$ do not depend on $t)$,
\begin{equation}\label{bgwx24c}
-\nabla\cdot\nabla p(\mathbf{x},t)+\frac{1}{c(\mathbf{x})^{2}}\frac{\partial^{2}p(\mathbf{x},t)}{\partial t^{2}}+\frac{\nabla\rho(\mathbf{x})}{\rho(\mathbf{x})}\cdot \nabla p(\mathbf{x},t)+\rho(\mathbf{x})
\nabla\cdot \mathbf{f}(\mathbf{x},t)=0~;~\forall \mathbf{x}\in\mathbb{R}^{3}
 ~.
\end{equation}
wherein $\mathbf{x}=(x_{1},x_{2},x_{3})$. This is the time domain version of the Bergman equation (BPDE} in sect. \ref{intro}) \cite{be46}.

Note that up till now we assumed the fluid to be lossless (inviscid) which means that $c$ and $\rho$ are real and independent of $t$. A convenient way to account for losses in the fluid is to consider $c$ and/or $\rho$ to be complex and dispersive, i.e.,
\begin{equation}\label{bgwx24d}
c=c(\mathbf{x},\omega)=c'(\mathbf{x},\omega)+ic''(\mathbf{x},\omega)~~,~~\rho(\mathbf{x},\omega)=
\rho'(\mathbf{x},\omega)+i\rho''(\mathbf{x},\omega)
 ~,
\end{equation}
wherein $c''<0$ and $\rho''>0$ for a passive (i.e., lossy ) fluid (the case considered hereafter) and $c''>0$ and $\rho''<0$ for an active fluid. This is what we shall do in the frequency-domain formulation of the wave equation.
%%%%%%%%%%%%%%%%%%%%%%%%%%%%%%%%%%%%%%%%%%%%%%%%%%%%%%%%%%%%%%%%%
\subsection{Frequency domain wave equation in an inviscid or viscous fluid}
By expanding the time domain pressure and force in the Fourier integrals
\begin{equation}\label{bgwx24e}
p(\mathbf{x},t)=\int_{-\infty}^{\infty}p(\mathbf{x},\omega)\exp(-i\omega t)d\omega
 ~,
\end{equation}
\begin{equation}\label{bgwx24f}
\mathbf{f}(\mathbf{x},t)=\int_{-\infty}^{\infty}\mathbf{f}(\mathbf{x},\omega)\exp(-i\omega t)d\omega
 ~,
\end{equation}
wherein $\omega=2\pi f$ is the angular frequency and $f$ the frequency, one obtains the frequency domain expression of the wave equation
\begin{equation}\label{bgwx24g}
\nabla\cdot\nabla p(\mathbf{x},\omega)+k^{2}(\mathbf{x},\omega)p(\mathbf{x},\omega)-\frac{\nabla\rho(\mathbf{x},\omega)}{\rho(\mathbf{x},\omega)}\cdot \nabla p(\mathbf{x},\omega)=\rho(\mathbf{x},\omega)
\nabla\cdot \mathbf{f}(\mathbf{x},\omega)~;~\forall \mathbf{x}\in\mathbb{R}^{3}
~,
\end{equation}
wherein
\begin{equation}\label{bgwx24h}
k(\mathbf{x},\omega):=\frac{\omega}{c(\mathbf{x},\omega)}
\end{equation}
is the wavenumber in the fluid, which is a generally-complex quantity due to the fact that $c$ is assumed to be generally-complex (to account for viscosity). Eq. (\ref{bgwx24g}) is the frequency domain version of the Bergman equation (BPDE).
%%%%%%%%%%%%%%%%%%%%%%%%%%%%%%%%%%%%%%%%%%%%%%%%%%%%%%%%%%%%%%%%%
\subsection{Getting rid of the gradient of pressure term via the $p-q$ transformation}\label{pq}
Eq. (\ref{bgwx24g}) was
\begin{equation}\label{gr01}
-p_{,ll}-k^{2}+\frac{\rho_{,l}}{\rho}p_{,l}+\rho f_{l,l}=0~;~l=1,2,3~;~\forall \mathbf{x}\in\mathbb{R}^{3}~.
\end{equation}
As we shall see later on, the presence of the gradient of pressure term (i.e., $p_{,ll}$) is a source of nuisances. To eliminate it, we attempt the change of variables \cite{be46,ma03}
\begin{equation}\label{gr02}
p=q\rho^{-m}:=q\eta^{-1}~,
\end{equation}
whence
\begin{equation}\label{gr03}
p_{,l}=(\eta^{-1})_{,l}q+\eta^{-1}q_{,l}~~,~~p_{,ll}=(\eta^{-1})_{,ll}q+2(\eta^{-1})_{,l}q_{l}+\eta^{-1}q_{,ll},
~,
\end{equation}
so that (\ref{gr01}) becomes
\begin{multline}\label{gr04}
-\eta^{-1}q_{,ll}+\left[-2(\eta^{-1})_{,l}+\frac{\rho_{,l}}{\rho}\eta^{-1}\right]q_{,l}+\left\{-(\eta^{-1})_{,ll}-k^{2}\eta^{-1}+
\frac{\rho_{,l}}{\rho}(\eta^{-1})_{,l}\right\}q+\rho f_{l,l}=0~;\\
~l=1,2,3~;~\forall \mathbf{x}\in\mathbb{R}^{3}~.
\end{multline}
Thus, the problem reduces to getting rid of $[~]$, i.e.,
\begin{equation}\label{gr05}
\left[-2(\eta^{-1})_{,l}+\frac{\rho_{,l}}{\rho}\eta^{-1}\right]=0~\Rightarrow~(2m+1)\rho^{-m-1}\rho_{,l}~;~l=1,2,3~;~\forall \mathbf{x}\in\mathbb{R}^{3}~,
\end{equation}
the only possible solution of which is $m=-1/2$. Thus, the proper transformation of variables is
\begin{equation}\label{gr06}
p=q\rho^{1/2}:=q\eta^{-1}~\Rightarrow~\eta^{-1}=\rho^{1/2}~,
\end{equation}
whence (\ref{gr04}) becomes
\begin{equation}\label{gr07}
q_{,ll}+\left[-\frac{3}{4}\left(\frac{\rho_{,l}}{\rho}\right)^{2}+\frac{1}{2}\frac{\rho_{,ll}}{\rho}+k^{2}\right]q-\rho^{1/2}f_{l,l}=0~;~l=1,2,3~;~\forall \mathbf{x}\in\mathbb{R}^{3}~,
\end{equation}
It can be shown that
\begin{equation}\label{gr08}
\left[-\frac{3}{4}\left(\frac{\rho_{,l}}{\rho}\right)^{2}+\frac{1}{2}\frac{\rho_{,ll}}{\rho}+k^{2}\right]=
-\rho^{1/2}\nabla\cdot\nabla\rho^{-1/2}~,
\end{equation}
so that we finally obtain the inhomogeneous Helmholtz equation for $q$
\begin{equation}\label{gr09}
\left[\nabla\cdot\nabla+\mathcal{K}^{2}(\mathbf{x},\omega)\right]q(\mathbf{x},\omega)=\rho^{1/2}(\mathbf{x},\omega)\nabla\cdot\mathbf{f}(\mathbf{x},\omega)~;~\forall \mathbf{x}\in\mathbb{R}^{3}~,
\end{equation}
wherein
\begin{equation}\label{gr10}
\mathcal{K}^{2}(\mathbf{x},\omega)=k^{2}-\rho^{1/2}(\mathbf{x},\omega)\nabla\cdot\nabla\rho^{-1/2}(\mathbf{x},\omega) ~.
\end{equation}
Thus, the problem of finding the pressure $p$ reduces to that of first finding $q$ via (\ref{gr09}) and then computing $p$ from $q$ via   (\ref{gr06}). This is apparently the method adopted in publications such as \cite{lo09}. We shall return to this issue later on in sect. \ref{dipq} of the present study.
%%%%%%%%%%%%%%%%%%%%%%%%%%%%%%%%%%%%%%%%%%%%%%%%%%%%%%%
\subsection{Specifics of 2D elastic and acoustic wave mechanics}\label{2D}
Since the remainder of this study  will be concerned mostly with two-dimensional wave-mechanical problems, we now consider the particularities of such problems.

 Let us return to the equation (\ref{bgwx10}) for wave motion in an elastic {\it solid} in which we replace the density $\rho$ and two Lam\'e parameters $\lambda$ and $\mu$ by the letters $R$, $L$ and $M$ whereas $u$ and $f$ are replaced by $U$ and $F$.
\begin{equation}\label{2D01}
(L+M) U_{k,kl}+M U_{l,kk}+
  L_{,l}U_{m,m}+
  M_{,k}(U_{k,l}+u_{l,k})+R(F_{l}-U_{l,tt})=0
 ~;~k,l,m=1,2,3~.
\end{equation}
In a general 3D situation, $R$, $L$ and $M$  and $\mathbf{F}$ depend on the three cartesian coordinates $x_{1},x_{2},x_{3}$ so that the three components of $\mathbf{U}$ depend as well on these three spatial coordinates.

We now consider the so-called 2D situation in which
\begin{equation}\label{2D02}
R_{,3}=0~~,~~L_{,3}=0~~,~~M_{,3}=0~~,~~F_{1,3}=F_{2,3}=F_{3,3}=0~.
\end{equation}
and we name the $x_{1}-x_{2}$ plane the sagittal plane.

It is then easy to show that
\begin{equation}\label{2D03}
U_{1,3}=U_{2,3}=U_{3,3}=0~,
\end{equation}
whence (\ref{2D01}) becomes
\begin{multline}\label{2D04}
(L+M)(U_{1,1l}+U_{2,2l})+M(U_{l,11}+U_{l,22})+
  L_{,l}(U_{1,1}+U_{2,2})+\\
  M_{,1}(U_{1,l}+U_{l,1})+M_{,2}(u_{2,l}+U_{l,2})+R(F_{l}-U_{l,tt})=0
 ~;~l=1,2,3~.
\end{multline}
from which we obtain the three relations:
\begin{equation}\label{2D05}
[(L+2M)U_{1,1}]_{,1}+[LU_{2,2}]_{,1}+[MU_{2,1}]_{,2}+[MU_{1,2}]_{,2}+
R(F_{1}-U_{1,tt})=0
~,
\end{equation}
\begin{equation}\label{2D06}
[(L+2M)U_{2,2}]_{,2}+[LU_{1,1}]_{,2}+[MU_{1,2}]_{,1}+[MU_{2,1}]_{,1}+
R(F_{2}-U_{2,tt})=0
~,
\end{equation}
\begin{equation}\label{2D07}
M[U_{3,11}+U_{3,22}]+M_{,1}U_{3,1}+M_{,2}U_{3,2}+
R(F_{3}-U_{3,tt})=0
~.
\end{equation}
The first two relations describe the wave motion (i.e., exclusively the $U_{1}$ and $U_{2}$ components of $\mathbf{U}$) in the sagittal plane so that this motion is termed 'in-plane' and is due to in-plane forces. The third relation describes the wave motion (i.e., exclusively the $U_{3}$ component of $\mathbf{U}$) along the $x_{3}$ direction perpendicular to the sagittal plane so that this motion is termed 'out-of plane' and is due to out-of-plane forces. Of utmost importance is to note that the out-of-plane motion is totally decoupled from the in-plane motion in this 2D situation.

We are especially interested here in this third relation (\ref{2D07}) which can be re-written as
\begin{equation}\label{2D08}
[U_{3,11}+U_{3,22}]+\frac{M_{,1}}{M}U_{3,1}+\frac{M_{,2}}{M}U_{3,2}+
\left(\frac{1}{C}\right)^{2}(F_{3}-U_{3,tt})=0
~.
\end{equation}
wherein
\begin{equation}\label{2D09}
\left(\frac{1}{C}\right)^{2}:=\frac{R}{M}~,
\end{equation}
and $C$ is the out-of-plane (or shear-horizontal) phase velocity (i.e., of $U_{3}$ motion) in the solid.

Until further notice, we make the following definitions: $\mathbf{x}:=(x_{1},x_{2})$,
$\nabla_{\mathbf{x}}:=(\frac{\partial}{\partial x_{1}},\frac{\partial}{\partial x_{2}})$, $U(\mathbf{x}):=U_{3}(\mathbf{x})$,  $F(\mathbf{x}):=F_{3}(\mathbf{x})$, whence (\ref{2D08}) can be written in the vectorial form
\begin{equation}\label{2D10}
\nabla_{\mathbf{x}}\cdot\nabla_{\mathbf{x}}U(\mathbf{x},t)-M(\mathbf{x})
\nabla_{\mathbf{x}}\left(\frac{1}{M(\mathbf{x})}\right)\cdot\nabla_{\mathbf{x}}U\mathbf{x},t)+
\left(\frac{1}{C(\mathbf{x})}\right)^{2}\left(F(\mathbf{x},t)-U_{,tt}(\mathbf{x},t)\right)=0~;~\forall \mathbf{x}\in\mathbb{R}^{2}~.
\end{equation}
The 2D version of the acoustic wave equation in inviscid {\it fluids} is, by virtue of (\ref{bgwx24c})
\begin{equation}\label{2D11}
\nabla_{\mathbf{x}}\cdot\nabla_{\mathbf{x}} p(\mathbf{x},t)-
\frac{\nabla_{\mathbf{x}}\rho(\mathbf{x})}{\rho(\mathbf{x})}\cdot \nabla_{\mathbf{x}} p(\mathbf{x},t)-\rho(\mathbf{x})
\nabla_{\mathbf{x}}\cdot \mathbf{f}(\mathbf{x},t)-\left(\frac{1}{c(\mathbf{x})}\right)^{2}p_{,tt}(\mathbf{x},t)=0~;~\forall \mathbf{x}\in\mathbb{R}^{2}
 ~.
\end{equation}
The important point is to notice that these two wave equations are exactly of the same form if the following associations are made: $U\leftrightarrow p$, $M\leftrightarrow 1/\rho$, $C\leftrightarrow c$ and $F/C^{2}\leftrightarrow-\rho\nabla\cdot \mathbf{f}$ which is consistent with the previous definitions $c^{2}=K/\rho$ and $C^{2}=M/R$ provided $1/R\leftrightarrow K$. This means that everything that will be found for the 2D  pressure wave in fluids will be applicable as well for the 2D out-of-plane particle displacement wave in solids provided that the acoustic density $\rho$ is associated with the reciprocal of the elastic shear modulus $1/M$. Thus, the assumption of constant density in 2D acoustics is equivalent to the assumption of constant shear modulus in 2D elasticity (in which case $c$ is not assumed spatially-constant in acoustics and  $C$ is not assumed spatially-constant in elasticity).
%%%%%%%%%%%%%%%%%%%%%%%%%%%%%%%%%%%%%%%%%%%%%%%%%%%%%%%%%%%%%%%%%%%%%%%%%%
\section{Acoustic radiation from applied sources in the absence of an obstacle in
unbounded space occupied by a homogeneous, lossless fluid}\label{acrad}
%
%%%%%%%%%%%%%%%%%%%%%%%%%%%%%%%%%%%%%%%%%%
\subsection{Description of the problem}\label{acraddp}
\begin{figure}[ht]
\begin{center}
\includegraphics[
height=2.5in, width=2.5in ] {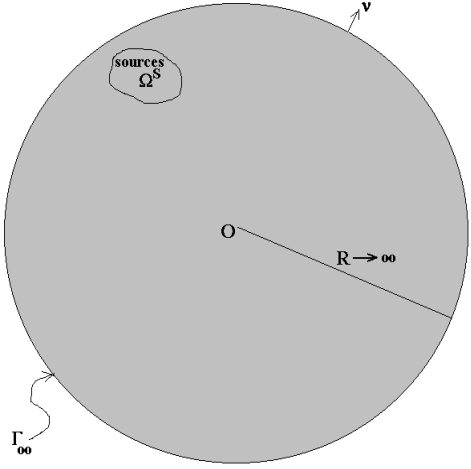} \caption{Source radiating in
free space filled by a homogeneous, lossless fluid.} \label{fig19}
\end{center}
\end{figure}

The configuration is depicted in fig. \ref{fig19}. Until further notice, it represents a 3D situation. The source, of
bounded support $\Omega^{s}$, is located in $\mathbb{R}^{3}$ which
is filled with a {\it homogeneous} medium
$M^{0}(\rho^{[0]},c^{[0]})$ such that $\rho^{[0]}$ and $c^{[0]}$
are position- and frequency-independent.
%%
%%%%%%%%%%%%%%%%%%%%%%%%%%%%%%%%%%%%%%%%%%%
\subsection{Governing equations}\label{acradge}
From now on, we place the analysis in the frequency domain, so that the governing equation (\ref{bgwx24g})is, due to the supposed-homogeneous nature of the fluid medium,
\begin{equation}\label{w2.3.2}
  \nabla^{2}p^{[0]}(\mathbf{x})+\left( k^{[0]}\right ) ^{2}p^{[0]}(\mathbf{x})=-s(\mathbf{x})
  ~~;~~\forall\mathbf{x}\in\mathbb{R}^{3}~,
\end{equation}
wherein $\omega$ is implicit in the arguments of the various
functions, $k^{[0]}=\omega/c^{[0]}$ is position-independent,
and $s:= -\rho \nabla\cdot \mathbf{f}$ represents the source
density associated with applied forces, whose support is assumed
to be the finite domain $\Omega^{s}$.

In addition, the pressure, which is bounded everywhere except in $\Omega^{s}$,
satisfies the radiation condition whose mathematical translation is
\begin{equation}\label{w2.3.2a}
   \lim_{r'\rightarrow\infty} \left[p_{,r'}^{[0]}(\mathbf{x'})-ik^{[0]}
p^{[0]}(\mathbf{x'})\right]=0~ ~,
\end{equation}
wherein $r'$ is the radial coordinate in the $\mathbf{x}'$ coordinate system.
%%%%%%%%%%%%%%%%%%%%%%%%%%%%%%%%%%
\subsection{Field representation}\label{acradfr}
The so-called {\it free-space Green's function}, which is bounded everywhere except for $\|\mathbf{x}-\mathbf{x'}\|=0$, satisfies
\begin{equation}\label{w2.3.3}
  \nabla_{\mathbf{x'}}^{2}G^{[0]}(\mathbf{x};\mathbf{x'})+(k^{0]})^{2}G^{0]}(\mathbf{x};\mathbf{x'})=
  -\delta(\mathbf{x}-\mathbf{x'})~,
\end{equation}
wherein $\delta(~)$ is the Dirac delta distribution whose sifting
property is expressed by
\begin{equation}\label{w2.3.4}
  \int_{\mathbb{R}^{3}}\mathcal{F}(\mathbf{x'})\delta(\mathbf{x}-\mathbf{x'})d\Omega(\mathbf{x'})=
  \mathcal{F}(\mathbf{x})~~;~~\forall~\mathbf{x}\in \mathbb{R}^{3}~,
\end{equation}
Furthermore, this Green's function satisfies the radiation condition
\begin{equation}\label{w2.3.4a}
  \lim_{r'\rightarrow\infty} \left[G_{,r'}^{[0]}(\mathbf{x};\mathbf{x'})-ik^{[0]}
G^{[0]}(\mathbf{x};\mathbf{x'})\right]=0~.
\end{equation}

It follows that
\begin{multline}\label{w2.3.4b}
  \int_{\mathbb{R}^{3}}\Big[G^{[0]}(\mathbf{x};\mathbf{x'})\nabla_\mathbf{x'}^{2}p^{[0]}(\mathbf{x'}-
  p^{[0]}(\mathbf{x'})\nabla_\mathbf{x'}^{2}G^{[0]}(\mathbf{x};\mathbf{x'})\Big]d\Omega(\mathbf{x'})=\\
  -\int_{\mathbb{R}^{3}}G^{[0]}(\mathbf{x};\mathbf{x'})
s(\mathbf{x})d\Omega(\mathbf{x'})+\int_{\mathbb{R}^{3}}p(\mathbf{x'}\delta(\mathbf{x}-\mathbf{x'})d\Omega\mathbf{x'})
~~;~~\forall~\mathbf{y}\in \mathbb{R}^{3}~,
\end{multline}
Applying Green's second identity  in
$\Omega=\mathbb{R}^{3}$, assumed to be "closed" by a hypersphere
of infinite radius whose boundary is $\Gamma_{\infty}$ (see fig.
\ref{fig19}), leads to
\begin{multline}\label{w2.3.5}
p^{0}(\mathbf{x})=\int_{\mathbb{R}^{3}}G^{[0]}(\mathbf{x};\mathbf{x'})
s(\mathbf{x'})d\Omega\mathbf{x'}) +
\\
\int_{\Gamma_{\infty}}\left [ G^{[0]}(\mathbf{x};\mathbf{x'})
\boldsymbol{\nu}(\mathbf{x'})\cdot\nabla_\mathbf{x'} p^{[0]}(\mathbf{x'})-
p^{[0]}(\mathbf{x})\boldsymbol{\nu}(\mathbf{x'})\cdot\nabla_\mathbf{x'}
G^{[0]}(\mathbf{x};\mathbf{x'})\right ] d\Gamma(\mathbf{x'}) ~~;
\\
~~\forall~\mathbf{y}\in \mathbb{R}^{3}~,
\end{multline}
The second integral can be written as
\begin{multline}\label{w2.3.6}
\int_{\Gamma_{\infty}}\left [ G^{[0]}(\mathbf{x};\mathbf{x'})
\boldsymbol{\nu}(\mathbf{x'})\cdot\nabla_\mathbf{x'} p^{[0]}(\mathbf{x'})-
p^{[0]}(\mathbf{x'})\boldsymbol{\nu}(\mathbf{x'})\cdot\nabla_\mathbf{x'}
G^{[0]}(\mathbf{x};\mathbf{x'})\right ] d\Gamma(\mathbf{x'})=
\\
\int_{\Gamma_{\infty}} G^{[0]}(\mathbf{x};\mathbf{x'})\left [
\boldsymbol{\nu}(\mathbf{x'})\cdot\nabla_\mathbf{x'} p^{[0]}(\mathbf{x'})-ik^{[0]}
p^{[0]}(\mathbf{x'})\right ] d\Gamma(\mathbf{x'})-
\\
\int_{\Gamma_{\infty}}p^{[0]}(\mathbf{x'})\left[
\boldsymbol{\nu}(\mathbf{x'})\cdot\nabla_\mathbf{x'}
G^{[0]}(\mathbf{x};\mathbf{x'})-ik^{[0]}
G^{[0]}(\mathbf{x},x'\right] d\Gamma (\mathbf{x'})
~~;~~\forall~\mathbf{x}\in \mathbb{R}^{3}~,
\end{multline}
and since $p^{[0]}$ and $G^{[0]}$ are bounded on $\Gamma_{\infty}$ and these two functions satisfy the radiations conditions (\ref{w2.3.2a}) and (\ref{w2.3.4a}), both of the integrals on the right
hand side of this expression vanish. Thus,
\begin{equation}\label{w2.3.7}
p^{[0]}(\mathbf{y})=\int_{\mathbb{R}^{3}}G^{[0]}(\mathbf{x};\mathbf{x'})
s(\mathbf{x'})d\Omega(\mathbf{x'})~~;~~\forall~\mathbf{x}\in
\mathbb{R}^{3}~,
\end{equation}
or, using the fact that the support of the applied sources is
$\Omega^{s}$
\begin{equation}\label{w2.3.8}
p^{[0]}(\mathbf{x})=\int_{\Omega^{s}}G^{[0]}(\mathbf{x};\mathbf{x'})
s(\mathbf{x'})d\Omega(\mathbf{x'})~~;~~\forall~\mathbf{x}\in
\mathbb{R}^{3}~,
\end{equation}
Since $G^{[0]}$ is  a function of $\mathbf{x}-\mathbf{x'}$, one sees
that (\ref{w2.3.8}) expresses the rather intuitive fact that the
pressure field radiated by applied sources of density $s$ is
expressed by a convolution
 between the latter and the pressure field radiated by a "point" source.

It is readily verified  by means of (\ref{w2.3.8}) and
(\ref{w2.3.4}) that when $s$ corresponds to a "point" source, the
radiated field is the free-space Green's function.
\newline
\newline
{\it Remark}
\newline
An important feature of (\ref{w2.3.8}) is that not only does it
provide a domain integral {\it representation} of the radiated
field, but also the {\it solution} to the  forward problem of
radiation by sources of {\it known} density $s$ and {\it known}
support $\Omega^{s}$, the kernel of the integral by which this
solution is computed being {\it known} as well.

The inverse radiation problem (also called the inverse source
problem), which is to determine the sources from the radiated
field, cannot be solved in such a simple manner. However, this inverse problem is linear in that the pressure depends linearly on the pressure field data.
%%%%%%%%%%%%%%%%%%%%%%%%%%%%%%%%%%%%%%
\subsection{The 2D free-space Green's function}
From this point on, everything will concern the 2D situation ($x_{3}$ being the ignorable coordinate due to the invariance of the constitutive parameters of the medium, impressed force and consequently the pressure field with respect to $x_{3}$) in which  $\mathbf{x}=(x_{1},x_{2})$ and $\mathbf{x'}=(x'_{1},x'_{2})$ (in cartesian coordinates).

As shown previously in this study, the free-space Green's function (GF)constitutes  a fundamental tool for the study of wave propagation. Rather than derive the 2D GF from the governing equations, we shall assume known the  expression (\cite{mf53}, p. 822, 891) for this GF and show that it satisfies the governing equation and radiation condition.

The 2D GF is
\begin{equation}\label{gf1}
G^{[0]}(\mathbf{x};\mathbf{x'})=\frac{i}{4}H_{0}^{(1)}(k^{[0]}(\mathbf{x};\mathbf{x'}))~,
\end{equation}
wherein $H_{0}^{(1)}$ is the zeroth-order Hankel function of the first kind \cite{as68}. This function admits the following representation in polar coordinates ((\cite{mf53}, p. 827) in which $\mathbf{x}=(r,\theta)$ and $\mathbf{x'}=(r',\theta')$:
\begin{multline}\label{gf2}
G^{[0]}(\mathbf{x};\mathbf{x'})=\\
\frac{i}{4}\sum_{n=-\infty}^{\infty}\left[H(r-r')H_{n}^{(1)}(k^{[0]}r)J_{n}(k^{[0]}r')+
H(r'-r)J_{n}(k^{[0]}r)H_{n}^{(1)}(k^{[0]}r')\right]\exp[in(\theta-\theta')]~,
\end{multline}
wherein $J_{n}(.)$ is the $n$-th order Bessel function and $H(\zeta)$ the Heaviside distribution ($H(\zeta>0)=1$, $H(\zeta<0)=0$).
%%%%%%%%%%%%%%%%%%%%%%%%%%%%%%%%%%%%%%%%%%%
\subsubsection{Demonstration that the 2D GF satisfies the 2D Helmholtz equation}\label{acradgf}
We first want to show that(\ref{gf2}) satisfies the Helmholtz equation in polar coordinates (\cite{mf53}, p. 825; \cite{fr60})
\begin{multline}\label{gf3}
\left(\nabla_{\mathbf{x}}\cdot\nabla_{\mathbf{x}}+\big(k^{[0]}\big)^{2}\right)G^{[0]}(\mathbf{x};\mathbf{x'})=\\
\left[\frac{1}{r}\frac{\partial}{\partial r}\left(r\frac{\partial}{\partial r}\right)+\frac{1}{r^{2}}\frac{\partial^{2}}{\partial\theta^{2}}+\big(k^{[0]}\big)^{2}\right]G^{[0]}(r,\theta;r',\theta)=
-\frac{1}{r}\delta(r-r')\delta(\theta-\theta')=-\delta(\mathbf{x}-\mathbf{x'})~,
\end{multline}
From the facts that
\begin{equation}\label{gf4}
H_{,\zeta}=\delta(\zeta)~~,~~\delta(-\zeta)=\delta(\zeta)~~,~~f(\zeta)\delta(\zeta)=f(0)\delta(\zeta)~,
\end{equation}
we obtain:
\begin{multline}\label{gf5}
G_{,r}^{[0]}(r,\theta;r',\theta)=\frac{i}{4}\sum_{n=-\infty}^{\infty}\Big\{\big[\delta(r-r')H_{n}^{(1)}(k^{[0]}r)J_{n}(k^{[0]}r')-
\delta(r'-r)J_{n}(k^{[0]}r)H_{n}^{(1)}(k^{[0]}r')\big]+\\
k^{[0]}\big[H(r-r')\dot{H}_{n}^{(1)}(k^{[0]}r)J_{n}(k^{[0]}r')+
H(r'-r)\dot{J}_{n}(k^{[0]}r)H_{n}^{(1)}(k^{[0]}r')\big]\Big\}\exp[in(\theta-\theta')]=\\
\frac{ik^{[0]}}{4}\sum_{n=-\infty}^{\infty}\big[H(r-r')\dot{H}_{n}^{(1)}(k^{[0]}r)J_{n}(k^{[0]}r')+
H(r'-r)\dot{J}_{n}(k^{[0]}r)H_{n}^{(1)}(k^{[0]}r')\big]\Big\}\exp[in(\theta-\theta')]~,
\end{multline}
wherein $\dot{H}_{n}(\zeta):=\frac{d}{d\zeta}H_{n}(\zeta)$, $\ddot{H}_{n}(\zeta):=\frac{d^{2}}{d\zeta^{2}}H_{n}(\zeta)$, $\dot{J}_{n}(\zeta):=\frac{d}{d\zeta}J_{n}(\zeta)$, $\ddot{J}_{n}(\zeta):=\frac{d^{2}}{d\zeta^{2}}J_{n}(\zeta)$.
\begin{multline}\label{gf6}
G_{,rr}^{[0]}(r,\theta;r',\theta)=\frac{ik^{[0]}}{4}\sum_{n=-\infty}^{\infty}\Big\{\big[\delta(r-r')\dot{H}_{n}^{(1)}(k^{[0]}r)J_{n}(k^{[0]}r')-
\delta(r'-r)\dot{J}_{n}(k^{[0]}r)H_{n}^{(1)}(k^{[0]}r')\big]+\\
k^{[0]}\big[H(r-r')\ddot{H}_{n}^{(1)}(k^{[0]}r)J_{n}(k^{[0]}r')+
H(r'-r)\ddot{J}_{n}(k^{[0]}r)H_{n}^{(1)}(k^{[0]}r')\big]\Big\}\exp[in(\theta-\theta')]=\\
\frac{ik^{[0]}}{4}\delta(r'-r)\sum_{n=-\infty}^{\infty}\big[\dot{H}_{n}^{(1)}(k^{[0]}r)J_{n}(k^{[0]}r)-
\dot{J}_{n}(k^{[0]}r)H_{n}^{(1)}(k^{[0]}r)\big]+\\
k^{[0]}\big[H(r-r')\ddot{H}_{n}^{(1)}(k^{[0]}r)J_{n}(k^{[0]}r')+
H(r'-r)\ddot{J}_{n}(k^{[0]}r)H_{n}^{(1)}(k^{[0]}r')\big]\Big\}\exp[in(\theta-\theta')]~,
\end{multline}
But (\cite{as68}, p. 360)
\begin{equation}\label{gf7}
~\dot{H}_{n}^{(1)}(k^{[0]}r)J_{n}(k^{[0]}r)-
\dot{J}_{n}(k^{[0]}r)H_{n}^{(1)}(k^{[0]}r)=\frac{2i}{\pi k^{[0]}r}
,
\end{equation}
so that
\begin{multline}\label{gf8}
G_{,rr}^{[0]}(r,\theta;r',\theta)=\frac{ik^{[0]}}{4}\frac{2i}{\pi k^{[0]}r}\delta(r-r')\sum_{n=-\infty}^{\infty}\exp[in(\theta-\theta')]+\\
\frac{i\big(k^{[0]}\big)^{2}}{4}\sum_{n=-\infty}^{\infty}\big[H(r-r')\ddot{H}_{n}^{(1)}(k^{[0]}r)J_{n}(k^{[0]}r')+
H(r'-r)\ddot{J}_{n}(k^{[0]}r)H_{n}^{(1)}(k^{[0]}r')\big]\exp[in(\theta-\theta')]~.
\end{multline}
The Poisson sum formula (\cite{mf53}, p. 483) tells us that
\begin{equation}\label{gf9}
\sum_{n=-\infty}^{\infty}\exp[in(\zeta-\zeta')]=2\pi\sum_{n=-\infty}^{\infty}\delta(\zeta-\zeta'+2n\pi)~;~\zeta\in\mathbb{R},~\zeta'\in\mathbb{R}
,
\end{equation}
whence
\begin{equation}\label{gf10}
\sum_{n=-\infty}^{\infty}\exp[in(\theta-\theta')]=2\pi\delta(\theta-\theta')~;~\theta\in[0,2\pi[,~\theta'\in[0,2\pi[
.
\end{equation}
Consequently
\begin{multline}\label{gf11}
G_{,rr}^{[0]}(r,\theta;r',\theta)=-\frac{1}{r}\delta(r-r')\delta(\theta-\theta')+\\
\frac{i\big(k^{[0]}\big)^{2}}{4}\sum_{n=-\infty}^{\infty}\big[H(r-r')\ddot{H}_{n}^{(1)}(k^{[0]}r)J_{n}(k^{[0]}r')+
H(r'-r)\ddot{J}_{n}(k^{[0]}r)H_{n}^{(1)}(k^{[0]}r')\big]\exp[in(\theta-\theta')]
~.
\end{multline}
It ensues that
\begin{multline}\label{gf12}
\nabla_{\mathbf{x}}^{2}G^{[0]}(r,\theta;r',\theta)=-\frac{1}{r}\delta(r-r')\delta(\theta-\theta')+\\
\frac{i}{4}\sum_{n=-\infty}^{\infty}\Big\{
H(r-r')\Big[-\frac{n^{2}}{r^{2}}H_{n}^{(1)}(k^{[0]}r)+
\frac{k^{[0]}}{r}\dot{H}_{n}^{(1)}(k^{[0]}r)+\big(k^{[0]}\big)^{2}\ddot{H}_{n}^{(1)}(k^{[0]}r)\Big]J_{n}(k^{[0]}r')\\+
H(r'-r)\Big[-\frac{n^{2}}{r^{2}}J_{n}(k^{[0]}r)+
\frac{k^{[0]}}{r}\dot{J}_{n}(k^{[0]}r)+
\big(k^{[0]}\big)^{2}\ddot{J}_{n}(k^{[0]}r)\big]H_{n}^{(1)}(k^{[0]}r')\Big\}\exp[in(\theta-\theta')]
~.
\end{multline}
But, by means of (\cite{as68}, p. 358), and for $w_{n}=J_{n}$ or $w_{n}=H_{n}^{(1)}$:
\begin{equation}\label{gf13}
\zeta^{2}w_{n,\zeta\zeta}(\zeta)+\zeta w_{n,\zeta}+(\zeta^{2}-n^{2})w_{n}(\zeta)=0
.
\end{equation}
whence
\begin{multline}\label{gf14}
\nabla_{\mathbf{x}}^{2}G^{[0]}(r,\theta;r',\theta)=-\frac{1}{r}\delta(r-r')\delta(\theta-\theta')-\\
\frac{i\big(k^{[0]}\big)^{2}}{4}\sum_{n=-\infty}^{\infty}\Big\{
H(r-r')H_{n}^{(1)}(k^{[0]}r)J_{n}(k^{[0]}r')+
H(r'-r)J_{n}(k^{[0]}r)+H_{n}^{(1)}(k^{[0]}r')\Big\}\exp[in(\theta-\theta')]
~.
\end{multline}
or
\begin{multline}\label{gf15}
\Big(\nabla_{\mathbf{x}}^{2}+\big(k^{[0]}\big)^{2}\Big)G^{[0]}(r,\theta;r',\theta)=-\frac{1}{r}\delta(r-r')\delta(\theta-\theta')-\\
-\frac{i\big(k^{[0]}\big)^{2}}{4}\sum_{n=-\infty}^{\infty}\Big\{
H(r-r')H_{n}^{(1)}(k^{[0]}r)J_{n}(k^{[0]}r')+
H(r'-r)J_{n}(k^{[0]}r)+H_{n}^{(1)}(k^{[0]}r')\Big\}\exp[in(\theta-\theta')]+\\
\frac{i\big(k^{[0]}\big)^{2}}{4}\sum_{n=-\infty}^{\infty}\Big\{
H(r-r')H_{n}^{(1)}(k^{[0]}r)J_{n}(k^{[0]}r')+
H(r'-r)J_{n}(k^{[0]}r)+H_{n}^{(1)}(k^{[0]}r')\Big\}\exp[in(\theta-\theta')]=\\
-\frac{1}{r}\delta(r-r')\delta(\theta-\theta')
~,
\end{multline}
which demonstrates (\ref{gf3}).
%%%%%%%%%%%%%%%%%%%%%%%%%%%%%%%%%%%%%%%%%%%%%%
\subsubsection{Demonstration that the 2D GF satisfies the radiation condition}\label{acradrc}
Our second task is to show that our 2D GF  satisfies the radiation condition;
\begin{equation}\label{gf16}
\lim_{r\rightarrow\infty}\left[G_{,r}^{[0]}(r,\theta;r',\theta')-ik^{[0]}G^{[0]}(r,\theta;r',\theta')\right]=0~;~r\rightarrow\infty
~,
\end{equation}
assuming that $r'$ is bounded. This means that we are interested in the case $r>>r'$ for which $H(r-r')=1$ and $H(r'-r)=0$, so that
\begin{equation}\label{gf17}
G^{[0]}(r,\theta;r',\theta')=
\frac{i}{4}\sum_{n=-\infty}^{\infty}H_{n}^{(1)}(k^{[0]}r)J_{n}(k^{[0]}r')\exp[in(\theta-\theta')]~;~r>>r'~,
\end{equation}
\begin{equation}\label{gf18}
G_{,r}^{[0]}(r,\theta;r',\theta')=
\frac{ik^{[0]}}{4}\sum_{n=-\infty}^{\infty}\dot{H}_{n}^{(1)}(k^{[0]}r)J_{n}(k^{[0]}r')\exp[in(\theta-\theta')]~;~r>>r'~.
\end{equation}
We make use of the asymptotic forms of the Hankel functions (\cite{as68}, p. 364)
\begin{equation}\label{gf19}
H_{n}^{(1)}(\zeta)\sim(\sqrt{\frac{2}{\pi}}\zeta^{-1/2}\exp[i(\zeta-n\pi/2-\pi/4)]~;~\|\zeta\|\rightarrow\infty
~,
\end{equation}
whence
\begin{equation}\label{gf20}
\dot{H}_{n}^{(1)}(\zeta)\sim(\sqrt{\frac{2}{\pi}}\Big[-\frac{1}{2}\zeta^{-3/2}+i\zeta^{-1/2}\exp[i(\zeta-n\pi/2-\pi/4)]~;~\|\zeta\|\rightarrow\infty
~.
\end{equation}
it follows that
\begin{multline}\label{gf21}
G_{,r}^{[0]}(r,\theta;r',\theta')-ik^{[0]}G^{[0]}(r,\theta;r',\theta')\sim\\
\frac{-k^{[0]}}{4}\sqrt{\frac{2}{\pi k^{[0]}r}}\exp[i(k^{[0]}r-\pi/4)]\sum_{n=-\infty}^{\infty}(1-1)\exp[in(\theta-\theta'-\pi/2)]=0~;~r\rightarrow\infty
~.
\end{multline}
so that (\ref{gf16}) is satisfied.
%%%%%%%%%%%%%%%%%%%%%%%%%%%%%%%%%%%%%%%%%%%%%%%%%%%%%%%%%%%%%%%%%%%%%%%%%%%%%%%%%%%%%%%%%%%%%%%
%%%%%%%%%%%%%%%%%%%%%%%%%%%%%%%%%%%%%%%%%%%%%%%%%%%%%%%%%%%%%%%%%%%%%%%%%%%%%%%%%
\section{Scattering of the sound radiated from applied sources
by a generally-heterogeneous fluid-like obstacle within an unbounded,
homogeneous, background fluid medium}\label{obst}
%
%%%%%%%%%%%%%%%%%%%%%%%%%%%%%%%%%%%
\subsection{Description of the problem}
\begin{figure}[ht]
\begin{center}
\includegraphics[
height=3in, width=3in ] {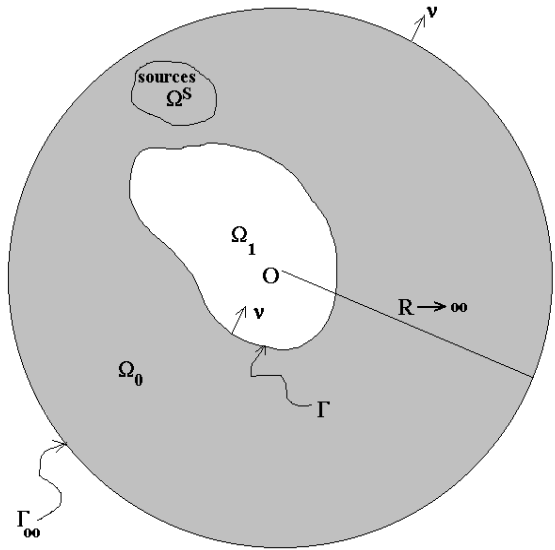} \caption{Configuration for the
scattering of the acoustic wave radiated by applied sources from a fluid-like
obstacle located in free space.} \label{fig20}
\end{center}
\end{figure}
The problem differs from the previous one in that the
formerly-homogeneous fluid medium now contains a generally-{\it heterogeneous},
spatially-bounded or unbounded (in one direction only) obstacle
(see fig. \ref{fig20} for the case of a bounded obstacle). This
obstacle can be composed of a fluid-like material with either
continuously-varying or discontinuously-varying physical
properties. Thus, the domain  of the obstacle can be simply- or
multiply-connected. In the latter case, the obstacle is actually a
set of smaller obstacles, all contained within the virtual
boundary of the larger finite-domain obstacle.

The beauty of the domain integral (DI) method, which will be discussed in detail further on, is that it can cope with a great variety of topologies, all with the same formalism. This feature is particularly-useful in the inverse problem context.

As will be shown, the obstacle, by its very presence, gives rise
to a field outside (and inside) itself that is different from what
it was in the the previous section. The difference between the two
total fields in these two problems is termed the scattered or {\it diffracted
field} and designated by $p^{d}$, and what was formerly called the
field due to applied sources is here called the {\it incident
field} and designated by $p^{i}$.

The unbounded portion of space outside of the obstacle and support of the sources is called
$\Omega_{0}$ and is occupied by the homogeneous medium
$M^{[0]}(\rho^{[0]},c^{[0]})$, with $\rho^{[0]},~c^{[0]}$ constants
with respect to  position. The obstacle forms the domain $\Omega_{1}$
and is occupied by the generally-heterogeneous fluid-like medium
$M(\rho^{[1]},c^{[1]})$ such that $\rho^{[1]}$ and $c^{[1]}$ are
generally position-dependent, complex and dispersive in the sense of (\label{bgwx24d}). The domain $\Omega_{s}$ of the sources is assumed to be wholly exterior to $\Omega_{1}$.
%%%%%%%%%%%%%%%%%%%%%%%%%%%%%%%%%%%%%%%%%%%%%%%%%%%%%%%%%%%%%%%%%%%%%%%
\subsection{The 2D scattering problem for a macroscopically-homogeneous, isotropic circular cylindrical obstacle by the domain decomposition separation of variables (DD-SOV) technique}\label{ddsov}
This scattering problem is canonical in that it serves to introduce many of the concepts that are employed in more general scattering problems, including those in which the constant density assumption is not made. However, it is important to stress that the DD-SOV technique that we shall employ to solve this problem is not applicable to obstacles with more general (than circular cylindrical, spherical, etc.) shapes in which the Helmholtz equation is not separable.

Let $(r,\theta,z)$ be the cylindrical coordinates. The obstacle is a circular cylinder, infinite in the $z$-direction and  whose cross-section is the disk $\varpi_1=\{r<a~;~\forall \theta\in[0,2\pi]\}$ the  boundary of which is the (circular) curve $\gamma=\{r=a~;~\forall \theta\in[0,2\pi]\}$. The incident field is assumed to not depend on $z$, so that, due to the homogeneous, isotropic nature of the obstacle and the shape-invariance of the latter with respect to $z$, it ensues that the scattered field is also invariant with respect to $z$. In the sagittal (orthogonal to the $z$ axis) plane, the region outside of $\varpi_1$ is designated by $\varpi_0$, so that the problem is to determine the scattered fields in $\varpi_0$, $\varpi_1$, and on $\gamma$, it being understood that the media in the two 2D regions $\varpi_0$, $\varpi_1$ have the same properties as those in the previous 3D regions $\Omega_0$, $\Omega_1$ respectively. The case addressed here assumes that both the mass density of, and wavespeed in, the obstacle are  constants as a function of position, i.e.,
\begin{equation}\label{ddsov0a}
\rho(\mathbf{x})=\rho^{[1]}={\text const.}~,~~c(\mathbf{x})=c^{[1]}={\text const.}~,~;~\forall \mathbf{x}\in\varpi_{1}~,
\end{equation}
the consequence of which (as concerns the wavespeed) is
\begin{equation}\label{ddsov0b}
k(\mathbf{x})=k^{[1]}={\text const.}~;~\forall \mathbf{x}\in\varpi_{1}~,
\end{equation}
it being recalled that $\rho^{[0]}$ and $k^{[0]}$ are also (different) constants.
%%%%%%%%%%%%%%%%%%%%%%%%%%%%%%%%%%%%%%%%%
\subsubsection{Governing equations via DD}
We search for the solution of the following problem:
\begin{multline}\label{ddsov1}
\left(\nabla_{\mathbf{x}}\cdot\nabla_{\mathbf{x}}+\big(k^{[0]}\big)^{2}\right)p^{d[l]}(\mathbf{x})=\\
\left[\frac{1}{r}\frac{\partial}{\partial r}\left(r\frac{\partial}{\partial r}\right)+\frac{1}{r^{2}}\frac{\partial^{2}}{\partial\theta^{2}}+\big(k^{[0]}\big)^{2}\right]p^{d[l]}(r,\theta;r',\theta)=0~;~\mathbf{x}\in\Omega_{l},~l=0,1
~,
\end{multline}
\begin{equation}\label{ddsov2}
\lim_{r\rightarrow\infty}\left[p_{,r}^{[0]}(\mathbf{x})-ik^{[0]}p^{[0]}(\mathbf{x})\right]=0~,
\end{equation}
\begin{equation}\label{ddsov3}
p^{[0]}(\mathbf{x})-p^{[1]}(\mathbf{x})=0~;~\mathbf{x}\in\Gamma~,
\end{equation}
\begin{equation}\label{ddsov4}
\frac{1}{\rho^{[0]}}p_{,r}^{[0]}(\mathbf{x})-\frac{1}{\rho^{[1]}}p_{,r}^{[1]}(\mathbf{x})=0~;~\mathbf{x}\in\Gamma~,
\end{equation}
wherein
\begin{equation}\label{ddsov5}
p^{d[0]}(\mathbf{x})=p^{[0]}(\mathbf{x})-p^{i}(\mathbf{x})~~,~~p^{d[1]}(\mathbf{x})=p^{[1]}(\mathbf{x})~,
\end{equation}
%%
%%%%%%%%%%%%%%%%%%%%%%%%%%%%%%%%%%%%%%%%%%%%%%%
\subsubsection{Field representations via SOV}
We had:
\begin{equation}\label{ddsov6}
p^{i}(\mathbf{x})=\int_{\Omega^{s}}G^{[0]}(\mathbf{x};\mathbf{x'})
s(\mathbf{x'})d\Omega(\mathbf{x'})~~;~~\forall~\mathbf{x}\in
\mathbb{R}^{3}~,
\end{equation}
which, after the introduction of the SOV representation of the 2D GF becomes
\begin{multline}\label{ddsov7}
p^{i}(\mathbf{x})=
\frac{i}{4}\int_{\Omega^{s}}
\sum_{n=-\infty}^{\infty}\left[H(r-r')H_{n}^{(1)}(k^{[0]}r)J_{n}(k^{[0]}r')+
H(r'-r)J_{n}(k^{[0]}r)H_{n}^{(1)}(k^{[0]}r')\right]\times\\
\exp[in(\theta-\theta')]~
s(\mathbf{x'})d\Omega(\mathbf{x'})~~;~~\forall~\mathbf{x}\in
\mathbb{R}^{3}~.
\end{multline}
Suppose the support of the source is between $r=r^{-}$ and $r=r^{+}>r^{-}$ in terms of $r$ and between $\theta=\theta^{-}$ and $\theta=\theta^{+}>\theta{-}$ in terms of $\theta$, with the understanding that $r^{-}>a$, $a$ being the largest value of $r$ along the boundary $\Gamma$ of the obstacle. Then
\begin{multline}\label{ddsov8}
p^{i}(\mathbf{x})=\frac{i}{4}
\sum_{n=-\infty}^{\infty}\int_{\theta^{-}}^{\theta^{+}}d\theta'\exp[in(\theta-\theta')]\times\\
\int_{r^{-}}^{r^{+}}dr'r'\left[H(r-r')H_{n}^{(1)}(k^{[0]}r)J_{n}(k^{[0]}r')+
H(r'-r)J_{n}(k^{[0]}r)H_{n}^{(1)}(k^{[0]}r')\right]~
s(r',\theta')~~;~~\forall~\mathbf{x}\in
\mathbb{R}^{3}~,
\end{multline}
so that, from the definition of the Heaviside distribution,
\begin{multline}\label{ddsov9}
p^{i}(r>r^{+},\theta)=\\
\sum_{n=-\infty}^{\infty}H_{n}^{(1)}(k^{[0]}r)\exp[in\theta]\frac{i}{4}\int_{\theta^{-}}^{\theta^{+}}d\theta'\exp[-in\theta']
\int_{r^{-}}^{r^{+}}dr'r'J_{n}(k^{[0]}r')~
s(r',\theta'):=\\
\sum_{n=-\infty}^{\infty}A_{n}^{i[0]}H_{n}^{(1)}(k^{[0]}r)\exp[in\theta]~~;~~\forall~\theta\in[0,2\pi[~,
\end{multline}
with
\begin{equation}\label{ddsov10}
A_{n}^{i[0]}=
\frac{i}{4}\int_{\theta^{-}}^{\theta^{+}}d\theta'\exp[-in\theta']
\int_{r^{-}}^{r^{+}}dr'r'J_{n}(k^{[0]}r')~
s(r',\theta')
\mathbb{R}^{3}~.
\end{equation}
and
\begin{multline}\label{ddsov11}
p^{i}(r<r^{-},\theta)=\\
\sum_{n=-\infty}^{\infty}J_{n}(k^{[0]}r)\exp[in\theta]\frac{i}{4}\int_{\theta^{-}}^{\theta^{+}}d\theta'\exp[-in\theta']
\int_{r^{-}}^{r^{+}}dr'r'H_{n}^{(1)}(k^{[0]}r')~
s(r',\theta'):=\\
\sum_{n=-\infty}^{\infty}B_{n}^{i[0]}J_{n}(k^{[0]}r)\exp[in\theta]~~;~~\forall~\theta\in[0,2\pi[~,
\end{multline}
with
\begin{equation}\label{ddsov12}
B_{n}^{i[0]}=
\frac{i}{4}\int_{\theta^{-}}^{\theta^{+}}d\theta'\exp[-in\theta']
\int_{r^{-}}^{r^{+}}dr'r'H_{n}^{(1)}(k^{[0]}r')~
s(r',\theta')
\mathbb{R}^{3}~.
\end{equation}
Consequently, the representation of the incident wave due to applied sources in $\Omega^{s}$ that we have to take into account in the transmission conditions (\ref{ddsov3})-(\ref{ddsov4}) is
\begin{equation}\label{ddsov13}
p^{i}(r<r^{-},\theta)=\sum_{n=-\infty}^{\infty}B_{n}^{[0]}J_{n}(k^{[0]}r)\exp[in\theta]~~;~~\forall~\theta\in[0,2\pi[~.
\end{equation}
wherein we have replaced $B_{n}^{i[0]}$ simply by $B_{n}^{[0]}$.

Note that even in the case of a plane incident wave (incident angle $\theta^{i}$),
\begin{equation}\label{ddsov14}
p^{i}(r,\theta)=B^{i}\exp[-ik^{[0]}r\cos(\theta-\theta^{i})]~,
\end{equation}
the SOV expression of $p^{i}(r,\theta)$, now valid for all $r$, is the same (\cite{as68}, p. 361) as in (\ref{ddsov13}), but with
\begin{equation}\label{ddsov15}
B_{n}^{[0]}=B^{i}\exp[-in(\theta^{i}-\pi/2)].
\end{equation}

Note that the $B_{n}^{[0]}$ are known since $s(\mathbf{x})$ was assumed known in the case of source wave incidence and $B^{i}$ and $\theta^{i}$ are assumed to be known in the case of plane wave incidence.

The SOV representations of the scattered fields satisfying the boundedness condition (notably within the cylinder) and the radiation condition (outside and far from the cylinder) are:
\begin{equation}\label{ddsov16}
p^{d[0]}(r,\theta)=\sum_{n=-\infty}^{\infty}A_{n}^{[0]}H_{n}^{(1)}(k^{[0]}r)\exp(in\theta)~~;~~\forall~r\ge a,~\forall~\theta\in[0,2\pi[~,
\end{equation}
\begin{equation}\label{ddsov17}
p^{d[1]}(r,\theta)=\sum_{n=-\infty}^{\infty}B_{n}^{[1]}J_{n}(k^{[1]}r)\exp(in\theta)~~;~~\forall~r\le a,~\forall~\theta\in[0,2\pi[~,
\end{equation}
with the understanding that the coefficients $A_{n}^{[0]}$ and  $B_{n}^{[1]}$ are unknown.
%%%%%%%%%%%%%%%%%%%%%%%%%%%%%%%%%%%
\subsubsection{Determination of the unknown coefficients via the transmission conditions}
A consequence of the transmission conditions is:
\begin{equation}\label{ddsov18}
\int_{0}^{2\pi}\left[p^{[0]}(a,\theta)-p^{[1]}(a,\theta)\right]\exp(-im\theta)d\theta=0~;~\forall m\in\mathbb{Z}~,
\end{equation}
\begin{equation}\label{ddsov19}
\int_{0}^{2\pi}\left[\frac{1}{\rho^{[0]}}p_{,r}^{[0]}(a,\theta)-\frac{1}{\rho^{[1]}}p_{,r}^{[1]}(a,\theta)\right]\exp(-im\theta)d\theta=0~;~\forall m\in\mathbb{Z}~,
\end{equation}
or
\begin{equation}\label{ddsov20}
\sum_{n=-\infty}^{\infty}\left[B_{n}^{[0]}J_{n}(k^{[0]}a)+A_{n}^{[0]}H_{n}^{(1)}(k^{[0]}a)-B_{n}^{[1]}J_{n}(k^{[1]}a)\right]\int_{0}^{2\pi}\exp[i(n-m)\theta]d\theta=0~;~\forall m\in\mathbb{Z}~,
\end{equation}
\begin{multline}\label{ddsov21}
\sum_{n=-\infty}^{\infty}\left[\left(\frac{k^{[0]}}{\rho^{[0]}}\right)\left(B_{n}^{[0]}\dot{J}_{n}(k^{[0]}a)+A_{n}^{[0]}\dot{H}_{n}^{(1)}(k^{[0]}a)\right)-
\left(\frac{k^{[1]}}{\rho^{[1]}}\right)B_{n}^{[1]}\dot{J}_{n}(k^{[1]}a)\right]\times\\
\int_{0}^{2\pi}\exp[i(n-m)\theta]d\theta=0~;~\forall m\in\mathbb{Z}~,
\end{multline}
so that, from the identity
\begin{equation}\label{ddsov22}
\int_{0}^{2\pi}\exp[i(n-m)\theta]d\theta=2\pi\delta_{mn}~,
\end{equation}
(wherein $\delta_{mn}$ is the Kronecker delta symbol), it ensues
\begin{equation}\label{ddsov23}
A_{n}^{[0]}H_{n}^{(1)}(k^{[0]}a)-B_{n}^{[1]}\dot{J}_{n}(k^{[1]}a)=-B_{n}^{[0]}\dot{J}_{n}(k^{[0]}a)~;~\forall n\in\mathbb{Z}~,
\end{equation}
\begin{equation}\label{ddsov24}
\left(\frac{k^{[0]}a}{\rho^{[0]}}\right)A_{n}^{[0]}\dot{H}_{n}^{(1)}(k^{[0]}a)-
\left(\frac{k^{[1]}a}{\rho^{[1]}}\right)B_{n}^{[1]}\dot{J}_{n}(k^{[1]}a)=-\left(\frac{k^{[0]}a}{\rho^{[0]}}\right)B_{n}^{[0]}\dot{J}_{n}(k^{[0]}a)~;~\forall n\in\mathbb{Z}~,
\end{equation}
the solution of which, on account of (\cite{as68}, p. 360), is
\begin{equation}\label{ddsov25}
A_{n}^{[0]}=B_{n}^{[0]}\left[\frac{\left(\frac{k^{[1]}a}{\rho^{[1]}}\right)\dot{J}_{n}(k^{[1]}a)J_{n}(k^{[0]}a)-
\left(\frac{k^{[0]}a}{\rho^{[0]}}\right)\dot{J}_{n}(k^{[0]}a)J_{n}(k^{[1]}a)}
{\left(\frac{k^{[0]}a}{\rho^{[0]}}\right)\dot{H}^{(1)}_{n}(k^{[0]}a)J_{n}(k^{[1]}a)-
\left(\frac{k^{[1]}a}{\rho^{[1]}}\right)\dot{J}_{n}(k^{[1]}a)H^{(1)}_{n}(k^{[0]}a)}\right]~;~n\in\mathbb{Z}
~,
\end{equation}
\begin{equation}\label{ddsov26}
B_{n}^{[1]}=B_{n}^{[0]}\left[\frac{\frac{2i}{\pi\rho^{[0]}}}
{\left(\frac{k^{[0]}a}{\rho^{[0]}}\right)\dot{H}^{(1)}_{n}(k^{[0]}a)J_{n}(k^{[1]}a)-
\left(\frac{k^{[1]}a}{\rho^{[1]}}\right)\dot{J}_{n}(k^{[1]}a)H^{(1)}_{n}(k^{[0]}a)}\right]~;~n\in\mathbb{Z}
~.
\end{equation}
This solves the problem of acoustic scattering by a macroscopically-homogeneous, isotropic circular cylinder.

Note that (\ref{ddsov25})-(\ref{ddsov26}) constitute the exact reference solutions for the cylindrical wave amplitudes inside and outside the homogeneous cylinder whose constitutive properties (i.e., mass density and wavespeed) are generally-different from those of the host medium.

Note further that it is not easy to discern in (\ref{ddsov25})-(\ref{ddsov26}) how the constant density  assumption (i.e.,$\rho^{[1]}=\rho^{[0]}$) affects the scattering amplitudes, nor, more generally, how the latter (and the scattered wavefield) depend on the contrast of mass densities $\epsilon=\frac{\rho^{[1]}-\rho^{[0]}}{\rho^{[1]}}$. We shall return to this issue in sect. \ref{di}.
%%%%%%%%%%%%%%%%%%%%%%%%%%%%%%%%%%%%%%%%%%%%%%%%%%%%%%%%%%%%%%%%%%%%%%%%%%%%%%%%%%%%%%%%%%%
\subsection{The coupled boundary integral (BI-BI) method applicable to the special case in which both the mass density of, and wavespeed in, the fluid-like obstacle do not depend on position}\label{constrhoonecone}
This case, which is treated in publications such as \cite{gu00},  again assumes that both the mass density of, and wavespeed in, the obstacle are  constants as a function of position, i.e.,
\begin{equation}\label{konstrhoonecone1}
\rho(\mathbf{x})=\rho^{[1]}={\text const.}~,~~c(\mathbf{x})=c^{[1]}={\text const.}~,~;~\forall \mathbf{x}\in\Omega_{1}~,
\end{equation}
the consequence of which (as concerns the wavespeed) is
\begin{equation}\label{konstrhoonecone2}
k(\mathbf{x})=k^{[1]}={\text const.}~;~\forall \mathbf{x}\in\Omega_{1}~,
\end{equation}
it being recalled that $\rho^{[0]}$ and $k^{[0]}$ are also (different) constants.
%%%%%%%%%%%%%%%%%%%%%%%%%%%%%%%%%%%%%%%%%%%%%%%%%%%%%
\subsubsection{BI-BI method: domain decomposition}\label{constrhooneconedc}
Instead of employing the frequency domain wave equation in $\mathbb{R}^{3}$ (\ref{bgwx24c}), we decompose the latter into two wave equations, one for $\Omega_{0}$ in which the total pressure is $p^{[0]}$ and the other for $\Omega_{1}$ in which the total pressure is $p^{[1]}$, these two domains being separated by $\Gamma$, through which the pressure is continuous and the velocity potential is continuous. Thus, assuming once again that the support of the source is in $\Omega_{0}$:
\begin{equation}\label{konstrhoonecone3}
\left(\nabla_{\mathbf{x}}\cdot\nabla_{\mathbf{x}}+\big(k^{[0]}\big)^{2}\right)p^{[0]}(\mathbf{x})=-s(\mathbf{x})~;~\forall \mathbf{x}\in\Omega_{0}~.
\end{equation}
\begin{equation}\label{konstrhoonecone4}
\left(\nabla_{\mathbf{x}}\cdot\nabla_{\mathbf{x}}+\big(k^{[1]}\big)^{2}\right)p^{[1]}(\mathbf{x})=
0~;~\forall \mathbf{x}\in\Omega_{1}~.
\end{equation}
By employing Green's second identity and the radiation condition we obtain, from (\ref{konstrhoonecone3}) and the wave equation satisfied by the free space Green's function:
\begin{equation}\label{konstrhoonecone5}
\mathcal{H}_{\Omega_{0}}(\mathbf{x})p^{[0]}(\mathbf{x})=p^{i}(\mathbf{x})+\int_{\Gamma}\left[G^{[0]}(\mathbf{x};\mathbf{x'})\boldsymbol{\nu(\mathbf{x'})}\cdot \nabla_{\mathbf{x'}}p^{[0]}(\mathbf{x'})-p^{[0]}(\mathbf{x'})\boldsymbol{\nu(\mathbf{x'})}\cdot \nabla_{\mathbf{x'}}G^{[0]}(\mathbf{x};\mathbf{x'})\right]d\Gamma(\mathbf{x'})~.
\end{equation}

Similarly, by employing Green's second identity,  we obtain, from (\ref{konstrhoonecone4}) and the wave equation satisfied by the free space Green's function $G^{[1]}$ (satisfying the same wave equation as $G^{[0]}$ in which $k^{[0]}$  is replaced by $k^{[1]}$):
\begin{equation}\label{konstrhoonecone6}
\mathcal{H}_{\Omega_{1}}(\mathbf{x})p^{[1]}(\mathbf{x})=-\int_{\Gamma}\left[G^{[1]}(\mathbf{x};\mathbf{x'})\boldsymbol{\nu(\mathbf{x'})}\cdot \nabla_{\mathbf{x'}}p^{[1]}(\mathbf{x'})-p^{[1]}(\mathbf{x'})\boldsymbol{\nu(\mathbf{x'})}\cdot \nabla_{\mathbf{x'}}G^{[1]}(\mathbf{x};\mathbf{x'})\right]d\Gamma(\mathbf{x'})~.
\end{equation}
wherein $\boldsymbol{\nu}$ is the unit vector normal to $\Gamma$ pointing into $\Omega_{1}$, $d\Gamma$ the differential area  element of the boundary $\Gamma$, and we have made use of the fact that
\begin{equation}\label{konstrhoonecone7}
\int_{\Omega}f(\mathbf{x'})\delta(\mathbf{x}-\mathbf{x'})d\Omega(\mathbf{x'})=\mathcal{H}_{\Omega}(\mathbf{x})f(\mathbf{x})~,
\end{equation}
with $\mathcal{H}_{\Omega}$  the domain Heaviside distribution equal to 1 for $\mathbf{x}\in\Omega$ and to 0 for $\mathbf{x}\in(\mathbf{R}^{3}-\Omega)$.

If we assume that $\mathbf{x}$ is a point on $\Gamma$, $\mathbf{x^{+}}$  a point on $\Gamma^{+}$ (the surface just outside, and homothetic to, $\Gamma$), and $\mathbf{x^{-}}$ a point on $\Gamma^{-}$ (the surface just inside, and homothetic to, $\Gamma$), all three points lying along the line associated with $\boldsymbol{\nu(\mathbf{x})}$, then
\begin{equation}\label{konstrhoone8}
\mathbf{x^{\pm}}=\mathbf{x}\pm\varepsilon\boldsymbol{\nu(\mathbf{x})}~.
\end{equation}
wherein $\varepsilon$ is a small, positive number. It follows that
\begin{equation}\label{konstrhoone9}
\mathcal{H}_{\Omega_{0}}(\mathbf{x^{+}})=1~~,~~\mathcal{H}_{\Omega_{1}}(\mathbf{x^{-}})=1~,
\end{equation}
whence
\begin{equation}\label{konstrhoonecone10}
p^{[0]}(\mathbf{x^{+}})=p^{i}(\mathbf{x^{+}})+\int_{\Gamma}\left[G^{[0]}(\mathbf{x^{+}};\mathbf{x'})\boldsymbol{\nu(\mathbf{x'})}\cdot \nabla_{\mathbf{x'}}p^{[0]}(\mathbf{x'})-p^{[0]}(\mathbf{x'})\boldsymbol{\nu(\mathbf{x'})}\cdot \nabla_{\mathbf{x'}}G^{[0]}(\mathbf{x^{+}};\mathbf{x'})\right]d\Gamma(\mathbf{x'})~,
\end{equation}
\begin{equation}\label{konstrhoonecone11}
p^{[1]}(\mathbf{x^{-}})=-\int_{\Gamma}\left[G^{[1]}(\mathbf{x^{-}};\mathbf{x'})\boldsymbol{\nu(\mathbf{x'})}\cdot \nabla_{\mathbf{x'}}p^{[1]}(\mathbf{x'})-p^{[1]}(\mathbf{x'})\boldsymbol{\nu(\mathbf{x'})}\cdot \nabla_{\mathbf{x'}}G^{[1]}(\mathbf{x^{-}};\mathbf{x'})\right]d\Gamma(\mathbf{x'})~.
\end{equation}
With the notations:
\begin{multline}\label{konstrhoonecone12}
\lim_{\varepsilon\rightarrow 0}
p^{[0]}(\mathbf{x^{+}})=p^{[0]}(\mathbf{x})~~,~~\lim_{\varepsilon\rightarrow 0}
p^{i}(\mathbf{x^{+}})=p^{i}(\mathbf{x})~~,~~\\
\lim_{\varepsilon\rightarrow 0}
G^{[0]}(\mathbf{x^{+}};\mathbf{x})=G^{[0]+}(\mathbf{x};\mathbf{x})~~,~~
\lim_{\varepsilon\rightarrow 0}\boldsymbol{\nu(\mathbf{x'})}\cdot \nabla_{\mathbf{x'}}G^{[1]}(\mathbf{x^{-}};\mathbf{x'})=
\boldsymbol{\nu(\mathbf{x'})}\cdot \nabla_{\mathbf{x'}}G^{[1]-}(\mathbf{x};\mathbf{x'})
\end{multline}
we obtain
\begin{multline}\label{konstrhoonecone13}
p^{[0]}(\mathbf{x})=p^{i}(\mathbf{x})+\\
\int_{\Gamma}\left[G^{[0]+}(\mathbf{x};\mathbf{x'})\boldsymbol{\nu(\mathbf{x'})}\cdot \nabla_{\mathbf{x'}}p^{[0]}(\mathbf{x'})-p^{[0]}(\mathbf{x'})\boldsymbol{\nu(\mathbf{x'})}\cdot \nabla_{\mathbf{x'}}G^{[0]+}(\mathbf{x};\mathbf{x'})\right]d\Gamma(\mathbf{x'})~;~\forall \mathbf{x}\in\Gamma~,
\end{multline}
\begin{equation}\label{konstrhoonecone14}
p^{[1]}(\mathbf{x})=-\int_{\Gamma}\left[G^{[1]-}(\mathbf{x};\mathbf{x'})\boldsymbol{\nu(\mathbf{x'})}\cdot \nabla_{\mathbf{x'}}p^{[1]}(\mathbf{x'})-p^{[1]}(\mathbf{x'})\boldsymbol{\nu(\mathbf{x'})}\cdot \nabla_{\mathbf{x'}}G^{[1]-}(\mathbf{x};\mathbf{x'})\right]d\Gamma(\mathbf{x'})~;~\forall \mathbf{x}\in\Gamma~.
\end{equation}
Recall that the fields satisfy the continuity conditions:
\begin{equation}\label{konstrhoonecone15}
p^{[0]}(\mathbf{x})-p^{[1]}(\mathbf{x})=0~~,~~
\frac{1}{\rho^{[0]}}\boldsymbol{\nu(\mathbf{x})}\cdot\nabla_{\mathbf{x}}p^{[0]}(\mathbf{x})-
\frac{1}{\rho^{[1]}}\boldsymbol{\nu(\mathbf{x})}\cdot\nabla_{\mathbf{x}}p^{[1]}(\mathbf{x})=0~;~\forall \mathbf{x}\in\Gamma~,
\end{equation}
whence (\ref{konstrhoonecone13}) takes the form
\begin{multline}\label{konstrhoonecone16}
p^{[1]}(\mathbf{x})=p^{i}(\mathbf{x})+\\
\int_{\Gamma}\left[G^{[0]+}(\mathbf{x};\mathbf{x'})\frac{\rho^{[0]}}{\rho^{[1]}}
\boldsymbol{\nu(\mathbf{x'})}\cdot \nabla_{\mathbf{x'}}p^{[1]}(\mathbf{x'})-p^{[1]}(\mathbf{x'})\boldsymbol{\nu(\mathbf{x'})}\cdot \nabla_{\mathbf{x'}}G^{[0]+}(\mathbf{x};\mathbf{x'})\right]d\Gamma(\mathbf{x'})~;~\forall \mathbf{x}\in\Gamma~,
\end{multline}
so that, with the notations
\begin{equation}\label{konstrhoonecone17}
U(\mathbf{x})=p^{[1]}(\mathbf{x}\in\Gamma)~~,~~U^{i}(\mathbf{x})=p^{i}(\mathbf{x}\in\Gamma)~~,~~V(\mathbf{x})=
-\frac{1}{k^{[1]}}\boldsymbol{\nu(\mathbf{x})}\cdot \nabla_{\mathbf{x}}p^{[1]}(\mathbf{x}\in\Gamma),
\end{equation}
we are finally confronted with the following {\it  system of two coupled boundary integral equations} in the two unknown functions $U(\mathbf{x})$ and $V(\mathbf{x})$:
\begin{equation}\label{konstrhoonecone18}
U(\mathbf{x})=U^{i}(\mathbf{x})+\int_{\Gamma}\left[-G^{[0]+}(\mathbf{x};\mathbf{x'})\frac{\rho^{[0]}}{\rho^{[1]}}k^{[1]}V(\mathbf{x'})
-U(\mathbf{x'})\boldsymbol{\nu(\mathbf{x'})}\cdot \nabla_{\mathbf{x'}}G^{[0]+}(\mathbf{x};\mathbf{x'})\right]d\Gamma(\mathbf{x'})~;~\forall \mathbf{x}\in\Gamma~,
\end{equation}
\begin{equation}\label{konstrhoonecone19}
U(\mathbf{x})=-\int_{\Gamma}\left[-G^{[1]-}(\mathbf{x};\mathbf{x'})k^{[1]}V(\mathbf{x'})-U(\mathbf{x'})\boldsymbol{\nu}(\mathbf{x'})\cdot \nabla_{\mathbf{x'}}G^{[1]-}(\mathbf{x};\mathbf{x'})\right]d\Gamma(\mathbf{x'})~;~\forall \mathbf{x}\in\Gamma~.
\end{equation}
%%
%%%%%%%%%%%%%%%%%%%%%%%%%%%%%%%%%%%%%%%%%%%
\subsubsection{BI-BI method: iterative method for obtaining the solutions for $U$ and $V$}
Let $U^{(j)}$ and $V^{(j)}$ be the $j$-th order approximations of the coupled BI-BI equations (\ref{konstrhoonecone18})-(\ref{konstrhoonecone19}). Then, after the intialization
\begin{equation}\label{konstrhoonecone19a}
U^{(0)}(\mathbf{x})=U^{i}(\mathbf{x})~,
\end{equation}
the iterative method proceeds (in the indicated order) as follows:
\begin{multline}\label{konstrhoonecone19b}
U^{(j-1)}(\mathbf{x})=-\int_{\Gamma}\left[-G^{[1]-}(\mathbf{x};\mathbf{x'})k^{[1]}V^{(j)}(\mathbf{x'})-U^{(j-1)}(\mathbf{x'})\boldsymbol{\nu}(\mathbf{x'})\cdot \nabla_{\mathbf{x'}}G^{[1]-}(\mathbf{x};\mathbf{x'})\right]d\Gamma(\mathbf{x'})~;\\
~j=1,2,...~;~\forall \mathbf{x}\in\Gamma~,
\end{multline}
\begin{multline}\label{konstrhoonecone19c}
U^{(j)}(\mathbf{x})=U^{i}(\mathbf{x})+\int_{\Gamma}\left[-G^{[0]+}(\mathbf{x};\mathbf{x'})\frac{\rho^{[0]}}{\rho^{[1]}}k^{[1]}V^{(j)}(\mathbf{x'})
-U^{(j-1)}(\mathbf{x'})\boldsymbol{\nu(\mathbf{x'})}\cdot \nabla_{\mathbf{x'}}G^{[0]+}(\mathbf{x};\mathbf{x'})\right]d\Gamma(\mathbf{x'});\\
~j=1,2,...~;~\forall \mathbf{x}\in\Gamma~.
\end{multline}
Note that this method requires the resolution of a single BIE of the second kind (\ref{konstrhoonecone19b}), followed by the computation of a single  integral transform (\ref{konstrhoonecone19c}), for each approximation order $j\geq 1$.
%%%%%%%%%%%%%%%%%%%%%%%%%%%%%%%%%%%%%%%%%%%
\subsubsection{BI-BI method: the 2D case (of cylindrical symmetry)}
Let the origin $O$ be located within the obstacle, whence the cylindrical coordinates $r,\theta,z$ are appropriate, with $z$ the ignorable coordinate. Then, $\mathbf{x}=(r,\theta)$, $\mathbf{x'}=(r',\theta')$, the closed boundary $\Gamma$ becomes the closed curve $\gamma$ and $d\Gamma$ becomes $d\gamma$. $\gamma$ is assumed to be describable by the single-valued function
\begin{equation}\label{konstrhoonecone20}
r=r_{\gamma}(\theta)~,
\end{equation}
and
\begin{multline}\label{konstrhoonecone21}
\nabla_{\mathbf{x}}=\Big(\frac{\partial}{\partial r},\frac{1}{r}\frac{\partial}{\partial \theta}\Big)~~,~~\boldsymbol{\nu}(\mathbf{x})=\frac{1}{\sigma_{\gamma}(\theta)}\big(-r_{\gamma},\dot{r}_{\gamma}\big)~~,
~~d\gamma(\mathbf{x})=\sigma_{\gamma}(\theta)d\theta~~,\\~~\sigma_{\gamma}=
\sigma_{\gamma}(\theta)=\sqrt{\big(\dot{r}_{\gamma}\big)^{2}+\big(r_{\gamma}\big)^{2}}~~~,
~~\boldsymbol{\nu}\cdot \nabla_{\mathbf{x}}=\frac{\frac{\dot{r}_{\gamma}}{r_{\gamma}}\frac{\partial }{\partial\theta}-r_{\gamma}\frac{\partial }
{\partial r}}{\sigma_{\gamma}(\theta)}~.
\end{multline}
The $2\pi$-periodicity (in terms of $\theta$) of the boundary makes it legitimate to assume
\begin{equation}\label{konstrhoonecone22}
U(\mathbf{x})=\sum_{m=-\infty}^{\infty}U_{m}(r)e^{im\theta}~~,~~
U^{i}(\mathbf{x})=\sum_{m=-\infty}^{\infty}U^{i}_{m}(r)e^{im\theta}~~,~~V(\mathbf{x})=\sum_{m=-\infty}^{\infty}V_{m}(r)e^{im\theta}~.
\end{equation}
Furthermore, the Greens' functions admit the representations
\begin{equation}\label{konstrhoonecone23}
G^{[l]}(\mathbf{x};\mathbf{x'})=\sum_{n=-\infty}^{\infty}G_{n}^{[l]}(r;r')e^{in(\theta-\theta')}~~,~~
\boldsymbol\nu(\mathbf{x'})\cdot\nabla_{\mathbf{x'}}G^{[l]}(\mathbf{x};\mathbf{x'})=
\frac{1}{k^{[l]}}\sum_{n=-\infty}^{\infty}F_{n}^{[l]}(r;r')e^{in(\theta-\theta')}~,
\end{equation}
so that the two boundary integral equations (BIE) lead (with $r'_{\gamma}=r_{\gamma}(\theta')$) to
\begin{multline}\label{konstrhoonecone24}
\sum_{m=-\infty}^{\infty}U_{m}(r_{\gamma})e^{im\theta}=\sum_{m=-\infty}^{\infty}U_{m}^{i}(r_{\gamma})e^{im\theta}+\\
\sum_{m=-\infty}^{\infty}\sum_{n=-\infty}^{\infty}e^{in\theta}\int_{0}^{2\pi}d\theta'\sigma'_{\gamma}
\left[-G_{n}^{[0]+}(r_{\gamma};r'_{\gamma})\frac{\rho^{[0]}}{\rho^{[1]}}k^{[1]}V_{m}(r'_{\gamma})
-k^{[0]}U_{m}(r'_{\gamma})F_{n}^{[0]+}(r_{\gamma};r'_{\gamma})\right]e^{i(m-n)\theta'}~;\\
~\forall\theta\in[0,2\pi[~,
\end{multline}
\begin{multline}\label{konstrhoonecone25}
\sum_{m=-\infty}^{\infty}U_{m}(r_{\gamma})e^{im\theta}=\\
-\sum_{m=-\infty}^{\infty}\sum_{n=-\infty}^{\infty}e^{in\theta}\int_{0}^{2\pi}d\theta'\sigma'_{\gamma}
\left[-G_{n}^{[1]-}(r_{\gamma};r'_{\gamma})k^{[1]}V_{m}(r'_{\gamma})
-k^{[1]}U_{m}(r'_{\gamma})F_{n}^{[1]-}(r_{\gamma};r'_{\gamma})\right]e^{i(m-n)\theta'}~;\\
~\forall\theta\in[0,2\pi[~.
\end{multline}
We project these relations so as to obtain the coupled system of two matrix integral equations (for the two unknown vector functions $\{U_{m};\forall m\in\mathbb{Z};\forall\theta\in[0,2\pi[\}$, $\{V_{m};\forall m\in\mathbb{Z};\forall\theta\in[0,2\pi[\}$)
\begin{multline}\label{konstrhoonecone27}
\sum_{m=-\infty}^{\infty}\int_{0}^{2\pi}d\theta U_{l}(r_{\gamma})e^{i(m-l)\theta}=
\sum_{m=-\infty}^{\infty}\int_{0}^{2\pi}d\theta U_{l}^{i}(r_{\gamma})e^{i(m-l)\theta}+\\
\sum_{m=-\infty}^{\infty}\sum_{n=-\infty}^{\infty}\int_{0}^{2\pi}d\theta e^{i(n-l)\theta}\int_{0}^{2\pi}d\theta'\sigma'_{\gamma}
\left[-G_{n}^{[0]+}(r_{\gamma};r'_{\gamma})\frac{\rho^{[0]}}{\rho^{[1]}}k^{[1]}V_{m}(r'_{\gamma})
-k^{[0]}U_{m}(r'_{\gamma})F_{n}^{[0]+}(r_{\gamma};r'_{\gamma})\right]\times\\
e^{i(m-n)\theta'}~;~\forall l\in\mathbb{Z}~,~\forall\theta\in[0,2\pi[~,
\end{multline}
\begin{multline}\label{konstrhoonecone28}
\sum_{m=-\infty}^{\infty}\int_{0}^{2\pi}d\theta U_{l}(r_{\gamma})e^{i(m-l)\theta}=\\
-\sum_{m=-\infty}^{\infty}\sum_{n=-\infty}^{\infty}\int_{0}^{2\pi}d\theta e^{i(n-l)\theta}\int_{0}^{2\pi}d\theta'\sigma'_{\gamma}\times\\
\left[-G_{n}^{[1]-}(r_{\gamma};r'_{\gamma})k^{[1]}V_{m}(r'_{\gamma})
-k^{[1]}U_{m}(r'_{\gamma})F_{n}^{[1]-}(r_{\gamma};r'_{\gamma})\right]\times\\
e^{i(m-n)\theta'}~;~\forall l\in\mathbb{Z}~,~\forall\theta\in[0,2\pi[~.
\end{multline}
%%
%%%%%%%%%%%%%%%%%%%%%%%%%%%%%%%%%%%%%%%%%%%%%%%
\subsubsection{BI-BI method: the case of a homogeneous, circular cylindrical obstacle}\label{chco}
The origin $O$ is at the center of the circular boundary of radius $a$, so that
\begin{equation}\label{chco1}
r_{\gamma}=a~~,~~\sigma_{\gamma}=a~~,~~\boldsymbol{\nu}(\mathbf{x})\cdot\nabla_{\mathbf{x}}=-\frac{\partial}{\partial r}~~,~~d\gamma=ad\theta~,
\end{equation}
whereupon the previous two matrix integral equations become
\begin{multline}\label{chco2}
\sum_{m=-\infty}^{\infty}U_{m}(a)\int_{0}^{2\pi}d\theta e^{i(m-l)\theta}=\sum_{m=-\infty}^{\infty}U_{m}^{i}(a)\int_{0}^{2\pi}d\theta e^{i(m-l)}+\\
a\sum_{m=-\infty}^{\infty}\sum_{n=-\infty}^{\infty}
\left[-G_{n}^{[0]+}(a,a)\frac{\rho^{[0]}}{\rho^{[1]}}k^{[1]}V_{m}(a)
-k^{[0]}U_{m}(a)F_{n}^{[0]+}(a;a)\right]\times\\
\int_{0}^{2\pi}d\theta e^{i(n-l)\theta}\int_{0}^{2\pi}d\theta'e^{i(m-n)\theta'}~;~\forall l\in\mathbb{Z}~,
\end{multline}
\begin{multline}\label{chco3}
\sum_{m=-\infty}^{\infty}U_{m}(a)\int_{0}^{2\pi}d\theta e^{i(m-l)\theta}=\\
-a\sum_{m=-\infty}^{\infty}\sum_{n=-\infty}^{\infty}
\left[-G_{n}^{[1]-}(a;a)k^{[1]}V_{m}(a)
-k^{[1]}U_{m}(a)F_{n}^{[1]-}(a;a)\right]\times\\
\int_{0}^{2\pi}d\theta e^{i(n-l)\theta}\int_{0}^{2\pi}d\theta'e^{i(m-n)\theta'}~;~\forall l\in\mathbb{Z}~.
\end{multline}
We now make use of the identity
\begin{equation}\label{chco4}
\int_{0}^{2\pi}d\theta e^{i(n-l)\theta}=2\pi\delta_{nl}~,
\end{equation}
so as to find
\begin{multline}\label{chco5}
\sum_{m=-\infty}^{\infty}U_{m}(a)\delta_{ml}=\sum_{m=-\infty}^{\infty}U_{m}^{i}(a)\delta_{ml}+\\
2\pi a\sum_{m=-\infty}^{\infty}\sum_{n=-\infty}^{\infty}
\left[-G_{n}^{[0]+}(a,a)\frac{\rho^{[0]}}{\rho^{[1]}}k^{[1]}V_{m}(a)
-k^{[0]}U_{m}(a)F_{n}^{[0]+}(a;a)\right]
\delta_{nl}\delta_{mn}~;~\forall l\in\mathbb{Z}~,
\end{multline}
\begin{multline}\label{chco6}
\sum_{m=-\infty}^{\infty}U_{m}(a)\delta_{ml}=\\
-2\pi a\sum_{m=-\infty}^{\infty}\sum_{n=-\infty}^{\infty}
\left[G_{n}^{[1]-}(a;a)k^{[1]}V_{m}(a)
-k^{[1]}U_{m}(a)F_{n}^{[1]-}(a;a)\right]
\delta_{nl}\delta_{mn}~;~\forall l\in\mathbb{Z}~,
\end{multline}
or simply,
\begin{equation}\label{chco7}
U_{l}(a)=U_{l}^{i}(a)+
2\pi a
\left[-G_{l}^{[0]+}(a,a)\frac{\rho^{[0]}}{\rho^{[1]}}k^{[1]}V_{l}(a)
-k^{[0]}U_{l}(a)F_{l}^{[0]+}(a;a)\right]
~;~\forall l\in\mathbb{Z}~,
\end{equation}
\begin{equation}\label{chco8}
U_{l}(a)=
-2\pi a
\left[-G_{l}^{[1]-}(a;a)k^{[1]}V_{l}(a)
-k^{[1]}U_{l}(a)F_{l}^{[1]-}(a;a)\right]
~;~\forall l\in\mathbb{Z}~,
\end{equation}
which represent two coupled systems of linear equations for the two unknown vectors $\{U_{l}(a);\forall l\in\mathbb{Z}\}$, $\{V_{l}(a);\forall l\in\mathbb{Z}\}$. Note that:
\begin{equation}\label{chco9}
F_{l}^{[j]\pm}(a;a)=-\frac{1}{k^{[j]}}G_{l,r'}^{[j]\pm}(a;a)
~;~l\in\mathbb{Z}~.
\end{equation}

To go a step further, we must recall that
\begin{equation}\label{chco10}
G_{l}^{[j]}(r;r')=\frac{i}{4}\left[H(r-r')H^{(1)}_{l}(k^{[j]}r)J_{l}(k^{[j]}r')+H(r'-r)H^{(1)}_{l}(k^{[j]}r')J_{l}(k^{[j]}r)\right]
~;~l\in\mathbb{Z}~,
\end{equation}
so that
\begin{equation}\label{chco11}
G_{l,r'}^{[j]}(r;r')=\frac{ik^{[j]}}{4}\left[H(r-r')H^{(1)}_{l}(k^{[j]}r)\dot{J}_{l}(k^{[j]}r')+
H(r'-r)\dot{H}^{(1)}_{l}(k^{[j]}r')J_{l}(k^{[j]}r)\right]
~;~l\in\mathbb{Z}~,
\end{equation}
wherein $H(\zeta>0)=1$ and $H(\zeta<0)=0$. Consequently,
\begin{equation}\label{chco10}
G_{l}^{[j]+}(a;a')=G_{l}^{[j]-}(a;a')=\frac{i}{4}H^{(1)}_{l}(k^{[j]}a)J_{l}(k^{[j]}a)
~;~l\in\mathbb{Z}~,
\end{equation}
\begin{equation}\label{chco11}
G_{l,r'}^{[j]+}(a;a)=\frac{ik^{[j]}}{4}H^{(1)}_{l}(k^{[j]}a)\dot{J}_{l}(k^{[j]}a)~~,~~G_{l,r'}^{[j]-}(a;a)=\frac{ik^{[j]}}{4}
\dot{H}^{(1)}_{l}(k^{[j]}a)J_{l}(k^{[j]}a)
~;~l\in\mathbb{Z}~,
\end{equation}
from which it follows that;
\begin{equation}\label{chco12}
U_{l}(a)=U_{l}^{i}(a)+
\frac{i\pi}{2}
\left[-\frac{\rho^{[0]}}{\rho^{[1]}}k^{[1]}aH^{(1)}_{l}(k^{[0]}a)J_{l}(k^{[0]}a)V_{l}(a)
+k^{[0]}aH^{(1)}_{l}(k^{[0]}a)\dot{J}_{l}(k^{[0]}a)U_{l}(a)\right]
~;~\forall l\in\mathbb{Z}~,
\end{equation}
\begin{equation}\label{chco13}
U_{l}(a)=
-\frac{i\pi}{2}
\left[-k^{[1]}aH^{(1)}_{l}(k^{[1]}a)J_{l}(k^{[1]}a)V_{l}(a)
+k^{[1]}a\dot{H}^{(1)}_{l}(k^{[1]}a)J_{l}(k^{[1]}a)U_{l}(a)\right]
~;~\forall l\in\mathbb{Z}~,
\end{equation}
or
\begin{equation}\label{chco14}
\begin{array}{c}
X_{l}^{11}Y_{l}^{1}+X_{l}^{12}Y_{l}^{2}=Z_{l}^{1}\\
X_{l}^{21}Y_{l}^{1}+X_{l}^{22}Y_{l}^{2}=Z_{l}^{2}
\end{array}
~;~\forall l\in\mathbb{Z}~,
\end{equation}
in which
\begin{equation}\label{chco15}
\begin{array}{cc}
X_{l}^{11}=1-\frac{i\pi}{2}k^{[0]}aH^{(1)}_{l}(k^{[0]}a)\dot{J}_{l}(k^{[0]}a)~~,~~ X_{l}^{12}=\frac{i\pi}{2}\frac{\rho^{[0]}}{\rho^{[1]}}k^{[1]}aH^{(1)}_{l}(k^{[0]}a)J_{l}(k^{[0]}a)\\
X_{l}^{21}=1+\frac{i\pi}{2}k^{[1]}a\dot{H}^{(1)}_{l}(k^{[1]}a)J_{l}(k^{[1]}a)~~,~~ X_{l}^{22}=-\frac{i\pi}{2}k^{[1]}aH^{(1)}_{l}(k^{[1]}a)J_{l}(k^{[1]}a)\\
Y_{l}^{1}=U_{l}(a)~~,~~Y_{l}^{2}=V_{l}(a)\\
Z_{l}^{1}=U_{l}^{i}(a)~~,~~Z_{l}^{2}=0
\end{array}
~.
\end{equation}
We now make use of the Wronskian formula for the Bessel and Hankel functions (\cite{as68}, p. 360)
\begin{equation}\label{chco16}
H^{(1)}_{l}(k^{[j]}a)\dot{J}_{l}(k^{[j]}a)-\dot{H}^{(1)}_{l}(k^{[j]}a)J_{l}(k^{[j]}a)=\frac{-2i}{\pi k^{[j]}a}~~,~~
~,
\end{equation}
to obtain:
\begin{equation}\label{chco17}
\begin{array}{cc}
X_{l}^{11}=-\frac{i\pi}{2}k^{[0]}a\dot{H}^{(1)}_{l}(k^{[0]}a)J_{l}(k^{[0]}a)~~,~~ X_{l}^{12}=\frac{i\pi}{2}\frac{\rho^{[0]}}{\rho^{[1]}}k^{[1]}aH^{(1)}_{l}(k^{[0]}a)J_{l}(k^{[0]}a)\\
X_{l}^{21}=\frac{i\pi}{2}k^{[1]}aH^{(1)}_{l}(k^{[1]}a)\dot{J}_{l}(k^{[1]}a)~~,~~ X_{l}^{22}=-\frac{i\pi}{2}k^{[1]}aH^{(1)}_{l}(k^{[1]}a)J_{l}(k^{[1]}a)\\
\end{array}
~.
\end{equation}
We showed previously that
\begin{equation}\label{chco18}
U_{l}^{i}(a)=B_{l}^{[0]}J_{l}(k^{[0]}a)~,
\end{equation}
so that the solution of the system of linear equations turns out to be:
\begin{equation}\label{chco19}
\begin{array}{c}
Y_{l}^{1}=U_{l}(a)=B_{l}^{[1]}J_{l}(k^{[1]}a)~~,~~B_{l}^{[1]}=B_{l}^{[0]}\left[\frac{\frac{2i}{\pi\rho^{[0]}}}
{\frac{k^{[0]}a}{\rho^{[0]}}\dot{H}^{(1)}_{l}(k^{[0]}a)J_{l}(k^{[1]}a)-
\frac{k^{[1]}a}{\rho^{[1]}}H^{(1)}_{l}(k^{[0]}a)\dot{J}_{l}(k^{[1]}a)}\right] \\
Y_{l}^{2}=V_{l}(a)=B_{l}^{[1]}\dot{J}_{l}(k^{[1]}a)
\end{array}
~.
\end{equation}
The last step is to obtain explicit expressions for the fields within and outside the obstacle from $U_{l}(a)~,~V_{l}(a)$.
We found previously  (\ref{konstrhoonecone6})-(\ref{konstrhoonecone7}) that:
\begin{equation}\label{chco20}
p^{[0]}(\mathbf{x})=p^{i}(\mathbf{x})+\int_{\gamma}\left[G^{[0]}(\mathbf{x};\mathbf{x'})\boldsymbol{\nu(\mathbf{x'})}\cdot \nabla_{\mathbf{x'}}p^{[0]}(\mathbf{x'})-p^{[0]}(\mathbf{x'})\boldsymbol{\nu(\mathbf{x'})}\cdot \nabla_{\mathbf{x'}}G^{[0]}(\mathbf{x};\mathbf{x'})\right]d\gamma(\mathbf{x'})~;~\forall \mathbf{x}\in \varpi^{[0]}~,
\end{equation}
\begin{equation}\label{chco21}
p^{[1]}(\mathbf{x})=-\int_{\gamma}\left[G^{[1]}(\mathbf{x};\mathbf{x'})\boldsymbol{\nu(\mathbf{x'})}\cdot \nabla_{\mathbf{x'}}p^{[1]}(\mathbf{x'})-p^{[1]}(\mathbf{x'})\boldsymbol{\nu(\mathbf{x'})}\cdot \nabla_{\mathbf{x'}}G^{[1]}(\mathbf{x};\mathbf{x'})\right]d\gamma(\mathbf{x'})~;~\forall \mathbf{x}\in \varpi^{[1]}~,
\end{equation}
wherein $\varpi^{[1]}$ is the disk $\{r<a~;~\forall \theta\in[0,2\pi[\}$ and $\varpi^{[0]}=\mathbb{R}^{2}-\varpi^{[1]}$. Because of the $2\pi$-angular periodicity of the fields, we have
\begin{equation}\label{chco22}
p^{[j]}(\mathbf{x})=\sum_{m=0}^{\infty}p_{m}^{[j]}(r)e^{im\theta}~;~j=0,1~~,~~p^{i}(\mathbf{x})=\sum_{m=0}^{\infty}p_{m}^{i}(r)e^{im\theta}~,
\end{equation}
so that, after evaluation of $\int_{0}^{2\pi}p^{[j]}(\mathbf{x})e^{-il\theta}d\theta$, we obtain the following explicit representations of the $r-$ dependent components of the fields:
\begin{equation}\label{chco23}
p_{l}^{[0]}(r>a)=p_{l}^{i}(r>a)-2\pi a\left[G_{l}^{[0]}(r>a;a)k^{[1]}\frac{\rho^{[0]}}{\rho^{[1]}}V_{l}(a)-G_{l,r'}^{[0]}(r>a;a)U_{l}(a)\right]~,
\end{equation}
\begin{equation}\label{chco24}
p_{l}^{[1]}(r<a)=2\pi a\left[G_{l}^{[1]}(r<a;a)k^{[1]}V_{l}(a)-G_{l,r'}^{[0]}(r<a;a)U_{l}(a)\right]~,
\end{equation}
or, more explicitly:
\begin{equation}\label{chco25}
p_{l}^{[0]}(r>a)=p_{l}^{i}(r>a)-\frac{i\pi}{2}\left[
k^{[1]}a\frac{\rho^{[0]}}{\rho^{[1]}}H^{(1)}_{l}(k^{[0]}r)J_{l}(k^{[0]}a) V_{l}(a)-
k^{[0]}aH^{(1)}_{l}(k^{[0]}r)\dot{J}_{l}(k^{[0]}a)U_{l}(a)
\right]~,
\end{equation}
\begin{equation}\label{chco26}
p_{l}^{[1]}(r<a)=\frac{i\pi}{2}\left[
k^{[1]}aJ_{l}(k^{[1]}r) H^{(1)}_{l}(k^{[1]}a)V_{l}(a)-k^{[1]}aJ_{l}(k^{[1]}r)\dot{H}^{(1)}_{l}(k^{[1]}a)U_{l}(a)
\right]~,
\end{equation}
or, even more explicitly;
\begin{equation}\label{chco27}
p_{l}^{[0]}(r>a)=p_{l}^{i}(r>a)-B_{l}^{[1]}H^{(1)}_{l}(k^{[0]}r)\frac{i\pi}{2}\left[
k^{[1]}a\frac{\rho^{[0]}}{\rho^{[1]}}J_{l}(k^{[0]}a) \dot{J}_{l}(k^{[1]}a)-
k^{[0]}a\dot{J}_{l}(k^{[0]}a)J^{(1)}_{l}(k^{[1]}a)
\right]~,
\end{equation}
\begin{equation}\label{chco28}
p_{l}^{[1]}(r<a)=B_{l}^{[1]}J_{l}(k^{[1]}r)\frac{i\pi}{2}k^{[1]}a\left[
 H^{(1)}_{l}(k^{[1]}a)\dot{J}_{l}(k^{[1]}a)-\dot{H}^{(1)}_{l}(k^{[1]}a)J_{l}(k^{[1]}a)
\right]~,
\end{equation}
which, on account of (\ref{chco16}) and (\ref{chco19}), become:
\begin{multline}\label{chco29}
p_{l}^{[0]}(r>a)=p_{l}^{i}(r>a)+A_{l}^{[0]}H^{(1)}_{l}(k^{[0]}r)~~,\\
~~A_{l}^{[0]}=B_{l}^{[0]}\left[\frac{
\frac{k^{[1]}a}{\rho^{[1]}}J_{l}(k^{[0]}a) \dot{J}_{l}(k^{[1]}a)-
\frac{k^{[0]}a}{\rho^{[0]}}\dot{J}_{l}(k^{[0]}a)J^{(1)}_{l}(k^{[1]}a)}{\frac{k^{[0]}a}{\rho^{[0]}}\dot{H}^{(1)}_{l}(k^{[0]}a)J_{l}(k^{[1]}a)-
\frac{k^{[1]}a}{\rho^{[1]}}H^{(1)}_{l}(k^{[0]}a)\dot{J}_{l}(k^{[1]}a)}\right]
~,
\end{multline}
\begin{equation}\label{chco30}
p_{l}^{[1]}(r<a)=B_{l}^{[1]}J_{l}(k^{[1]}r)
~,
\end{equation}
these being identical to the exact, reference solutions (\ref{ddsov25})-(\ref{ddsov26}) obtained previously by the DD-SOV method.
%%%%%%%%%%%%%%%%%%%%%%%%%%%%%%%%%%%%%%%%%%%%%%%%%%%%%%%%%%%%%%%%%%%%%%%%%%%%%%%%%%%%%%%%%%%
\subsection{The  coupled domain-boundary (DI-BI) method applicable to the special case in which the mass density of the fluid-like obstacle does not depend on position}\label{constrhoone}
This case, which is treated in publications such as \cite{gu00},  differs from the most general one only by the fact that the mass density of the obstacle is a constant as a function of position, i.e.,
\begin{equation}\label{constrhoone1}
\rho(\mathbf{x})=\rho^{[1]}={\text const.}~;~\forall \mathbf{x}\in\Omega_{1}~,
\end{equation}
it being implicitly assumed that
\begin{equation}\label{constrhoone2}
c(\mathbf{x})=c^{[1]}(\mathbf{x})~;~\forall \mathbf{x}\in\Omega_{1}~.
\end{equation}
%%
%%%%%%%%%%%%%%%%%%%%%%%%%%%%%%%%%%%%%%%%%%%%%%%%%%%%%
\subsubsection{Domain decomposition}
Instead of employing the frequency domain wave equation in $\mathbb{R}^{3}$, we decompose the latter into two wave equations, one for $\Omega_{0}$ in which the total pressure is $p^{[0]}$ and the other for $\Omega_{1}$ in which the total pressure is $p^{[1]}$, these two domains being separated by $\Gamma$, through which the pressure is continuous and the velocity potential is continuous. Thus:
\begin{equation}\label{constrhoone3}
\left(\nabla_{\mathbf{x}}\cdot\nabla_{\mathbf{x}}+\big(k^{[0]}\big)^{2}\right)p^{[0]}(\mathbf{x})=-s(\mathbf{x})~;~\forall \mathbf{x}\in\Omega_{0}~.
\end{equation}
\begin{equation}\label{constrhoone4}
\left(\nabla_{\mathbf{x}}\cdot\nabla_{\mathbf{x}}+\big(k^{[0]}\big)^{2}\right)p^{[1]}(\mathbf{x})=
\Big(\big(k^{[0]}\big)^{2}-\big(k^{[1]}(\mathbf{x})\big)^{2}\Big)p^{[1]}~;~\forall \mathbf{x}\in\Omega_{1}~.
\end{equation}
By employing Green's second identity and the radiation condition we obtain, from (\ref{constrhoone3}) and the wave equation satisfied by the free space Green's function:
\begin{equation}\label{constrhoone5}
\mathcal{H}_{\Omega_{0}}(\mathbf{x})p^{[0]}(\mathbf{x})=p^{i}(\mathbf{x})+\int_{\Gamma}\left[G^{[0]}(\mathbf{x};\mathbf{x'})\boldsymbol{\nu}\cdot \nabla_{\mathbf{x'}}p^{[0]}(\mathbf{x'})-p^{[0]}(\mathbf{x'})\boldsymbol{\nu}\cdot \nabla_{\mathbf{x'}}G^{[0]}(\mathbf{x};\mathbf{x'})\right]d\Gamma(\mathbf{x'})~,
\end{equation}
wherein we have made use of the fact that
\begin{equation}\label{constrhoone6}
\int_{\Omega}f(\mathbf{x'})\delta(\mathbf{x}-\mathbf{x'})d\Omega(\mathbf{x'})=\mathcal{H}_{\Omega}(\mathbf{x})f(\mathbf{x})~,
\end{equation}
with $\mathcal{H}_{\Omega}(\mathbf{x})$  the domain Heaviside distribution equal to 1 for $\mathbf{x}\in\Omega$ and to 0 for $\mathbf{x}\in(\mathbf{R}^{3}-\Omega)$.

Similarly, by employing Green's second identity,  we obtain, from (\ref{constrhoone4}) and the wave equation satisfied by the free space Green's function:
\begin{multline}\label{constrhoone7}
\mathcal{H}_{\Omega_{1}}(\mathbf{x})p^{[1]}(\mathbf{x})=-\int_{\Gamma}\left[G^{[0]}(\mathbf{x};\mathbf{x'})\boldsymbol{\nu}\cdot \nabla_{\mathbf{x'}}p^{[1]}(\mathbf{x'})-p^{[1]}(\mathbf{x'})\boldsymbol{\nu}\cdot \nabla_{\mathbf{x'}}G^{[0]}(\mathbf{x};\mathbf{x'})\right]d\Gamma(\mathbf{x'})+\\
\int_{\Omega_{1}}G^{[0]}(\mathbf{x};\mathbf{x'})\Big(\big(k^{[1]}(\mathbf{x'})\big)^{2}-\big(k^{[0]}\big)^{2}\Big)p^{[1]}(\mathbf{x'})
d\Omega(\mathbf{x'})~.
\end{multline}
Eq. (\ref{constrhoone5}) yields
\begin{multline}\label{constrhoone8}
p^{[0]}(\mathbf{x})=p^{i}(\mathbf{x})+\int_{\Gamma}\Big[G^{[0]+}(\mathbf{x};\mathbf{x'})\boldsymbol{\nu}\cdot \nabla_{\mathbf{x'}}p^{[0]}(\mathbf{x'})d\Gamma(\mathbf{x'})-\\
p^{[0]}(\mathbf{x'})\boldsymbol{\nu}\cdot \nabla_{\mathbf{x'}}G^{[0]+}(\mathbf{x};\mathbf{x'})\Big]d\Gamma(\mathbf{x'})~;~\forall \mathbf{x}\in\Gamma~,
\end{multline}
whereas (\ref{constrhoone7}) gives rise to
\begin{multline}\label{constrhoone9}
p^{[1]}(\mathbf{x})=-\int_{\Gamma}\Big[G^{[0]-}(\mathbf{x};\mathbf{x'})\boldsymbol{\nu}\cdot \nabla_{\mathbf{x'}}p^{[1]}(\mathbf{x'}-p^{[1]}(\mathbf{x'})\boldsymbol{\nu}\cdot \nabla_{\mathbf{x'}}G^{[0]-}(\mathbf{x};\mathbf{x'})\Big]d\Gamma(\mathbf{x'})+\\
\int_{\Omega_{1}}G^{[0]}(\mathbf{x};\mathbf{x'})\Big(\big(k^{[1]}(\mathbf{x'})\big)^{2}-
\big(k^{[0]}\big)^{2}\Big)p^{[1]}(\mathbf{x'})d\Omega(\mathbf{x'})~;~\forall \mathbf{x}\in\Gamma~.
\end{multline}
wherefrom it ensues (by linear superposition, and on account of the position-free nature of the two mass densities and the transmission condition on pressure and velocity potential):
\begin{multline}\label{constrhoone10}
\left(1+\frac{\rho^{[0]}}{\rho^{[1]}}\right)p^{[1]}(\mathbf{x})=p^{i}(\mathbf{x})+
\int_{\Gamma}\Big\{\frac{\rho^{[0]}}{\rho^{[1]}}\Big[G^{[0]+}(\mathbf{x};\mathbf{x'})-
G^{[0]-}(\mathbf{x};\mathbf{x'})\Big]
\boldsymbol{\nu}\cdot \nabla_{\mathbf{x'}}p^{[1]}(\mathbf{x'})-\\
p^{[1]}(\mathbf{x'})\Big[\boldsymbol{\nu}\cdot \nabla_{\mathbf{x'}}G^{[0]+}(\mathbf{x};\mathbf{x'})-\frac{\rho^{[0]}}{\rho^{[1]}}\boldsymbol{\nu}\cdot \nabla_{\mathbf{x'}}G^{[0]-}(\mathbf{x};\mathbf{x'})\Big]\Big\}d\Gamma(\mathbf{x'})+\\
\frac{\rho^{[0]}}{\rho^{[1]}}\int_{\Omega_{1}}G^{[0]}(\mathbf{x};\mathbf{x'})\Big(\big(k^{[1]}(\mathbf{x'})\big)^{2}-
\big(k^{[0]}\big)^{2}\Big)p^{[1]}(\mathbf{x'})d\Omega(\mathbf{x'})~;~\forall \mathbf{x}\in\Gamma~
~.
\end{multline}
Eq. (\ref{constrhoone5}) gives rise to
\begin{multline}\label{constrhoone11}
0=p^{i}(\mathbf{x})+\int_{\Gamma}\left[G^{[0]}(\mathbf{x};\mathbf{x'})\boldsymbol{\nu}\cdot \nabla_{\mathbf{x'}}p^{[0]}(\mathbf{x'})d\Gamma(\mathbf{x'})-p^{[0]}(\mathbf{x'})\boldsymbol{\nu}\cdot \nabla_{\mathbf{x'}}G^{[0]}(\mathbf{x};\mathbf{x'})\right]d\Gamma(\mathbf{x'})~;\\
~\forall \mathbf{x}\in\Omega_{1}~,
\end{multline}
whereas  (\ref{constrhoone7}) yields
\begin{multline}\label{constrhoone12}
p^{[1]}(\mathbf{x})=-\int_{\Gamma}\left[G^{[0]}(\mathbf{x};\mathbf{x'})\boldsymbol{\nu}\cdot \nabla_{\mathbf{x'}}p^{[1]}(\mathbf{x'})-p^{[1]}(\mathbf{x'})\boldsymbol{\nu}\cdot \nabla_{\mathbf{x'}}G^{[0]}(\mathbf{x};\mathbf{x'})\right]d\Gamma(\mathbf{x'})+\\
\int_{\Omega_{1}}G^{[0]}(\mathbf{x};\mathbf{x'})\Big(\big(k^{[1]}(\mathbf{x'})\big)^{2}-
\big(k^{[0]}\big)^{2}\Big)p^{[1]}(\mathbf{x'})d\Omega(\mathbf{x'})~;~\forall \mathbf{x}\in\Omega_{1}~,
\end{multline}
so that their linear superposition results (also on account of the position-free nature of the two mass densities and the transmission condition on pressure and velocity potential) in
\begin{multline}\label{constrhoone13}
p^{[1]}(\mathbf{x})=\frac{\rho^{[1]}}{\rho^{[0]}}p^{i}(\mathbf{x})+\left(1-\frac{\rho^{[1]}}{\rho^{[0]}}\right)\int_{\Gamma}p^{[1]}(\mathbf{x'})\boldsymbol{\nu}\cdot \nabla_{\mathbf{x'}}G^{[0]}(\mathbf{x};\mathbf{x'})d\Gamma(\mathbf{x'})+\\
\int_{\Omega_{1}}G^{[0]}(\mathbf{x};\mathbf{x'})\Big(\big(k^{[1]}(\mathbf{x'})\big)^{2}-
\big(k^{[0]}\big)^{2}\Big)p^{[1]}(\mathbf{x'})d\Omega(\mathbf{x'})~;~\forall \mathbf{x}\in\Omega_{1}~.
\end{multline}

To retrieve $p^{[0]}(\mathbf{x}\in\Omega_{0})$ from these two functions, we proceed as follows. Eq. (\ref{constrhoone5}) yields
\begin{multline}\label{constrhoone14}
p^{[0]}(\mathbf{x})=p^{i}(\mathbf{x})+\int_{\Gamma}\left[G^{[0]}(\mathbf{x};\mathbf{x'})\boldsymbol{\nu}\cdot \nabla_{\mathbf{x'}}p^{[0]}(\mathbf{x'})d\gamma(\mathbf{x'})-p^{[0]}(\mathbf{x'})\boldsymbol{\nu}\cdot \nabla_{\mathbf{x'}}G^{[0]}(\mathbf{x};\mathbf{x'})\right]d\Gamma(\mathbf{x'})~;\\
~\forall \mathbf{x}\in\Omega_{0}~,
\end{multline}
whereas (\ref{constrhoone7}) gives rise to
\begin{multline}\label{constrhoone15}
0=-\int_{\Gamma}\left[G^{[0]}(\mathbf{x};\mathbf{x'})\boldsymbol{\nu}\cdot \nabla_{\mathbf{x'}}p^{[1]}(\mathbf{x'})-p^{[1]}(\mathbf{x'})\boldsymbol{\nu}\cdot \nabla_{\mathbf{x'}}G^{[0]}(\mathbf{x};\mathbf{x'})\right]d\Gamma(\mathbf{x'})+\\
\int_{\Omega_{1}}G^{[0]}(\mathbf{x};\mathbf{x'})\Big(\big(k^{[1]}(\mathbf{x'})\big)^{2}-
\big(k^{[0]}\big)^{2}\Big)p^{[1]}(\mathbf{x'})d\Omega(\mathbf{x'})~;~\forall \mathbf{x}\in\Omega_{0}~,
\end{multline}
so that their linear superposition results (also on account of the position-free nature of the two mass densities and the transmission condition on pressure and velocity potential) in
\begin{multline}\label{constrhoone15}
p^{[0]}(\mathbf{x})=p^{i}(\mathbf{x})-
\left(1-\frac{\rho^{[0]}}{\rho^{[1]}}\right)\int_{\Gamma}p^{[1]}(\mathbf{x'})\boldsymbol{\nu}\cdot \nabla_{\mathbf{x'}}G^{[0]}(\mathbf{x};\mathbf{x'})d\Gamma(\mathbf{x'})+\\
\frac{\rho^{[0]}}{\rho^{[1]}}\int_{\Omega_{1}}G^{[0]}(\mathbf{x};\mathbf{x'})\Big(\big(k^{[1]}(\mathbf{x'})\big)^{2}-
\big(k^{[0]}\big)^{2}\Big)p^{[1]}(\mathbf{x'})d\Omega(\mathbf{x'})~;~\forall \mathbf{x}\in\Omega_{0}~.
\end{multline}
Let:
\begin{multline}\label{constrhoone15a}
U(\mathbf{x})=p^{[1]}(\mathbf{x}\in\Gamma)~~,~~U^{i}(\mathbf{x})=p^{i}(\mathbf{x}\in\Gamma)~~,~~
W(\mathbf{x})=p^{[1]}(\mathbf{x}\in\Omega_{1})~~,~~W^{i}(\mathbf{x})=p^{i}(\mathbf{x}\in\Omega_{1})~.
\end{multline}
Because of the continuity of $G^{[0]}(\mathbf{x};\mathbf{x'})$ (illustrated hereafter), the term in $\boldsymbol{\nu}\cdot \nabla_{\mathbf{x'}}p^{[1]}(\mathbf{x'})$ disappears in (\ref{constrhoone10}), so that the latter becomes:
\begin{multline}\label{constrhoone15b}
\left(1+\frac{\rho^{[0]}}{\rho^{[1]}}\right)U(\mathbf{x})=U^{i}(\mathbf{x})
-\int_{\Gamma}U(\mathbf{x'})\Big[\boldsymbol{\nu}\cdot \nabla_{\mathbf{x'}}G^{[0]+}(\mathbf{x};\mathbf{x'})-\frac{\rho^{[0]}}{\rho^{[1]}}\boldsymbol{\nu}\cdot \nabla_{\mathbf{x'}}G^{[0]-}(\mathbf{x};\mathbf{x'})\Big]d\Gamma(\mathbf{x'})+\\
\frac{\rho^{[0]}}{\rho^{[1]}}\int_{\Omega_{1}}G^{[0]}(\mathbf{x};\mathbf{x'})\Big(\big(k^{[1]}\mathbf{x'})\big)^{2}-
\big(k^{[0]}\big)^{2}\Big)W(\mathbf{x'})d\Omega(\mathbf{x'})~;~\forall \mathbf{x}\in\Gamma~
~.
\end{multline}
whereas (\ref{constrhoone13}) becomes:
\begin{multline}\label{constrhoone15c}
W(\mathbf{x})=\frac{\rho^{[1]}}{\rho^{[0]}}W^{i}(\mathbf{x})+
\left(1-\frac{\rho^{[1]}}{\rho^{[0]}}\right)\int_{\Gamma}U(\mathbf{x'})\boldsymbol{\nu}\cdot \nabla_{\mathbf{x'}}G^{[0]}(\mathbf{x};\mathbf{x'})d\Gamma(\mathbf{x'})+\\
\int_{\Omega_{1}}G^{[0]}(\mathbf{x};\mathbf{x'})\Big(\big(k^{[1]}(\mathbf{x'})\big)^{2}-
\big(k^{[0]}\big)^{2}\Big)W(\mathbf{x'})d\Omega(\mathbf{x'})~;~\forall \mathbf{x}\in\Omega_{1}~.
\end{multline}
  Eqs. (\ref{constrhoone15b}) and (\ref{constrhoone15c}) constitute a system of two coupled (one BI, the other DI) integral  equations  for the two unknown functions $U(\mathbf{x}$ and $W(\mathbf{x})$. Once these two functions are determined, $p^{[0]}$ follows by simple integration via
\begin{multline}\label{constrhoone15d}
p^{[0]}(\mathbf{x})=p^{i}(\mathbf{x})-
\left(1-\frac{\rho^{[0]}}{\rho^{[1]}}\right)\int_{\Gamma}U(\mathbf{x'})\boldsymbol{\nu}\cdot \nabla_{\mathbf{x'}}G^{[0]}(\mathbf{x};\mathbf{x'})d\Gamma(\mathbf{x'})+\\
\frac{\rho^{[0]}}{\rho^{[1]}}\int_{\Omega_{1}}G^{[0]}(\mathbf{x};\mathbf{x'})\Big(\big(k^{[1]}(\mathbf{x'})\big)^{2}-
\big(k^{[0]}\big)^{2}\Big)W(\mathbf{x'})d\Omega(\mathbf{x'})~;~\forall \mathbf{x}\in\Omega_{0}~.
\end{multline}
%%
%%%%%%%%%%%%%%%%%%%%%%%%%%%%%%%%%%%%%%%%%%%
\subsubsection{DI-BI method: iterative methods for obtaining the solutions for $U$ and $W$}
Let $U^{(j)}$ and $W^{(j)}$ be the $j$-th order approximations of the coupled DI-BI equations (\ref{constrhoone15b})-(\ref{constrhoone15c}).

A first iterative method proceeds, after the initialization
\begin{equation}\label{constrhoone15e}
U^{(0)}(\mathbf{x})=\frac{U^{i}(\mathbf{x})}{1+\frac{\rho^{[0]}}{\rho^{[1]}}}~,
\end{equation}
as follows (in the indicated order):
\begin{multline}\label{constrhoone15f}
W^{(j)}(\mathbf{x})=\frac{\rho^{[1]}}{\rho^{[0]}}W^{i}(\mathbf{x})+
\left(1-\frac{\rho^{[1]}}{\rho^{[0]}}\right)\int_{\Gamma}U^{(j-1)}(\mathbf{x'})\boldsymbol{\nu}\cdot \nabla_{\mathbf{x'}}G^{[0]}(\mathbf{x};\mathbf{x'})d\Gamma(\mathbf{x'})+\\
\int_{\Omega_{1}}G^{[0]}(\mathbf{x};\mathbf{x'})\Big(\big(k^{[1]}(\mathbf{x'})\big)^{2}-
\big(k^{[0]}\big)^{2}\Big)W^{(j)}(\mathbf{x'})d\Omega(\mathbf{x'})~;~\forall \mathbf{x}\in\Omega_{1}~;~j=1,2,...~,
\end{multline}
\begin{multline}\label{constrhoone15g}
\left(1+\frac{\rho^{[0]}}{\rho^{[1]}}\right)U^{(j)}(\mathbf{x})=U^{i}(\mathbf{x})
-\int_{\Gamma}U^{(j-1)}(\mathbf{x'})\Big[\boldsymbol{\nu}\cdot \nabla_{\mathbf{x'}}G^{[0]+}(\mathbf{x};\mathbf{x'})-\frac{\rho^{[0]}}{\rho^{[1]}}\boldsymbol{\nu}\cdot \nabla_{\mathbf{x'}}G^{[0]-}(\mathbf{x};\mathbf{x'})\Big]d\Gamma(\mathbf{x'})+\\
\frac{\rho^{[0]}}{\rho^{[1]}}\int_{\Omega_{1}}G^{[0]}(\mathbf{x};\mathbf{x'})\Big(\big(k^{[1]}(\mathbf{x'})\big)^{2}-
\big(k^{[0]}\big)^{2}\Big)W^{(j-1)}(\mathbf{x'})d\Omega(\mathbf{x'})~;~\forall \mathbf{x}\in\Gamma~;~j=1,2,...~
~.
\end{multline}
Note that this iterative method requires the resolution of a single   domain integral equation (DIE) (\ref{constrhoone15f}) of the second kind for $W^{(j)}(\mathbf{x})$, followed by the computation  of $U^{(j)}(\mathbf{x})$, via  the simple integral transform (\ref{constrhoone15g}), for each approximation order $j\geq 1$..

A perhaps more efficient iterative method proceeds, after the initialization
\begin{equation}\label{constrhoone15h}
W^{(0)}(\mathbf{x})=\frac{\rho^{[1]}}{\rho^{[0]}}W^{i}(\mathbf{x})~,
\end{equation}
as follows (in the indicated order):
\begin{multline}\label{constrhoone15i}
\left(1+\frac{\rho^{[0]}}{\rho^{[1]}}\right)U^{(j)}(\mathbf{x})=U^{i}(\mathbf{x})
-\int_{\Gamma}U^{(j)}(\mathbf{x'})\Big[\boldsymbol{\nu}\cdot \nabla_{\mathbf{x'}}G^{[0]+}(\mathbf{x};\mathbf{x'})-\frac{\rho^{[0]}}{\rho^{[1]}}\boldsymbol{\nu}\cdot \nabla_{\mathbf{x'}}G^{[0]-}(\mathbf{x};\mathbf{x'})\Big]d\Gamma(\mathbf{x'})+\\
\frac{\rho^{[0]}}{\rho^{[1]}}\int_{\Omega_{1}}G^{[0]}(\mathbf{x};\mathbf{x'})\Big(\big(k^{[1]}(\mathbf{x'})\big)^{2}-
\big(k^{[0]}\big)^{2}\Big)W^{(j-1)}(\mathbf{x'})d\Omega(\mathbf{x'})~;~\forall \mathbf{x}\in\Gamma~;~j=1,2,...~
~,
\end{multline}
\begin{multline}\label{constrhoone15j}
W^{(j)}(\mathbf{x})=\frac{\rho^{[1]}}{\rho^{[0]}}W^{i}(\mathbf{x})+
\left(1-\frac{\rho^{[1]}}{\rho^{[0]}}\right)\int_{\Gamma}U^{(j)}(\mathbf{x'})\boldsymbol{\nu}\cdot \nabla_{\mathbf{x'}}G^{[0]}(\mathbf{x};\mathbf{x'})d\Gamma(\mathbf{x'})+\\
\int_{\Omega_{1}}G^{[0]}(\mathbf{x};\mathbf{x'})\Big(\big(k^{[1]}(\mathbf{x'})\big)^{2}-
\big(k^{[0]}\big)^{2}\Big)W^{(j-1)}(\mathbf{x'})d\Omega(\mathbf{x'})~;~\forall \mathbf{x}\in\Omega_{1}~;~j=1,2,...~.
\end{multline}
Note that this iterative method requires the resolution of a single   boundary integral equation (\ref{constrhoone15i}) of the second kind for $U^{(j)}(\mathbf{x})$ (in principle, less computationally-intensive than a DIE), followed by the computation  of $W^{(j)}(\mathbf{x})$, via  the simple integral transform (\ref{constrhoone15j}), for each approximation order $j\geq 1$..
%%%%%%%%%%%%%%%%%%%%%%%%%%%%%%%%%%%%%%%%%%%%%%%%%%%%%%%%%%%%%%%%%%%%%%%%%%%%%%
\subsubsection{DI-BI method: the 2D (i.e., cylindrical symmetry) case in polar coordinates when $k^{[1]}(\mathbf{x})=k^{[1]}(r)$}
Let the origin $O$ be located within the obstacle. The constitutive parameters as well as the pressure are $2\pi-$periodic functions of $\theta$ and the boundary (formerly $\Gamma$, now $\gamma$) of the obstacle is a circle of radius $a$ in the sagittal plane, so that from (\ref{constrhoone10}) it ensues that
\begin{multline}\label{constrhoone16}
\left(1+\frac{\rho^{[0]}}{\rho^{[1]}}\right)\int_{0}^{2\pi}d\theta e^{-il\theta}p^{[1]}(a,\theta)=\int_{0}^{2\pi}d\theta e^{-il\theta} p^{i}(a,\theta)-\\
\int_{0}^{2\pi}d\theta e^{-il\theta}\int_{0}^{2\pi}d\theta'a \Big\{\frac{\rho^{[0]}}{\rho^{[1]}}\Big[G^{[0]+}(a,\theta;a,\theta')-
G^{[0]-}(a,\theta;a,\theta')\Big]
p_{,r'}^{[1]}(a,\theta')-\\
p^{[1]}(a,\theta')\Big[G_{,r'}^{[0]+}(a,\theta;a,\theta')-\frac{\rho^{[0]}}{\rho^{[1]}}
G_{,r'}^{[0]-}(a,\theta;a,\theta')\Big]\Big\}+\\
\frac{\rho^{[0]}}{\rho^{[1]}}\int_{0}^{2\pi}d\theta e^{-il\theta}\int_{0}^{a}dr'~r' \int_{0}^{2\pi}d\theta'G^{[0]}(a,\theta;r',\theta')\Big(\big(k^{[1]}(r')\big)^{2}-
\big(k^{[0]}\big)^{2}\Big)p^{[1]}(r',\theta')~;~\forall l\in\mathbb{Z}~,
\end{multline}
whereas (\ref{constrhoone13}) gives rise to
\begin{multline}\label{constrhoone17}
\int_{0}^{2\pi}d\theta e^{-il\theta}p^{[1]}(r,\theta)=\frac{\rho^{[1]}}{\rho^{[0]}}\int_{0}^{2\pi}d\theta e^{-il\theta}p^{i}(r,\theta)-\\
\left(1-\frac{\rho^{[1]}}{\rho^{[0]}}\right)
\int_{0}^{2\pi}d\theta e^{-il\theta}\int_{0}^{\pi}d\theta'ap^{[1]}(a,\theta')G_{,r'}^{[0]}(r,\theta;a,\theta')+\\
\int_{0}^{2\pi}d\theta e^{-il\theta}\int_{0}^{a}dr'~r' \int_{0}^{2\pi}d\theta' G^{[0]}(r,\theta;r',\theta')\Big(\big(k^{[1]}(r')\big)^{2}-
\big(k^{[0]}\big)^{2}\Big)p^{[1]}(r',\theta')~;~\forall r<a~;~\forall l\in\mathbb{Z}~,
\end{multline}
and (\label{constrhoone15}) yields
\begin{multline}\label{constrhoone17a}
\int_{0}^{2\pi}d\theta e^{-il\theta}p^{[0]}(r,\theta)=\int_{0}^{2\pi}d\theta e^{-il\theta}p^{i}(r,\theta)+\\
\left(1-\frac{\rho^{[0]}}{\rho^{[1]}}\right)
\int_{0}^{2\pi}d\theta e^{-il\theta}\int_{0}^{\pi}d\theta'ap^{[1]}(a,\theta')G_{,r'}^{[0]}(r,\theta;a,\theta')+\\
\frac{\rho^{[0]}}{\rho^{[1]}}\int_{0}^{2\pi}d\theta e^{-il\theta}\int_{0}^{a}dr'~r' \int_{0}^{2\pi}d\theta' G^{[0]}(r,\theta;r',\theta')\Big(\big(k^{[1]}(r')\big)^{2}-
\big(k^{[0]}\big)^{2}\Big)p^{[1]}(r',\theta')~;~\forall r>a~;~\forall l\in\mathbb{Z}~.
\end{multline}
We make the following separation-of-variables (SOV) ansatzes (suggested by the $2\pi$-periodicity of the involved functions:
\begin{equation}\label{constrhoone18}
p^{[1]}(r,\theta)=\sum_{m\in\mathbb{Z}}p_{m}^{[1]}(r)e^{im\theta}~,~p^{i}(r,\theta)=\sum_{m\in\mathbb{Z}}p_{m}^{i}(r)e^{im\theta}~,
\end{equation}
\begin{equation}\label{constrhoone19}
 G^{[0]}(r,\theta;r',\theta')=\sum_{n\in\mathbb{Z}}G_{n}^{[0]}(r;r')e^{in(\theta-\theta')}~,~
 G_{,r'}^{[0]\pm}(a,\theta;a,\theta')=\sum_{n\in\mathbb{Z}}G_{n,r'}^{[0]\pm}(a;a)e^{in(\theta-\theta')}~,
\end{equation}
and make use of the identity
\begin{equation}\label{constrhoone20}
\int_{0}^{2\pi}d\theta e^{i(m-l)\theta}=2\pi\delta_{lm}~;~\l,m\in\mathbb{Z}~,
\end{equation}
wherein $\delta_{lm}$ is the Kronecker delta symbol, to find:
\begin{multline}\label{constrhoone21}
\left(1+\frac{\rho^{[0]}}{\rho^{[1]}}\right)p_{l}^{[1]}(a)=p_{l}^{i}(a)-
2\pi a \Big\{\frac{\rho^{[0]}}{\rho^{[1]}}\Big[G_{l}^{[0]+}(a;a)-
G^{[0]-}(a;a)\Big]
p_{l,r'}^{[1]}(a)-\\
p_{l}^{[1]}(a)\Big[G_{l,r'}^{[0]+}(a;a)-\frac{\rho^{[0]}}{\rho^{[1]}}
G_{l,r'}^{[0]-}(a;a)\Big]\Big\}+
2\pi\frac{\rho^{[0]}}{\rho^{[1]}}\int_{0}^{a}dr'~r' G_{l}^{[0]}(a;a)\Big(\big(k^{[1]}(r')\big)^{2}-
\big(k^{[0]}\big)^{2}\Big)p_{l}^{[1]}(r')~;\\
~\forall l\in\mathbb{Z}~,
\end{multline}
\begin{multline}\label{constrhoone22}
p_{l}^{[1]}(r)=\frac{\rho^{[1]}}{\rho^{[0]}}p_{l}^{i}(r)-
2\pi a\left(1-\frac{\rho^{[1]}}{\rho^{[0]}}\right)
p_{l}^{[1]}(a)G_{l,r'}^{[0]}(r;a)+\\
2\pi\int_{0}^{a}dr'~r' G_{l}^{[0]}(r;r')\Big(\big(k^{[1]}(r')\big)^{2}-
\big(k^{[0]}\big)^{2}\Big)p_{l}^{[1]}(r')~;~\forall r<a~;~\forall l\in\mathbb{Z}~,
\end{multline}
\begin{multline}\label{constrhoone22a}
p_{l}^{[0]}(r)=p_{l}^{i}(r)+2\pi a \left(1-\frac{\rho^{[0]}}{\rho^{[1]}}\right)
p_{l}^{[1]}(a)G_{l,r'}^{[0]}(r;a)+\\
2\pi\frac{\rho^{[0]}}{\rho^{[1]}}\int_{0}^{a}dr'~r' G_{l}^{[0]}(r;r')\Big(\big(k^{[1]}(r')\big)^{2}-
\big(k^{[0]}\big)^{2}\Big)p_{l}^{[1]}(r')~;~\forall r>a~;~\forall l\in\mathbb{Z}~,
\end{multline}
Recall that
\begin{equation}\label{constrhoone23}
G_{l}^{[0]}(r,r')=\frac{i}{4}\left[H(r-r')H_{l}^{(1)}(k^{[0]}r)J_{l}(k^{[0]}r')+H(r'-r)H_{l}^{(1)}(k^{[0]}r')J_{l}(k^{[0]}r)\right]~,
\end{equation}
\begin{equation}\label{constrhoone24}
G_{l,r'}^{[0]}(r;r')=\frac{ik^{[0]}}{4}\big[H(r-r')H_{l}^{(1)}(k^{[0]}r)\dot{J}_{l}(k^{[0]}r')+
H(r'-r)J_{l}(k^{[0]}r)\dot{H}_{l}^{(1)}(k^{[0]}r')\big]~,
\end{equation}
so that
\begin{equation}\label{constrhoone25}
G_{l}^{[0]+}(a;a)=G_{l}^{[0]-}(a;a)=\frac{i}{4}H_{l}^{(1)}(k^{[0]}a)\dot{J}_{l}(k^{[0]}a)
~,
\end{equation}
which fact demonstrates the previously-mentioned  continuity of $G_{l}^{[0]}$, and
\begin{equation}\label{constrhoone26}
G_{l,r'}^{[0]+}(a;a)=\frac{ik^{[0]}}{4}H_{l}^{(1)}(k^{[0]}a)\dot{J}_{l}(k^{[0]}a)~~,~~G_{l,r'}^{[0]-}(a;a)=\frac{ik^{[0]}}{4}
J_{l}(k^{[0]}a)\dot{H}_{l}^{(1)}(k^{[0]}a)\big]~.
\end{equation}
Moreover,
\begin{equation}\label{constrhoone27}
G_{l}^{[0]}(r;r'<r)=\frac{i}{4}H_{l}^{(1)}(k^{[0]}r)J_{l}(k^{[0]}r')~~,~~
G_{l}^{[0]}(r;r'>r)=\frac{i}{4}H_{l}^{(1)}(k^{[0]}r')J_{l}(k^{[0]}r)
~,
\end{equation}
the consequences of which are:
\begin{multline}\label{constrhoone28}
\left(1+\frac{\rho^{[0]}}{\rho^{[1]}}\right)U_{l}(a)=U_{l}^{i}(a)+
U_{l}(a)\frac{i\pi}{2}k^{[0]}a\Big[H_{l}^{(1)}(k^{[0]}a)\dot{J}_{l}(k^{[0]}a)-
\frac{\rho^{[0]}}{\rho^{[1]}}J_{l}(k^{[0]}a)\dot{H}_{l}^{(1)}(k^{[0]}a)\Big]+\\
\frac{i\pi}{2}\frac{\rho^{[0]}}{\rho^{[1]}}H_{l}^{(1)}(k^{[0]}a)\int_{0}^{a}dr'~r'J_{l}(k^{[0]}r') \Big(\big(k^{[1]}(r')\big)^{2}-
\big(k^{[0]}\big)^{2}\Big)W_{l}(r')~;\\
~\forall l\in\mathbb{Z}~,
\end{multline}
\begin{multline}\label{constrhoone29}
W_{l}(r)=\frac{\rho^{[1]}}{\rho^{[0]}}W_{l}^{i}(r)-
U_{l}(a)\frac{i\pi}{2} k^{[0]}a\left(1-\frac{\rho^{[1]}}{\rho^{[0]}}\right)\dot{H}_{l}^{(1)}(k^{[0]}a)J_{l}(k^{[0]}r)
+\\
\frac{i\pi}{2}\Big[\int_{0}^{r}dr'~r'H_{l}^{(1)}(k^{[0]}r)J_{l}(k^{[0]}r')\Big(\big(k^{[1]}(r')\big)^{2}-
\big(k^{[0]}\big)^{2}\Big)W_{l}(r')+\\
\int_{r}^{a}dr'~r'J_{l}(k^{[0]}r)H_{l}^{(1)}(k^{[0]}r') \Big(\big(k^{[1]}(r')\big)^{2}-
\big(k^{[0]}\big)^{2}\Big)W_{l}(r')
\Big]~;~\forall r<a~;~\forall l\in\mathbb{Z}~,
\end{multline}
\begin{multline}\label{constrhoone29a}
p_{l}^{[0]}(r)=p_{l}^{i}(r)+
U_{l}(a)\frac{i\pi}{2} k^{[0]}a\left(1-\frac{\rho^{[0]}}{\rho^{[1]}}\right)\dot{J}_{l}(k^{[0]}a)H_{l}^{(1)}(k^{[0]}r)
+\\
\frac{i\pi}{2}\frac{\rho^{[0]}}{\rho^{[1]}}\int_{0}^{a}dr'~r'H_{l}^{(1)}(k^{[0]}r)J_{l}(k^{[0]}r')\Big(\big(k^{[1]}(r')\big)^{2}-
\big(k^{[0]}\big)^{2}\Big)W_{l}(r')~;~\forall r>a~;~\forall l\in\mathbb{Z}~,
\end{multline}
wherein
\begin{equation}\label{constrhoone30}
U_{l}(a)=p^{[1]}(a)~~,~~W_{l}(r)=p^{[1]}(r)~.
\end{equation}
%%
%%%%%%%%%%%%%%%%%%%%%%%%%%%%%%%%%%%%%%%%%%%%%%%%%%%%%%%%%%%%%%%%%%%%%%%%%%%%%%
\subsubsection{DI-BI method: the 2D (i.e., cylindrical symmetry) case in polar coordinates when $k^{[1]}(\mathbf{x})=k^{[1]}$}
In the case $k^{[1]}=$const., the  relations (\ref{constrhoone28})-(\ref{constrhoone29}) become:
\begin{multline}\label{constrhoone31}
\left(1+\frac{\rho^{[0]}}{\rho^{[1]}}\right)U_{l}(a)=U_{l}^{i}(a)+
U_{l}(a)\frac{i\pi}{2}k^{[0]}a\Big[H_{l}^{(1)}(k^{[0]}a)\dot{J}_{l}(k^{[0]}a)-
\frac{\rho^{[0]}}{\rho^{[1]}}J_{l}(k^{[0]}a)\dot{H}_{l}^{(1)}(k^{[0]}a)\Big]+\\
\frac{i\pi}{2}\frac{\rho^{[0]}}{\rho^{[1]}}\Big(\big(k^{[1]}\big)^{2}-
\big(k^{[0]}\big)^{2}\Big)H_{l}^{(1)}(k^{[0]}a)\int_{0}^{a}dr'~r'J_{l}(k^{[0]}r') W_{l}(r')~;\\
~\forall l\in\mathbb{Z}~,
\end{multline}
\begin{multline}\label{constrhoone32}
W_{l}(r)=\frac{\rho^{[1]}}{\rho^{[0]}}W_{l}^{i}(r)-
U_{l}(a)\frac{i\pi}{2} k^{[0]}a\left(1-\frac{\rho^{[1]}}{\rho^{[0]}}\right)\dot{H}_{l}^{(1)}(k^{[0]}a)J_{l}(k^{[0]}r)
+\\
\frac{i\pi}{2}\Big(\big(k^{[1]}\big)^{2}-
\big(k^{[0]}\big)^{2}\Big[H_{l}^{(1)}(k^{[0]}r)\int_{0}^{r}dr'~r'J_{l}(k^{[0]}r')W_{l}(r')+\\
J_{l}(k^{[0]}r)\int_{r}^{a}dr'~r'H_{l}^{(1)}(k^{[0]}r')W_{l}(r')
\Big]~;~\forall r<a~;~\forall l\in\mathbb{Z}~,
\end{multline}
\begin{multline}\label{constrhoone32a}
p_{l}^{[0]}(r)=p_{l}^{i}(r)+
U_{l}(a)\frac{i\pi}{2} k^{[0]}a\left(1-\frac{\rho^{[0]}}{\rho^{[1]}}\right)\dot{J}_{l}(k^{[0]}a)H_{l}^{(1)}(k^{[0]}r)
+\\
\frac{i\pi}{2}\frac{\rho^{[0]}}{\rho^{[1]}}\Big(\big(k^{[1]}\big)^{2}-
\big(k^{[0]}\big)^{2}\Big)H_{l}^{(1)}(k^{[0]}r)\int_{0}^{a}dr'~r'J_{l}(k^{[0]}r')W_{l}(r')~;~\forall r>a~;~\forall l\in\mathbb{Z}~.
\end{multline}
Let:
\begin{multline}\label{constrhoone33}
K_{l}(r)=\Big(\big(k^{[1]}\big)^{2}-\big(k^{[0]}\big)^{2}\Big)\int_{0}^{r}dr'~r'J_{l}(k^{[0]}r') W_{l}(r')~~ ,\\
L_{l}(r)=\Big(\big(k^{[1]}\big)^{2}-\big(k^{[0]}\big)^{2}\Big)\int_{r}^{a}dr'~r'H_{l}^{(1)}(k^{[0]}r')W_{l}(r')~.
\end{multline}
Then (\ref{constrhoone31})-(\ref{constrhoone32a}) become:
\begin{multline}\label{constrhoone34}
\left(1+\frac{\rho^{[0]}}{\rho^{[1]}}\right)U_{l}(a)=U_{l}^{i}(a)+
U_{l}(a)\frac{i\pi}{2}k^{[0]}a\Big[H_{l}^{(1)}(k^{[0]}a)\dot{J}_{l}(k^{[0]}a)-
\frac{\rho^{[0]}}{\rho^{[1]}}J_{l}(k^{[0]}a)\dot{H}_{l}^{(1)}(k^{[0]}a)\Big]+\\
\frac{i\pi}{2}\frac{\rho^{[0]}}{\rho^{[1]}}H_{l}^{(1)}(k^{[0]}a)K_{l}(a)~;~\forall l\in\mathbb{Z}~,
\end{multline}
\begin{multline}\label{constrhoone35}
W_{l}(r)=\frac{\rho^{[1]}}{\rho^{[0]}}W_{l}^{i}(r)-
U_{l}(a)\frac{i\pi}{2} k^{[0]}a\left(1-\frac{\rho^{[1]}}{\rho^{[0]}}\right)\dot{H}_{l}^{(1)}(k^{[0]}a)J_{l}(k^{[0]}r)
+\\
\frac{i\pi}{2}\Big[H_{l}^{(1)}(k^{[0]}r)K_{l}(r)+
J_{l}(k^{[0]}r)L_{l}(r)\Big]~;~\forall r<a~;~\forall l\in\mathbb{Z}~,
\end{multline}
\begin{multline}\label{constrhoone35a}
p_{l}^{[0]}(r)=p_{l}^{i}(r)+
U_{l}(a)\frac{i\pi}{2} k^{[0]}a\left(1-\frac{\rho^{[0]}}{\rho^{[1]}}\right)\dot{J}_{l}(k^{[0]}a)H_{l}^{(1)}(k^{[0]}r)
+\\
\frac{i\pi}{2}\frac{\rho^{[0]}}{\rho^{[1]}}H_{l}^{(1)}(k^{[0]}r)K_{l}(r)~;~\forall r>a~;~\forall l\in\mathbb{Z}~.
\end{multline}

We first consider in more detail the  boundary integral equation (\ref{constrhoone34}) and the domain integral equation (\ref{constrhoone35}). We make the following pressure field ansatz, dictated notably by the boundedness condition of the pressure field for $r\le a$
\begin{equation}\label{constrhoone36}
W_{l}(r)=B_{l}^{[1]}J_{l}(k^{[1]}r)~;~r\le a~,
\end{equation}
so that we are faced with  the two integrals
\begin{multline}\label{constrhoone37}
K_{l}(r)=\Big(\big(k^{[1]}\big)^{2}-\big(k^{[0]}\big)^{2}\Big)\int_{0}^{r}dr'~r'J_{l}(k^{[0]}r') J_{l}(^{[1]}r')~~ ,\\
L_{l}(r)=\Big(\big(k^{[1]}\big)^{2}-\big(k^{[0]}\big)^{2}\Big)\int_{r}^{a}dr'~r'H_{l}^{(1)}(k^{[0]}r')J_{l}(k^{[1]}r') \Big)W_{l}(r')~.
\end{multline}
which can be evaluated with the help of (\cite{as68}, p. 484) so as to yield;
\begin{multline}\label{constrhoone38}
K_{l}(r)=\left[-k^{[1]}r\dot{J}_{l}(k^{[1]}r)J_{l}(k^{[0]}r)+k^{[0]}r\dot{J}_{l}(k^{[0]}r)J_{l}(k^{[1]}r)\right]  ~~ ,\\
L_{l}(r)=\left[-k^{[1]}a\dot{J}_{l}(k^{[1]}a)H_{l}^{(1)}(k^{[0]}a)+k^{[0]}a\dot{H}_{l}^{(1)}(k^{[0]}a)J_{l}(k^{[1]}a)\right]-\\
\left[-k^{[1]}r\dot{J}_{l}(k^{[1]}r)H_{l}^{(1)}(k^{[0]}r)+k^{[0]}r\dot{H}_{l}^{(1)}(k^{[0]}r)J_{l}(k^{[1]}r)\right]~,
\end{multline}
whence, with the help of the Wronskian (\cite{as68}, p. 360)
\begin{multline}\label{constrhoone39}
H_{l}^{(1)}(k^{[0]}r)K_{l}(r)+J_{l}(k^{[0]}r)L_{l}(r)=\\
-\frac{2i}{\pi}J_{l}(k^{[1]}r)+
J_{l}(k^{[0]}r)\left[-k^{[1]}a\dot{J}_{l}(k^{[1]}a)H_{l}^{(1)}(k^{[0]}a)+k^{[0]}a\dot{H}_{l}^{(1)}(k^{[0]}a)J_{l}(k^{[1]}a)\right]~.
\end{multline}
It follows that:
\begin{multline}\label{constrhoone40}
\left(1+\frac{\rho^{[0]}}{\rho^{[1]}}\right)U_{l}(a)=U_{l}^{i}(a)+
U_{l}(a)\frac{i\pi}{2}k^{[0]}a\Big[H_{l}^{(1)}(k^{[0]}a)\dot{J}_{l}(k^{[0]}a)-
\frac{\rho^{[0]}}{\rho^{[1]}}J_{l}(k^{[0]}a)\dot{H}_{l}^{(1)}(k^{[0]}a)\Big]+\\
B_{l}^{[1]}\frac{i\pi}{2}\frac{\rho^{[0]}}{\rho^{[1]}}
H_{l}^{(1)}(k^{[0]}a)\left[-k^{[1]}r\dot{J}_{l}(k^{[1]}r)J_{l}(k^{[0]}r)+
k^{[0]}r\dot{J}_{l}(k^{[0]}r)J_{l}(k^{[1]}r)\right]~;~\forall l\in\mathbb{Z}~,
\end{multline}
\begin{multline}\label{constrhoone41}
W_{l}(r)=\frac{\rho^{[1]}}{\rho^{[0]}}W_{l}^{i}(r)-
U_{l}(a)\frac{i\pi}{2} k^{[0]}a\left(1-\frac{\rho^{[1]}}{\rho^{[0]}}\right)\dot{H}_{l}^{(1)}(k^{[0]}a)J_{l}(k^{[0]}r)
+\\
B_{l}^{[1]}\frac{i\pi}{2}\Big\{-\frac{2i}{\pi}J_{l}(k^{[1]}r)+
J_{l}(k^{[0]}r)\left[-k^{[1]}a\dot{J}_{l}(k^{[1]}a)H_{l}^{(1)}(k^{[0]}a)+k^{[0]}a\dot{H}_{l}^{(1)}(k^{[0]}a)J_{l}(k^{[1]}a)\right] \Big\}~;\\
~\forall r<a~;~\forall l\in\mathbb{Z}~.
\end{multline}
First, consider (\ref{constrhoone40}) in more detail. With the help of the previously-mentioned Wronskian, it is easily  shown that:
\begin{multline}\label{constrhoone41}
\left(1+\frac{\rho^{[0]}}{\rho^{[1]}}\right)-\frac{i\pi}{2}k^{[0]}a\Big[H_{l}^{(1)}(k^{[0]}a)\dot{J}_{l}(k^{[0]}a)-
\frac{\rho^{[0]}}{\rho^{[1]}}J_{l}(k^{[0]}a)\dot{H}_{l}^{(1)}(k^{[0]}a)\Big]=\\
1-\frac{i\pi}{2}k^{[0]}a\left(1-\frac{\rho^{[0]}}{\rho^{[1]}}\right)H_{l}^{(1)}(k^{[0]}a)\dot{J}_{l}(k^{[0]}a)
\end{multline}
whence (\ref{constrhoone40}) becomes
\begin{multline}\label{constrhoone42}
U_{l}(a)\left[1-\frac{i\pi}{2}k^{[0]}a H_{l}^{(1)}(k^{[0]}a)\dot{J}_{l}(k^{[0]}a)\right]+
B_{l}^{[1]}\frac{i\pi}{2}k^{[1]}a \frac{\rho^{[0]}}{\rho^{[1]}}H_{l}^{(1)}(k^{[0]}a)\dot{J}_{l}(k^{[1]}a)J_{l}(k^{[0]}a)+\\
\frac{i\pi}{2}k^{[0]}a\frac{\rho^{[0]}}{\rho^{[1]}}H_{l}^{(1)}(k^{[0]}a)\dot{J}_{l}(k^{[0]}a)
\left[U_{l}(a)-B_{l}^{[1]}J_{l}(k^{[1]}a)\right]=U_{l}^{i}(a)
~;~\forall l\in\mathbb{Z}~,
\end{multline}
Again, employing the Wronskian, we find
\begin{equation}\label{constrhoone43}
1-\frac{i\pi}{2}k^{[0]}a H_{l}^{(1)}(k^{[0]}a)\dot{J}_{l}(k^{[0]}a)=-\frac{i\pi}{2}k^{[0]}a J_{l}(k^{[0]}a)\dot{H}_{l}^{(1)}(k^{[0]}a)
~,
\end{equation}
so that
\begin{multline}\label{constrhoone44}
\frac{\pi\rho^{[0]}}{2i}J_{l}(k^{[0]}a)\left[
U_{l}(a)\frac{k^{[0]}a}{\rho^{[0]}}\dot{H}_{l}^{(1)}(k^{[0]}a)-B_{l}^{[1]} \frac{k^{[1]}a}{\rho^{[1]}}H_{l}^{(1)}(k^{[0]}a)\dot{J}_{l}(k^{[1]}a)
\right]+\\
\frac{i\pi}{2}k^{[0]}a\frac{\rho^{[0]}}{\rho^{[1]}}H_{l}^{(1)}(k^{[0]}a)\dot{J}_{l}(k^{[0]}a)
\left[U_{l}(a)-B_{l}^{[1]}J_{l}(k^{[1]}a)\right]=U_{l}^{i}(a)
~;~\forall l\in\mathbb{Z}~.
\end{multline}
On account of (\ref{constrhoone36}) it seems reasonable to assume
\begin{equation}\label{constrhoone45}
U_{l}(a)=B_{l}^{[1]}J_{l}(k^{[1]}a)~,
\end{equation}
whence
\begin{equation}\label{constrhoone46}
B_{l}^{[1]} \frac{\pi\rho^{[0]}}{2i}J_{l}(k^{[0]}a)
\left[
\frac{k^{[0]}a}{\rho^{[0]}}J_{l}(k^{[1]}a)\dot{H}_{l}^{(1)}(k^{[0]}a)- \frac{k^{[1]}a}{\rho^{[1]}}H_{l}^{(1)}(k^{[0]}a)\dot{J}_{l}(k^{[1]}a)
\right]=U_{l}^{i}(a)
~;~\forall l\in\mathbb{Z}~.
\end{equation}
and after recalling that $U_{l}^{i}(a)=B_{l}^{[0]}J_{l}(k^{[0]}a)$, it  finally ensues that:
\begin{equation}\label{constrhoone47}
B_{l}^{[1]}= B_{l}^{[0]}\left[\frac{\frac{2i}{\pi\rho^{[0]}}}{\frac{k^{[0]}a}{\rho^{[0]}}J_{l}(k^{[1]}a)\dot{H}_{l}^{(1)}(k^{[0]}a)- \frac{k^{[1]}a}{\rho^{[1]}}H_{l}^{(1)}(k^{[0]}a)\dot{J}_{l}(k^{[1]}a)}
\right]~;~\forall l\in\mathbb{Z}~,
\end{equation}
which agrees with the DD-SOV solution (\ref{ddsov26}) for $B_{l}^{[1]}$.

To see if the ansatzes (\ref{constrhoone36}) and (\ref{constrhoone45}) are admissible we return to (\ref{constrhoone41}) which must yield the same expression for $B_{l}^{[1]}$ as previously. Thus, under the hypotheses (\ref{constrhoone36}) and (\ref{constrhoone45}), (\ref{constrhoone41}) yields
\begin{multline}\label{constrhoone48}
B_{l}^{[1]}J_{l}(k^{[1]}r)=\frac{\rho^{[1]}}{\rho^{[0]}}W_{l}^{i}(r)-
B_{l}^{[1]}J_{l}(k^{[1]}a)\frac{i\pi}{2} k^{[0]}a\left(1-\frac{\rho^{[1]}}{\rho^{[0]}}\right)\dot{H}_{l}^{(1)}(k^{[0]}a)J_{l}(k^{[0]}r)
+\\
B_{l}^{[1]}\frac{i\pi}{2}\Big\{-\frac{2i}{\pi}J_{l}(k^{[1]}r)+
J_{l}(k^{[0]}r)\left[-k^{[1]}a\dot{J}_{l}(k^{[1]}a)H_{l}^{(1)}(k^{[0]}a)+k^{[0]}a\dot{H}_{l}^{(1)}(k^{[0]}a)J_{l}(k^{[1]}a)\right] \Big\}~;\\
~\forall r<a~;~\forall l\in\mathbb{Z}~.
\end{multline}
whence
\begin{multline}\label{constrhoone48}
B_{l}^{[1]}\frac{\pi\rho^{[1]}}{2i}J_{l}(k^{[0]}r)\left[
\frac{k^{[0]}a}{\rho^{[0]}}J_{l}(k^{[1]}a)\dot{H}_{l}^{(1)}(k^{[0]}a)-
\frac{k^{[1]}a}{\rho^{[1]}}H_{l}^{(1)}(k^{[0]}a)\dot{J}_{l}(k^{[1]}a)
\right]=\\
\frac{\rho^{[1]}}{\rho^{[0]}}W_{l}^{i}(r)
~\forall r<a~;~\forall l\in\mathbb{Z}~.
\end{multline}
the only possible solution of which is again (\ref{constrhoone47}) if it is recalled that $W_{l}^{i}(r)=B_{l}^{[0]}J_{l}(k^{[0]}r)$. Thus, (\ref{constrhoone36}) and (\ref{constrhoone45}) are admissible ansatzes in that the same expression for $B_{l}^{[1]}$ is obtained via both the first and second coupled BIE+DIE.

To obtain an explicit expression for the pressure field in $\varpi_{0}$, we return to (\ref{constrhoone35a}) in which we make use of the now-known facts that
\begin{equation}\label{constrhoone49}
U_{l}(a)=B_{l}^{[1]}J_{l}(k^{[1]}a)~~,~~W_{l}(r)=B_{l}^{[1]}J_{l}(k^{[1]}r)~,
\end{equation}
together with the expression for $K_{l}(r)$ (\ref{constrhoone38}), to obtain
\begin{multline}\label{constrhoone50}
p_{l}^{[0]}(r)=p_{l}^{i}(r)+
B_{l}^{[1]}\frac{i\pi}{2} k^{[0]}a\left(1-\frac{\rho^{[0]}}{\rho^{[1]}}\right)J_{l(k^{[1]}a}\dot{J}_{l}(k^{[0]}a)H_{l}^{(1)}(k^{[0]}r)
+\\
\frac{i\pi}{2}\frac{\rho^{[0]}}{\rho^{[1]}}H_{l}^{(1)}(k^{[0]}r)\left[-k^{[1]}r\dot{J}_{l}(k^{[1]}r)J_{l}(k^{[0]}r)+k^{[0]}r\dot{J}_{l}(k^{[0]}r)J_{l}(k^{[1]}r)\right]~;~\forall r>a~;~\forall l\in\mathbb{Z}~,
\end{multline}
which simplifies to:
\begin{multline}\label{constrhoone51}
p_{l}^{[0]}(r)=p_{l}^{i}(r)+
B_{l}^{[1]}\frac{i\pi\rho^{[0]}}{2} \left[\frac{k^{[0]}a}{\rho^{[0]}}J_{l}(k^{[1]}a)\dot{J}_{l}(k^{[0]}a)-
\frac{k^{[1]}a}{\rho^{[1]}}J_{l}(k^{[0]}a)\dot{J}_{l}(k^{[1]}a)\right]H_{l}^{(1)}(k^{[0]}r)
~;~\forall r>a~;~\forall l\in\mathbb{Z}~,
\end{multline}
whence finally:
\begin{multline}\label{constrhoone52}
p_{l}^{[0]}(r)=p_{l}^{i}(r)+A_{l}^{[0]}H_{l}^{(1)}(k^{[0]}r)~;~\forall r>a~;~\forall l\in\mathbb{Z}~~,\\
~~A_{l}^{[0]}=B_{l}^{[0]}\left[\frac{
\frac{k^{[1]}a}{\rho^{[1]}}J_{l}(k^{[0]}a)\dot{J}_{l}(k^{[1]}a)-
\frac{k^{[0]}a}{\rho^{[0]}}J_{l}(k^{[1]}a)\dot{J}_{l}(k^{[0]}a)}
{\frac{k^{[0]}a}{\rho^{[0]}}J_{l}(k^{[1]}a)\dot{H}_{l}^{(1)}(k^{[0]}a)-
 \frac{k^{[1]}a}{\rho^{[1]}}\dot{J}_{l}(k^{[1]}a)H_{l}^{(1)}(k^{[0]}a)}\right]
~.
\end{multline}
This, of course, agrees with the previously-obtained exact, reference DD-SOV (\ref{ddsov25}), as well as BI-BI, solutions  of this problem.
%%%%%%%%%%%%%%%%%%%%%%%%%%%%%%%%%%%%%%%%%%%%%%%%%%%%%%%%%%%%%%%%%%%%%%%%%%%%%%%%
\subsection{The BI-DI method}\label{bidi}
\subsubsection{Governing equations}
We again make the following assumptions:
\begin{equation}\label{bidi1}
\rho(\mathbf{x})=\rho^{[l]}=const.~;~\forall \mathbf{x}\in\Omega_{j}~,~j=0,1~,
\end{equation}
wherein $\rho^{[1]}$ is generally different from $\rho^{[0]}$ and
\begin{equation}\label{bidi2}
\begin{array}{c}
c(\mathbf{x})=c^{[0]}=const.~;~\forall \mathbf{x}\in\Omega_{0}~,\\
c(\mathbf{x})=c^{[1]}(\mathbf{x})~;~\forall \mathbf{x}\in\Omega_{1}~.
\end{array}
\end{equation}
and appeal to the following results found in sect. \ref{constrhoone}:
\begin{equation}\label{bidi3}
\mathcal{H}_{\Omega_{0}}(\mathbf{x})p^{[0]}(\mathbf{x})=p^{i}(\mathbf{x})+\int_{\Gamma}\left[G^{[0]}(\mathbf{x};\mathbf{x'})\boldsymbol{\nu}\cdot \nabla_{\mathbf{x'}}p^{[0]}(\mathbf{x'})-p^{[0]}(\mathbf{x'})\boldsymbol{\nu}\cdot \nabla_{\mathbf{x'}}G^{[0]}(\mathbf{x};\mathbf{x'})\right]d\Gamma(\mathbf{x'})~,
\end{equation}
\begin{multline}\label{bid4}
\mathcal{H}_{\Omega_{1}}(\mathbf{x})p^{[1]}(\mathbf{x})=-\int_{\Gamma}\left[G^{[0]}(\mathbf{x};\mathbf{x'})\boldsymbol{\nu}\cdot \nabla_{\mathbf{x'}}p^{[1]}(\mathbf{x'})-p^{[1]}(\mathbf{x'})\boldsymbol{\nu}\cdot \nabla_{\mathbf{x'}}G^{[0]}(\mathbf{x};\mathbf{x'})\right]d\Gamma(\mathbf{x'})+\\
\int_{\Omega_{1}}G^{[0]}(\mathbf{x};\mathbf{x'})\Big(\big(k^{[1]}(\mathbf{x'})\big)^{2}-\big(k^{[0]}\big)^{2}\Big)p^{[1]}(\mathbf{x'})
d\Omega(\mathbf{x'})~.
\end{multline}
Adding these two relations, and making use of the properties of the Heaviside distributions results in:
\begin{multline}\label{bidi5}
p(\mathbf{x})=p^{i}(\mathbf{x})+\\
\int_{\Gamma}\Big(G^{[0]}(\mathbf{x};\mathbf{x'})
\left[\boldsymbol{\nu}\cdot \nabla_{\mathbf{x'}}p^{[0]}(\mathbf{x'})-\boldsymbol{\nu}\cdot \nabla_{\mathbf{x'}}p^{[1]}(\mathbf{x'})\right]-
\left[p^{[0]}(\mathbf{x'})-p^{[1]}(\mathbf{x'})\right]\boldsymbol{\nu}\cdot \nabla_{\mathbf{x'}}G^{[0]}(\mathbf{x};\mathbf{x'})\Big)d\Gamma(\mathbf{x'})+\\
\int_{\Omega_{1}}G^{[0]}(\mathbf{x};\mathbf{x'})\Big(\big(k^{[1]}(\mathbf{x'})\big)^{2}-\big(k^{[0]}\big)^{2}\Big)p^{[1]}(\mathbf{x'})
d\Omega(\mathbf{x'})~;~\forall \mathbf{x}\in \mathbf{R}^{3}~,
\end{multline}
or, after employment of the transmission conditions across $\Gamma$
\begin{multline}\label{bidi6}
p(\mathbf{x})=p^{i}(\mathbf{x})+
\int_{\Gamma}G^{[0]}(\mathbf{x};\mathbf{x'})
\left(\frac{\rho^{[0]}}{\rho^{[1]}}-1\right)\boldsymbol{\nu}\cdot \nabla_{\mathbf{x'}}p^{[1]}(\mathbf{x'})d\Gamma(\mathbf{x'})+\\
\int_{\Omega_{1}}G^{[0]}(\mathbf{x};\mathbf{x'})\Big(\big(k^{[1]}(\mathbf{x'})\big)^{2}-\big(k^{[0]}\big)^{2}\Big)p^{[1]}(\mathbf{x'})
d\Omega(\mathbf{x'})~;~\forall \mathbf{x}\in \mathbf{R}^{3}~.
\end{multline}
This relation forms the basis of what we term the coupled BI-DI method. We note that (\ref{bidi6}) gives rise to two relations:
\begin{multline}\label{bidi7}
p(\mathbf{x})=p^{[1]}(\mathbf{x})=p^{i}(\mathbf{x})+
\int_{\Gamma}G^{[0]}(\mathbf{x};\mathbf{x'})
\left(\frac{\rho^{[0]}}{\rho^{[1]}}-1\right)\boldsymbol{\nu}\cdot \nabla_{\mathbf{x'}}p^{[1]}(\mathbf{x'})d\Gamma(\mathbf{x'})+\\
\int_{\Omega_{1}}G^{[0]}(\mathbf{x};\mathbf{x'})\Big(\big(k^{[1]}(\mathbf{x'})\big)^{2}-\big(k^{[0]}\big)^{2}\Big)p^{[1]}(\mathbf{x'})
d\Omega(\mathbf{x'})~;~\forall \mathbf{x}\in \Omega_{1}~,
\end{multline}
\begin{multline}\label{bidi8}
p(\mathbf{x})=p^{[0]}(\mathbf{x})=p^{i}(\mathbf{x})+
\int_{\Gamma}G^{[0]}(\mathbf{x};\mathbf{x'})
\left(\frac{\rho^{[0]}}{\rho^{[1]}}-1\right)\boldsymbol{\nu}\cdot \nabla_{\mathbf{x'}}p^{[1]}(\mathbf{x'})d\Gamma(\mathbf{x'})+\\
\int_{\Omega_{1}}G^{[0]}(\mathbf{x};\mathbf{x'})\Big(\big(k^{[1]}(\mathbf{x'})\big)^{2}-\big(k^{[0]}\big)^{2}\Big)p^{[1]}(\mathbf{x'})
d\Omega(\mathbf{x'})~;~\forall \mathbf{x}\in \Omega_{0}~,
\end{multline}
so that we must first solve for $p^{[1]}$ via the integral equation (\ref{bidi7}) and then solve for $p^{[0]}$ via the integral transform (\ref{bidi8}). We now proceed to do this for a circular cylindrical obstacle.
%%%%%%%%%%%%%%%%%%%%%%%%%%%%%%%%%%%%%%%%%%%%%%%%
\subsubsection{BI-DI method: the integral equation and integral transform for a circular cylindrical obstacle}
For a circular cylindrical obstacle we proceed as in sect. \ref{constrhoone} to obtain the integral equation
\begin{multline}\label{bidi9}
    p_{l}(r)=p_{l}^{[1]}(r)=p_{l}^{i}(r)-\left(\frac{\rho^{[0]}}{\rho^{[1]}}-1\right)2\pi a G_{l}^{[0]}(r;a)p_{l,r}^{[1]}(a)+\\
   2\pi\int_{0}^{a}dr'r'\Big(\big(k^{[1]}(\mathbf{x'})\big)^{2}-\big(k^{[0]}\big)^{2}\Big)G_{l}^{[0]}(r;r')p_{l}^{[1]}(r')
  ~;~\forall r\in[0,a[~,
\end{multline}
and the integral transform
\begin{multline}\label{bidi10}
    p_{l}(r)=p_{l}^{[0]}(r)=p_{l}^{i}(r)-\left(\frac{\rho^{[0]}}{\rho^{[1]}}-1\right)2\pi a G_{l}^{[0]}(r;a)p_{l,r}^{[1]}(a)+\\
   2\pi\int_{0}^{a}dr'r'\Big(\big(k^{[1]}(\mathbf{x'})\big)^{2}-\big(k^{[0]}\big)^{2}\Big)G_{l}^{[0]}(r;r')p_{l}^{[1]}(r')
  ~;~\forall r>a~.
\end{multline}
Employing the previously-obtained polar coordinate representations of the Green's function results in:
\begin{multline}\label{bidi11}
    p_{l}^{[1]}(r)=p_{l}^{i}(r)-\left(\frac{\rho^{[0]}}{\rho^{[1]}}-1\right)\frac{i\pi}{2} a J_{l}(k^{[0]}r)H_{l}^{(1)}(k^{[0]}a)p_{l,r}^{[1]}(a)+\\
   \frac{i\pi}{2}\Big[\int_{0}^{r}dr'r'\Big(\big(k^{[1]}(\mathbf{x'})\big)^{2}-\big(k^{[0]}\big)^{2}\Big)
   H_{l}^{(1)}(k^{[0]}r)J_{l}(k^{[0]}r')p_{l}^{[1]}(r')+\\
   \int_{r}^{a}dr'r'\Big(\big(k^{[1]}(\mathbf{x'})\big)^{2}-\big(k^{[0]}\big)^{2}\Big)
   J_{l}^{(1)}(k^{[0]}r)H_{l}^{(1)}(k^{[0]}r')p_{l}^{[1]}(r')\Big]
   ~;~\forall r\in [0,a[~,
\end{multline}
\begin{multline}\label{bidi12}
    p_{l}^{[0]}(r)=p_{l}^{i}(r)-\left(\frac{\rho^{[0]}}{\rho^{[1]}}-1\right)\frac{i\pi}{2} a H_{l}^{(1)}(k^{[0]}r)J_{l}(k^{[0]}a)p_{l,r}^{[1]}(a)+\\
   \frac{i\pi}{2}\int_{0}^{a}dr'r'\Big(\big(k^{[1]}(\mathbf{x'})\big)^{2}-\big(k^{[0]}\big)^{2}\Big)
   H_{l}^{(1)}(k^{[0]}r)J_{l}(k^{[0]}r')p_{l}^{[1]}(r')
   ~;~\forall r>a~.
\end{multline}
%%
%%%%%%%%%%%%%%%%%%%%%%%%%%%%%%%%%%%%%%%%%%%%%%%%
\subsubsection{BI-DI method: the integral equation and integral transform for a circular cylindrical obstacle in which the density and wavespeed are constants with respect to the space coordinates}
Now $k^{[1]}$ is a constant so that the term $\Big(\big(k^{[1]}(\mathbf{x'})\big)^{2}-\big(k^{[0]}\big)^{2}\Big)$ can  be placed outside the integrals. Moreover, we can adopt the ansatzes
\begin{equation}\label{bidi13}
    p_{l}^{[1]}(r)=B_{l}^{[1]}J_{l}(k^{[1]}r)~\Rightarrow~p_{l,r}^{[1]}(r)=B_{l}^{[1]}k^{[1]}\dot{J}_{l}(k^{[1]}r)~,
\end{equation}
\begin{equation}\label{bidi14}
 p_{l}^{[0]}(r)= p_{l}^{i}(r)+A_{l}^{[0]}H_{l}^{(1)}(k^{[0]}r)
\end{equation}
and recall that
\begin{equation}\label{bidi15}
 p_{l}^{i}(r)= B_{l}^{[0]}J_{l}(k^{[0]}r)~;~\forall r\in[0,a[~,
\end{equation}
so that (\ref{bidi11})-(\ref{bidi12}) become
\begin{multline}\label{bidi14}
    B_{l}^{[1]}J_{l}(k^{[1]}r)=B_{l}^{[0]}(r)J_{l}(k^{[0]}r)-B_{l}^{[1]}\left(\frac{\rho^{[0]}}{\rho^{[1]}}-1\right)\frac{i\pi}{2} k^{[1]}a J_{l}(k^{[0]}r)H_{l}^{(1)}(k^{[0]}a)\dot{J}_{l}(k^{[1]}a)+\\
   B_{l}^{[1]}\frac{i\pi}{2}\Big[H_{l}^{(1)}(k^{[0]}r)\Big(\big(k^{[1]}(\mathbf{x'})\big)^{2}-\big(k^{[0]}\big)^{2}\Big)\int_{0}^{r}dr'r'
   J_{l}(k^{[0]}r')J_{l}(k^{[1]}r')+\\
   J_{l}(k^{[0]}r)\Big(\big(k^{[1]}(\mathbf{x'})\big)^{2}-\big(k^{[0]}\big)^{2}\Big)\int_{r}^{a}dr'r'
   H_{l}^{(1)}(k^{[0]}r')J_{l}^{(1)}(k^{[1]}r')\Big]
   ~;~\forall r\in [0,a[~,
\end{multline}
\begin{multline}\label{bidi15}
    p_{l}^{i}(r)+A_{l}^{[0]}H_{l}^{(1)}(k^{[0]}r)=p_{l}^{i}(r)-B_{l}^{[1]}\left(\frac{\rho^{[0]}}{\rho^{[1]}}-1\right)\frac{i\pi}{2} k^{[1]}a H_{l}^{(1)}(k^{[0]}r)J_{l}(k^{[0]}a)\dot{J}_{l}(k^{[1]}a)+\\
    B_{l}^{[1]}\frac{i\pi}{2}H_{l}^{(1)}(k^{[0]}r\Big(\big(k^{[1]}(\mathbf{x'})\big)^{2}-\big(k^{[0]}\big)^{2}\Big)\int_{0}^{a}dr'r'
   )J_{l}(k^{[0]}r')J_{l}(k^{[1]}r')
   ~;~\forall r>a~,
\end{multline}
from which we obtain, via (\ref{constrhoone37})-(\ref{constrhoone39}):
\begin{equation}\label{bidi16}
B_{l}^{[1]}= B_{l}^{[0]}\left[\frac{\frac{2i}{\pi\rho^{[0]}}}{\frac{k^{[0]}a}{\rho^{[0]}}J_{l}(k^{[1]}a)\dot{H}_{l}^{(1)}(k^{[0]}a)- \frac{k^{[1]}a}{\rho^{[1]}}H_{l}^{(1)}(k^{[0]}a)\dot{J}_{l}(k^{[1]}a)}
\right]~,
\end{equation}
\begin{equation}\label{bidi17}
A_{l}^{[0]}=B_{l}^{[0]}\left[\frac{
\frac{k^{[1]}a}{\rho^{[1]}}J_{l}(k^{[0]}a)\dot{J}_{l}(k^{[1]}a)-
\frac{k^{[0]}a}{\rho^{[0]}}J_{l}(k^{[1]}a)\dot{J}_{l}(k^{[0]}a)}
{\frac{k^{[0]}a}{\rho^{[0]}}J_{l}(k^{[1]}a)\dot{H}_{l}^{(1)}(k^{[0]}a)-
 \frac{k^{[1]}a}{\rho^{[1]}}\dot{J}_{l}(k^{[1]}a)H_{l}^{(1)}(k^{[0]}a)}\right]
~.
\end{equation}
which agrees, once again, with the exact, reference DD-SOV solutions (\ref{ddsov25})-(\ref{ddsov26}) as well as with (\ref{constrhoone47}) and (\ref{constrhoone52}).
%%%%%%%%%%%%%%%%%%%%%%%%%%%%%%%%%%%%%%%%%%%%%%%%%%%%%%%%%%%%
\subsection{The Domain Integral (DI) method}\label{DI}\label{di}
%
%%%%%%%%%%%%%%%%%%%%%%%%%%%%%%%%%%%%
\subsubsection{Governing equations}
The governing (BPDE) equation for this problem is  re-written here for
convenience (in simplified form with the $\omega$-dependence being
implicit in $p$, $s$, $\rho$ and  $c$).
\begin{equation}\label{2.3.9}
\nabla_{\mathbf{x}}\cdot\nabla_{\mathbf{x}} p(\mathbf{x})+\big( k(\mathbf{x})\big)^{2}p(\mathbf{x})-\frac{\nabla_{\mathbf{x}}\rho(\mathbf{x})}{\rho(\mathbf{x})}\cdot \nabla p(\mathbf{x})=-s(\mathbf{x})=\rho(\mathbf{x})
\nabla_{\mathbf{x}}\cdot \mathbf{f}(\mathbf{x})~;~\forall \mathbf{x}\in\mathbb{R}^{3}
~,
\end{equation}
In addition, the field is bounded everywhere except in the support of the sources and satisfies the radiation condition.

Recall that the field radiated by the sources, henceforth termed the "incident field" and designated by $p^{i}$, satisfies
\begin{equation}\label{w2.3.10}
\nabla_{\mathbf{x}}\cdot\nabla_{\mathbf{x}} p^{i}(\mathbf{x})+\big(k^{[0]}\big)^{2}p^{i}(\mathbf{x})=-s(\mathbf{x})~~;~~\forall\mathbf{x}\in
\mathbb{R}^{3}~.
\end{equation}
Subtracting these two equations gives
\begin{equation}\label{w2.3.11}
  \nabla_{\mathbf{x}}^{2}p^{d}(\mathbf{x})+
  \left(k^{[0]}\right ) ^{2}p^{d}(\mathbf{x})=-s^{d}(\mathbf{x})~~;~~\forall\mathbf{x}\in
\mathbb{R}^{3}~,
\end{equation}
wherein
\begin{equation}\label{w2.3.12}
 p^{d}(\mathbf{x}):= p(\mathbf{x})-p^{i}(\mathbf{x})~,
\end{equation}
is the diffracted pressure field  and
\begin{equation}\label{w2.3.14}
 s^{d}(\mathbf{x})=\left[ \big(k(\mathbf{x})\big)^{2}-
 \big( k^{0}\big ) ^{2}\right ] p(\mathbf{x})-
 \frac{\nabla_{\mathbf{x}}\rho(\mathbf{x})}{\rho(\mathbf{x})}\cdot \nabla_{\mathbf{x}} p(\mathbf{x})
 ~.
\end{equation}
the induced source density.
%%%%%%%%%%%%%%%%%%%%%%%%%%%%%%%%%%%%%%%%%%
\subsubsection{DI field representation}
The important feature of (\ref{w2.3.11}) is that it is exactly of
the same form as the partial differential equation relative to
radiation from a bounded source of sect. \ref{acradfr}. Moreover, since $p$ and $p^{i}$
satisfy the radiation condition, their difference $p^{d}$ must
also satisfy the radiation condition, which fact authorizes (with
the appropriate substitutions) the use of the previous result
(\ref{w2.3.7}), i.e.
\begin{equation}\label{w2.3.15}
p^{d}(\mathbf{x})=\int_{\mathbb{R}^{3}}G^{[0]}(\mathbf{x};\mathbf{x'})
s^{d}(\mathbf{x'})d\Omega(\mathbf{x'})~~;~~\forall~\mathbf{x}\in
\mathbb{R}^{3}~.
\end{equation}
\newline
{\it Remark a}
\newline
Contrary to what was stated in the remark at the end of sect. \ref{acradfr},
here the domain integral does not solve the
forward problem at hand, which is to determine the pressure field,
given the applied sources, spatial distributions of density and
wavespeed, and geometry of the configuration. The reason for this
is that the term $s^{d}$ under the integral sign in
(\ref{w2.3.15}) is a function of $p$ and $\nabla p$, both of which
are unknown functions. This is seen from:
\begin{multline}\label{w2.3.17}
p(\mathbf{x})=p^{i}(\mathbf{x})+\\
\int_{\mathbb{R}^{3}}G^{[0]}(\mathbf{x};\mathbf{x'})
\left(\left[ \big(k(\mathbf{x'})\big)^{2}-
 \big( k^{0}\big ) ^{2}\right] p(\mathbf{x'})-
 \frac{\nabla_{\mathbf{x'}}\rho(\mathbf{x'})}{\rho(\mathbf{x'})}\cdot \nabla_{\mathbf{x'}} p(\mathbf{x'})\right)d\Omega(\mathbf{x'})~~;~~\forall~\mathbf{x}\in
\mathbb{R}^{3}~,
\end{multline}
wherein
\begin{equation}\label{w2.3.17a}
p^{i}(\mathbf{x})=\int_{\Omega^{s}}G^{[0]}(\mathbf{x};\mathbf{x'})
s^{i}(\mathbf{x'})d\Omega(\mathbf{x'})~~;~~\forall~\mathbf{x}\in
\mathbb{R}^{3}~.
\end{equation}
We shall return to this expression, which is often termed the {\it Bergman integral representation} of the pressure, further on.
\newline
\newline
{\it Remark b}
\newline
If both the wavespeed $c$ (and therefore $k$) in the first term containing $k^{2}-(k^{[0]})^{2}$ is everywhere the same, i.e., equal to $k^{[0]}$, and the mass density $\rho$ in the term containing $\nabla\rho$ is everywhere the same, i.e., equal to $\rho^{[0]}$, then both these terms vanish so that the total pressure field $p$ reduces to the incident pressure field $p^{i}$, as one would expect in the case of absence of the obstacle.
\newline
\newline
{\it Remark c}
\newline
If the mass density $\rho$ in the term containing $\nabla\rho$ is everywhere the same, i.e., equal to $\rho^{[0]}$, then this term vanishes, and, as we shall see hereafter, the determination of the pressure field within (and outside) the obstacle is straightforward via the domain integral representation
\begin{equation}\label{w2.3.17b}
p(\mathbf{x})=p^{i}(\mathbf{x})+
\int_{\mathbb{R}^{3}}G^{[0]}(\mathbf{x};\mathbf{x'})
\left[ \big(k(\mathbf{x'})\big)^{2}-
 \big( k^{0}\big ) ^{2}\right] p(\mathbf{x'})d\Omega(\mathbf{x'})~~;~~\forall~\mathbf{x}\in
\mathbb{R}^{3}~,
\end{equation}
which is what is often termed the {\it Lippmann-Schwinger integral representation} of the pressure field.
%%%%%%%%%%%%%%%%%%%%%%%%%%%%%%%%%%%%
\subsubsection{DI method: a two-step scheme deriving from the $p-q$ transformation}\label{dipq}
The until-now-considered (standard) version of the DI method seeks to obtain the pressure $p$ of the equation
\begin{equation}\label{dipq01}
\nabla_{\mathbf{x}}\cdot\nabla_{\mathbf{x}} p(\mathbf{x})+\big( k(\mathbf{x})\big)^{2}p(\mathbf{x})-\frac{\nabla_{\mathbf{x}}\rho(\mathbf{x})}{\rho(\mathbf{x})}\cdot \nabla p(\mathbf{x})=-s(\mathbf{x})=\rho(\mathbf{x})
\nabla_{\mathbf{x}}\cdot \mathbf{f}(\mathbf{x})~;~\forall \mathbf{x}\in\mathbb{R}^{3}
~,
\end{equation}
whereas the $p-q$ transform version of the DI method first seeks to obtain the transformed pressure $q$ of the equation
\begin{equation}\label{dipq01}
\left[\nabla_{\mathbf{x}}\cdot\nabla_{\mathbf{x}}+\left(\mathcal{K}(\mathbf{x})\right)^{2}\right]q(\mathbf{x})=-S(\mathbf{x})=\rho^{1/2}(\mathbf{x})\nabla_{\mathbf{x}}\cdot\mathbf{f}(\mathbf{x})~;~\forall \mathbf{x}\in\mathbb{R}^{3}~,
\end{equation}
wherein
\begin{equation}\label{dipq02}
\mathcal{K}^{2}(\mathbf{x},\omega)=k^{2}-\rho^{1/2}(\mathbf{x},\omega)\nabla\cdot\nabla\rho^{-1/2}(\mathbf{x},\omega) ~.
\end{equation}
followed by the obtention of $p$ from $q$ from
\begin{equation}\label{dipq03}
p=q\rho^{1/2}~.
\end{equation}
In the same manner that we obtained the standard version DI equation (\ref{w2.3.17}) from (\ref{dipq01}) we find the $p-q$ version DI equation
\begin{multline}\label{dipq04}
q(\mathbf{x})=q^{i}(\mathbf{x})+\\
\int_{\mathbb{R}^{3}}G^{[0]}(\mathbf{x};\mathbf{x'})
\left(\left[ \big(k(\mathbf{x'})\big)^{2}-
 \big( k^{0}\big ) ^{2}\right]-
 \rho^{1/2}(\mathbf{x'})\nabla_{\mathbf{x'}}\cdot\nabla_{\mathbf{x'}}\rho^{-1/2}(\mathbf{x'})\right) q(\mathbf{x'})d\Omega(\mathbf{x'})~~;~~\forall~\mathbf{x}\in
\mathbb{R}^{3}~,
\end{multline}
wherein
\begin{equation}\label{dipq05}
q^{i}(\mathbf{x})=\int_{\mathbb{R}^{3}}G^{[0]}(\mathbf{x};\mathbf{x'})S(\mathbf{x'})d\Omega(\mathbf{x'})~~;~~\forall~\mathbf{x}\in
\mathbb{R}^{3}~.
\end{equation}
This shows that the $p-q$ transform has enabled the elimination of the $\nabla_{\mathbf{x'}} p$ term in the integral but this 'advantage' is obtained at the expense of requiring the computation of $\nabla_{\mathbf{x'}}\cdot\nabla_{\mathbf{x'}}\rho^{-1/2}(\mathbf{x'})$ which may be difficult if there exists a discontinuity of $\rho$ in the medium (this actually occurring on the boundary of an obstacle in which the density is constant and different from the density in the host medium). Our opinion on this matter is that the $p-q$ method cannot be employed in cases in which the density is discontinuous across the boundary of the obstacle, and if such be the case, a smoothing strategy must be employed to render $\rho$ artificially-continuous, such as is done in \cite{pa19}.

Consequently, we prefer to employ the standard version of the DI method from now on.
%%%%%%%%%%%%%%%%%%%%%%%%%%%%%%%%%%%%
\subsubsection{DI method: an iterative scheme}
With the so-called Born approximation initialization:
\begin{equation}\label{it1}
p^{(0)}(\mathbf{x})=p^{i}(\mathbf{x})~,
\end{equation}
the iterative scheme becomes
\begin{multline}\label{it2}
p^{(j)}(\mathbf{x})=p^{i}(\mathbf{x})+\\
\int_{\mathbb{R}^{3}}G^{[0]}(\mathbf{x};\mathbf{x'})
\left(\left[ \big(k(\mathbf{x'})\big)^{2}-
 \big( k^{0}\big ) ^{2}\right] p^{(j-1)}(\mathbf{x'})-
 \frac{\nabla_{\mathbf{x'}}\rho(\mathbf{x'})}{\rho(\mathbf{x'})}\cdot \nabla_{\mathbf{x'}} p^{(j-1)}(\mathbf{x'})\right)d\Omega(\mathbf{x'})~~;\\
 ~~\forall~\mathbf{x}\in
\mathbb{R}^{3}~~;~~j=1,2,....~.
\end{multline}
Owing to the fact that this scheme is not very informative concerning the constant-density assumption issue, it will be replaced by another, more appropriate,  iterative scheme in a subsequent section.
%%%%%%%%%%%%%%%%%%%%%%%%%%%%%%%%%%%%
\subsubsection{DI method: the 2D case of cylindrical symmetry}
Let the origin $O$ be located within the obstacle, whence the cylindrical coordinates $r,\theta,z$ are appropriate, with $z$ the ignorable coordinate. Then, $\mathbf{x}=(r,\theta)$, $\mathbf{x'}=(r',\theta')$, the closed boundary $\Gamma$ becomes the closed curve $\gamma$ and $d\Omega$ becomes $d\varpi$. $\gamma$ is assumed to be describable by the single-valued function
\begin{equation}\label{di10}
r=r_{\gamma}(\theta)~,
\end{equation}
and
\begin{equation}\label{di11}
\nabla_{\mathbf{x}}=\Big(\frac{\partial}{\partial r},\frac{1}{r}\frac{\partial}{\partial \theta}\Big)~~,~~
~~d\varpi(\mathbf{x})=rdrd\theta
~,
\end{equation}
so that
\begin{equation}\label{di11}
\frac{\nabla_{\mathbf{x}}\rho(\mathbf{x})}{\rho(\mathbf{x})}\cdot \nabla_{\mathbf{x}}p(\mathbf{x})=\frac{\rho_{,r}(r,\theta)}{\rho(r,\theta)}p_{,r}(r,\theta)+
\frac{1}{r^{2}}\frac{\rho_{,\theta}(r,\theta)}{\rho(r,\theta)}p_{,\theta}(r,\theta)
~.
\end{equation}
The $2\pi$-periodicity (in terms of $\theta$) of the boundary makes it legitimate to assume
\begin{equation}\label{di12}
p(\mathbf{x})=\sum_{m=-\infty}^{\infty}p_{m}(r)e^{im\theta}~~,~~
p^{i}(\mathbf{x})=\sum_{m=-\infty}^{\infty}U^{i}_{m}(r)e^{im\theta}~,
\end{equation}
whence
\begin{equation}\label{di13}
\frac{\nabla_{\mathbf{x}}\rho(\mathbf{x})}{\rho(\mathbf{x})}\cdot \nabla_{\mathbf{x}}p(\mathbf{x})=\sum_{m=-\infty}^{\infty}\left[\frac{\rho_{,r}(r,\theta)}{\rho(r,\theta)}p_{m,r}(r)+
\frac{im}{r^{2}}\frac{\rho_{,\theta}(r,\theta)}{\rho(r,\theta)}p_{m}(r)\right]e^{im\theta}
~.
\end{equation}
Henceforth, we assume that the constitutive properties of the cylindrical obstacle, depend, at worst, only on $r$ so that
\begin{equation}\label{di14}
\frac{\nabla_{\mathbf{x}}\rho(\mathbf{x})}{\rho(\mathbf{x})}\cdot \nabla_{\mathbf{x}}p(\mathbf{x})=\sum_{m=-\infty}^{\infty}\frac{\rho_{,r}(r)}{\rho(r)}p_{m,r}(r)e^{im\theta}=\sum_{m=-\infty}^{\infty}Q_{m}(r)e^{im\theta}
~,
\end{equation}
\begin{equation}\label{di15}
\left[\big(k(\mathbf{x'})\big)^{2}-
 \big( k^{0}\big ) ^{2}\right]p(\mathbf{x'})=\sum_{m=-\infty}^{\infty}\left[\big(k(r)\big)^{2}-
 \big( k^{0}\big ) ^{2}\right]p_{m}(r)e^{im\theta}=\sum_{m=-\infty}^{\infty}P_{m}(r)e^{im\theta}
~.
\end{equation}
Furthermore, the Green's function admits the representations
\begin{equation}\label{di16}
G^{[0]}(\mathbf{x};\mathbf{x'})=\sum_{n=-\infty}^{\infty}G_{n}^{[0]}(r;r')e^{in(\theta-\theta')}
~,
\end{equation}
wherein
\begin{equation}\label{di16a}
G_{n}^{[0]}(r;r')=\frac{i}{4}\left[H(r-r')H_{n}^{(1)}(k^{[0]}r)J_{n}(k^{[0]}r')+H(r'-r)H_{n}^{(1)}(k^{[0]}r')J_{n}(k^{[0]}r)\right]
~,
\end{equation}
so that the DI equation becomes, after projection
\begin{equation}\label{di17}
p_{l}(r)=p_{l}^{i}(r)+2\pi\int_{0}^{\infty}dr'r'G_{l}^{[0]}(r;r')P_{l}(r')+
2\pi\int_{0}^{\infty}dr'r'G_{l}^{[0]}(r;r')Q_{l}(r')~;~\forall r\in[0,\infty[~,~\forall l\in\mathbb{Z}~.
\end{equation}
%%
%%%%%%%%%%%%%%%%%%%%%%%%%%%%%%%%%%%%%%%%
\subsubsection{The DI method: case in which the boundary of the "obstacle" is a circle, $k^{[1]}=k^{[1]}(r)$ and $\rho^{[1]}$ is a constant as a function of $r$ and $\theta$}
We now assume that either $\gamma$ is a circle of radius $a$ or the obstacle is enclosed within a circular cylinder (now termed "obstacle") whose boundary (in the sagittal plane) is the circle $r=a=\max_{\theta\in[0,2\pi[}r_{\gamma}(\theta)$ and that the medium within this enclosure cylinder is now termed $M^{[1]}(\rho^{[1]},c^{[1]})$ (actually a mixture of the media of the host and that of the original obstacle ; see sect. \ref{mix} for more details on this issue) whence the fact that the medium outside this "obstacle"  is $M^{[0]}(\rho^{[0]},c^{[0]})$ whose constitutive parameters do not, as before, depend on position. Finally, we make the assumption that $\rho^{[1]}$ (but not $c^{[1]})$) does not depend on position. It ensues that;
\begin{equation}\label{di18}
k(r)=k^{[1]}(r)-\left[k^{[1]}(r)-k^{[0]}\right]H(r-a)~;~\forall r\in[0,\infty[~.
\end{equation}
\begin{equation}\label{di19}
\rho(r)=\rho^{[1]}-\left[\rho^{[1]}-\rho^{[0]}\right]H(r-a)~;~\forall r\in[0,\infty[~,
\end{equation}
wherein $H(\zeta)$ is the Heaviside distribution. The consequence of the last expression is
\begin{equation}\label{di20}
\rho_{,r}(r)=-\left[\rho^{[1]}-\rho^{[0]}\right]\delta(r-a)~;~\forall r\in[0,\infty[~,
\end{equation}
in which $\delta(\zeta)$ is the Dirac delta distribution. It follows that the DI equation takes the form:
\begin{multline}\label{di21}
p_{l}(r)=p_{l}^{i}(r)+2\pi\int_{0}^{a}dr'r'G_{l}^{[0]}(r;r')\left[\Big(k^{[1]}(r')\Big)^{2}-
 \Big( k^{[0]}\Big ) ^{2}\right]p_{l}(r')+\\
 2\pi\left[\rho^{[1]}-\rho^{[0]}\right]\int_{0}^{\infty}dr'r'G_{l}(^{[0]}r;r')\delta(r'-a)\frac{p_{l,r'}(r')}{\rho(r')}~;~\forall r\in[0,\infty[~,~\forall l\in\mathbb{Z}~,
\end{multline}
or, on account of the sifting property of the Dirac distribution, and the continuity of $\frac{p_{l,r'}(r')}{\rho(r')}$ for $r=a$,
\begin{multline}\label{di22}
p_{l}(r)=p_{l}^{i}(r)+2\pi\int_{0}^{a}dr'r'G_{l}^{[0]}(r;r')\left[\Big(k^{[1]}(r')\Big)^{2}-
 \Big( k^{[0]}\Big ) ^{2}\right]p_{l}(r')+\\
 2\pi\left[\rho^{[1]}-\rho^{[0]}\right]aG_{l}^{[0]}(r;a)\frac{p_{l,r'}(a)}{\rho(a)}:=p_{l}^{i}(r)+D_{l}(r)-E_{l}(r)~;~\forall r\in[0,\infty[~,~\forall l\in\mathbb{Z}~.
\end{multline}
%%
%%%%%%%%%%%%%%%%%%%%%%%%%%%%%%%%%%%%%%%%
\subsubsection{The DI method: case in which the boundary of the "obstacle" is a circle, and $k^{[1]}$ and $\rho^{[1]}$ are constants as a function of $r$ and $\theta$}
We had
\begin{equation}\label{di23}
E_{l}(r)=-2\pi\left[\rho^{[1]}-\rho^{[0]}\right]aG_{l}^{[0]}(r;a)\frac{p_{l,r'}(a)}{\rho(a)}~.
\end{equation}
 With $k^{[1]}(r)=k^{[1]}=$cte., $D_{l}(r)$ becomes
\begin{equation}\label{di23a}
D_{l}(r)=2\pi)\left[\Big(k^{[1]}\Big)^{2}-
 \Big( k^{[0]}\Big)^{2}\right]\int_{0}^{a}dr'r'G_{l}^{[0]}(r;r')p_{l}(r')~.
\end{equation}
First consider the internal field.
The boundedness condition on the field within the obstacle and the wave equation within the latter imply that it is reasonable to make the ansatzes
\begin{equation}\label{di24}
p_{l}(r'<a)=B_{l}^{[1]}J_{l}(k^{[1]}r')~,
\end{equation}
so that
\begin{equation}\label{di25}
D_{l}(r)=2\pi B_{l}^{[1]}\left[\Big(k^{[1]}\Big)^{2}-
 \Big( k^{[0]}\Big ) ^{2}\right]\int_{0}^{a}dr'r'G_{l}^{[0]}(r;r')J_{l}(k^{[1]}r')~.
\end{equation}
whence, on account of (\ref{di16a}),
\begin{multline}\label{di26}
D_{l}(r<a)=\frac{i\pi}{2}B_{l}^{[1]}\left[\Big(k^{[1]}\Big)^{2}-
 \Big( k^{[0]}\Big)^{2}\right]\times\\
 \left(\int_{0}^{r}dr'r'H_{l}^{(1)}(k^{[0]}r)J_{l}(k^{[0]}r')J_{l}(k^{[1]}r')+
 \int_{r}^{a}dr'r'J_{l}(k^{[0]}r)H_{l}^{(1)}(k^{[0]}r')J_{l}(k^{[1]}r')\right)~.
\end{multline}
Employing (\ref{constrhoone38})-(\ref{constrhoone39}) gives
\begin{multline}\label{di27}
D_{l}(r<a)=\\
B_{l}^{[1]}J_{l}(k^{[1]}r)+B_{l}^{[1]}\frac{i\pi}{2}J_{l}(k^{[0]}r)\left[
-k^{[1]}a\dot{J}_{l}(k^{[1]}a)H_{l}^{(1)}(k^{[0]}a)+k^{[0]}a J_{l}(k^{[1]}a)\dot{H}_{l}^{(1)}(k^{[0]}a)\right]=\\
B_{l}^{[1]}\left[J_{l}(k^{[1]}r)-J_{l}(k^{[0]}r)d_{l}^{<}\right]
~,
\end{multline}
wherein
\begin{equation}\label{di27a}
d_{l}^{<}=\frac{\pi\rho^{[0]}}{2i}\left[-\frac{k^{[1]}a}{\rho^{[0]}}\dot{J}_{l}(k^{[1]}a)H_{l}^{(1)}(k^{[0]}a)+\frac{k^{[0]}a}{\rho^{[0]}} J_{l}(k^{[1]}a)\dot{H}_{l}^{(1)}(k^{[0]}a)\right]
~.
\end{equation}
It follows from (\ref{di24}) that
\begin{equation}\label{di27b}
p_{l,r'}(r'<a)=B_{l}^{[1]}k^{[1]}\dot{J}_{l}(k^{[1]}r')
~,
\end{equation}
so that
\begin{equation}\label{di28}
E_{l}(r<a)=-B_{l}^{[1]}\frac{i\pi}{2}k^{[1]}a\left[\frac{\rho^{[1]}-\rho^{[0]}}{\rho^{[1]}}
\right]J_{l}(k^{[0]}r)H_{l}^{(1)}(k^{[0]}a)\dot{J}_{l}(k^{[1]}a)=-B_{l}^{[1]}J_{l}(k^{[0]}r)e_{l}^{<}
~,
\end{equation}
wherein
\begin{equation}\label{di28b}
e_{l}^{<}=-\frac{\pi}{2i}k^{[1]}a\left[\frac{\rho^{[1]}-\rho^{[0]}}{\rho^{[1]}}
\right]H_{l}^{(1)}(k^{[0]}a)\dot{J}_{l}(k^{[1]}a)
~.
\end{equation}
We found previously (\ref{di22}) that
\begin{equation}\label{di28c}
p_{l}(r<a)=p_{l}^{i}(r<a)+D_{l}(r<a)-E_{l}(r<a)~.
\end{equation}
whence
\begin{equation}\label{di29}
B_{l}^{[1]}J_{l}(k^{[1]}r)=B_{l}^{[0]}J_{l}(k^{[0]}r)+
B_{l}^{[1]}J_{l}(k^{[1]}r)+B_{l}^{[1]}J_{l}(k^{[0]}r)\left[-d_{l}^{<}+e_{l}^{<}\right]~;~\forall r<a
~,
\end{equation}
wherefrom
\begin{multline}\label{di29a}
0=B_{l}^{[0]}+B_{l}^{[1]}\left[-d_{l}^{<}+e_{l}^{<}\right]=
B_{l}^{[0]}+B_{l}^{[1]}\frac{i\pi}{2}\left[
-k^{[1]}a\dot{J}_{l}(k^{[1]}a)H_{l}^{(1)}(k^{[0]}a)+k^{[0]}a J_{l}(k^{[1]}a)\dot{H}_{l}^{(1)}(k^{[0]}a)\right]+\\
B_{l}^{[1]}\frac{i\pi}{2}k^{[1]}a\left[\frac{\rho^{[1]}-\rho^{[0]}}{\rho^{[1]}}
\right]H_{l}^{(1)}(k^{[0]}a)\dot{J}_{l}(k^{[1]}a)~,
\end{multline}
or
\begin{equation}\label{di30}
B_{l}^{[1]}= B_{l}^{[0]}\left[\frac{\frac{2i}{\pi\rho^{[0]}}}{\frac{k^{[0]}a}{\rho^{[0]}}J_{l}(k^{[1]}a)\dot{H}_{l}^{(1)}(k^{[0]}a)- \frac{k^{[1]}a}{\rho^{[1]}}H_{l}^{(1)}(k^{[0]}a)\dot{J}_{l}(k^{[1]}a)}
\right]~;~\forall l\in\mathbb{Z}~,
\end{equation}
which agrees, once again, with the exact, reference solution (\ref{ddsov26})and (\ref{constrhoone47}).

Next, consider the external field. Again, on account of (\ref{di24}):
\begin{equation}\label{di31}
D_{l}(r>a)=B_{l}^{[1]}\frac{i\pi}{2}\left[\Big(k^{[1]}\Big)^{2}-
 \Big( k^{[0]}\Big)^{2}\right]
 \int_{0}^{a}dr'r'H_{l}^{(1)}(k^{[0]}r)J_{l}(k^{[0]}r')J_{l}(k^{[1]}r')~,
\end{equation}
which yields, after the employment of (\ref{constrhoone38})-(\ref{constrhoone39})
\begin{multline}\label{di32}
D_{l}(r>a)=B_{l}^{[1]}\frac{i\pi}{2}H_{l}^{(1)}(k^{[0]}r)\left[-k^{[1]}a\dot{J}_{l}(k^{[1]}a)J_{l}(k^{[0]}a)+k^{[0]}a J_{l}(k^{[1]}a)\dot{J}_{l}(k^{[0]}a)\right]=\\-B_{l}^{[1]}H_{l}^{(1)}(k^{[0]}r)d_{l}^{>}~.
\end{multline}
Eq. (\ref{di28}) entails
\begin{equation}\label{di33}
E_{l}(r>a)=-B_{l}^{[1]}\frac{i\pi}{2}k^{[1]}a\left[\frac{\rho^{[1]}-\rho^{[0]}}{\rho^{[1]}}
\right]H_{l}^{(1)}(k^{[0]}r)J_{l}(k^{[0]}a)\dot{J}_{l}(k^{[1]}a)=-B_{l}^{[1]}H_{l}^{(1)}(k^{[0]}r)e_{l}^{>}
~.
\end{equation}
Moreover,
\begin{equation}\label{di34}
p_{l}(r>a)=p_{l}^{i}(r>a)+D_{l}(r>a)-E_{l}(r>a)~.
\end{equation}
whence
\begin{multline}\label{di35}
p_{l}(r>a)=p_{l}^{i}(r>a)+
B_{l}^{[1]}\frac{i\pi}{2}H_{l}^{(1)}(k^{[0]}r)\left[-k^{[1]}a\dot{J}_{l}(k^{[1]}a)J_{l}(k^{[0]}a)+k^{[0]}a J_{l}(k^{[1]}a)\dot{J}_{l}(k^{[0]}a)\right]+\\
B_{l}^{[1]}\frac{i\pi}{2}k^{[1]}a\left[\frac{\rho^{[1]}-\rho^{[0]}}{\rho^{[1]}}
\right]H_{l}^{(1)}(k^{[0]}r)J_{l}(k^{[0]}a)\dot{J}_{l}(k^{[1]}a)~,
\end{multline}
or
\begin{multline}\label{di36}
p_{l}(r>a)=p_{l}^{i}(r>a)+
A_{l}^{[0]}H_{l}^{(1)}(k^{[0]}r)~~,\\
A_{l}^{[0]}=B_{l}^{[0]}\left[\frac
{-\frac{k^{[0]}a}{\rho^{[0]}} J_{l}(k^{[1]}a)\dot{J}_{l}(k^{[0]}a)+\frac{k^{[1]}a)}{\rho^{[1]}}\dot{J}_{l}(k^{[1]}a)J_{l}(k^{[0]}a)}
{\frac{k^{[0]}a}{\rho^{[0]}}J_{l}(k^{[1]}a)\dot{H}_{l}^{(1)}(k^{[0]}a)- \frac{k^{[1]}a}{\rho^{[1]}}H_{l}^{(1)}(k^{[0]}a)\dot{J}_{l}(k^{[1]}a)}
\right]~,
\end{multline}
which agrees, once again, with the exact, reference DD-SOV solution (\ref{ddsov25}) and (\ref{constrhoone52}).
%%%%%%%%%%%%%%%%%%%%%%%%%%%%%%%%%%%
\subsubsection{DI method: the constant density solution for the "obstacle"}
Everything is the same as in the previous section except that we now assume that we are in the constant-density situation, i.e.,
\begin{equation}\label{di37}
\rho^{[1]}=\rho^{[0]}~,
\end{equation}
so that, assuming
\begin{equation}\label{di38}
p_{l}^{[1]}(r<a)=\mathcal{B}_{l}^{[1]}J_{l}(k^{[1]}r)~,
\end{equation}
we find, in the same manner as previously:
\begin{equation}\label{di39}
\mathcal{B}_{l}^{[1]}=\frac{B_{l}^{[0]}}{d_{l}^{<}}~,
\end{equation}
Recall by (\ref{di29}) that
\begin{equation}\label{di40}
B_{l}^{[1]}=\frac{B_{l}^{[0]}}{d_{l}^{<}-e_{l}^{<}}~,
\end{equation}
so that (\ref{di39}) is, of course, a consequence of (\ref{di40}) because $e_{l}^{<}=0$ when $\rho^{[1]}=\rho^{[0]}$.

The key feature of (\ref{di40}) is the separation, in its denominator, of the term ($e_{l}^{<}$) depending on the contrast of mass density density from the other term ($d_{l}^{<}$) which does not depend on the mass density. This crucial feature emerges clearly and naturally in the DI formulation, but not in the DD-SOV, BI-BI, DI-BI and BI-DI formulations.
%%%%%%%%%%%%%%%%%%%%%%%%%%%%%%%%%%%
\subsubsection{DI method: an iterative solution for the "obstacle" useful for small differences of density}\label{diit}
Eqs (\ref{di40}) and (\ref{di39}) entail
\begin{equation}\label{di41}
B_{l}^{[1]}=\frac{\frac{B_{l}^{[0]}}{d_{l}^{<}}}{1-\frac{e_{l}^{<}}{d_{l}^{<}}}=
\frac{\mathcal{B}_{l}^{[1]}~}{{1-\frac{e_{l}^{<}}{d_{l}^{<}}}},
\end{equation}
We have
\begin{equation}\label{di42}
\frac{e_{l}^{<}}{d_{l}^{<}}=\frac
{-\epsilon\frac{\pi}{2i}k^{[1]}a H_{l}^{(1)}(k^{[0]}a)\dot{J}_{l}(k^{[1]}a)}
{\frac{\pi\rho^{[0]}}{2i}\left[-\frac{k^{[1]}a}{\rho^{[0]}}\dot{J}_{l}(k^{[1]}a)H_{l}^{(1)}(k^{[0]}a)+
\frac{k^{[0]}a}{\rho^{[0]}}J_{l}(k^{[1]}a)\dot{H}_{l}^{(1)}(k^{[0]}a)\right]}=
\frac{\epsilon}
{1-\frac{k^{[0]}}{k^{[1]}}\frac{J_{l}(k^{[1]}a)\dot{H}_{l}^{(1)}(k^{[0]}a)}{\dot{J}_{l}(k^{[1]}a)H_{l}^{(1)}(k^{[0]}a)}}
~,
\end{equation}
or
\begin{equation}\label{di43}
\frac{e_{l}^{<}}{d_{l}^{<}}=\epsilon f_{l}
~,
\end{equation}
with
\begin{equation}\label{di44}
\epsilon=\frac{\rho^{[1]}-\rho^{[0]}}{\rho^{[1]}}~~,~~f_{l}=\frac{1}
{1-\frac{k^{[0]}}{k^{[1]}}\frac{J_{l}(k^{[1]}a)\dot{H}_{l}^{(1)}(k^{[0]}a)}{\dot{J}_{l}(k^{[1]}a)H_{l}^{(1)}(k^{[0]}a)}}
~.
\end{equation}
Note that $\epsilon$ is a dimensionless (potentially-) small parameter.

If the product $\|\epsilon f_{l}\|<1$ then it is appropriate to employ the geometric series
\begin{equation}\label{di45}
B_{l}^{[1]}=
\mathcal{B}_{l}^{[1]}\sum_{j=0}^{\infty}\left[\epsilon f_{l}\right]^{j},
\end{equation}
to represent $B_{l}^{[1]}$, whereby the iteration formula, initialized by
\begin{equation}\label{di46}
B_{l}^{[1](0)}=
\mathcal{B}_{l}^{[1]},
\end{equation}
is obtained:
\begin{equation}\label{di47}
B_{l}^{[1](j)}=B_{l}^{[1](j-1)}+\mathcal{B}_{l}^{[1]}\left[\epsilon f_{l}\right]^{j}~;~j=1,2,...
~.
\end{equation}
Thus, the smaller is $\epsilon$ (i.e., the closer  the two densities are to each other), the closer is $B_{l}^{[1](j)}$ to the constant-density solution $\mathcal{B}_{l}^{[1]}$, and the lesser the amount of iterates that are necessary to obtain a decent approximation to $B_{l}^{[1]}$. This will be verified numerically later on.

Let us now turn to $A_{l}^{[1]}$. We found previously from (\ref{di32})-(\ref{di33})that:
\begin{multline}\label{di48}
D_{l}(r>a)=-B_{l}^{[1]}H_{l}^{(1)}(k^{[0]}r)d_{l}^{>}~~,\\
~~d_{l}^{>}=
\frac{\pi\rho^{[0]}}{2i}\left[-\frac{k^{(1)}a}{\rho^{[0]}}\dot{J}_{l}(k^{[1]}a)J_{l}(k^{[0]}a)+
\frac{k^{(0)}a}{\rho^{[0]}}\dot{J}_{l}(k^{[0]}a)J_{l}(k^{[1]}a)\right]
~,
\end{multline}
\begin{equation}\label{di49}
E_{l}(r>a)=-B_{l}^{[1]}H_{l}^{(1)}(k^{[0]}r)e_{l}^{>}~~,\\
~~e_{l}^{>}=
\frac{-\pi}{2i}\epsilon k^{(1)}a \dot{J}_{l}(k^{[1]}a)J_{l}(k^{[0]}a)=\epsilon g_{l}
~,
\end{equation}
so that (\ref{di34}) becomes
\begin{equation}\label{di50}
p_{l}(r>a)=p_{l}^{i}(r>a)+B_{l}^{[1]}\left[-d_{l}^{>}+e_{l}^{>}\right]H_{l}^{(1)}(k^{[0]}r)=p_{l}^{i}(r>a)+A_{l}^{[1]}H_{l}^{(1)}(k^{[0]}r)~,
\end{equation}
with
\begin{equation}\label{di51}
A_{l}^{[0]}=B_{l}^{[1]}\left[-d_{l}^{>}+e_{l}^{>}\right]~,
\end{equation}
and, in the constant-density case
\begin{equation}\label{di52}
\mathcal{A}_{l}^{[0]}=-\mathcal{B}_{l}^{[1]}d_{l}^{>}~.
\end{equation}
Eq. (\ref{di41}) entails
\begin{equation}\label{di53}
A_{l}^{[0]}=\mathcal{B}_{l}^{[1]}\left[\frac
{-d_{l}^{>}+e_{l}^{>}}
{1-\frac{e_{l}^{<}}{d_{l}^{<}}}
\right]~,
\end{equation}
so that, on account of (\ref{di52}),
\begin{equation}\label{di54}
A_{l}^{[0]}=\mathcal{A}_{l}^{[1]}+\mathcal{B}_{l}^{[1]}\left[\frac
{-\frac{d_{l}^{>}e_{l}^{<}}{d_{l}^{<}}+e_{l}^{>}}
{1-\frac{e_{l}^{<}}{d_{l}^{<}}}
\right]=
\mathcal{A}_{l}^{[0]}+\mathcal{B}_{l}^{[1]}\epsilon\left[\frac
{-d_{l}^{>}f_{l}+g_{l}}
{1-\epsilon f_{l}}\right]~,
\end{equation}
or, employing once again the geometric series representation
\begin{equation}\label{di55}
A_{l}^{[0]}=
\mathcal{A}_{l}^{[1]}+\mathcal{B}_{l}^{[1]}\left[
-d_{l}^{>}f_{l}+g_{l}\right]\sum_{j=0}^{\infty}\left[\epsilon f_{l}\right]^{j+1}
~,
\end{equation}
whence the iterative scheme, with the intialization
\begin{equation}\label{di56}
A_{l}^{[0](0)}=
\mathcal{A}_{l}^{[0]}
~,
\end{equation}
for the obtention of successive approximations of $A_{l}^{[1]}$
\begin{equation}\label{di57}
A_{l}^{[0](j)}=
\mathcal{A}_{l}^{[0](j-1)}+\mathcal{B}_{l}^{[1]}\left[
-d_{l}^{>}f_{l}+g_{l}\right]\left[\epsilon f_{l}\right]^{j}~;~j=1,2,...
~,
\end{equation}
from which we notice that the $j$th-order  correction to $A_{l}^{[1]}$  is of order $\epsilon^{j}$, which, for small $\epsilon$, is the same as the $j$th-order correction to $B_{l}^{[1]}$, since the latter  is also of order $\epsilon^{j}$.
%%%%%%%%%%%%%%%%%%%%%%%%%%%%%%%%%%%%%%%%%%%%%%%%%%%%%%%%%%%%%%%%%%%%%%%%%%%%%%%%
\subsubsection{Reconciling various publications dealing with the DI method}\label{recon}
The publications \cite{mk67,lm95,ma03}, which apparently deal with the same problem of scattering by an obstacle having space variable-dependent constitutive properties, seem to differ from each other, and especially from our own formulation of the DI method. We show hereafter that they are all equivalent under certain prescribed conditions (actually the same as the ones assumed in our DI-BI method):
\begin{equation}\label{recon1}
\rho(\mathbf{x})=\rho^{[l]}=const.~;~\forall \mathbf{x}\in\Omega_{j}~,~j=0,1~,
\end{equation}
wherein $\rho^{[1]}$ is generally different from $\rho^{[0]}$ and
\begin{equation}\label{recon2}
\begin{array}{c}
c(\mathbf{x})=c^{[0]}=const.~;~\forall \mathbf{x}\in\Omega_{0}~,\\
c(\mathbf{x})=c^{[1]}(\mathbf{x})~;~\forall \mathbf{x}\in\Omega_{1}~.
\end{array}
\end{equation}
The starting point is the fundamental relation (\ref{bidi6}) of the BI-DI method, i.e.,
\begin{multline}\label{recon3}
p(\mathbf{x})=p^{i}(\mathbf{x})+
\int_{\Gamma}G^{[0]}(\mathbf{x};\mathbf{x'})
\left(\frac{\rho^{[0]}}{\rho^{[1]}}-1\right)\boldsymbol{\nu}\cdot \nabla_{\mathbf{x'}}p^{[1]}(\mathbf{x'})d\Gamma(\mathbf{x'})+\\
\int_{\Omega_{1}}G^{[0]}(\mathbf{x};\mathbf{x'})\Big(\big(k^{[1]}(\mathbf{x'})\big)^{2}-\big(k^{[0]}\big)^{2}\Big)p^{[1]}(\mathbf{x'})
d\Omega(\mathbf{x'}):=p^{i}(\mathbf{x})+I_{\Gamma}(\mathbf{x})+I_{\Omega}(\mathbf{x})~;~\forall \mathbf{x}\in \mathbf{R}^{3}~.
\end{multline}
From Green's first identity we obtain
\begin{multline}\label{recon7}
I_{\Gamma}(\mathbf{x})=\left(\frac{\rho^{[0]}}{\rho^{[1]}}-1\right)
\int_{\Gamma}G^{[0]}(\mathbf{x};\mathbf{x'})
\boldsymbol{\nu}\cdot \nabla_{\mathbf{x'}}p^{[1]}(\mathbf{x'})d\Gamma(\mathbf{x'})=\\
-\left(\frac{\rho^{[0]}}{\rho^{[1]}}-1\right)\int_{\Omega_{1}}\left[G^{[0]}(\mathbf{x};\mathbf{x'})\nabla_{\mathbf{x'}}\cdot\nabla_{\mathbf{x'}} p^{[1]}(\mathbf{x'})+\nabla_{\mathbf{x'}}G^{[0]}(\mathbf{x};\mathbf{x'})\cdot\nabla_{\mathbf{x'}} p^{[1]}(\mathbf{x'})\right]
d\Omega(\mathbf{x'})~;~\forall \mathbf{x}\in \mathbf{R}^{3}~.
\end{multline}
or, after recalling the partial differential equation (\ref{constrhoone4}) satisfied by $p^{[1]}$, and on account of (\ref{recon1})-(\ref{recon2}),
\begin{multline}\label{recon8}
I_{\Gamma}(\mathbf{x})=-\left(\frac{\rho^{[0]}}{\rho^{[1]}}-1\right)
\int_{\Omega_{1}}\left[-\left(k^{[1]}(\mathbf{x'})\right)^{2}G^{[0]}(\mathbf{x};\mathbf{x'}) p^{[1]}(\mathbf{x'})+\nabla_{\mathbf{x'}}G^{[0]}(\mathbf{x};\mathbf{x'})\cdot\nabla_{\mathbf{x'}} p^{[1]}(\mathbf{x'})\right]
d\Omega(\mathbf{x'})~;\\
~\forall \mathbf{x}\in \mathbf{R}^{3}~,
\end{multline}
whence
\begin{multline}\label{recon9}
p(\mathbf{x})=p^{i}(\mathbf{x})-(\alpha-1)
\int_{\Omega_{1}}\nabla_{\mathbf{x'}}G^{[0]}(\mathbf{x};\mathbf{x'})\cdot\nabla p^{[1]}(\mathbf{x'})
d\Omega(\mathbf{x'})+\\\big(k^{[0]}\big)^{2}\int_{\Omega_{1}}\Big(N(\mathbf{x'})\alpha-1\Big)G^{[0]}(\mathbf{x};\mathbf{x'}) p^{[1]}(\mathbf{x'})d\Omega(\mathbf{x'})~;
~\forall \mathbf{x}\in \mathbf{R}^{3}~,
\end{multline}
wherein
\begin{equation}\label{recon10}
    \alpha:=\frac{\rho^{[0]}}{\rho^{[1]}}~~,~~N(\mathbf{x'}):=\left(\frac{k^{[1]}(\mathbf{x'})}{k^{[0]}}\right)^{2}=
    \left(\frac{c^{[0]}}{c^{[1]}(\mathbf{x'})}\right)^{2}~.
\end{equation}
Eq. (\ref{recon9}) agrees with (3.5)-(3.7) of \cite{ma03} if we recognize the fact that Martin's Green's function is the negative of our Green's function. The 2D version of  (\ref{recon9}) also agrees with (2) of \cite{lm95} if we recall the correspondences of the constitutive parameters of the 2D acoustic problem with those of the 2D SH elastic wave problem (see sect. \ref{2D}). For the same reason, the 2D version of (\ref{recon9}) agrees with the 2D version of (12) in \cite{mk67}. However, it should be stressed that the authors of \cite{mk67} assume that their obstacle is {\it homogeneous} which means, in particular, that their equivalent of $\alpha$ is a constant with respect to the spatial coordinates (this being also our starting hypothesis), whereas the author of \cite{ma03} seems to conclude that his result holds also for $\alpha$ dependent on $\mathbf{x'}$.

We should mention that (\ref{recon9}) agrees with (2) of \cite{mc93} provided that it is understood that the authors of \cite{mc93} assume piecewise constant (in $\mathbb{R}^{2}$) compressibility $\kappa$ instead of our assumption of piecewise constant (in $\mathbb{R}^{2}$) wavespeed $c$.

Finally, it should be pointed out that the expression of the DI for 3D elastic wave scattering given in \cite{pa03} appears to derive from the Mal-Knopoff integral equation (attributed by the authors of \cite{pa03} to a study of Kupradze) for the particular case of variable $\rho$ assuming constant $\lambda$ and $\mu$ in $\mathbb{R}^{3}$.
%%%%%%%%%%%%%%%%%%%%%%%%%%%%%%%%%%%%%%%%%%%%%%%%%%%%%%%%%%%%%%%%%%%%%%%%%%%%%%%%
\subsection{The reasons, in general  2D acoustic scattering problems, for choosing the configuration of two homogeneous, isotropic media separated by a  boundary of canonical shape}\label{mix}
A general  acoustic scattering problem can take several elementary forms (this list is non-exhaustive):\\\\
(1) the obstacle is composed of several smaller, homogeneous objects of arbitrary shape separated one from the other,\\
(2) the obstacle is composed of only one object that is inhomogeneous and of arbitrary shape,\\
(3) the obstacle is composed of only one object that is homogeneous and of arbitrary shape,\\\\
the obstacle being, by definition, a subdomain of $\mathbb{R}^{n}~;~n=1,2,3$, within (i.e., its interior) which the constitutive properties are different from those of the region (i.e., its exterior of infinite extent) that prevail (i.e., the constitutive properties) in the absence of the obstacle.

In 3D acoustic scattering problems, it is convenient to situate the origin somewhere near the geometric midpoint  of the group of objects or within the boundary of the single object so that either the group, or single object, can be thought to be contained within a sphere (it is not necessary for this sphere to be the smallest one able to contain the group or single object. We then define $\Omega_{ext}$ and $\Omega_{int}$ as the regions exterior and interior to this sphere (henceforth termed the 'obstacle') and $\Gamma_{ext-int}$ as the boundary of the latter. By definition, the constitutive properties of the medium in $\Omega_{ext}$ are those of the homogeneous medium filling the host region  that prevails in the absence of the obstacle. Thus, the constitutive parameters of the medium in $\Omega_{ext}$ are $\rho^{[ext]}$, $c^{[ext]}$, which are both constants with respect to position in $\Omega_{ext}$. The question is: what mass density and wavespeed should we assign to the necessarily-inhomogeneous (except in case (3) when the boundary of the object is a sphere coinciding with the minimal sphere) medium occupying $\Omega_{int}$?

In 2D acoustic scattering problems (of more interest in this study), it is convenient to situate the origin somewhere near the geometric midpoint  of the group of cylindrical objects (i.e., infinitely long in the $z$ direction) or within the boundary of the single object so that either the group, or single object, can be thought to be contained within a circular cylinder (it is not necessary for this circular cylinder to be the smallest one able to contain the group or single object. We then define $\varpi_{ext}$ and $\varpi_{int}$ as the regions exterior and interior to this circular cylinder (henceforth termed the 'obstacle') in its cross section plane and $\gamma_{ext-int}$ as the boundary of the circular disk $\varpi_{int}$. By definition, the constitutive properties of the medium in $\varpi_{ext}$ are those of the homogeneous medium filling the host region  that prevails in the absence of the obstacle. Thus, the constitutive parameters of the medium in $\varpi_{ext}$ are $\rho^{[ext]}$, $c^{[ext]}$, which are both constants with respect to position in $\varpi_{ext}$. The question is again: what mass density and wavespeed should we assign to the necessarily-inhomogeneous (except in case (3) when the boundary of the object is a circular cylinder coinciding with the minimal circular cylinder) medium occupying $\varpi_{int}$?

Consider the circular cylinder in its sagittal plane, i.e., the disk $\varpi_{int}$, within which there are two regions, one ($\varpi_{int}-\varpi_{1}$) in which the constitutive parameters $\mathfrak{c}_{j}~;~j=1,2$ are  $\mathfrak{c}_{j}^{[0]}$ (constants with respect to $\mathbf{x}$) and the other ($\varpi_{1}$ in which the constitutive parameters are  $\mathfrak{c}_{j}^{[1]}(\mathbf{x})$ (i.e., generally position-dependent). A general mixing formula for the parameters $\mathfrak{c}_{j}$ is
\begin{equation}\label{mix-1}
    \bar{\mathfrak{c}}_{j}=\frac{\int_{\varpi_{int}}\mathfrak{c}_{j}(\mathbf{x})d\varpi(\mathbf{x})}{\int_{\varpi_{int}}d\varpi(\mathbf{x})}=
    \frac{
    \int_{\varpi_{int}-\varpi_{1}}\mathfrak{c}_{j}^{[0]}d\varpi(\mathbf{x})
    +\int_{\varpi_{1}}\mathfrak{c}_{j}^{[1]}(\mathbf{x})d\varpi(\mathbf{x})}
    {\int_{\varpi_{int}}d\varpi(\mathbf{x})}~.
\end{equation}
To simplify matters, we shall consider $\mathfrak{c}_{j}^{[1]}~;~j=1,2$ to also be constants with respect to position which amounts to the supposition that the composite medium within the disk is {\it binary}, i.e., composed of areas in which the physical properties $\mathfrak{c}^{[0]}$ are constants and other areas in which the physical properties $\mathfrak{c}^{[1]}$ are also constants, different from $\mathfrak{c}^{[0]}$. Consequently,
\begin{equation}\label{mix-2}
    \bar{\mathfrak{c}}_{j}=
    \frac{
    \mathfrak{c}_{j}^{[0]}\int_{\varpi_{int}-\varpi_{1}}d\varpi(\mathbf{x})
    +\mathfrak{c}_{j}^{[1]}\int_{\varpi_{1}}d\varpi(\mathbf{x})}
    {\int_{\varpi_{int}}d\varpi(\mathbf{x})}=
    \frac{
    \mathfrak{c}_{j}^{[0]}\big(\varpi_{int}-\varpi_{1}\big)+\mathfrak{c}_{j}^{[1]}\varpi_{1}}
    {\varpi_{int}}=
    \mathfrak{c}_{j}^{[0]}(1-\phi)+\mathfrak{c}_{j}^{[1]}\phi~.
\end{equation}
wherein $\phi=\frac{\varpi_{1}}{\varpi_{int}}$ is the {\it filling fraction}, or the ratio of the area $\varpi_{1}$ occupied by the material having   parameters $\mathfrak{c}^{[1]}~;~j=1,2$ to the total area in the disk $\varpi_{int}$.

Until now, we have been rather vague as to the physical nature of the parameters $\mathfrak{c}^{[1]}~;~j=1,2$. The most natural choice is:
\begin{equation}\label{mix-3}
    \mathfrak{c}_{1}=\rho~~,~~\mathfrak{c}_{2}=c=c'+ic''~,
\end{equation}
whence
\begin{equation}\label{mix-4}
    \bar{\mathfrak{c}}_{1}=\bar{\rho}=\rho^{[0]}(1-\phi)+\rho^{[1]}\phi~~,~~\bar{\mathfrak{c}}_{2}=\bar{c}=c^{[0]}(1-\phi)+c^{[1]}\phi~.
\end{equation}
Another choice is:
\begin{equation}\label{mix-5}
    \mathfrak{c}_{1}=(\rho)^{-1}~~,~~\mathfrak{c}_{2}=(c)^{-1}=(c'+ic'')^{-1}~,
\end{equation}
whence
\begin{equation}\label{mix-6}
    \bar{\mathfrak{c}}_{1}=(\bar{\rho})^{-1}=(\rho^{[0]})^{-1}(1-\phi)+(\rho^{[1]})^{-1}\phi~~,~~\
    \bar{\mathfrak{c}}_{2}=(\bar{c})^{-1}=(c^{[0]})^{-1}(1-\phi)+(c^{[1]})^{-1}\phi~.
\end{equation}
Eqs. (\ref{mix-3})-(\ref{mix-4}) and (\ref{mix-5})-(\ref{mix-6}) are the simplest (there exist more complex) versions of the so-called series   and parallel (respectively)  mixing formulae \cite{si02} for binary obstacles.

Now let us summarize what we have done. We began with a group of objects or single object of arbitrary shape, each object having properties $\mathfrak{c}_{j}^{[1]}$  different from those, $\mathfrak{c}_{j}^{[0]}$ (constants as a function of position), of the host. For the obvious reason, made evident in sect. \ref{obst}, that it is much simpler to treat a scattering problem involving a single, homogeneous obstacle of canonical shape than one involving several (or one) heterogeneous or homogeneous objects of arbitrary shape, we decided, aided by this possibility in the DI method, to enclose the original group of objects or single object of arbitrary shape by a virtual boundary of canonical shape (spherical in 3D, circular cylindrical in 2D). We termed the domain interior to this virtual boundary to be that of 'the obstacle'. This obstacle is generally-heterogeneous  because it comprises regions having the physical properties of the previous objects and another region having the physical properties of the host. We saw in sect. \ref{obst}, that the integral methods  turn out to be much more easily exploitable if this obstacle, enclosed by the virtual boundary of canonical shape, is homogeneous. Finally, we proposed a method for homogenizing the obstacle which appeals to a simple mixture formula. Thus, the original problem reduces to that of a (single) homogeneous obstacle enclosed within a virtual boundary of canonical shape, this problem having been treated in depth in various subsections of sect. \ref{obst}.

It can be objected that the problem we finally address is not the one we proposed to solve at the outset. This objection is founded, but we found no other way  to treat the original problem if the goal is to do this by other-than-purely-numerical means. In fact, this was precisely our goal as regards our quest for answers regarding the constant mass-density assumption question.

A second remark has to do with the 'closeness' of the solution of the scattering problem of  the 'new' obstacle (i.e., the homogeneous one with canonical-shaped virtual boundary) to that of the problem dealing with the 'old' obstacle (i.e., the one with several objects or one object with arbitrary shape). Homogenization, and, in particular, its mixture formula version, has been shown to give rise to very useful (in both the inverse and forward problem contexts \cite{qk11,wi16,wi18}), and often precise, approximations of the solution, especially at low frequencies. This fact constitutes another, important, justification of  the procedure adopted in our study.
%%%%%%%%%%%%%%%%%%%%%%%%%%%%%%%%%%%%%%%%%%%%%%%%%%%%%%%%%%%%%%%%%%%%%%%%%%%%%%%%%%%%%%%%%%%%%%%%%%%%%%%%%%%%%%%%%%%
\section{Numerical results}
%
%%%%%%%%%%%%%%%%%%%%%%%%%%%%%%%%%%%%%
\subsection{The numerical procedure and choice of parameters}
The aim of the computations was to compare numerically the acoustic scattering amplitudes $\mathcal{A}_{l}^{[0]},~\mathcal{B}_{l}^{[1]}$,   based on the constant-density assumption (i.e., $\epsilon=0$ in sect. \ref{diit}): (a) to the exact (i.e., DD-SOV) $A_{l}^{[0]},~B_{l}^{[1]}$ expressions (\ref{ddsov25})-(\ref{ddsov26}))respectively of the amplitudes, as well as (b) to the approximations $A_{l}^{[0](j)},~B_{l}^{[1](j)}~;~j=1,2,3$ of the amplitudes, all of  which do not rely on the constant-density assumption. $\mathcal{A}_{l}^{[0]},~\mathcal{B}_{l}^{[1]}$ are simply (\ref{ddsov25})-(\ref{ddsov26}), in which $\rho^{[1]}$ is taken to be equal to $\rho^{[0]}$. $A_{l}^{[0](j)},~B_{l}^{[1](j)}~;~j=1,2$ are given in (\ref{di46})-(\ref{di47}) and (\ref{di56})-(\ref{di57}) and $\epsilon$ is the mass density contrast parameter defined in (\ref{di44}).

The graphs are relative to the {\it external} amplitudes $\mathcal{A}_{l}^{[0]},~A_{l}^{[0]},~A_{l}^{[0](j)}$ and {\it internal} amplitudes $\mathcal{B}_{l}^{[1]},~B_{l}^{[1]},~B_{l}^{[1](j)}$ as a function of: frequency $f$, $\rho^{[1]}$ and $c^{[1]}$, the other parameters being fixed at the following values: $\rho^{[0]}=1000~Kgm^{-3}$, $c^{[0]}=1500~ms^{-1}$, $a=0.1~m$, $B_{l}^{[0]}=1$. Usually, the scattering amplitude order $l$ is taken to range from 0 to 3, but at high frequencies, we also give results for much larger $l$.
%\clearpage
%\newpage
%%%%%%%%%%%%%%%%%%%%%%%%%%%%%%%%%%%%%
\subsection{Variation of $\rho^{[1]}$ for various  orders $l$: case $\rho^{[1]}>\rho^{[0]}$ and low $f$}\label{fig01}
\begin{figure}[ht]
\begin{center}
\includegraphics[width=0.75\textwidth]{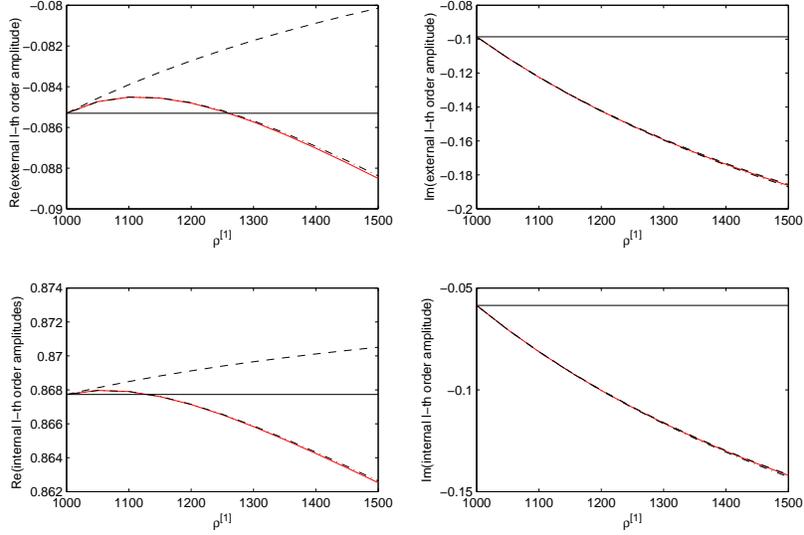}
\caption{Reflected and transmitted wavefield amplitudes as a function of $\rho^{[1]}>\rho^{[0]}$.
 The upper left(right) panels depict the real(imaginary) parts of  $A_{l}^{[0]}$ (red),
$\mathcal{A}_{l}^{[0]}$ (black ------),
 $A_{l}^{[0](1)}$ (black - - - -), $A_{l}^{[0](2)}$ (black -.-.-.-).
The lower left(right) panels depict the real(imaginary) parts of  $B_{l}^{[1]}$ (red), $\mathcal{B}_{l}^{[1]}$ (black ------), $B_{l}^{[1](1)}$ (black - - - -), $B_{l}^{[1](2)}$ (black -.-.-.-). Case $c^{[1]}=1700-i210~ms^{-1}$, $f=2000~Hz$, $l=0$.}
\label{fig101}
\end{center}
\end{figure}
\begin{figure}[ptb]
\begin{center}
\includegraphics[width=0.75\textwidth]{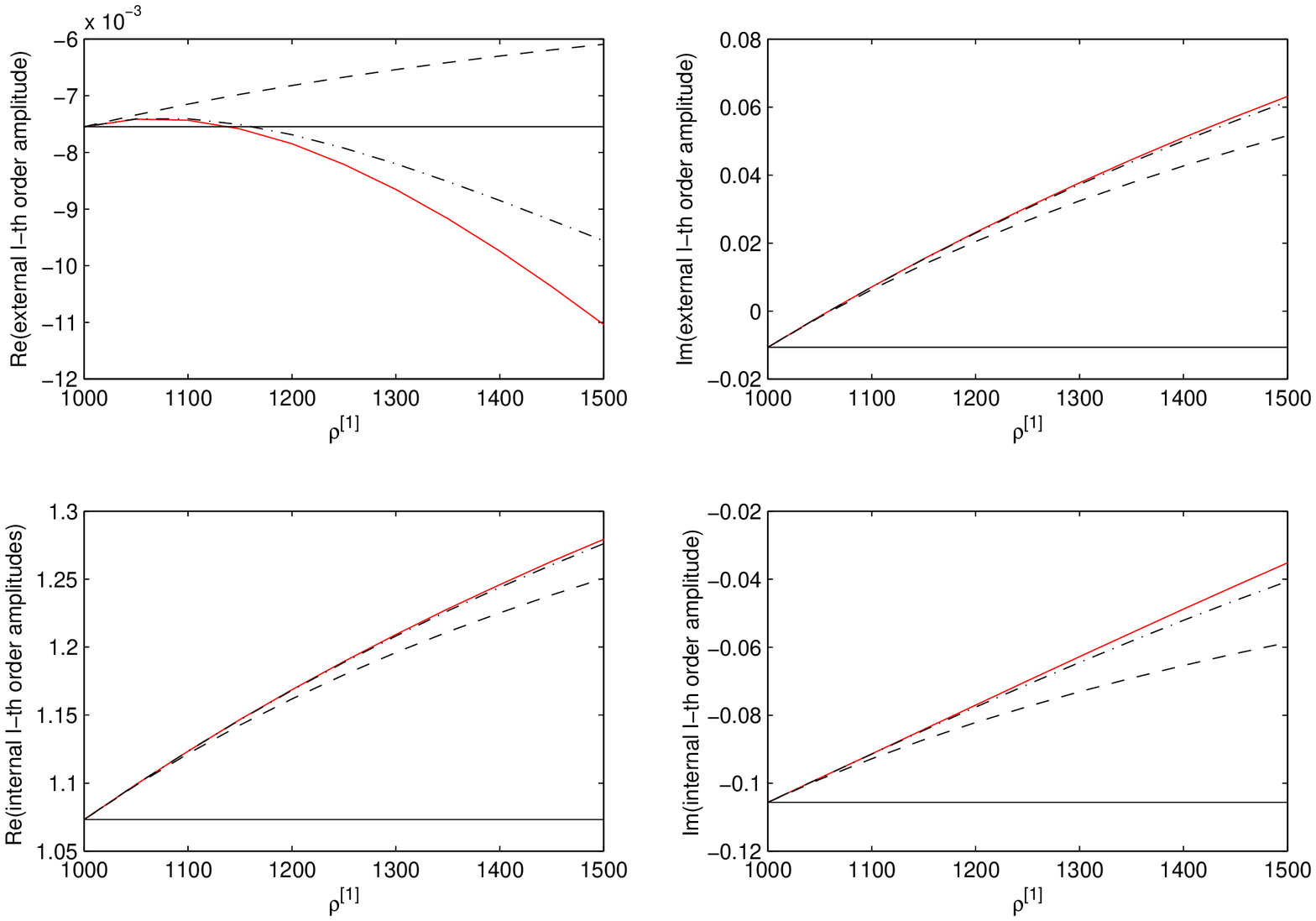}
\caption{Same as fig. \ref{fig101} except that $l=1$.}
\label{fig102}
\end{center}
\end{figure}
\begin{figure}[ptb]
\begin{center}
\includegraphics[width=0.75\textwidth]{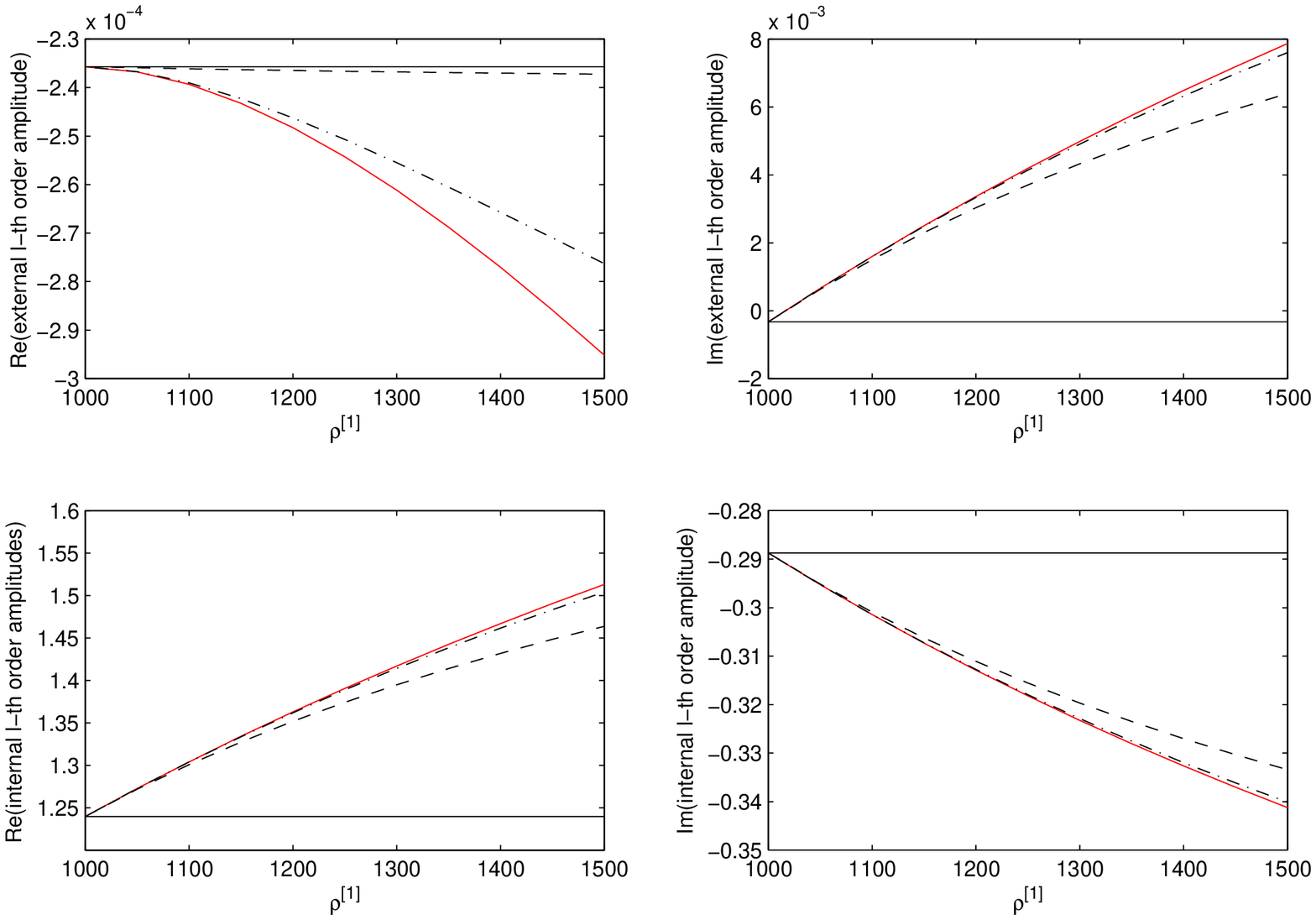}
\caption{Same as fig. \ref{fig101} except that $l=2$.}
\label{fig103}
\end{center}
\end{figure}
\begin{figure}[ptb]
\begin{center}
\includegraphics[width=0.75\textwidth]{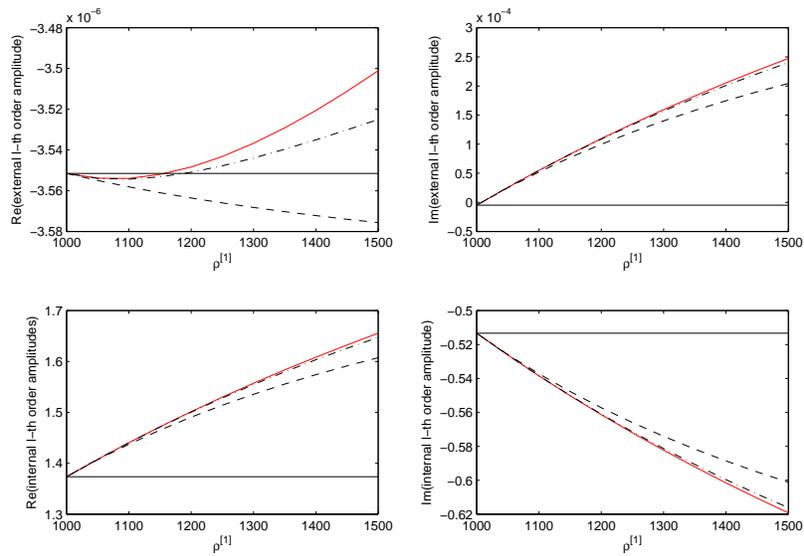}
\caption{Same as fig. \ref{fig101} except that $l=3$.}
\label{fig104}
\end{center}
\end{figure}
\clearpage
\newpage
We observe substantial qualitative differences, that increase with the mass density contrast, between $\mathcal{A}_{l}^{[0]},~\mathcal{B}_{l}^{[1]}$ and $A_{l}^{[0]},~B_{l}^{[1]}$  respectively except for very small differences of $\rho^{[1]}$ from $\rho^{[0]}$. Better agreement is attained, both quantitatively and qualitatively, by means of $A_{l}^{[0](1)},~B_{l}^{[1](1)}$, and near-coincidence with $A_{l}^{[0]},~B_{l}^{[1]}$ is obtained by means of $A_{l}^{[0](2)},~B_{l}^{[1](2)}$ respectively.

%%%%%%%%%%%%%%%%%%%%%%%%%%%%%%%%%%%%%
\subsection{Variation of $\rho^{[1]}$ for various  orders $l$: case $\rho^{[1]}<\rho^{[0]}$ and low $f$}
\begin{figure}[ht]
\begin{center}
\includegraphics[width=0.75\textwidth]{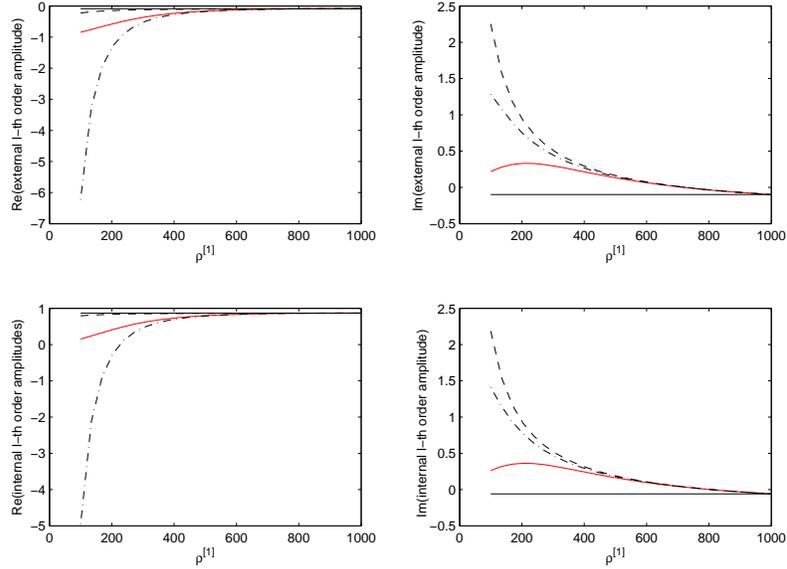}
\caption{Reflected and transmitted wavefield amplitudes as a function of $\rho^{[1]}<\rho^{[0]}$.  The upper left(right) panels depict the real(imaginary) parts of  $A_{l}^{[0]}$ (red),
$\mathcal{A}_{l}^{[0]}$ (black ------),
 $A_{l}^{[0](1)}$ (black - - - -), $A_{l}^{[0](2)}$ (black -.-.-.-).
The lower left(right) panels depict the real(imaginary) parts of  $B_{l}^{[1]}$ (red), $\mathcal{B}_{l}^{[1]}$ (black ------), $B_{l}^{[1](1)}$ (black - - - -), $B_{l}^{[1](2)}$ (black -.-.-.-). Case $c^{[1]}=1700-i210~ms^{-1}$, $f=2000~Hz$, $l=0$.}
\label{fig201}
\end{center}
\end{figure}
\begin{figure}[ptb]
\begin{center}
\includegraphics[width=0.75\textwidth]{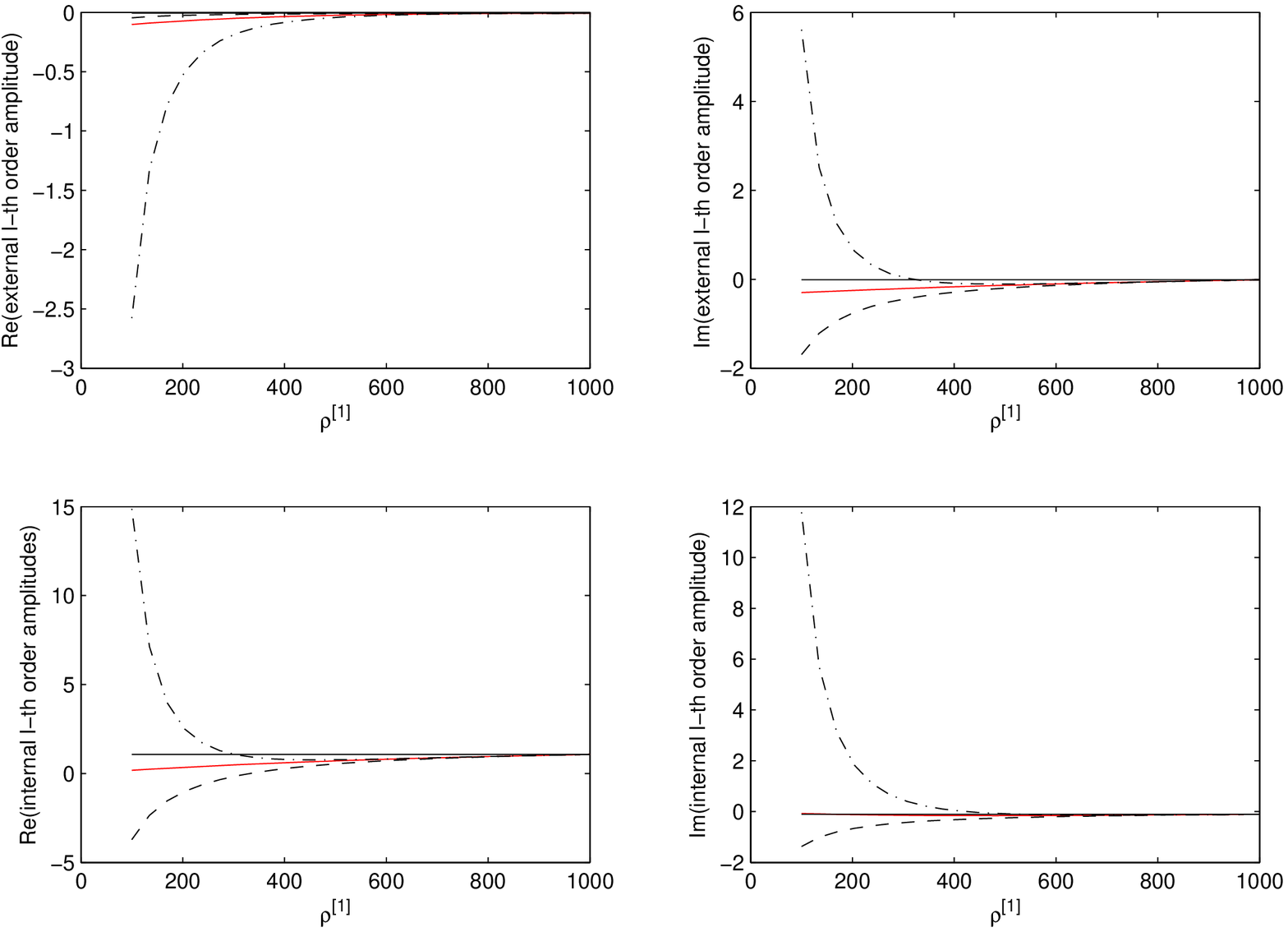}
\caption{Same as fig. \ref{fig201} except that $l=1$.}
\label{fig202}
\end{center}
\end{figure}
\begin{figure}[ptb]
\begin{center}
\includegraphics[width=0.75\textwidth]{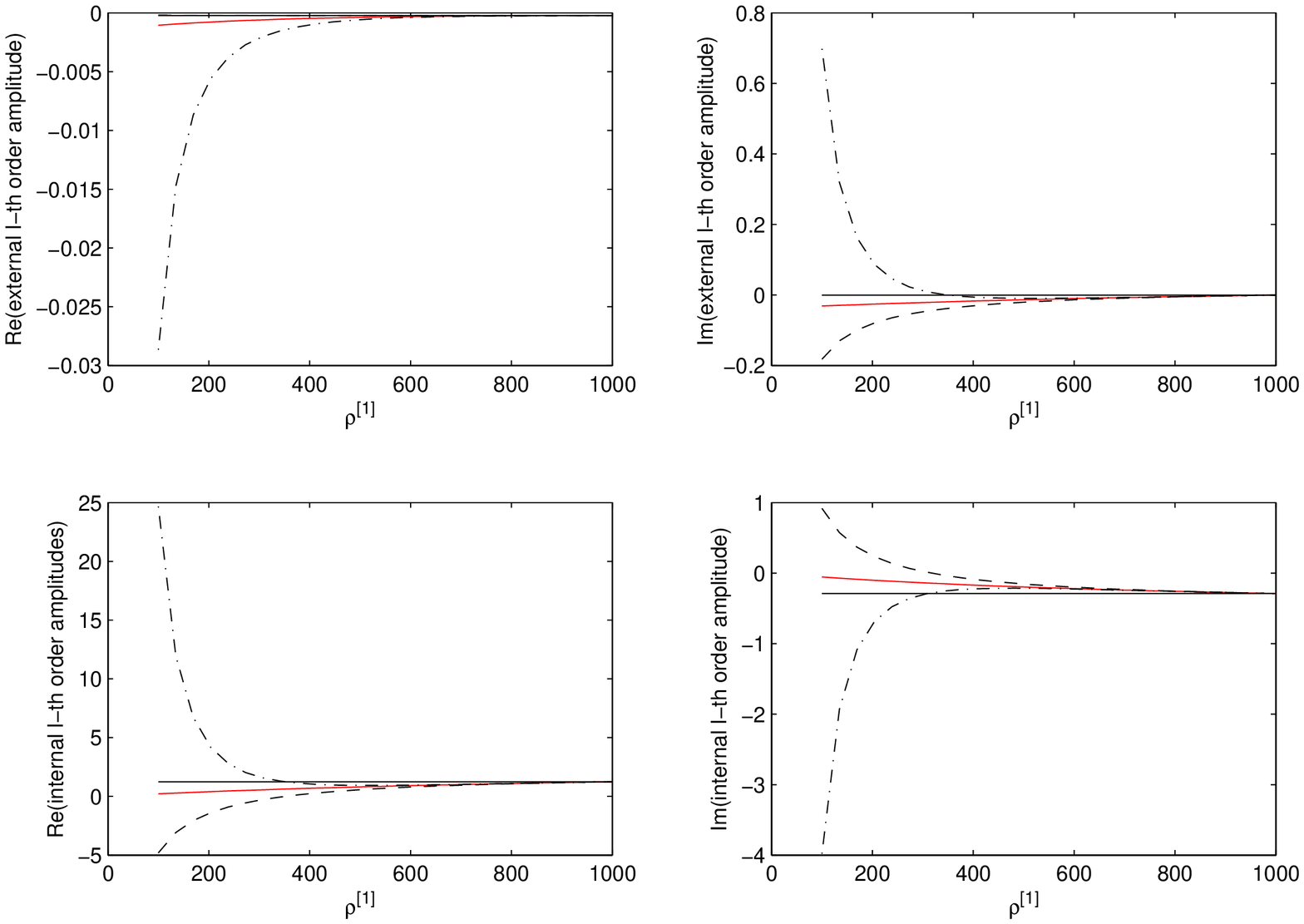}
\caption{Same as fig. \ref{fig201} except that $l=2$.}
\label{fig203}
\end{center}
\end{figure}
\begin{figure}[ptb]
\begin{center}
\includegraphics[width=0.75\textwidth]{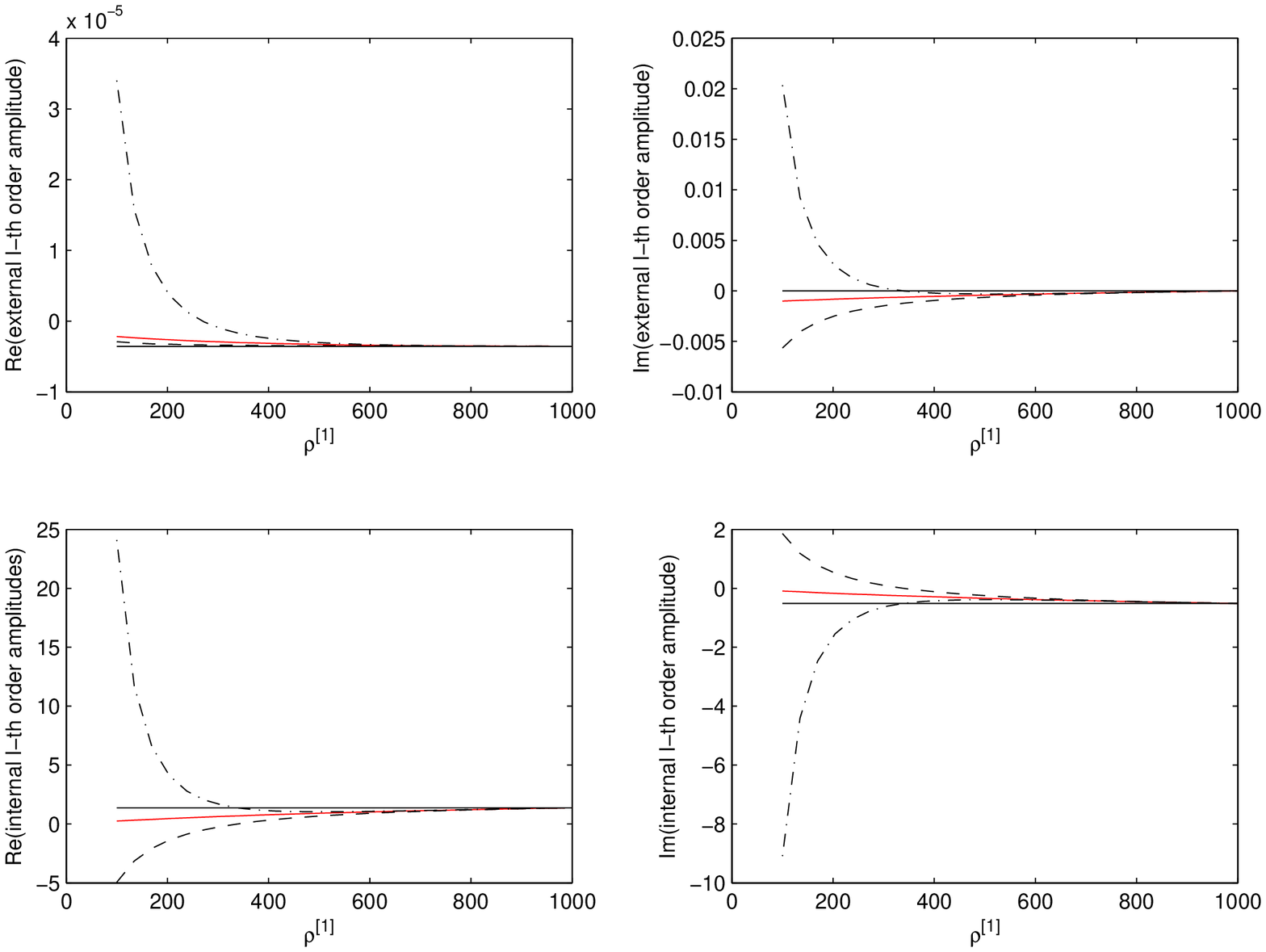}
\caption{Same as fig. \ref{fig201} except that $l=3$.}
\label{fig204}
\end{center}
\end{figure}
\clearpage
\newpage
The same remarks apply to this series of figures as in the  series of sect. \ref{fig01}.
%%%%%%%%%%%%%%%%%%%%%%%%%%%%%%%%%%%%%
\subsection{Variation of $\rho^{[1]}$ for various  orders $l$: case $\rho^{[1]}>\rho^{[0]}$ and higher $f$}\label{fig14}
\begin{figure}[ht]
\begin{center}
\includegraphics[width=0.75\textwidth]{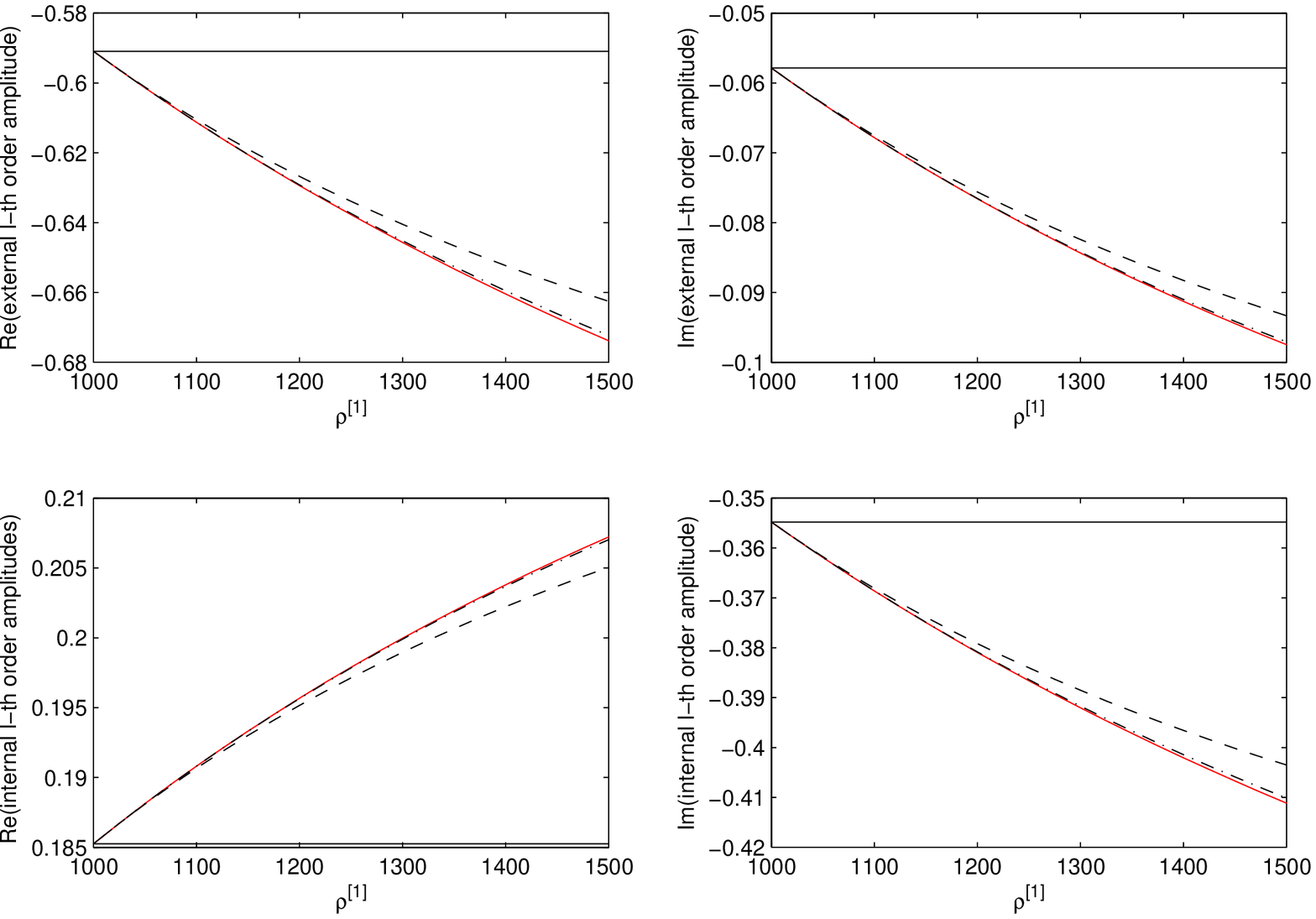}
\caption{Reflected and transmitted wavefield amplitudes as a function of $\rho^{[1]}>\rho^{[0]}$.
 The upper left(right) panels depict the real(imaginary) parts of  $A_{l}^{[0]}$ (red),
$\mathcal{A}_{l}^{[0]}$ (black ------),
 $A_{l}^{[0](1)}$ (black - - - -), $A_{l}^{[0](2)}$ (black -.-.-.-).
The lower left(right) panels depict the real(imaginary) parts of  $B_{l}^{[1]}$ (red), $\mathcal{B}_{l}^{[1]}$ (black ------), $B_{l}^{[1](1)}$ (black - - - -), $B_{l}^{[1](2)}$ (black -.-.-.-). Case $c^{[1]}=1700-i210~ms^{-1}$, $f=20000~Hz$, $l=0$.}
\label{fig1401}
\end{center}
\end{figure}
\begin{figure}[ptb]
\begin{center}
\includegraphics[width=0.75\textwidth]{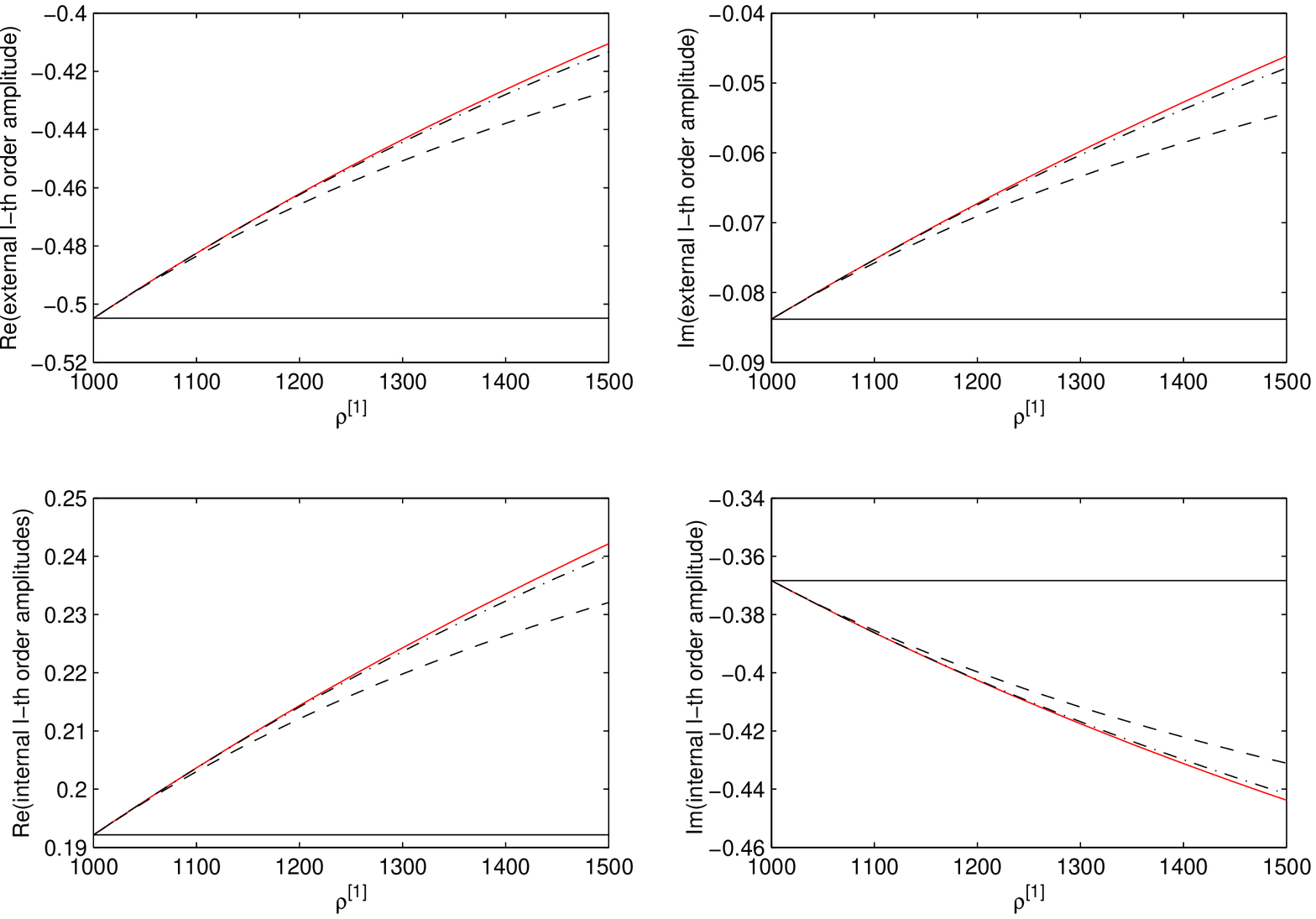}
\caption{Same as fig. \ref{fig1401} except that $l=1$.}
\label{fig1402}
\end{center}
\end{figure}
\begin{figure}[ptb]
\begin{center}
\includegraphics[width=0.75\textwidth]{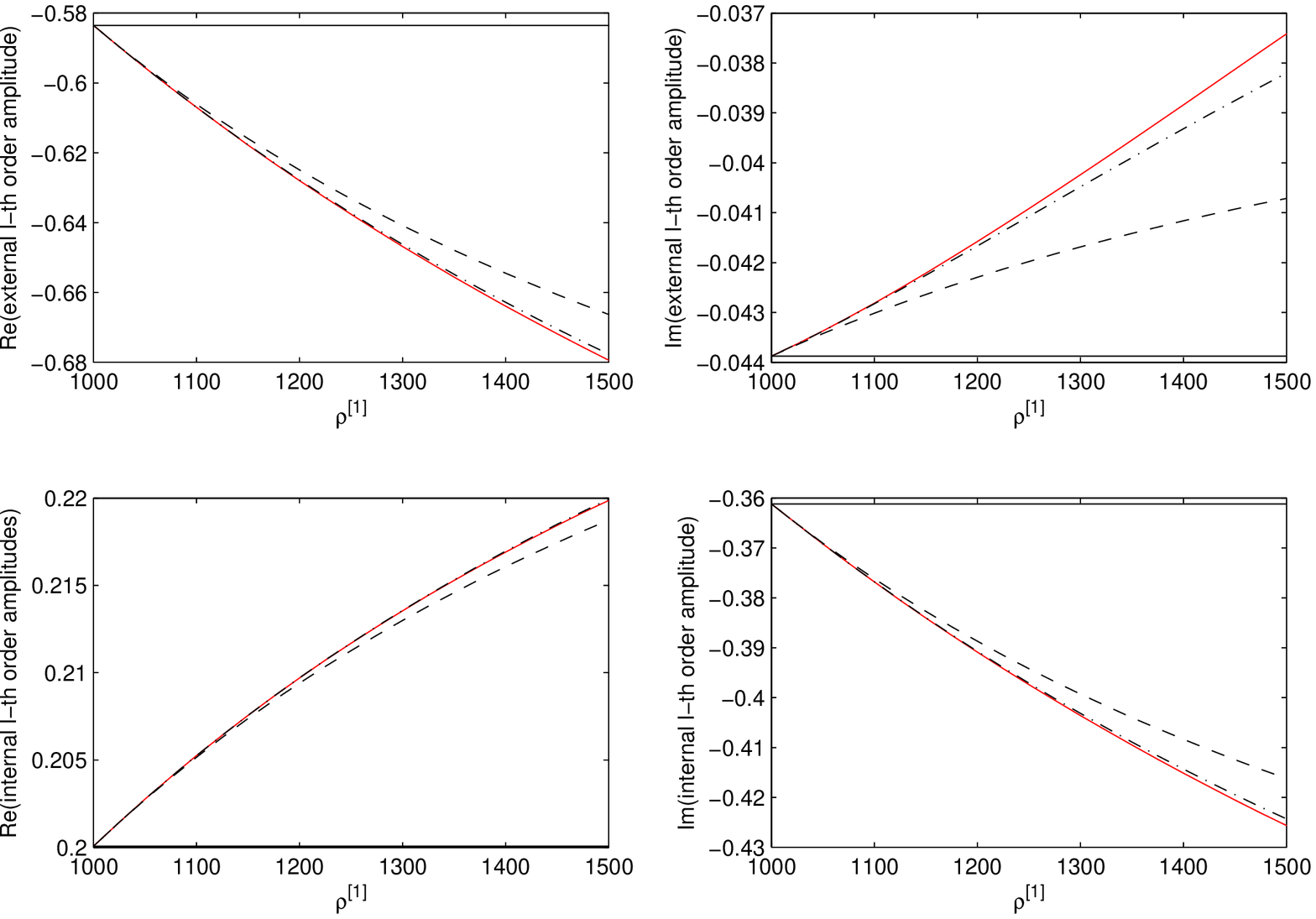}
\caption{Same as fig. \ref{fig1401} except that $l=2$.}
\label{fig1403}
\end{center}
\end{figure}
\begin{figure}[ptb]
\begin{center}
\includegraphics[width=0.75\textwidth]{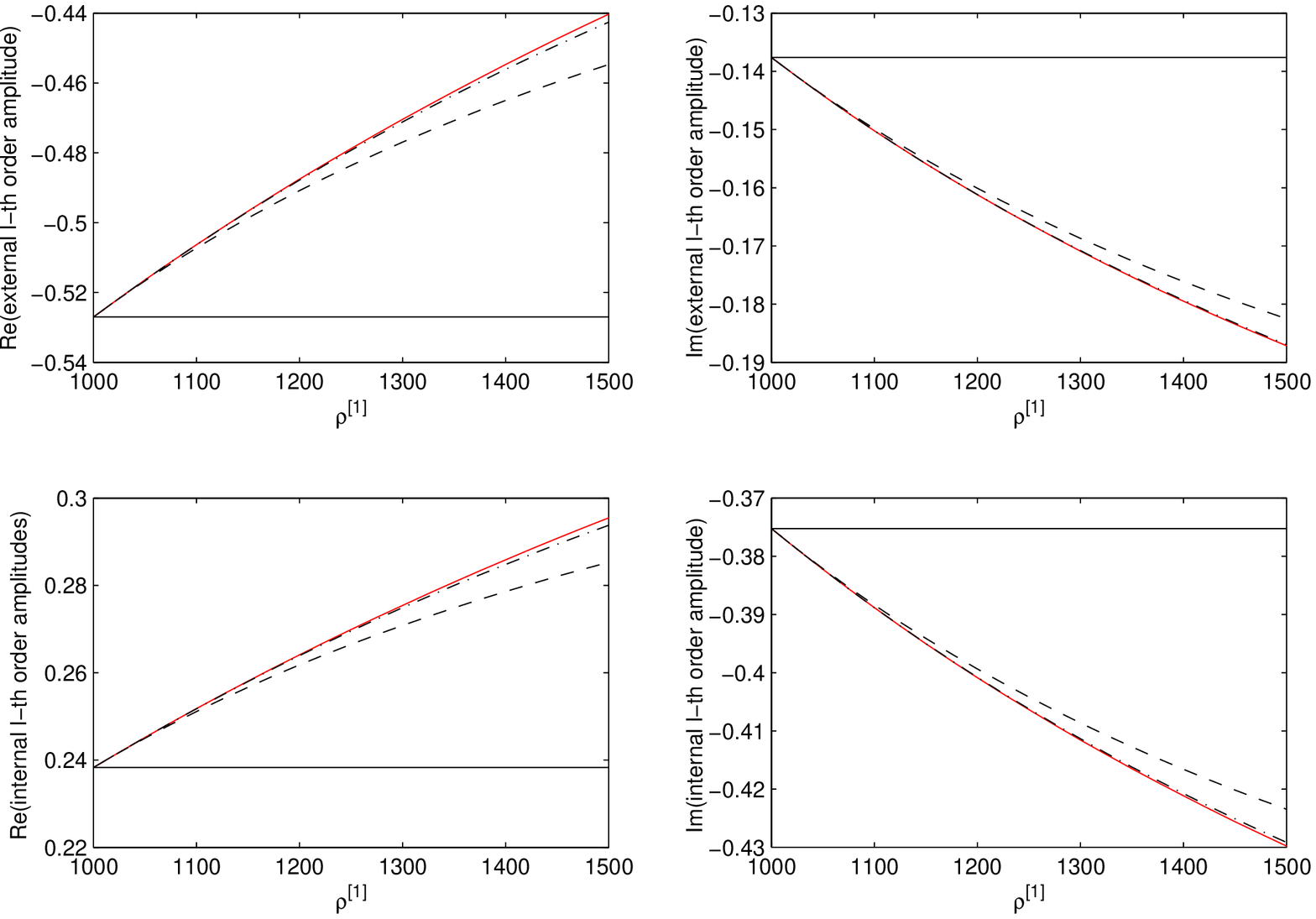}
\caption{Same as fig. \ref{fig1401} except that $l=3$.}
\label{fig1404}
\end{center}
\end{figure}
\clearpage
\newpage
The same remarks apply to this series of figures as in the  series of sect. \ref{fig01}.
%%%%%%%%%%%%%%%%%%%%%%%%%%%%%%%%%%%%%
\subsection{Variation of $\rho^{[1]}$ for various  orders $l$: case $\rho^{[1]}<\rho^{[0]}$ and higher $f$}\label{fig15}
\begin{figure}[ht]
\begin{center}
\includegraphics[width=0.75\textwidth]{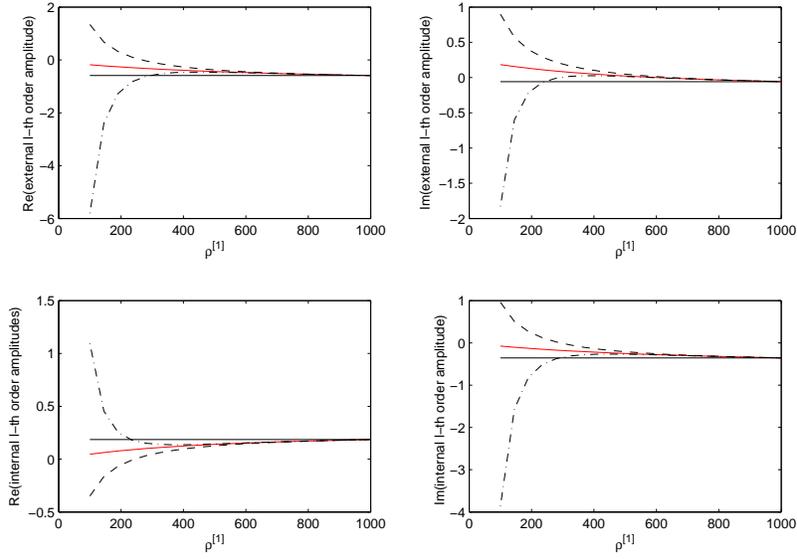}
\caption{Reflected and transmitted wavefield amplitudes as a function of $\rho^{[1]}<\rho^{[0]}$.  The upper left(right) panels depict the real(imaginary) parts of  $A_{l}^{[0]}$ (red),
$\mathcal{A}_{l}^{[0]}$ (black ------),
 $A_{l}^{[0](1)}$ (black - - - -), $A_{l}^{[0](2)}$ (black -.-.-.-).
The lower left(right) panels depict the real(imaginary) parts of  $B_{l}^{[1]}$ (red), $\mathcal{B}_{l}^{[1]}$ (black ------), $B_{l}^{[1](1)}$ (black - - - -), $B_{l}^{[1](2)}$ (black -.-.-.-). Case $c^{[1]}=1700-i210~ms^{-1}$, $f=20000~Hz$, $l=0$.}
\label{fig1501}
\end{center}
\end{figure}
\begin{figure}[ptb]
\begin{center}
\includegraphics[width=0.75\textwidth]{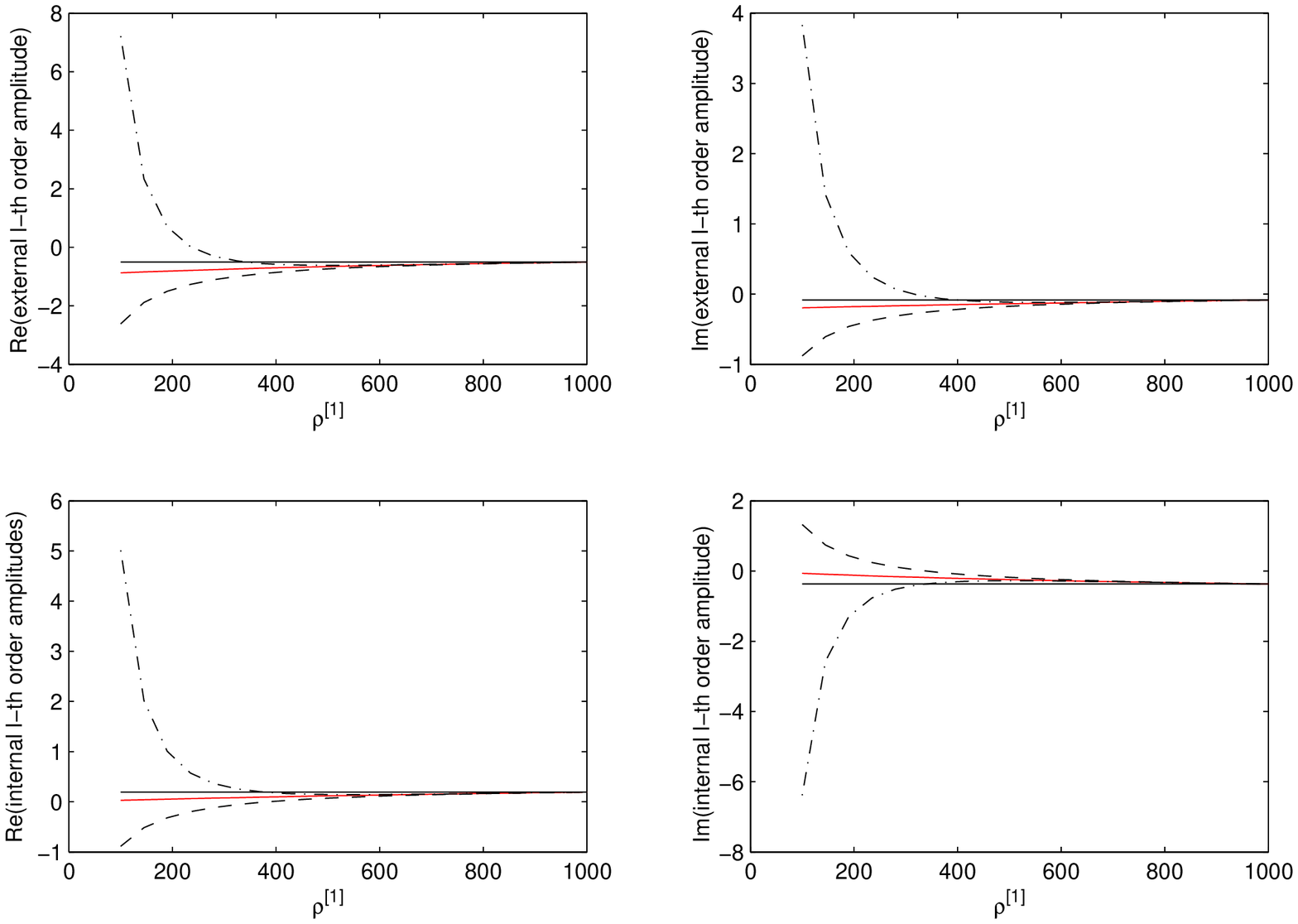}
\caption{Same as fig. \ref{fig1501} except that $l=1$.}
\label{fig1502}
\end{center}
\end{figure}
\begin{figure}[ptb]
\begin{center}
\includegraphics[width=0.75\textwidth]{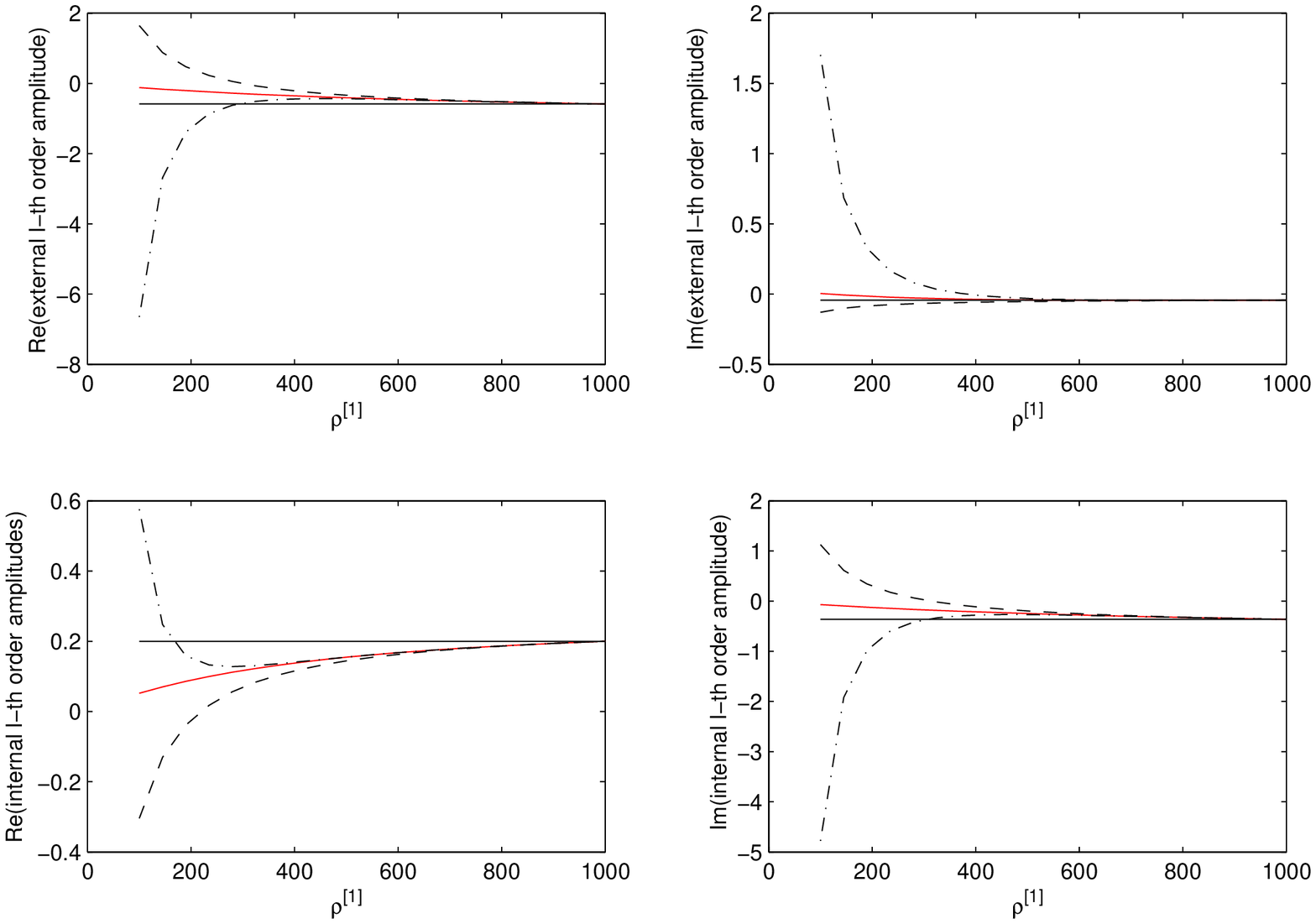}
\caption{Same as fig. \ref{fig1501} except that $l=2$.}
\label{fig1503}
\end{center}
\end{figure}
\begin{figure}[ptb]
\begin{center}
\includegraphics[width=0.75\textwidth]{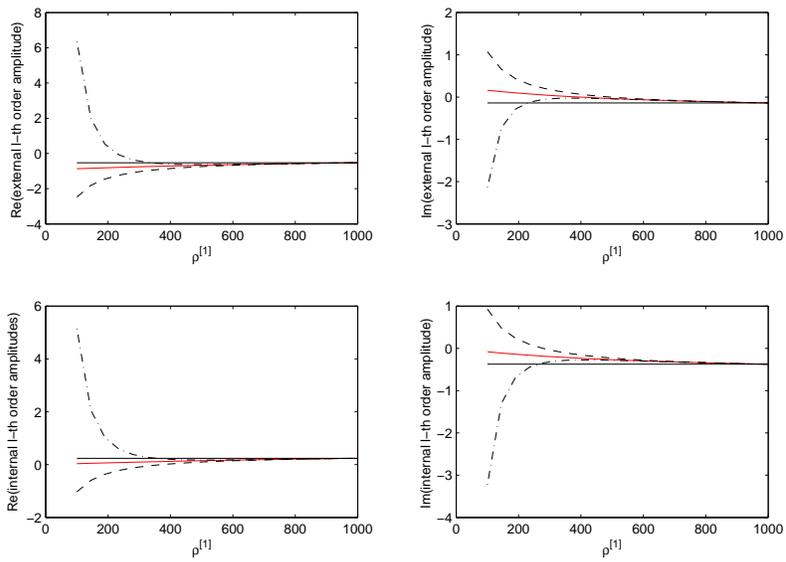}
\caption{Same as fig. \ref{fig1501} except that $l=3$.}
\label{fig1504}
\end{center}
\end{figure}
\clearpage
\newpage
The same remarks apply to this series of figures as in the  series of sect. \ref{fig01}.
%%%%%%%%%%%%%%%%%%%%%%%%%%%%%%%%%%%%%
\subsection{Variation of (low-to-medium) frequency $f$ for various orders $l$: case of fairly-large ($\epsilon=0.231$)  mass density contrast and large $\big|\Im(c^{[1]})\big|$}\label{fig03}
\begin{figure}[ht]
\begin{center}
\includegraphics[width=0.75\textwidth]{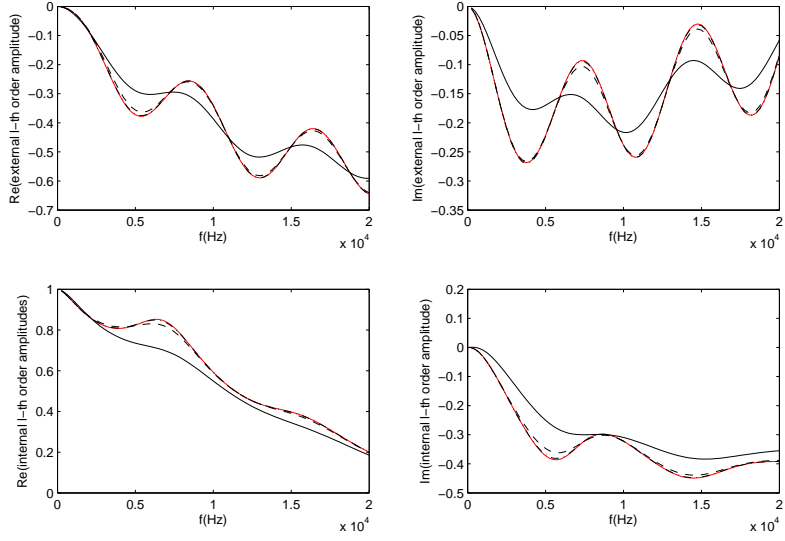}
\caption{Reflected and transmitted wavefield amplitudes as a function of $f$.  The upper left(right) panels depict the real(imaginary) parts of  $A_{l}^{[0]}$ (red),
$\mathcal{A}_{l}^{[0]}$ (black ------),
 $A_{l}^{[0](1)}$ (black - - - -), $A_{l}^{[0](2)}$ (black -.-.-.-).
The lower left(right) panels depict the real(imaginary) parts of  $B_{l}^{[1]}$ (red), $\mathcal{B}_{l}^{[1]}$ (black ------), $B_{l}^{[1](1)}$ (black - - - -), $B_{l}^{[1](2)}$ (black -.-.-.-). Case $\rho^{[1]}=1300~Kgm^{-3}$, $c^{[1]}=1700-i210~ms^{-1}$, $l=0$.}
\label{fig301}
\end{center}
\end{figure}
\begin{figure}[ptb]
\begin{center}
\includegraphics[width=0.75\textwidth]{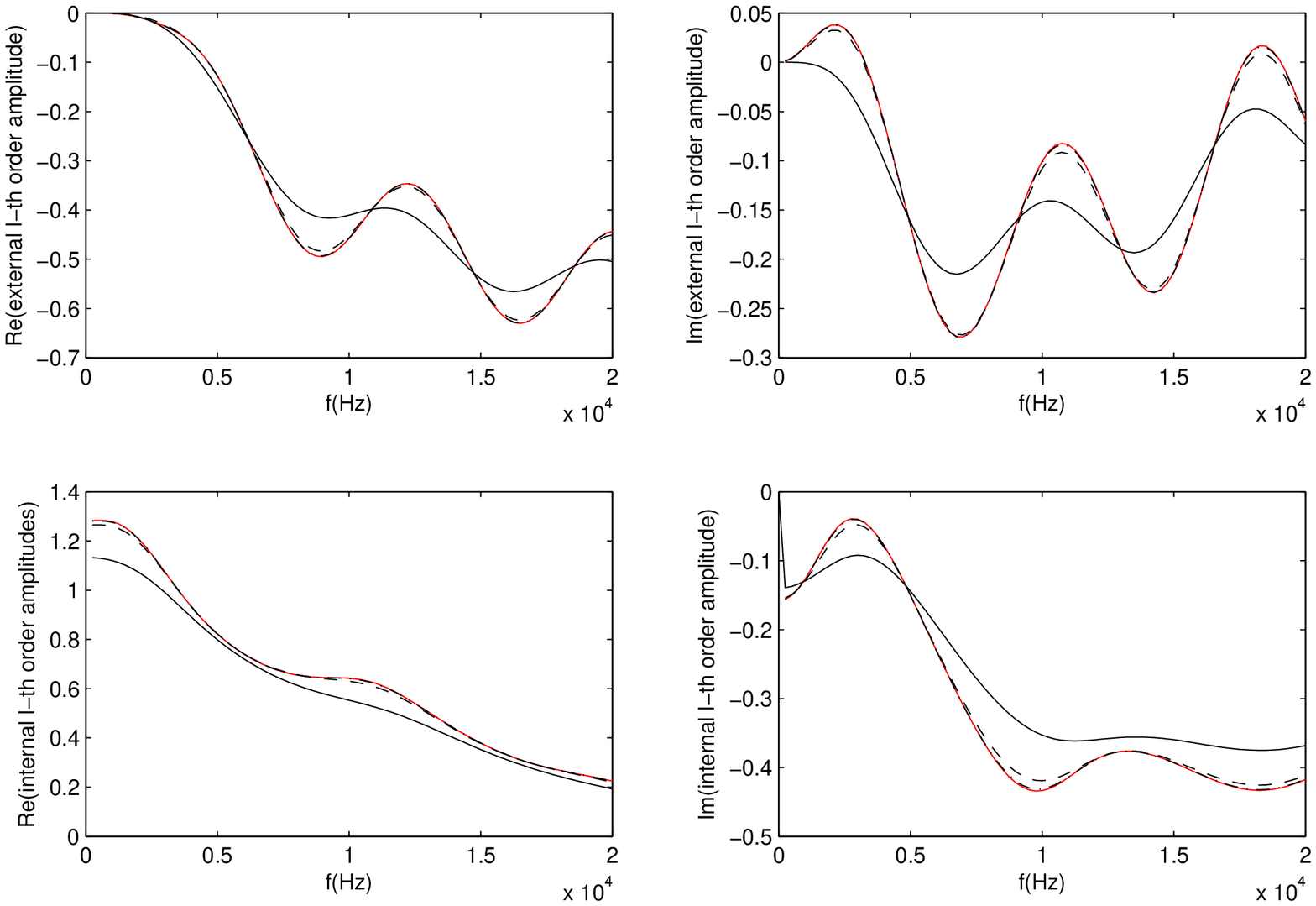}
\caption{Same as fig. \ref{fig301} except that $l=1$.}
\label{fig302}
\end{center}
\end{figure}
\begin{figure}[ptb]
\begin{center}
\includegraphics[width=0.75\textwidth]{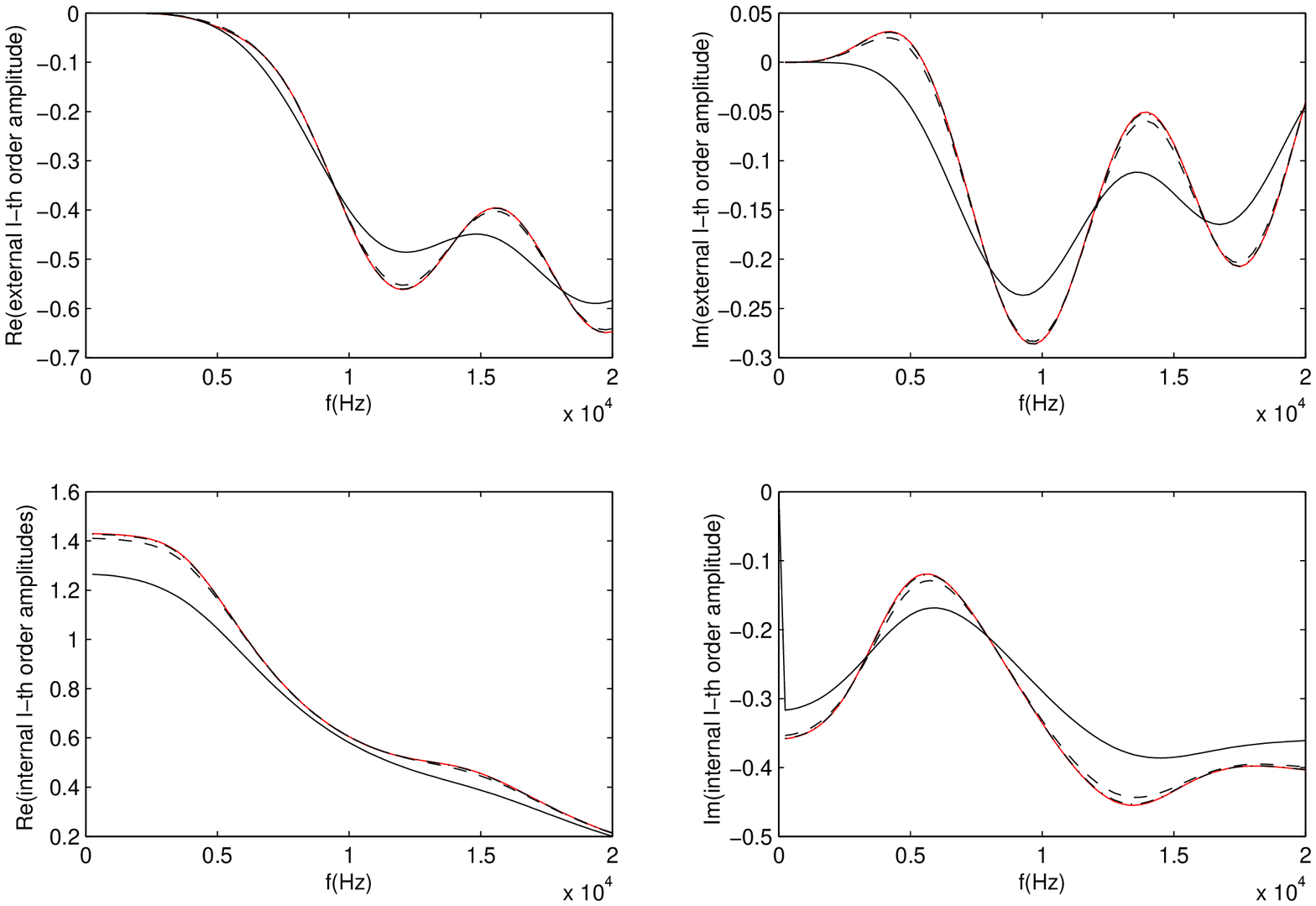}
\caption{Same as fig. \ref{fig301} except that $l=2$.}
\label{fig303}
\end{center}
\end{figure}
\begin{figure}[ptb]
\begin{center}
\includegraphics[width=0.75\textwidth]{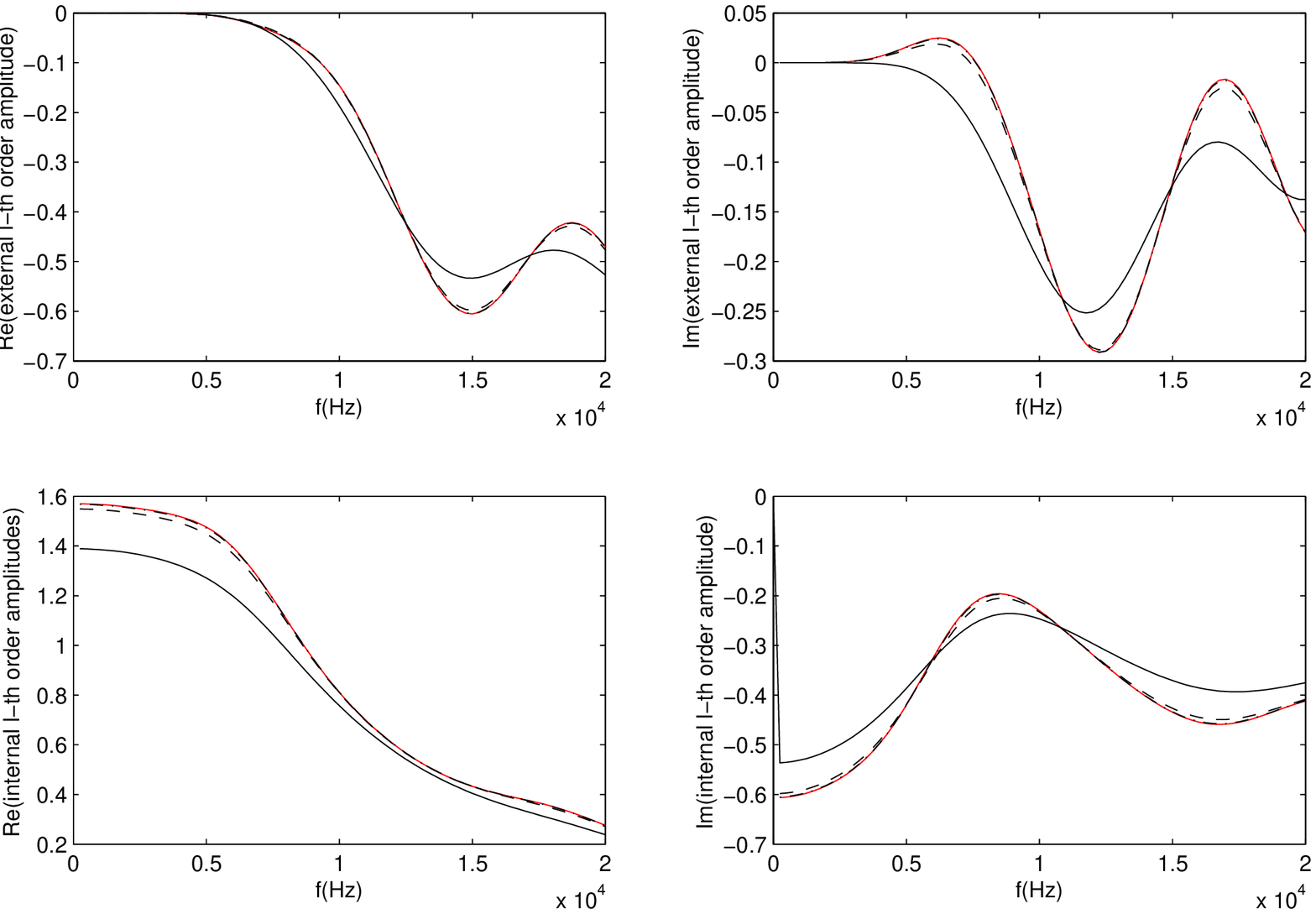}
\caption{Same as fig. \ref{fig301} except that $l=3$.}
\label{fig304}
\end{center}
\end{figure}
\clearpage
\newpage
We observe, between $\mathcal{A}_{l}^{[0]},~\mathcal{B}_{l}^{[1]}$ and $A_{l}^{[0]},~B_{l}^{[1]}$ respectively, substantial qualitative, and even quantitative, differences, that are particularly pronounced near the extrema of the amplitudes. Better agreement is attained, both quantitatively and qualitatively, by means of $A_{l}^{[0](1)},~B_{l}^{[1](1)}$, and near-coincidence with $A_{l}^{[0]},~B_{l}^{[1]}$ is obtained by means of $A_{l}^{[0](2)},~B_{l}^{[1](2)}$ respectively.
%%%%%%%%%%%%%%%%%%%%%%%%%%%%%%%%%%%%%%%%%%%%%%%%%%%%%
\subsection{Variation of frequency $f$ for various orders $l$: case of relatively-small ($\epsilon=0.091$) density contrast and large $\big|\Im(c^{[1]})\big|$}\label{fig04}
\begin{figure}[ht]
\begin{center}
\includegraphics[width=0.75\textwidth]{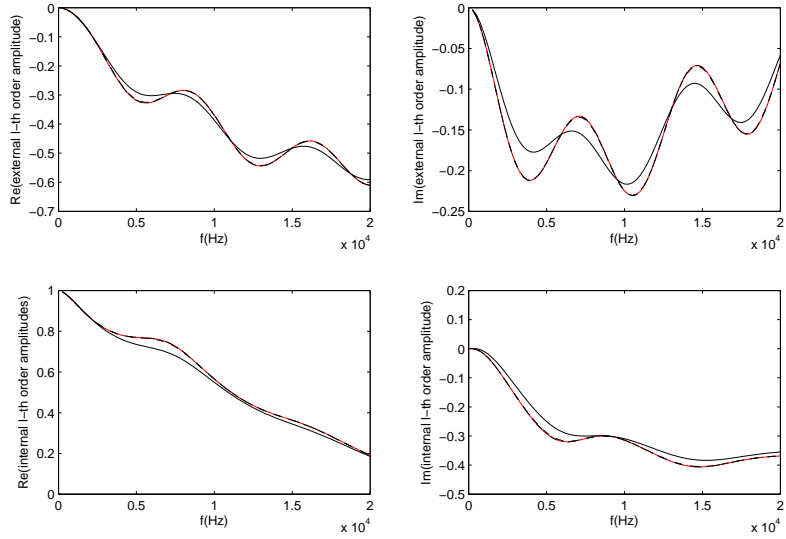}
\caption{Reflected and transmitted wavefield amplitudes as a function of $f$.  The upper left(right) panels depict the real(imaginary) parts of  $A_{l}^{[0]}$ (red),
$\mathcal{A}_{l}^{[0]}$ (black ------),
 $A_{l}^{[0](1)}$ (black - - - -), $A_{l}^{[0](2)}$ (black -.-.-.-).
The lower left(right) panels depict the real(imaginary) parts of  $B_{l}^{[1]}$ (red), $\mathcal{B}_{l}^{[1]}$ (black ------), $B_{l}^{[1](1)}$ (black - - - -), $B_{l}^{[1](2)}$ (black -.-.-.-). Case $\rho^{[1]}=1100~Kgm^{-3}$, $c^{[1]}=1700-i210~ms^{-1}$, $l=0$.}
\label{fig401}
\end{center}
\end{figure}
\begin{figure}[ptb]
\begin{center}
\includegraphics[width=0.75\textwidth]{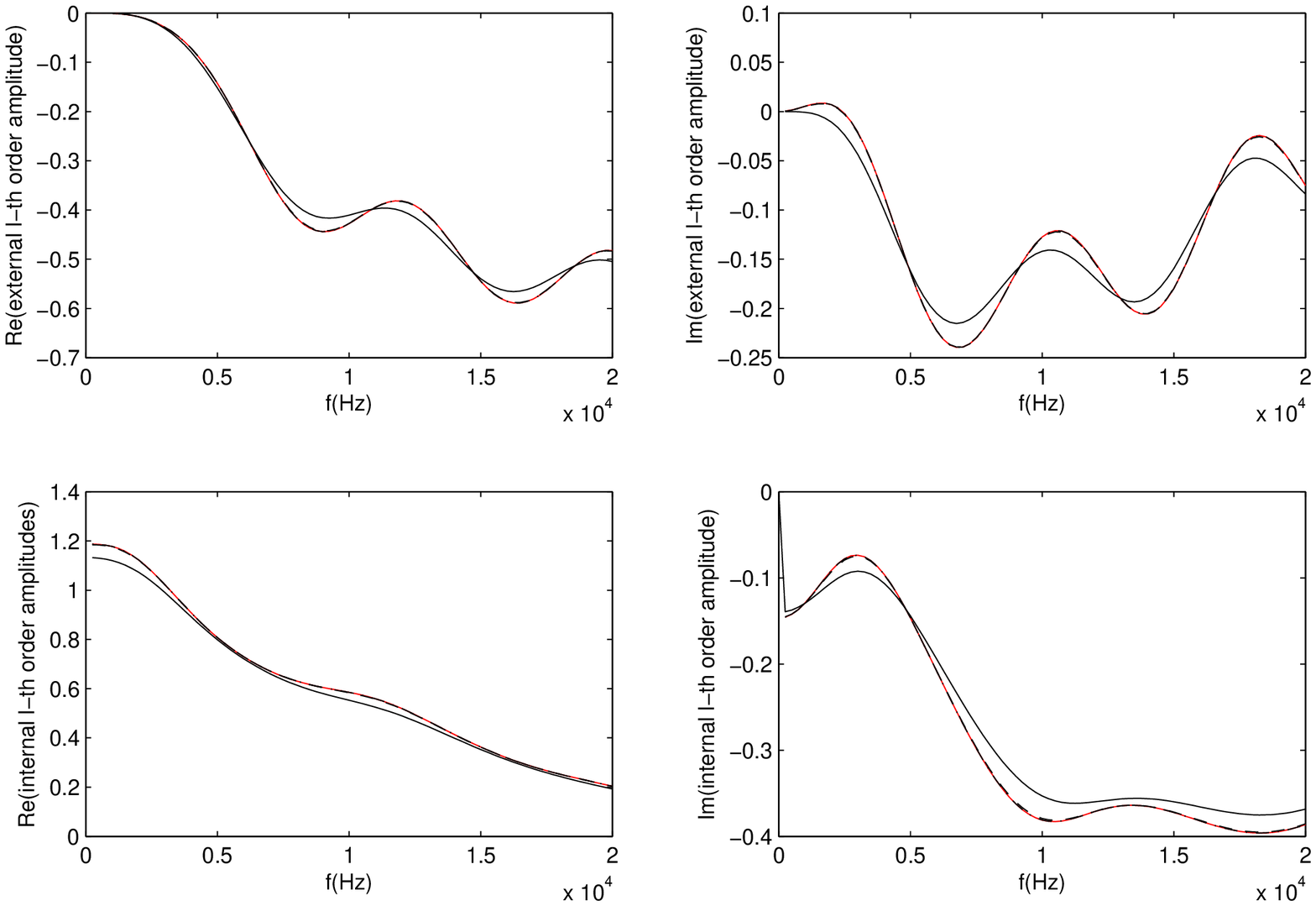}
\caption{Same as fig. \ref{fig401} except that $l=1$.}
\label{fig402}
\end{center}
\end{figure}
\begin{figure}[ptb]
\begin{center}
\includegraphics[width=0.75\textwidth]{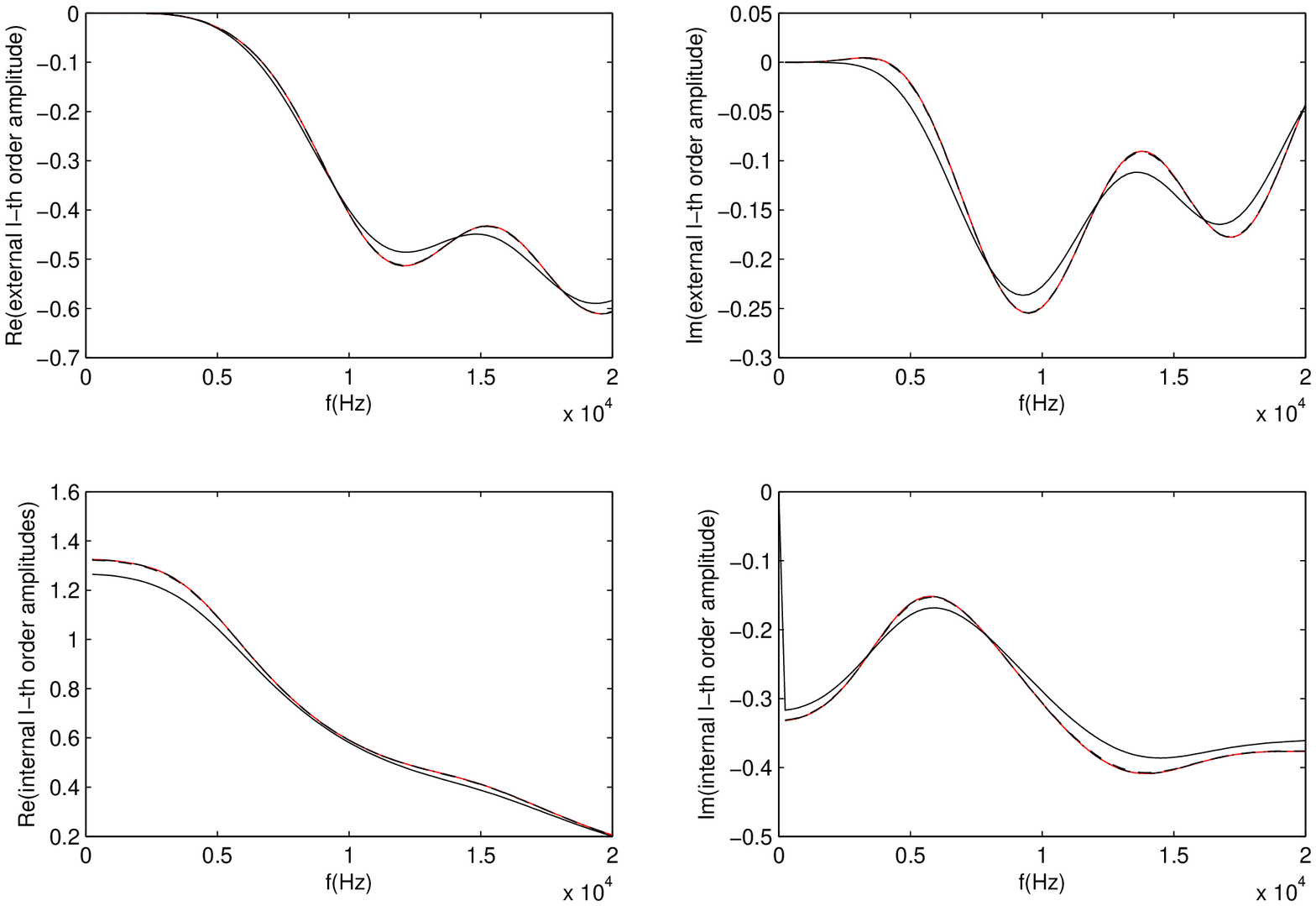}
\caption{Same as fig. \ref{fig401} except that $l=2$.}
\label{fig403}
\end{center}
\end{figure}
\begin{figure}[ptb]
\begin{center}
\includegraphics[width=0.75\textwidth]{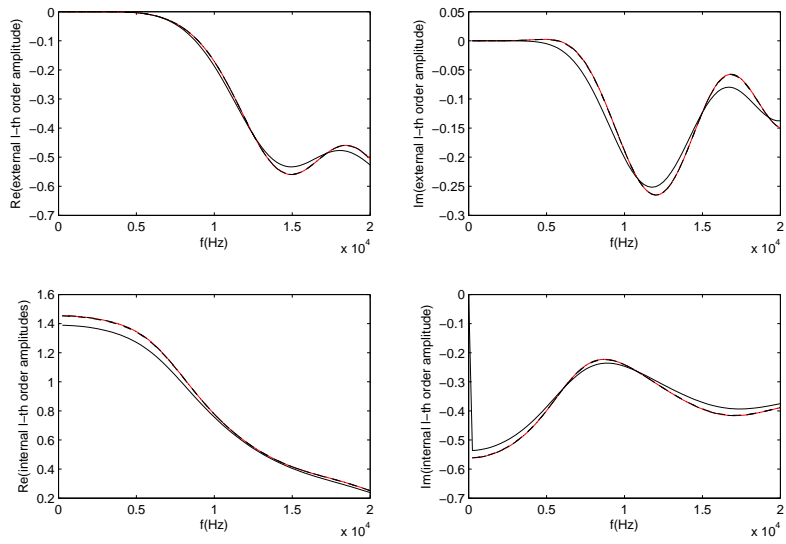}
\caption{Same as fig. \ref{fig401} except that $l=3$.}
\label{fig404}
\end{center}
\end{figure}
\clearpage
\newpage
As expected the differences between $\mathcal{A}_{l}^{[0]},~\mathcal{B}_{l}^{[1]}$ and $A_{l}^{[0]},~B_{l}^{[1]}$ respectively are less-pronounced than in sect. \ref{fig03}. Otherwise, the same remarks apply here than in the previous section.
%%%%%%%%%%%%%%%%%%%%%%%%%%%%%%%%%%%%%%%%%%%%%%%%%%%%%%%%%%%%%
\subsection{Variation of (low-to-medium) frequency $f$ for various orders $l$: case of relatively-small ($\epsilon=0.091$) small density contrast and small $\big|\Im(c^{[1]})\big|$}\label{fig05}
\begin{figure}[ht]
\begin{center}
\includegraphics[width=0.75\textwidth]{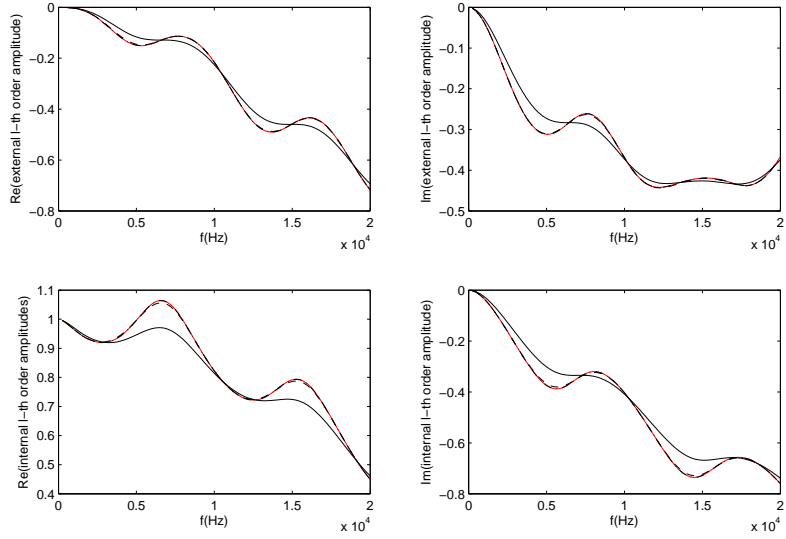}
\caption{Reflected and transmitted wavefield amplitudes as a function of $f$.  The upper left(right) panels depict the real(imaginary) parts of  $A_{l}^{[0]}$ (red),
$\mathcal{A}_{l}^{[0]}$ (black ------),
 $A_{l}^{[0](1)}$ (black - - - -), $A_{l}^{[0](2)}$ (black -.-.-.-).
The lower left(right) panels depict the real(imaginary) parts of  $B_{l}^{[1]}$ (red), $\mathcal{B}_{l}^{[1]}$ (black ------), $B_{l}^{[1](1)}$ (black - - - -), $B_{l}^{[1](2)}$ (black -.-.-.-). Case $\rho^{[1]}=1100~Kgm^{-3}$, $c^{[1]}=1700-i21~ms^{-1}$, $l=0$.}
\label{fig501}
\end{center}
\end{figure}
\begin{figure}[ptb]
\begin{center}
\includegraphics[width=0.75\textwidth]{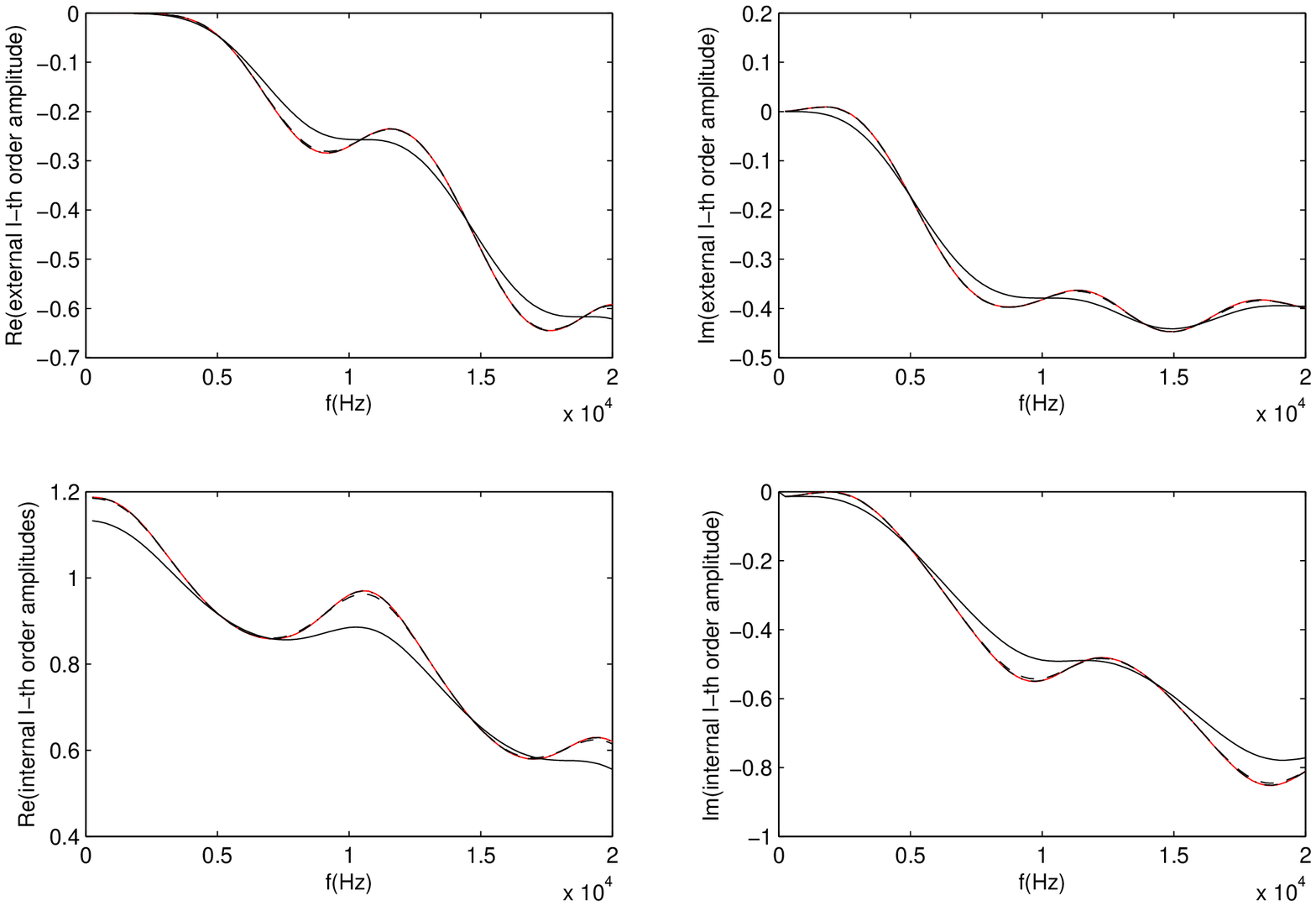}
\caption{Same as fig. \ref{fig501} except that $l=1$.}
\label{fig502}
\end{center}
\end{figure}
\begin{figure}[ptb]
\begin{center}
\includegraphics[width=0.75\textwidth]{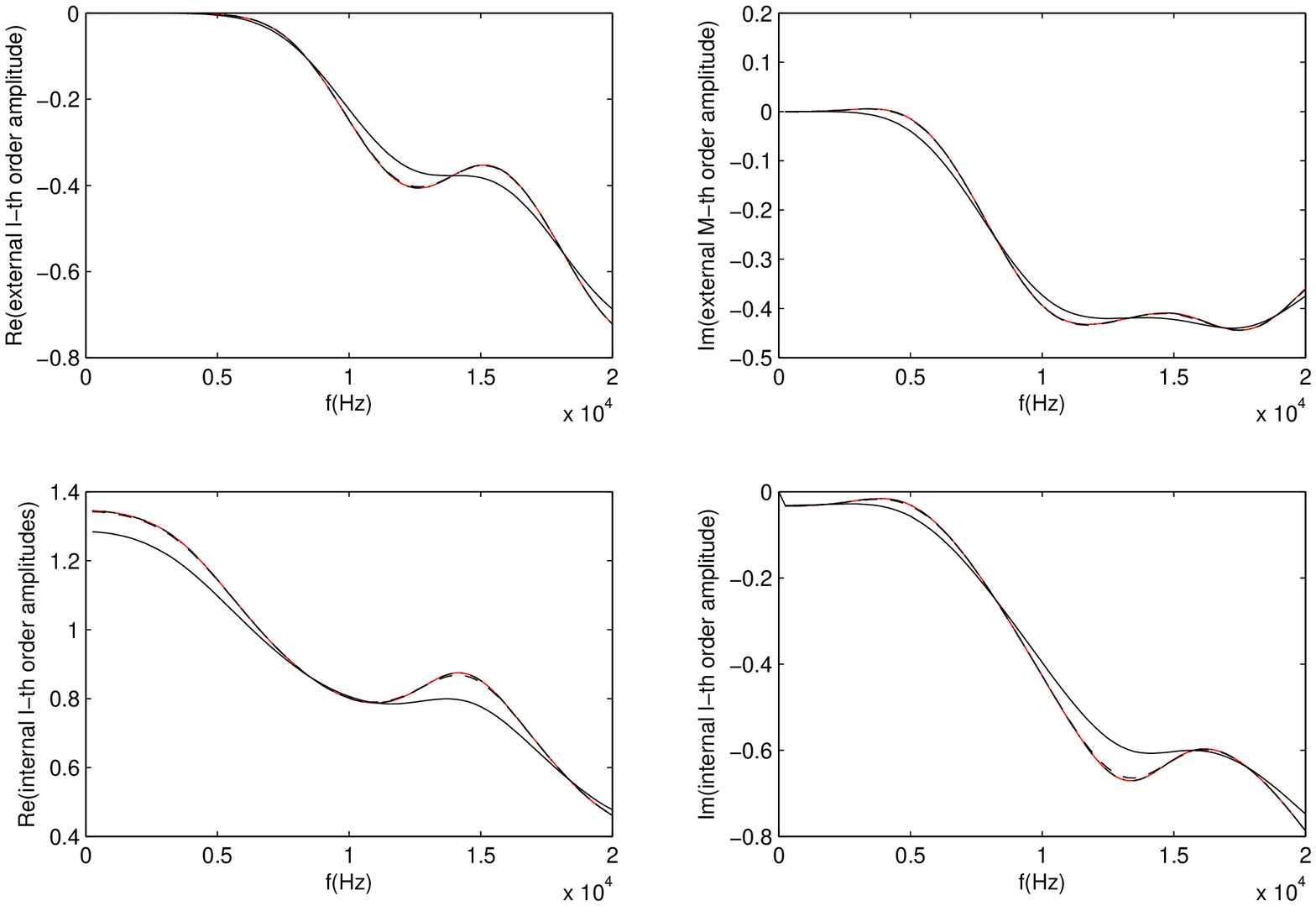}
\caption{Same as fig. \ref{fig501} except that $l=2$.}
\label{fig503}
\end{center}
\end{figure}
\begin{figure}[ptb]
\begin{center}
\includegraphics[width=0.75\textwidth]{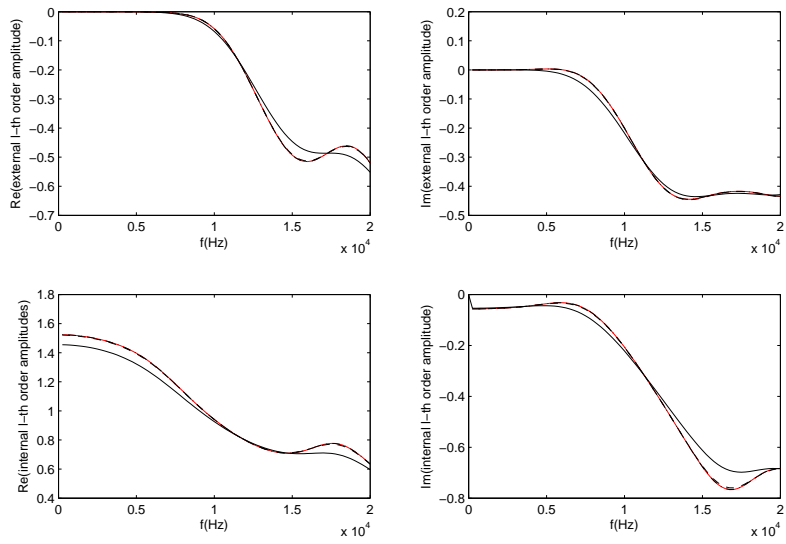}
\caption{Same as fig. \ref{fig501} except that $l=3$.}
\label{fig504}
\end{center}
\end{figure}
\clearpage
\newpage
Due to the lesser attenuation in the obstacle, the differences between $\mathcal{A}_{l}^{[0]},~\mathcal{B}_{l}^{[1]}$ and $A_{l}^{[0]},~B_{l}^{[1]}$ respectively are observed to be more-pronounced than in sect. \ref{fig04}. Otherwise, the same remarks apply here than in the previous section.
%%%%%%%%%%%%%%%%%%%%%%%%%%%%%%%%%%%%%%%%%%%%%%%%%%%%%%%%%%%%%
\subsection{Variation of low-to-medium frequency $f$ for various orders $l$: case of fairly-large ($\varepsilon=0.231$) mass density contrast and large  wavespeed contrast  for $\Re(c^{[1]})>c^{[0]}$}\label{fig06}
\begin{figure}[ht]
\begin{center}
\includegraphics[width=0.75\textwidth]{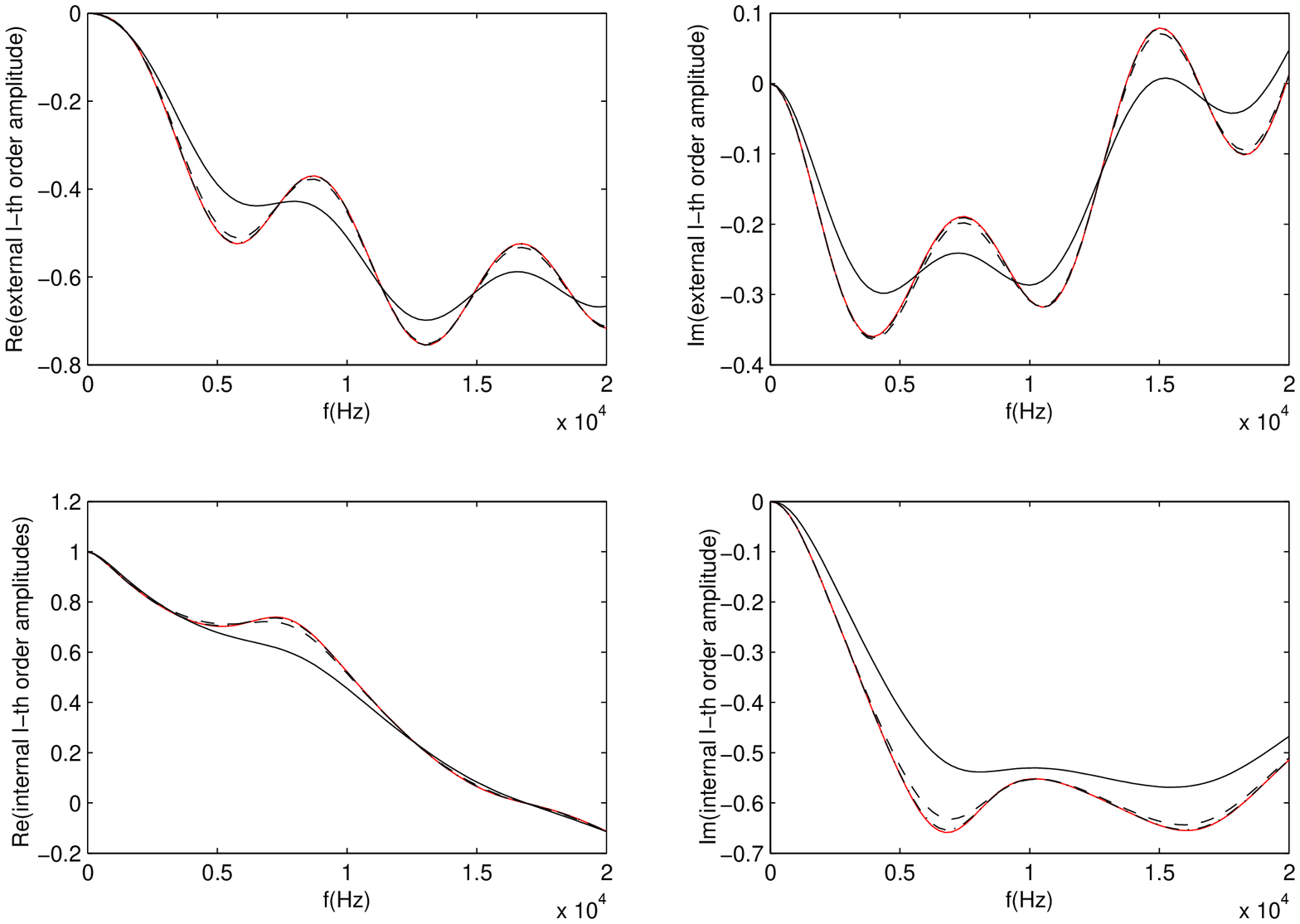}
\caption{Reflected and transmitted wavefield amplitudes as a function of $f$.  The upper left(right) panels depict the real(imaginary) parts of  $A_{l}^{[0]}$ (red),
$\mathcal{A}_{l}^{[0]}$ (black ------),
 $A_{l}^{[0](1)}$ (black - - - -), $A_{l}^{[0](2)}$ (black -.-.-.-).
The lower left(right) panels depict the real(imaginary) parts of  $B_{l}^{[1]}$ (red), $\mathcal{B}_{l}^{[1]}$ (black ------), $B_{l}^{[1](1)}$ (black - - - -), $B_{l}^{[1](2)}$ (black -.-.-.-). Case $\rho^{[1]}=1300~Kgm^{-3}$, $c^{[1]}=1900-i210~ms^{-1}$, $l=0$.}
\label{fig601}
\end{center}
\end{figure}
\begin{figure}[ptb]
\begin{center}
\includegraphics[width=0.75\textwidth]{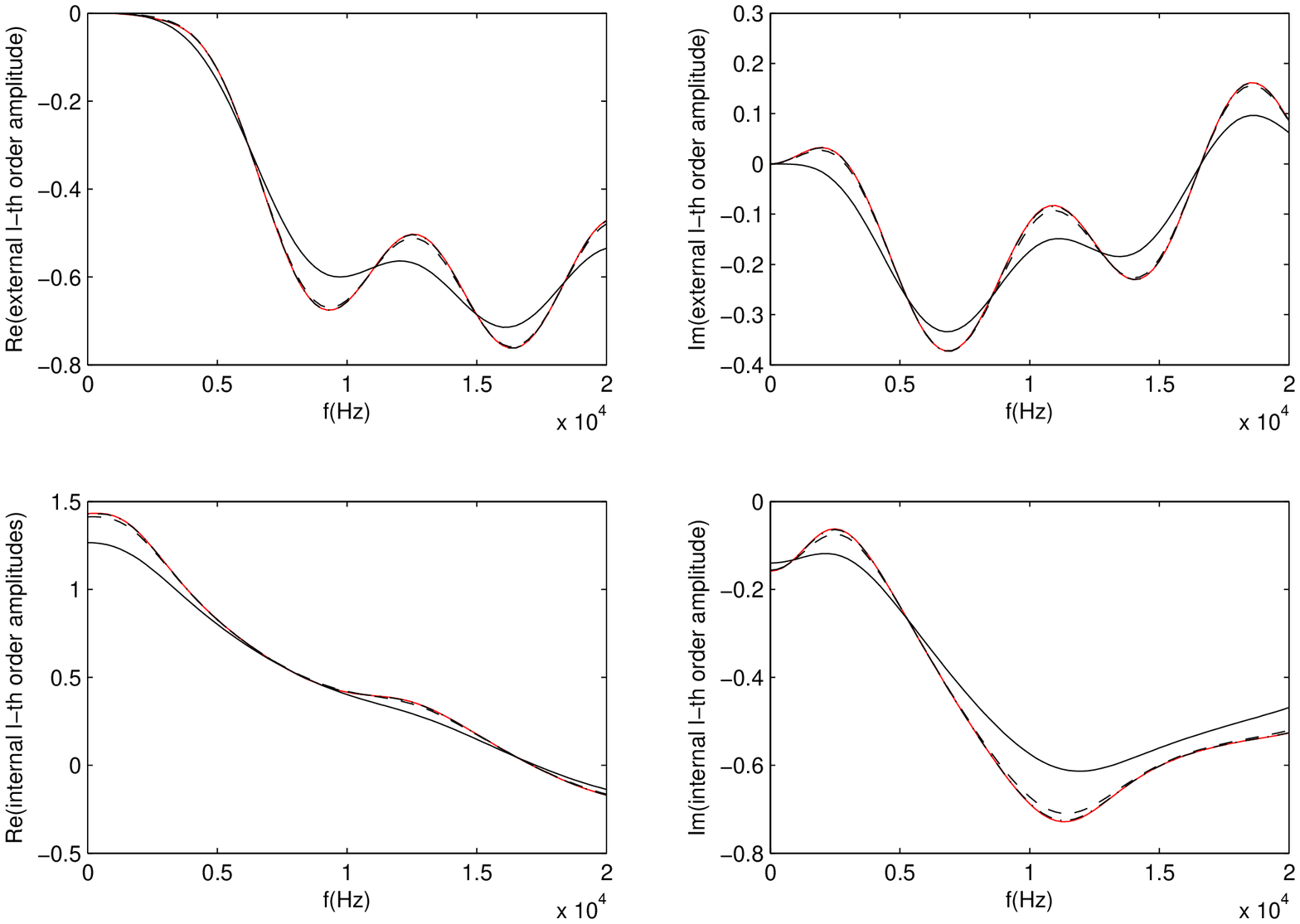}
\caption{Same as fig. \ref{fig601} except that $l=1$.}
\label{fig602}
\end{center}
\end{figure}
\begin{figure}[ptb]
\begin{center}
\includegraphics[width=0.75\textwidth]{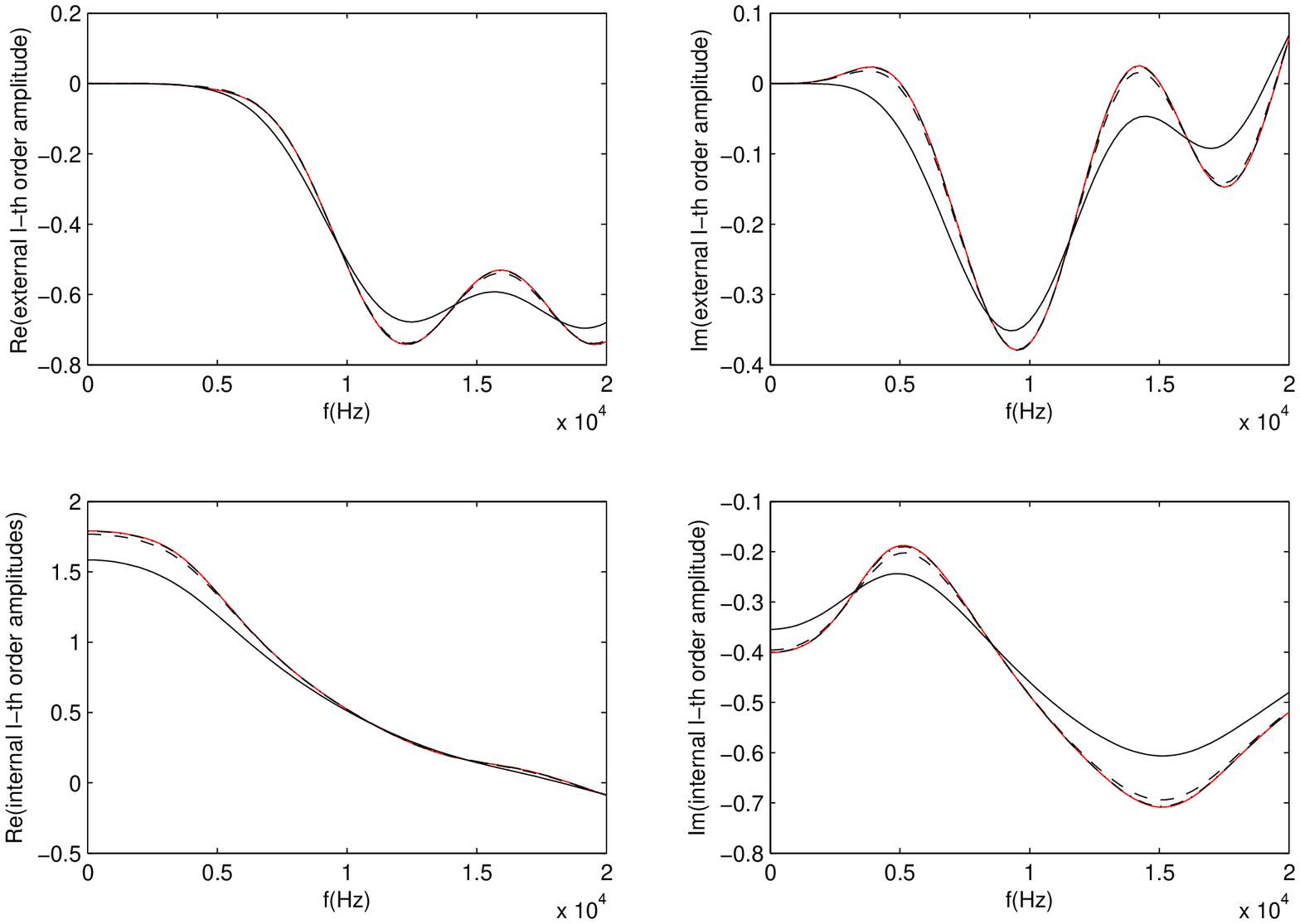}
\caption{Same as fig. \ref{fig601} except that $l=2$.}
\label{fig603}
\end{center}
\end{figure}
\begin{figure}[ptb]
\begin{center}
\includegraphics[width=0.75\textwidth]{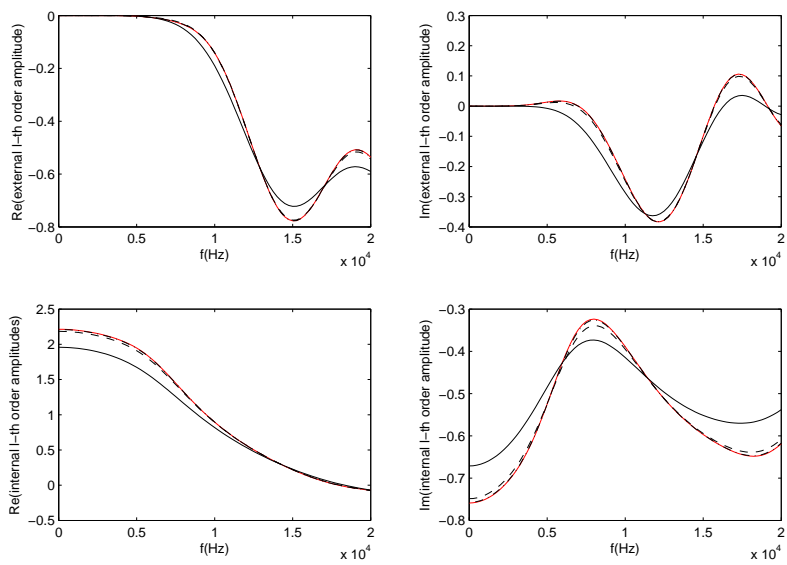}
\caption{Same as fig. \ref{fig601} except that $l=3$.}
\label{fig604}
\end{center}
\end{figure}
\clearpage
\newpage
We observe, between $\mathcal{A}_{l}^{[0]},~\mathcal{B}_{l}^{[1]}$ and $A_{l}^{[0]},~B_{l}^{[1]}$ respectively, substantial qualitative, and even quantitative, differences, that are particularly pronounced near the extrema of the amplitudes. Better agreement is attained,  both quantitatively and qualitatively, by means of $A_{l}^{[0](1)},~B_{l}^{[1](1)}$, and near-coincidence with $A_{l}^{[0]},~B_{l}^{[1]}$ is obtained by means of $A_{l}^{[0](2)},~B_{l}^{[1](2)}$ respectively.
%%%%%%%%%%%%%%%%%%%%%%%%%%%%%%%%%%%%%%%%%%%%%%%%%%%%%%%%%%%%%
\subsection{Variation of low-to-medium frequency $f$ for various orders $l$: case  of fairly-large ($\varepsilon=0.231$) mass density contrast and small  wavespeed contrast  for $\Re(c^{[1]})>c^{[0]}$}\label{fig07}
\begin{figure}[ht]
\begin{center}
\includegraphics[width=0.75\textwidth]{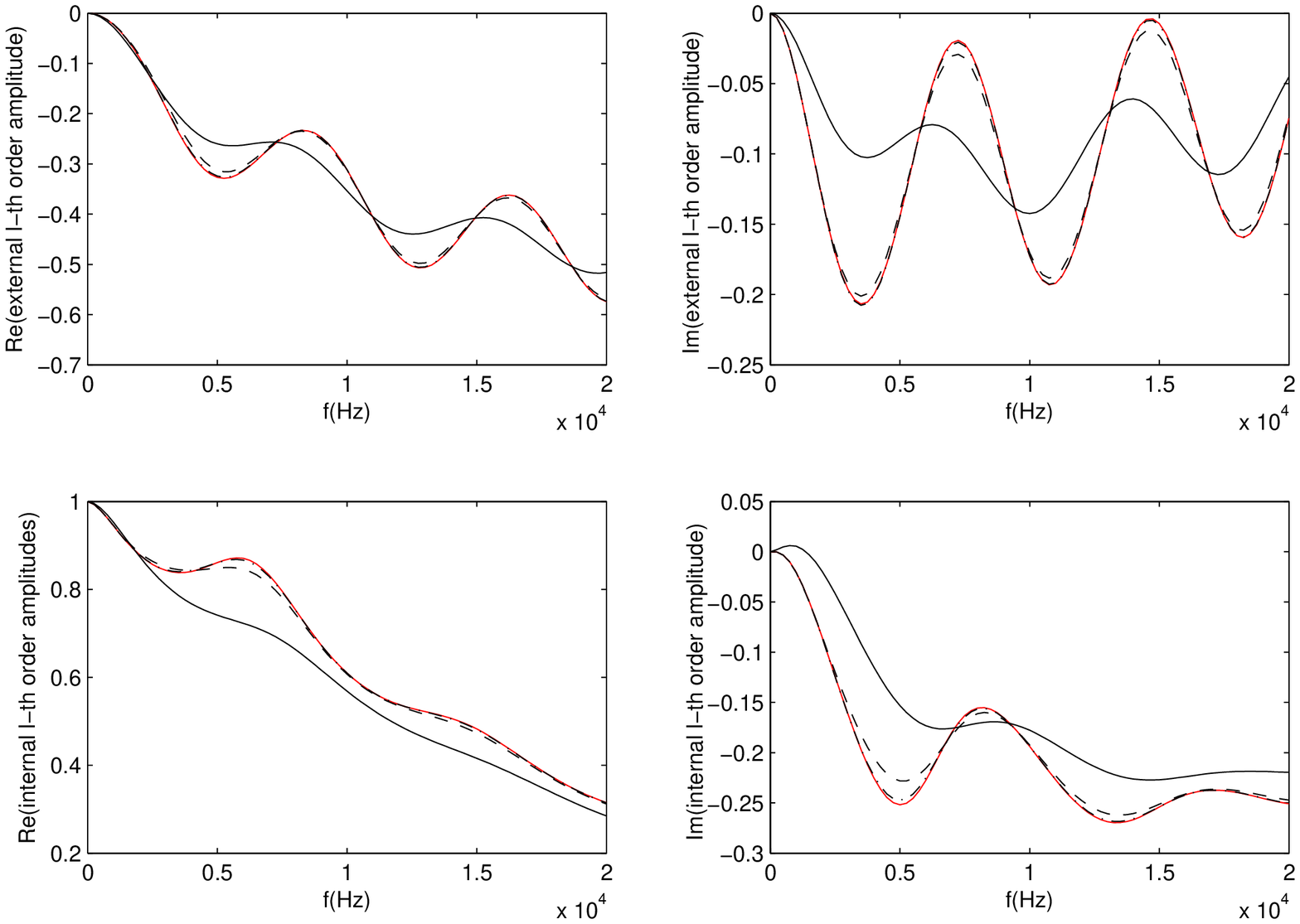}
\caption{Reflected and transmitted wavefield amplitudes as a function of $f$.  The upper left(right) panels depict the real(imaginary) parts of  $A_{l}^{[0]}$ (red),
$\mathcal{A}_{l}^{[0]}$ (black ------),
 $A_{l}^{[0](1)}$ (black - - - -), $A_{l}^{[0](2)}$ (black -.-.-.-).
The lower left(right) panels depict the real(imaginary) parts of  $B_{l}^{[1]}$ (red), $\mathcal{B}_{l}^{[1]}$ (black ------), $B_{l}^{[1](1)}$ (black - - - -), $B_{l}^{[1](2)}$ (black -.-.-.-). Case $\rho^{[1]}=1300~Kgm^{-3}$, $c^{[1]}=1600-i210~ms^{-1}$, $l=0$.}
\label{fig701}
\end{center}
\end{figure}
\begin{figure}[ptb]
\begin{center}
\includegraphics[width=0.75\textwidth]{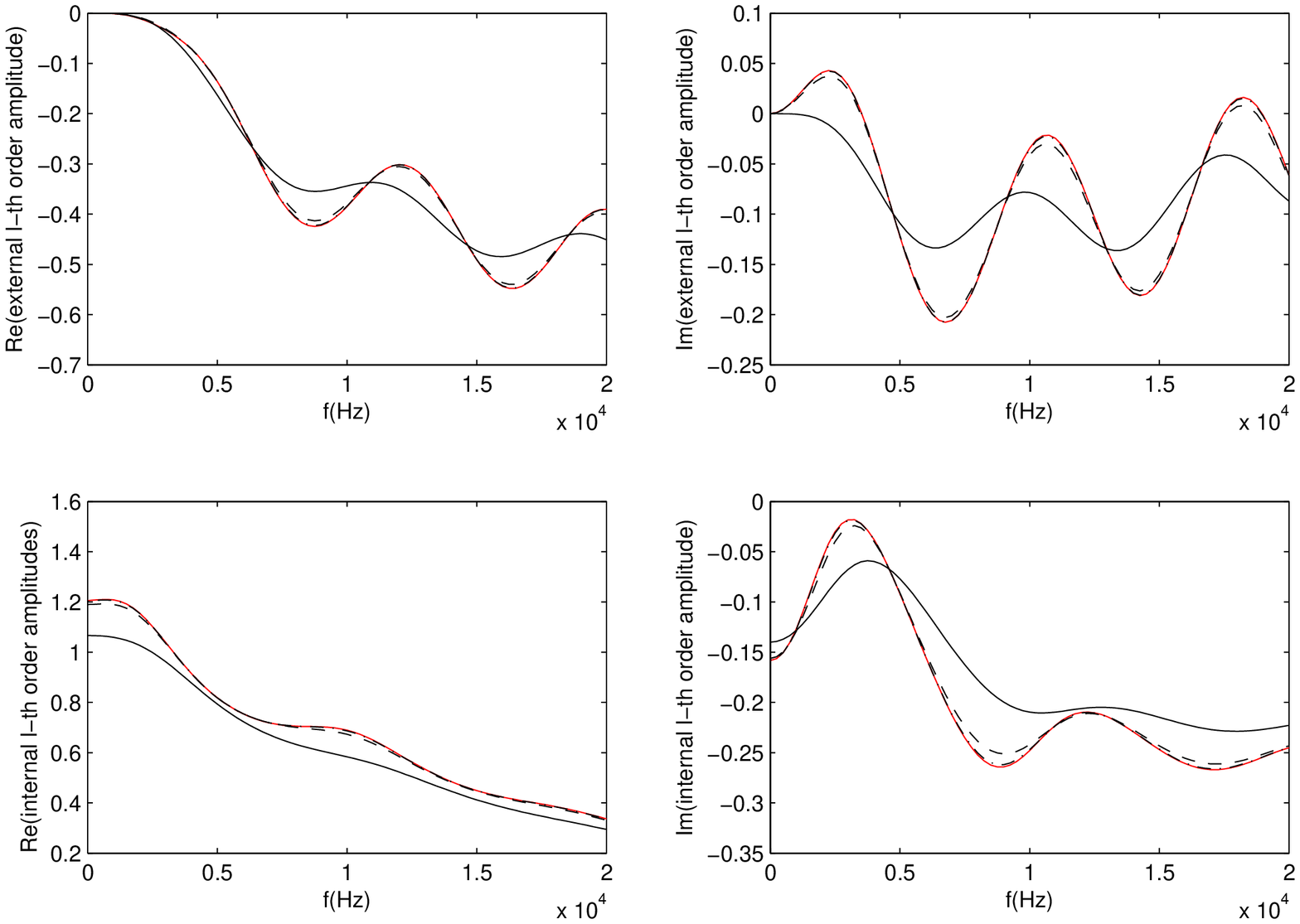}
\caption{Same as fig. \ref{fig701} except that $l=1$.}
\label{fig702}
\end{center}
\end{figure}
\begin{figure}[ptb]
\begin{center}
\includegraphics[width=0.75\textwidth]{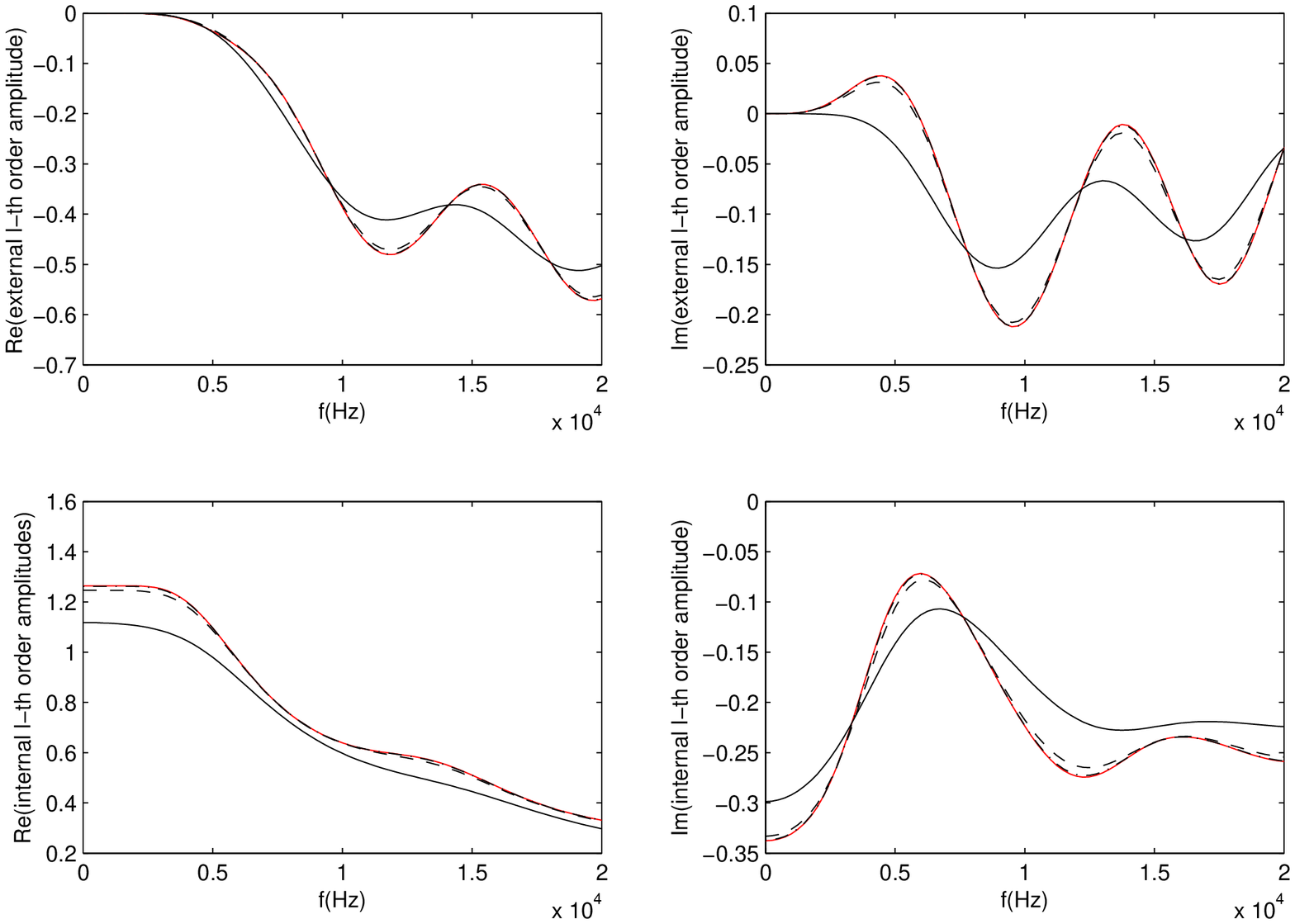}
\caption{Same as fig. \ref{fig701} except that $l=2$.}
\label{fig703}
\end{center}
\end{figure}
\begin{figure}[ptb]
\begin{center}
\includegraphics[width=0.75\textwidth]{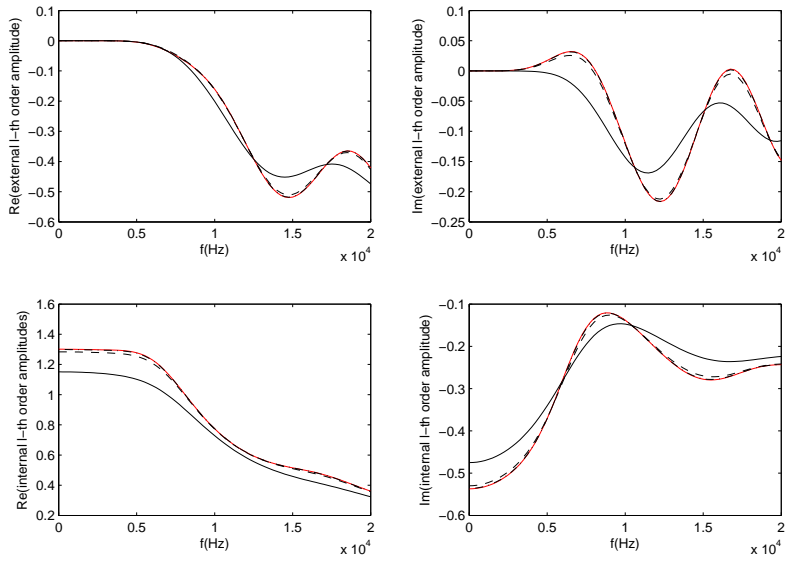}
\caption{Same as fig. \ref{fig701} except that $l=3$.}
\label{fig704}
\end{center}
\end{figure}
\clearpage
\newpage
We observe, between $\mathcal{A}_{l}^{[0]},~\mathcal{B}_{l}^{[1]}$ and $A_{l}^{[0]},~B_{l}^{[1]}$ the same qualitative, and somewhat aggravated quantitative, differences as in sect. \ref{fig06}. Otherwise, the same comments as in the previous section apply here.
%%%%%%%%%%%%%%%%%%%%%%%%%%%%%%%%%%%%%%%%%%%%%%%%%%%%%%%%%%%%%
\subsection{Variation of low-to-medium frequency $f$ for various orders $l$: case of fairly-large ($\varepsilon=0.231$) mass density contrast and small  wavespeed velocity contrast for $\Re(c^{[1]})<c^{[0]}$}\label{fig08}
\begin{figure}[ht]
\begin{center}
\includegraphics[width=0.75\textwidth]{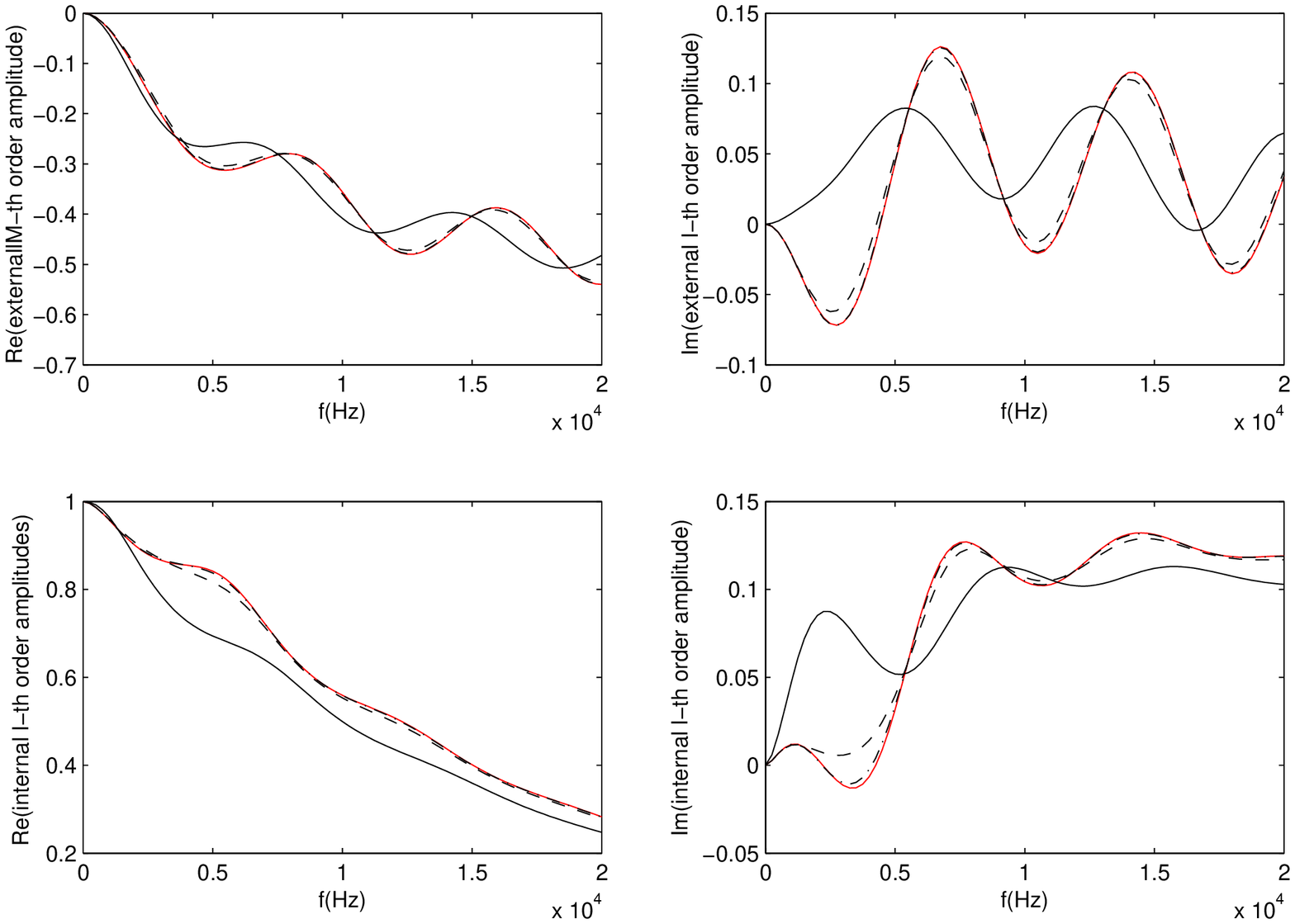}
\caption{Reflected and transmitted wavefield amplitudes as a function of $f$.  The upper left(right) panels depict the real(imaginary) parts of  $A_{l}^{[0]}$ (red),
$\mathcal{A}_{l}^{[0]}$ (black ------),
 $A_{l}^{[0](1)}$ (black - - - -), $A_{l}^{[0](2)}$ (black -.-.-.-).
The lower left(right) panels depict the real(imaginary) parts of  $B_{l}^{[1]}$ (red), $\mathcal{B}_{l}^{[1]}$ (black ------), $B_{l}^{[1](1)}$ (black - - - -), $B_{l}^{[1](2)}$ (black -.-.-.-). Case $\rho^{[1]}=1300~Kgm^{-3}$, $c^{[1]}=1400-i210~ms^{-1}$, $l=0$.}
\label{fig801}
\end{center}
\end{figure}
\begin{figure}[ptb]
\begin{center}
\includegraphics[width=0.75\textwidth]{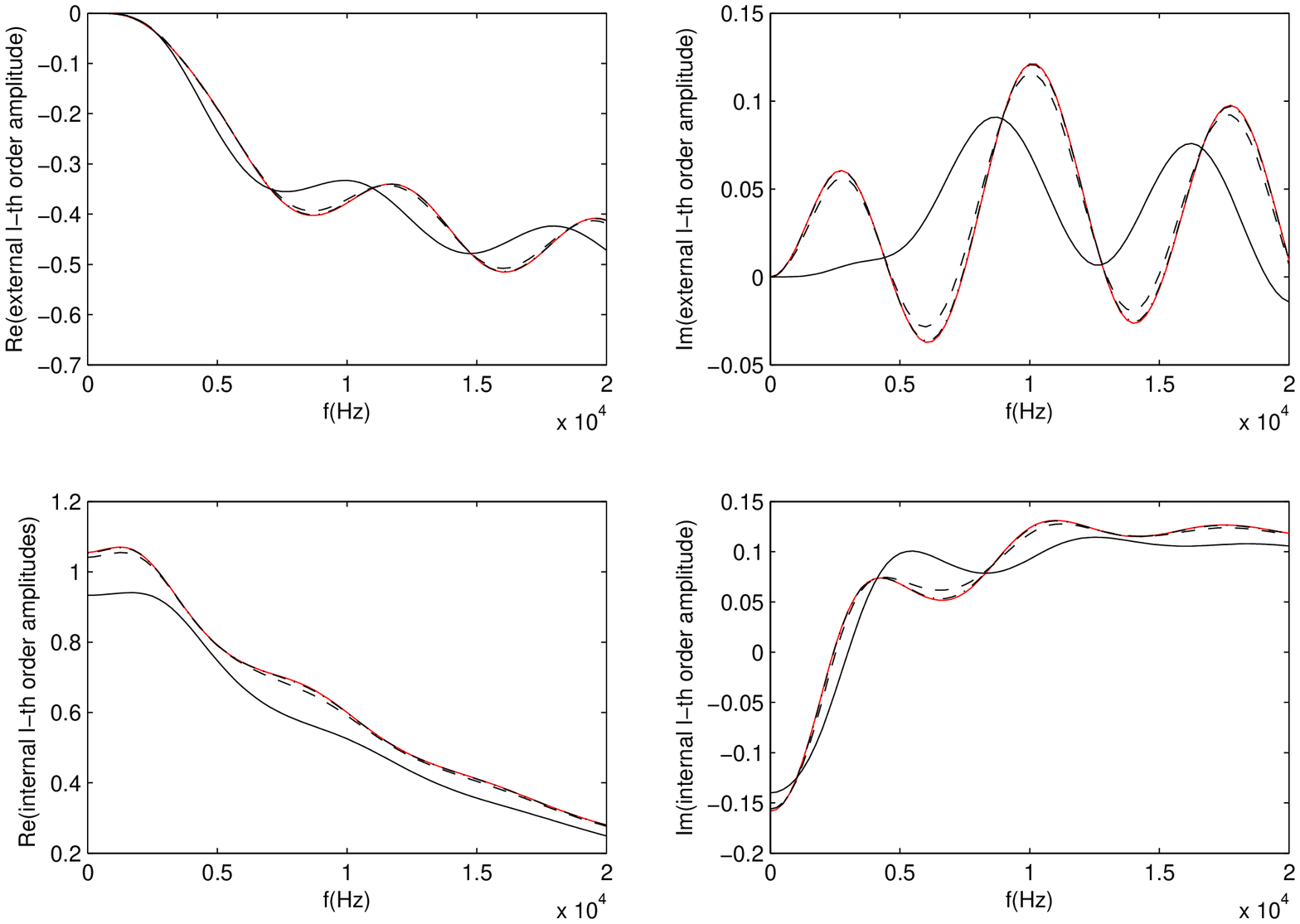}
\caption{Same as fig. \ref{fig801} except that $l=1$.}
\label{fig802}
\end{center}
\end{figure}
\begin{figure}[ptb]
\begin{center}
\includegraphics[width=0.75\textwidth]{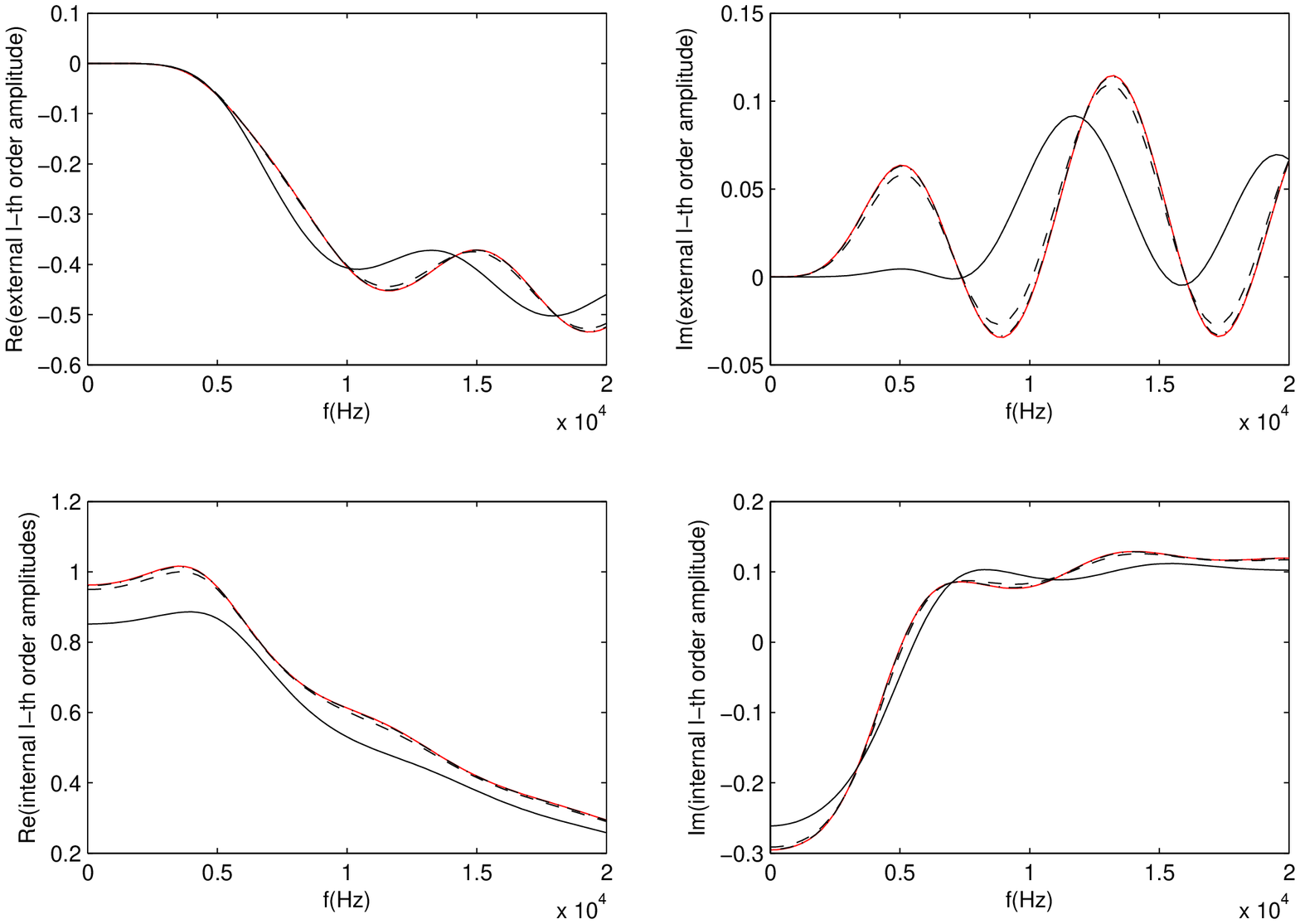}
\caption{Same as fig. \ref{fig801} except that $l=2$.}
\label{fig803}
\end{center}
\end{figure}
\begin{figure}[ptb]
\begin{center}
\includegraphics[width=0.75\textwidth]{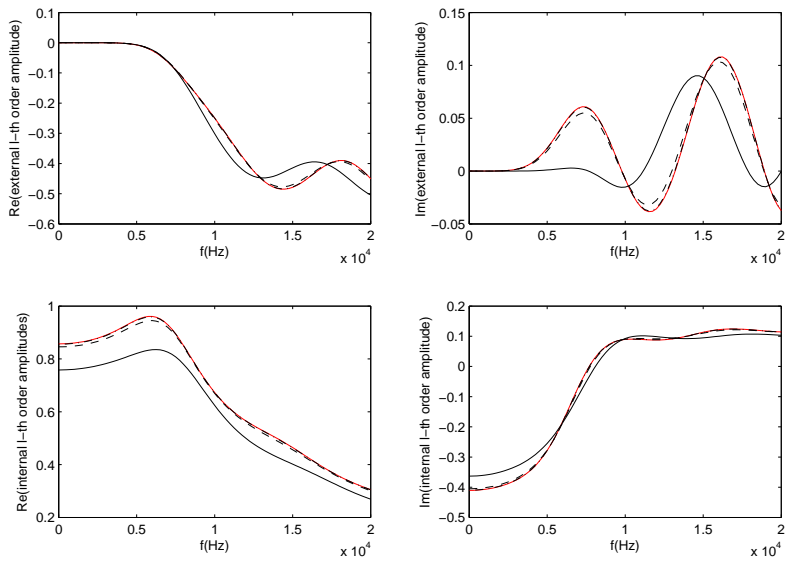}
\caption{Same as fig. \ref{fig801} except that $l=3$.}
\label{fig804}
\end{center}
\end{figure}
\clearpage
\newpage
We observe, between $\mathcal{A}_{l}^{[0]},~\mathcal{B}_{l}^{[1]}$ and $A_{l}^{[0]},~B_{l}^{[1]}$ the aggravated qualitative, and similar quantitative, differences as in sect. \ref{fig07}. Otherwise, the same comments as in the previous section apply here.
%%%%%%%%%%%%%%%%%%%%%%%%%%%%%%%%%%%%%%%%%%%%%%%%%%%%%%%%%%%%%
\subsection{Variation of high frequency $f$  for various orders $l$: case of fairly-large ($\varepsilon=0.231$) mass density contrast, small velocity contrast and  large $\big|\Im(c^{[1]})\big|$}\label{fig09}
\begin{figure}[ht]
\begin{center}
\includegraphics[width=0.75\textwidth]{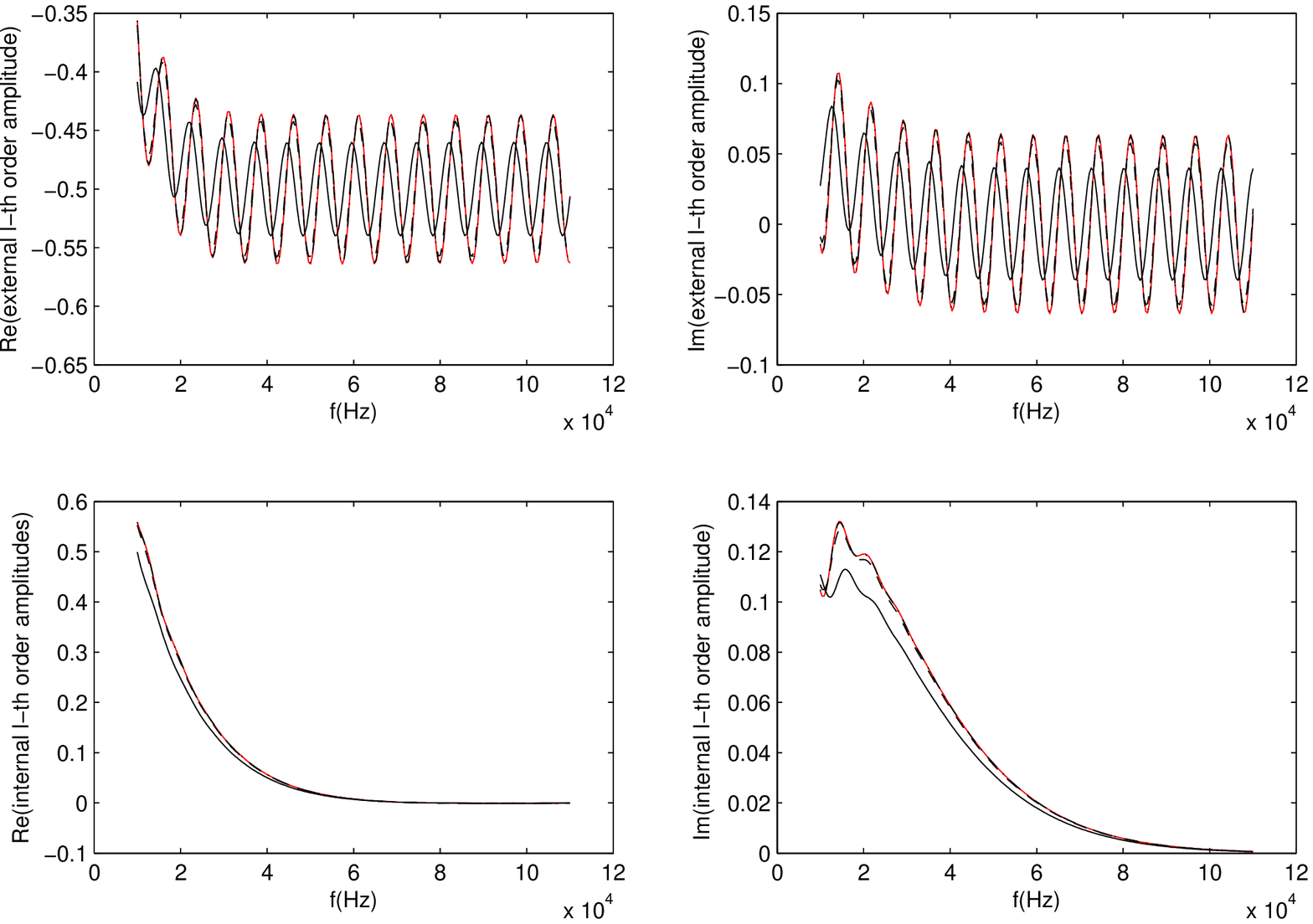}
\caption{Reflected and transmitted wavefield amplitudes as a function of $f$.  The upper left(right) panels depict the real(imaginary) parts of  $A_{l}^{[0]}$ (red),
$\mathcal{A}_{l}^{[0]}$ (black ------),
 $A_{l}^{[0](1)}$ (black - - - -), $A_{l}^{[0](2)}$ (black -.-.-.-).
The lower left(right) panels depict the real(imaginary) parts of  $B_{l}^{[1]}$ (red), $\mathcal{B}_{l}^{[1]}$ (black ------), $B_{l}^{[1](1)}$ (black - - - -), $B_{l}^{[1](2)}$ (black -.-.-.-). Case $\rho^{[1]}=1300~Kgm^{-3}$, $c^{[1]}=1400-i210~ms^{-1}$, $l=0$.}
\label{fig901}
\end{center}
\end{figure}
\begin{figure}[ptb]
\begin{center}
\includegraphics[width=0.75\textwidth]{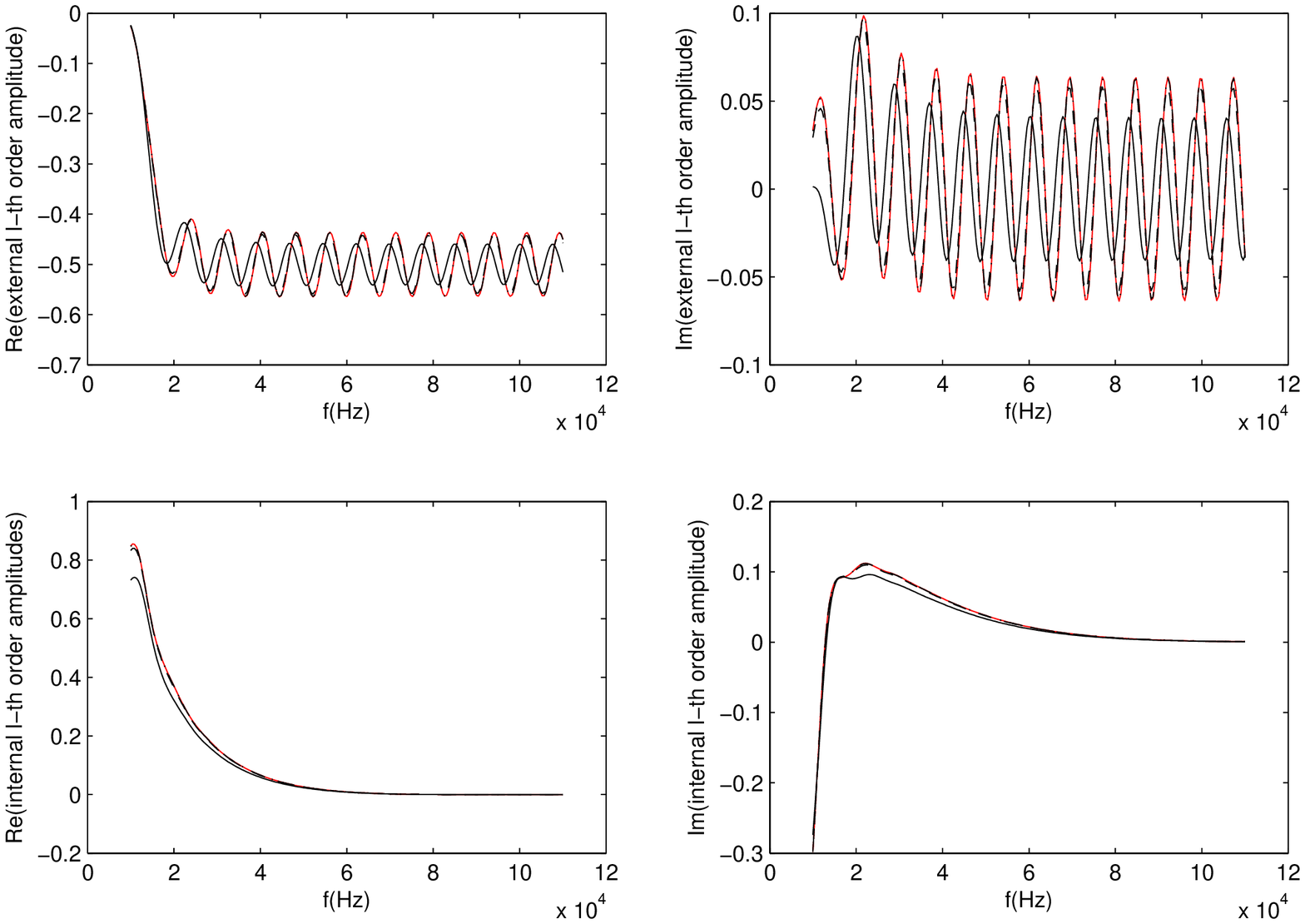}
\caption{Same as fig. \ref{fig901} except that $l=5$.}
\label{fig902}
\end{center}
\end{figure}
\begin{figure}[ptb]
\begin{center}
\includegraphics[width=0.75\textwidth]{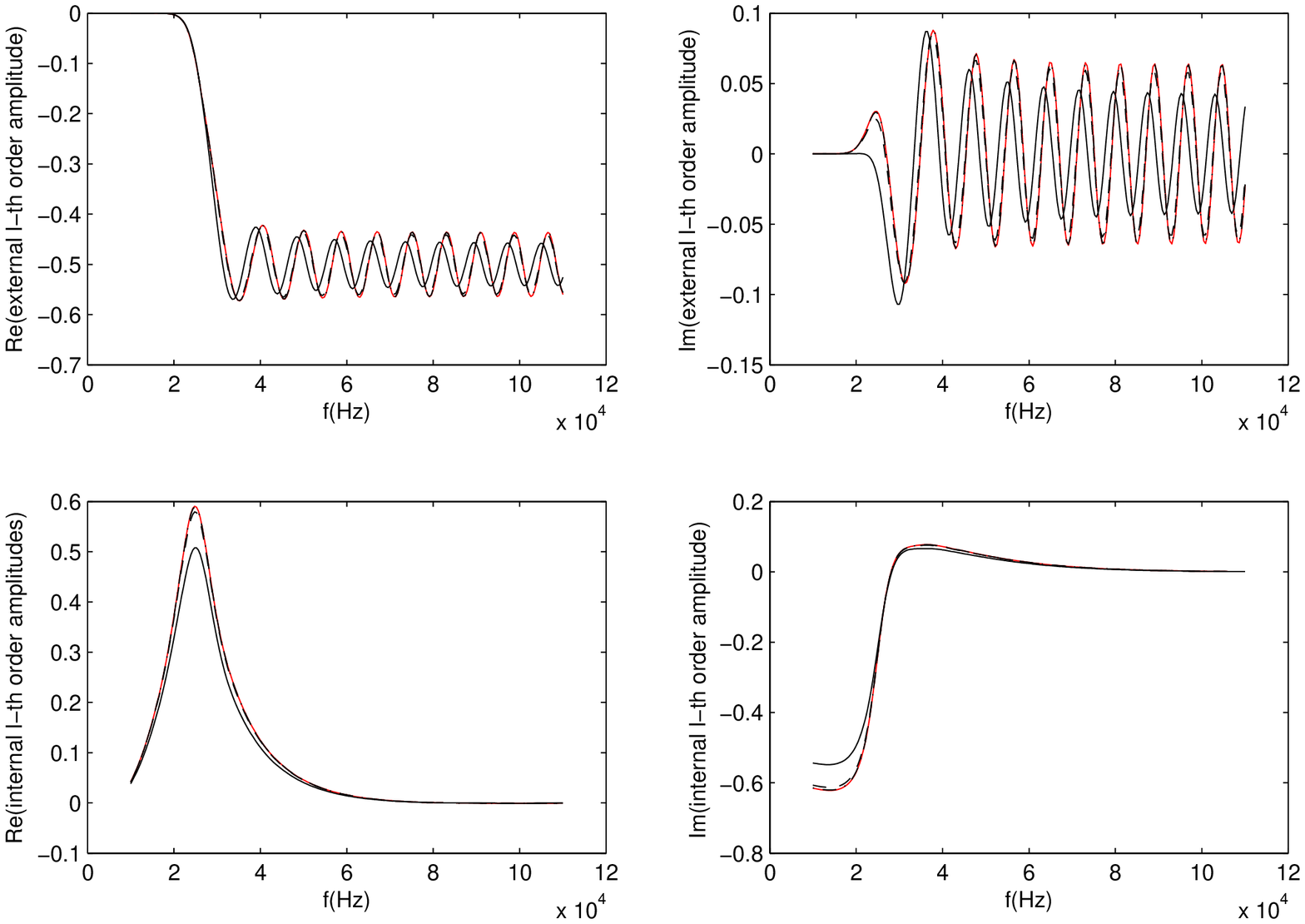}
\caption{Same as fig. \ref{fig901} except that $l=11$.}
\label{fig903}
\end{center}
\end{figure}
\begin{figure}[ptb]
\begin{center}
\includegraphics[width=0.75\textwidth]{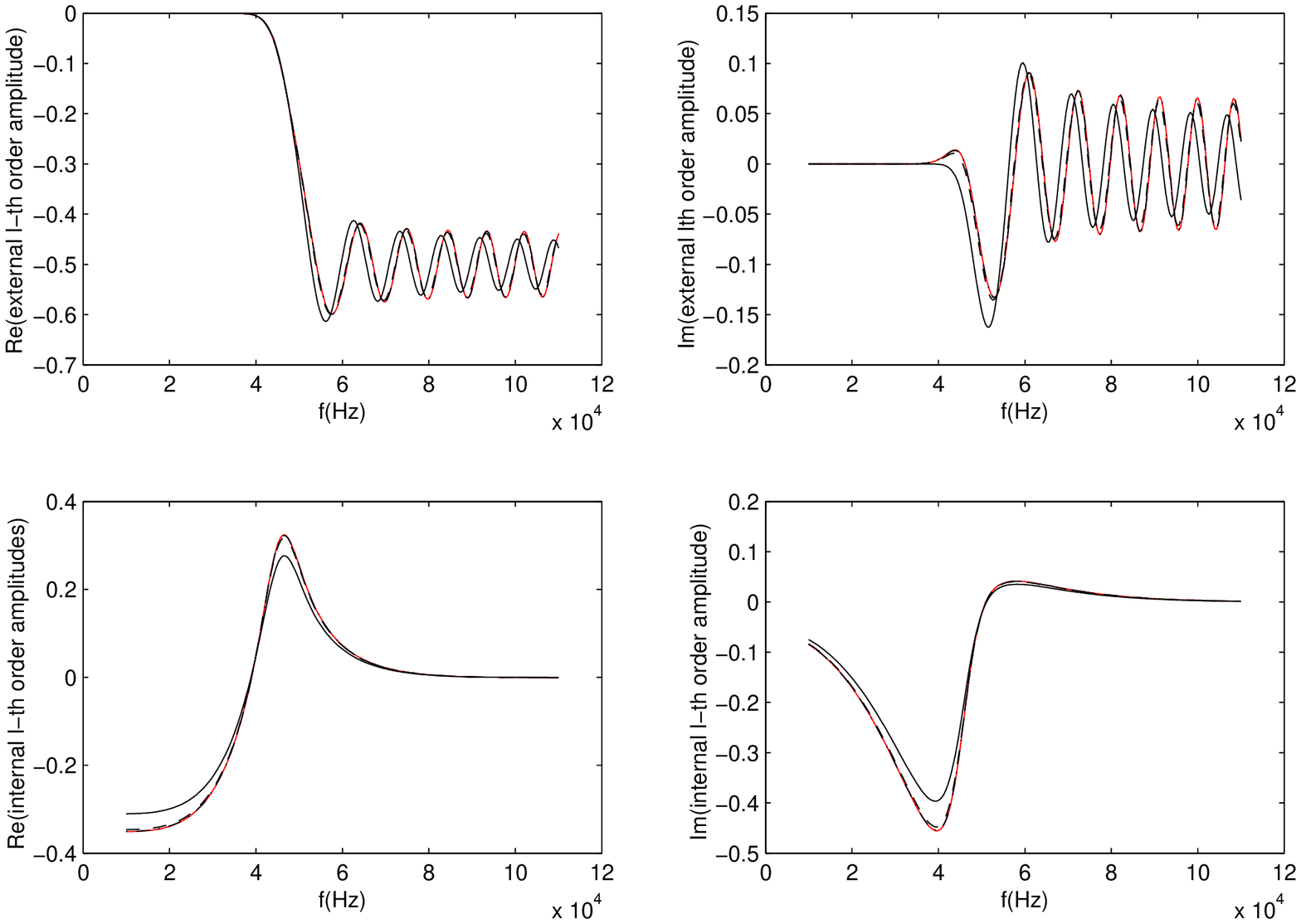}
\caption{Same as fig. \ref{fig901} except that $l=20$.}
\label{fig904}
\end{center}
\end{figure}
\clearpage
\newpage
We observe, between $\mathcal{A}_{l}^{[0]},~\mathcal{B}_{l}^{[1]}$ and $A_{l}^{[0]},~B_{l}^{[1]}$, rather large qualitative, and quantitative, differences. These differences are reduced between the exact amplitudes and  $A_{l}^{[0](1)},~B_{l}^{[1](1)}$, and even more so, between the exact amplitudes and  $A_{l}^{[0](2)},~B_{l}^{[1](2)}$.

A remarkable aspect of these figures is the evidence in the internal amplitudes of the occurrence of a resonance, followed by an asymptotic regime in which these amplitudes gently tend to zero. The resonances set in at $f=\sim20,\sim30,\sim50~KHz$ for $l=5,11,20$ respectively. These resonances are associated with the excitation of normal modes (as is testified by the fact that the denominator of $B_{l}^{[1]}$ vanishes or is nearly nil at the resonance frequencies), as explained in \cite{wi95}. The asymptotic regime seems to set in near $80~KHz$ for all $l$. Beyond this frequency, the external field is a quasi-periodic function of $f$, with slowly-decreasing envelope.
%%%%%%%%%%%%%%%%%%%%%%%%%%%%%%%%%%%%%%%%%%%%%%%%%%%%%%%%%%%%%
\subsection{Variation of frequency near the resonance frequency of the $l=20$ normal mode, for various $\Im(c^{[1]})$}\label{fig10}
\begin{figure}[ht]
\begin{center}
\includegraphics[width=0.75\textwidth]{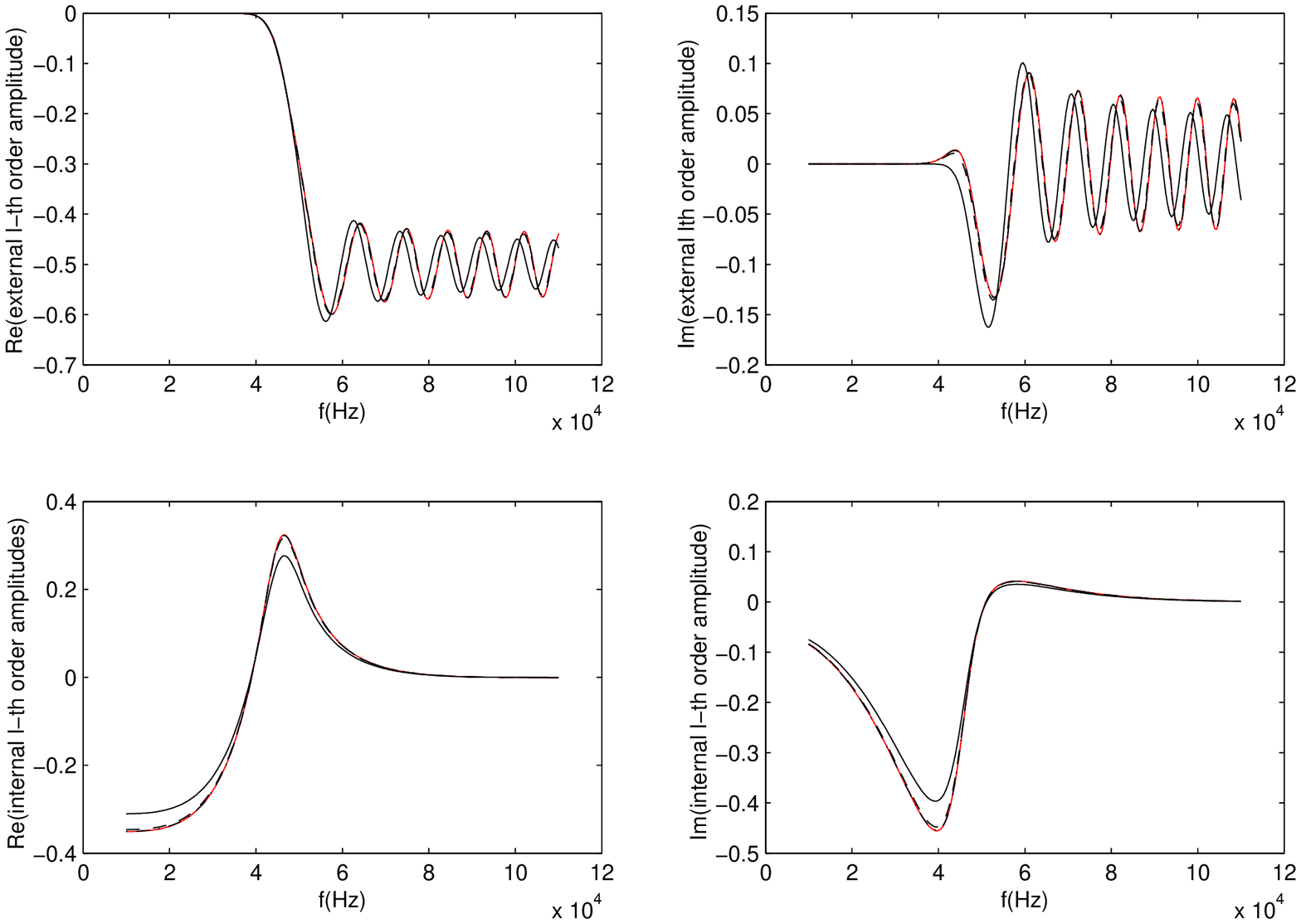}
\caption{Reflected and transmitted wavefield amplitudes as a function of $f$.  The upper left(right) panels depict the real(imaginary) parts of  $A_{l}^{[0]}$ (red),
$\mathcal{A}_{l}^{[0]}$ (black ------),
 $A_{l}^{[0](1)}$ (black - - - -), $A_{l}^{[0](2)}$ (black -.-.-.-).
The lower left(right) panels depict the real(imaginary) parts of  $B_{l}^{[1]}$ (red), $\mathcal{B}_{l}^{[1]}$ (black ------), $B_{l}^{[1](1)}$ (black - - - -), $B_{l}^{[1](2)}$ (black -.-.-.-). Case $\rho^{[1]}=1300~Kgm^{-3}$, $c^{[1]}=1400-i210~ms^{-1}$, $l=20$.}
\label{fig1001}
\end{center}
\end{figure}
\begin{figure}[ptb]
\begin{center}
\includegraphics[width=0.75\textwidth]{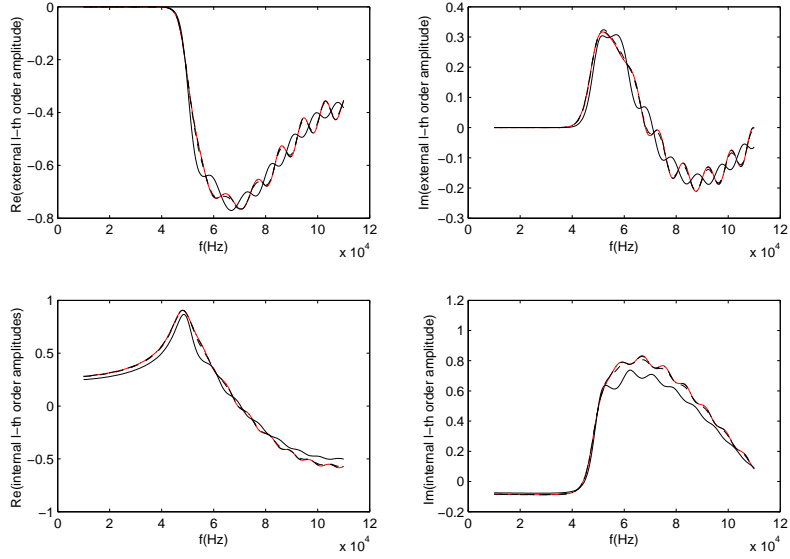}
\caption{Same as fig. \ref{fig1001} except that $c^{[1]}=1400-i21~ms^{-1}$.}
\label{fig1002}
\end{center}
\end{figure}
\begin{figure}[ptb]
\begin{center}
\includegraphics[width=0.75\textwidth]{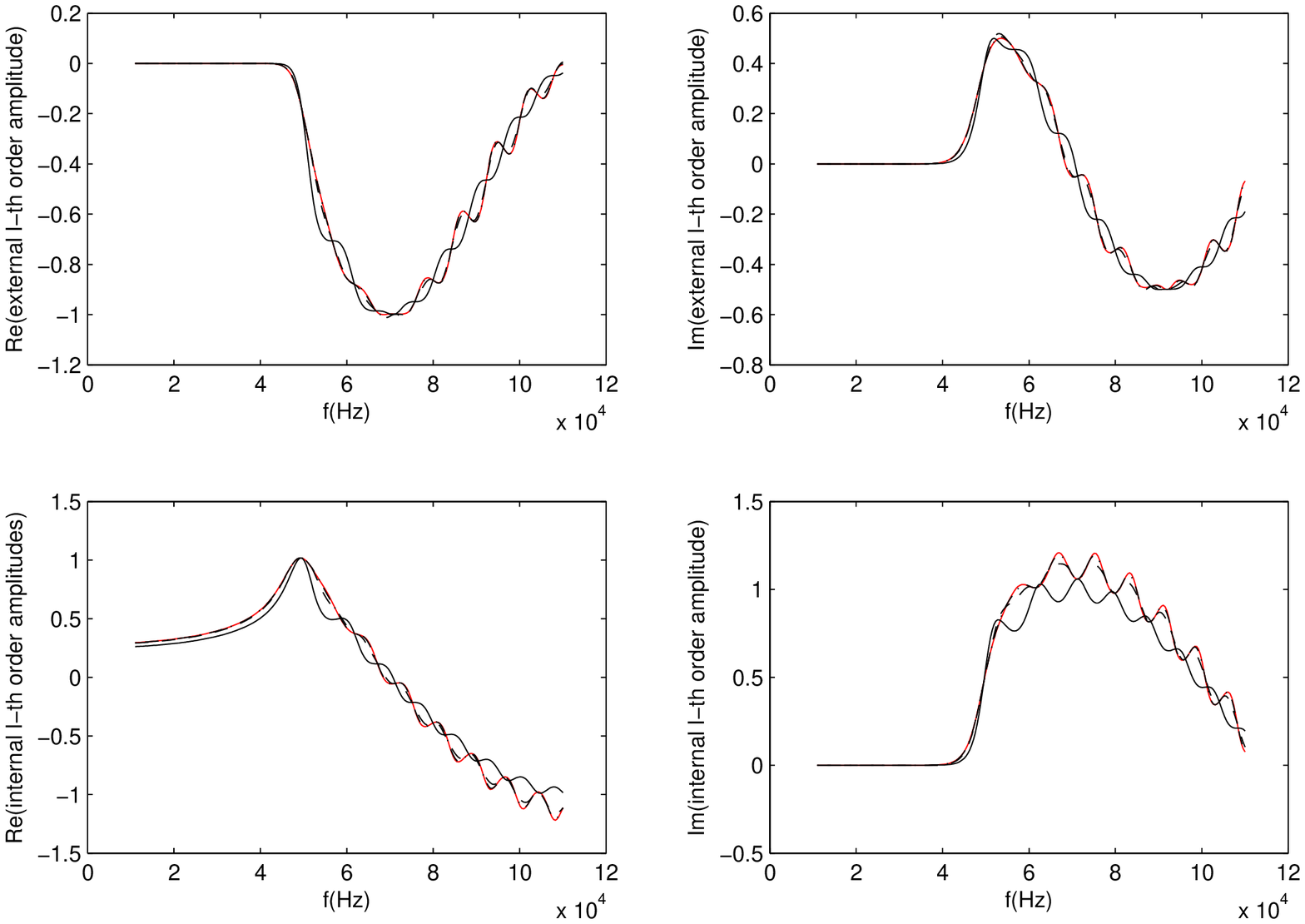}
\caption{Same as fig. \ref{fig1001} except that $c^{[1]}=1400-i0ms^{-1}$.}
\label{fig1003}
\end{center}
\end{figure}
\clearpage
\newpage
We again observe, between $\mathcal{A}_{l}^{[0]},~\mathcal{B}_{l}^{[1]}$ and $A_{l}^{[0]},~B_{l}^{[1]}$, rather large qualitative, and quantitative, differences. These differences are reduced between the exact amplitudes and  $A_{l}^{[0](1)},~B_{l}^{[1](1)}$, and even more so, between the exact amplitudes and  $A_{l}^{[0](2)},~B_{l}^{[1](2)}$.

To give further support to the normal mode resonance hypothesis, we progressively diminished $\big|\Im(c^{[1]})\big|$. This resulted, as expected, in the increase of the height of the resonance peak. This peak is not infinitely-high for $\Im(c^{[1]})=0$ because, even in the absence of material losses, there are pseudo-losses due to radiation damping (a part of the incident energy is always scattered to outer space (i.e., lost)) even when the obstacle acts like a closed cavity.
%%%%%%%%%%%%%%%%%%%%%%%%%%%%%%%%%%%%%%%%%%%%%%%%%%%%%%%%%%%%%
\subsection{Variation of frequency $f$ near the former $l=20$ normal mode resonance frequency  for various $\Re(c^{[1]})$}\label{fig11}
\begin{figure}[ht]
\begin{center}
\includegraphics[width=0.75\textwidth]{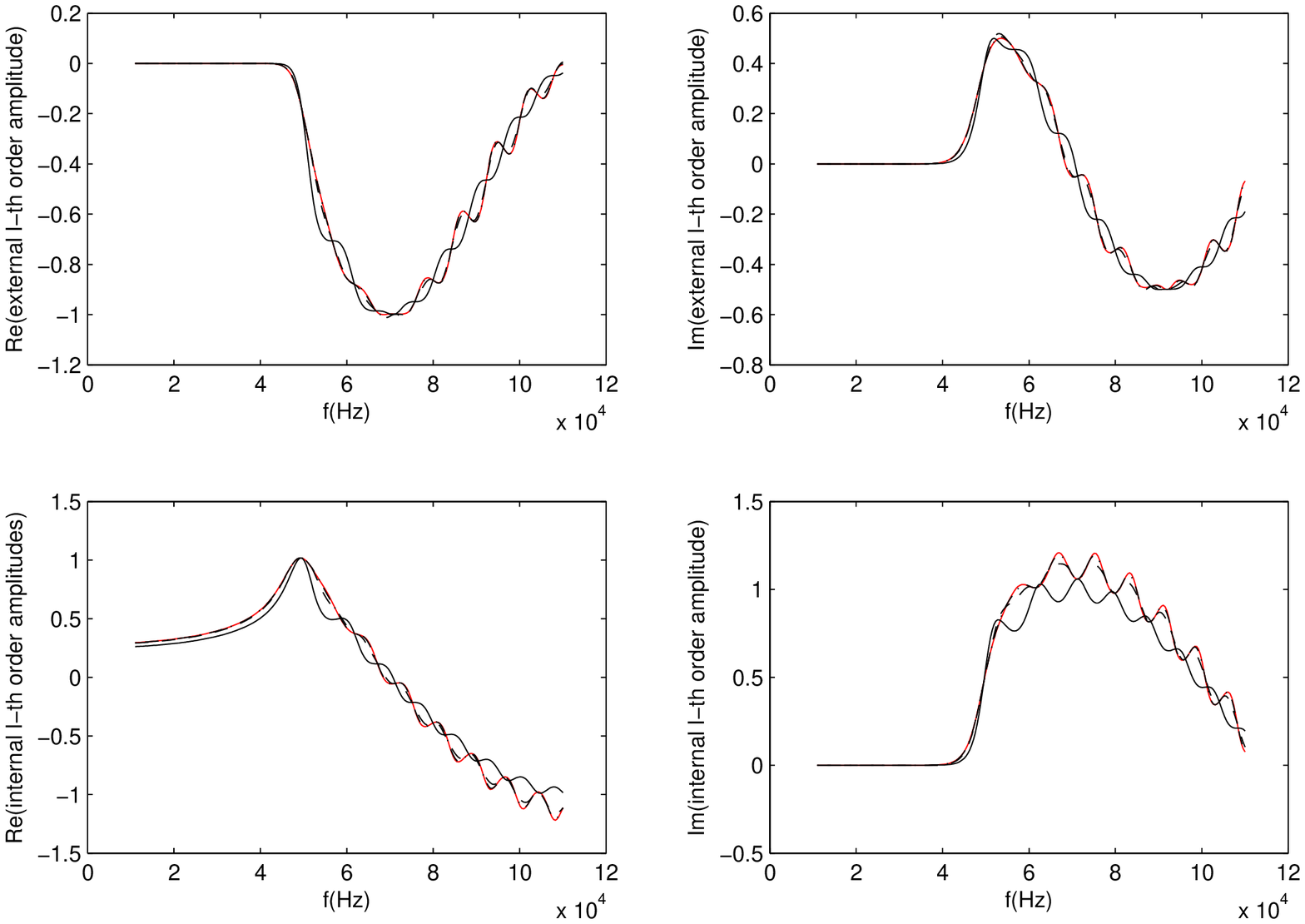}
\caption{Reflected and transmitted wavefield amplitudes as a function of $f$.  The upper left(right) panels depict the real(imaginary) parts of  $A_{l}^{[0]}$ (red),
$\mathcal{A}_{l}^{[0]}$ (black ------),
 $A_{l}^{[0](1)}$ (black - - - -), $A_{l}^{[0](2)}$ (black -.-.-.-).
The lower left(right) panels depict the real(imaginary) parts of  $B_{l}^{[1]}$ (red), $\mathcal{B}_{l}^{[1]}$ (black ------), $B_{l}^{[1](1)}$ (black - - - -), $B_{l}^{[1](2)}$ (black -.-.-.-). Case $\rho^{[1]}=1300~Kgm^{-3}$, $c^{[1]}=1400-i0~ms^{-1}$, $l=20$.}
\label{fig1101}
\end{center}
\end{figure}
\begin{figure}[ptb]
\begin{center}
\includegraphics[width=0.75\textwidth]{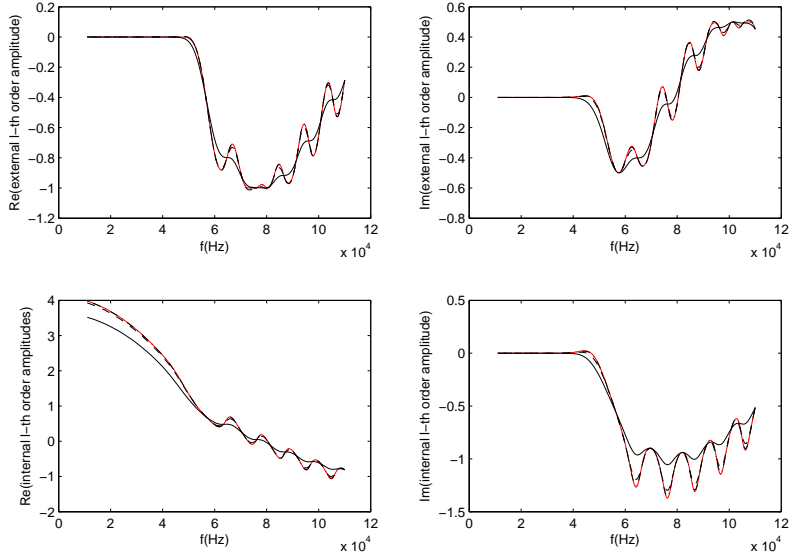}
\caption{Same as fig. \ref{fig1101} except that $c^{[1]}=1600-i0~ms^{-1}$.}
\label{fig1102}
\end{center}
\end{figure}
\begin{figure}[ptb]
\begin{center}
\includegraphics[width=0.75\textwidth]{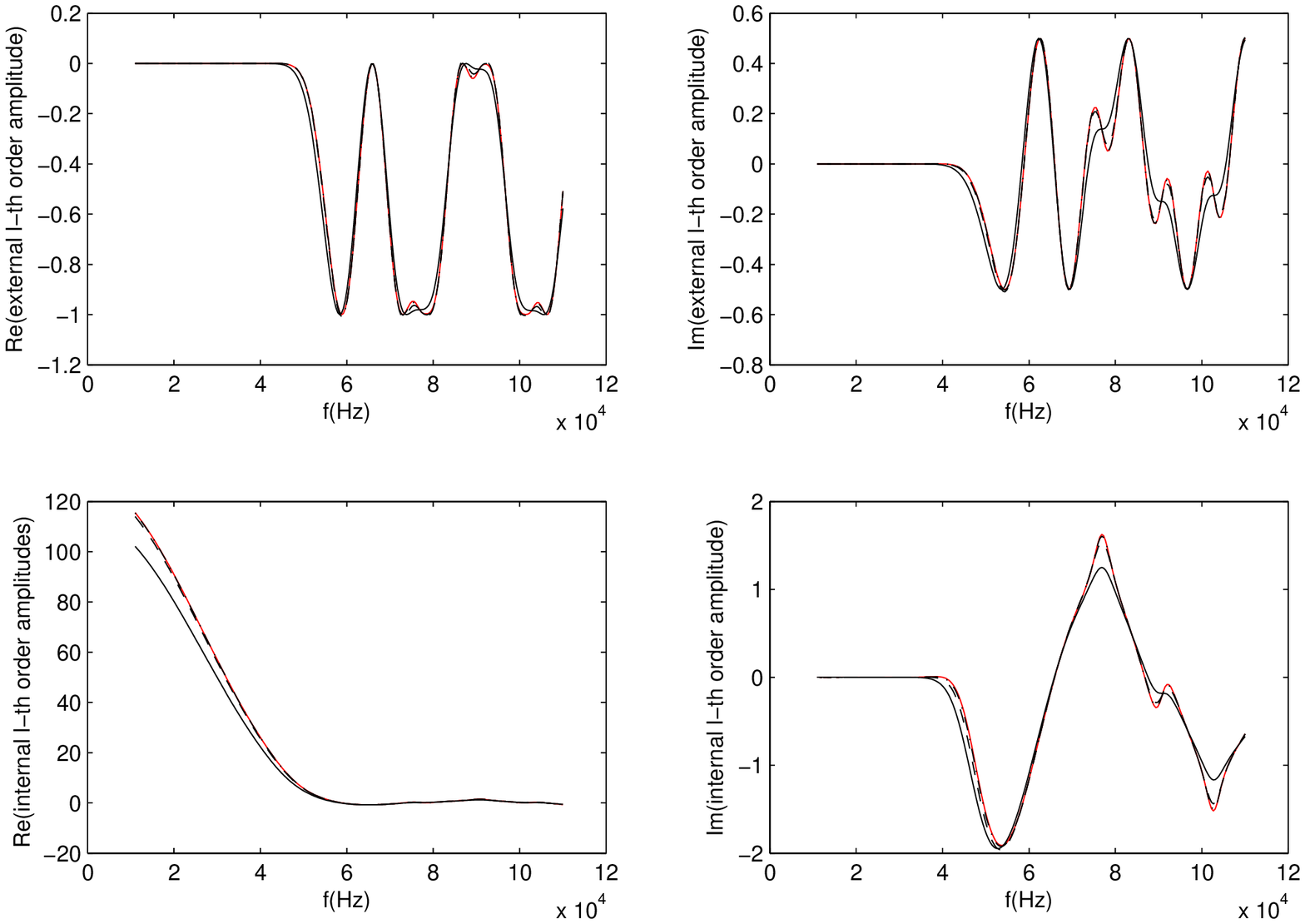}
\caption{Same as fig. \ref{fig1101} except that $c^{[1]}=1900-i0ms^{-1}$.}
\label{fig1103}
\end{center}
\end{figure}
\clearpage
\newpage
We  observe qualitative and quantitative differences between $\mathcal{A}_{l}^{[0]},~\mathcal{B}_{l}^{[1]}$ and $A_{l}^{[0]},~B_{l}^{[1]}$. These differences are reduced between the exact amplitudes and  $A_{l}^{[0](1)},~B_{l}^{[1](1)}$, and even more so, between the exact amplitudes and  $A_{l}^{[0](2)},~B_{l}^{[1](2)}$.

The increase of the velocity contrast is seen to progressively eliminate what was previously (when $c^{[1]}=1400~ms^{-1}$) identified as being the $l=20$ normal mode resonance near $f=50~KHz$. This does not necessarily mean that the resonances cease to exist althogehter, since they might show up for other $l$ at other frequencies when the velocity contrast is large (but we have not verified this).
%%%%%%%%%%%%%%%%%%%%%%%%%%%%%%%%%%%%%%%%%%%%%%%%%%%%%%%%%%%%%
\subsection{Variation of $\Re(c^{[1]})$ for various $l$: case of $\epsilon=0.231$ density contrast at $f=20~KHz$}\label{fig12}
\begin{figure}[ht]
\begin{center}
\includegraphics[width=0.75\textwidth]{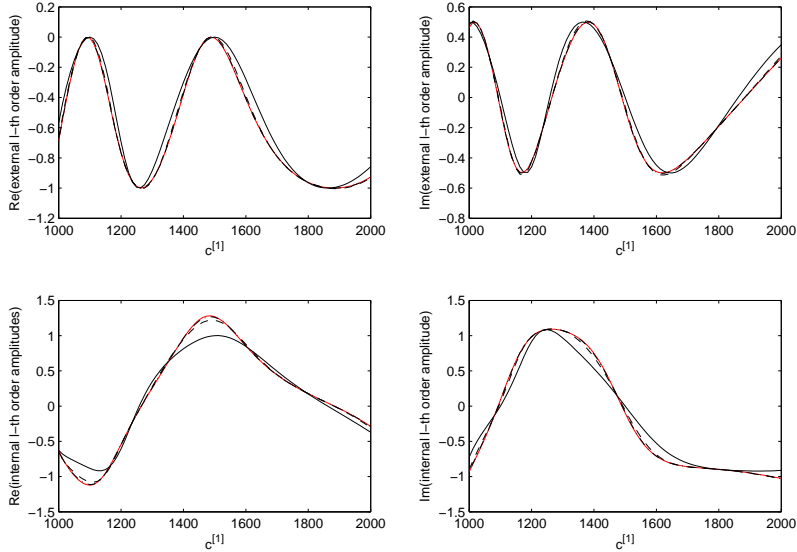}
\caption{Reflected and transmitted wavefield amplitudes as a function of $\Re(c^{[1]})$.  The upper left(right) panels depict the real(imaginary) parts of  $A_{l}^{[0]}$ (red),
$\mathcal{A}_{l}^{[0]}$ (black ------),
 $A_{l}^{[0](1)}$ (black - - - -), $A_{l}^{[0](2)}$ (black -.-.-.-).
The lower left(right) panels depict the real(imaginary) parts of  $B_{l}^{[1]}$ (red), $\mathcal{B}_{l}^{[1]}$ (black ------), $B_{l}^{[1](1)}$ (black - - - -), $B_{l}^{[1](2)}$ (black -.-.-.-). Case $\rho^{[1]}=1300~Kgm^{-3}$, $\Im c^{[1]}=0~ms^{-1}$, $f=20000~Hz$, $l=0$.}
\label{fig1201}
\end{center}
\end{figure}
\begin{figure}[ptb]
\begin{center}
\includegraphics[width=0.75\textwidth]{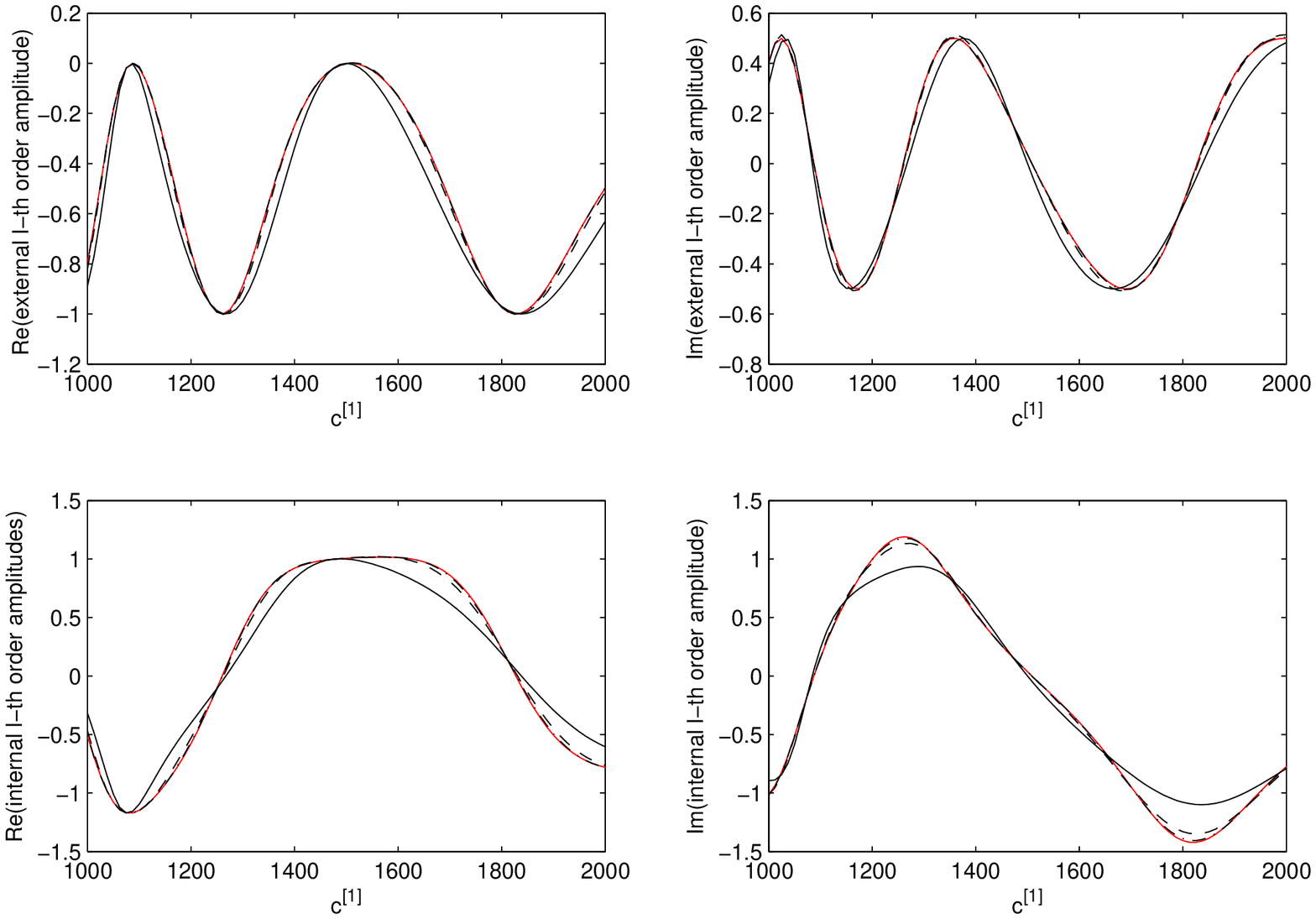}
\caption{Same as fig. \ref{fig1201} except that $l=1$.}
\label{fig1202}
\end{center}
\end{figure}
\begin{figure}[ptb]
\begin{center}
\includegraphics[width=0.75\textwidth]{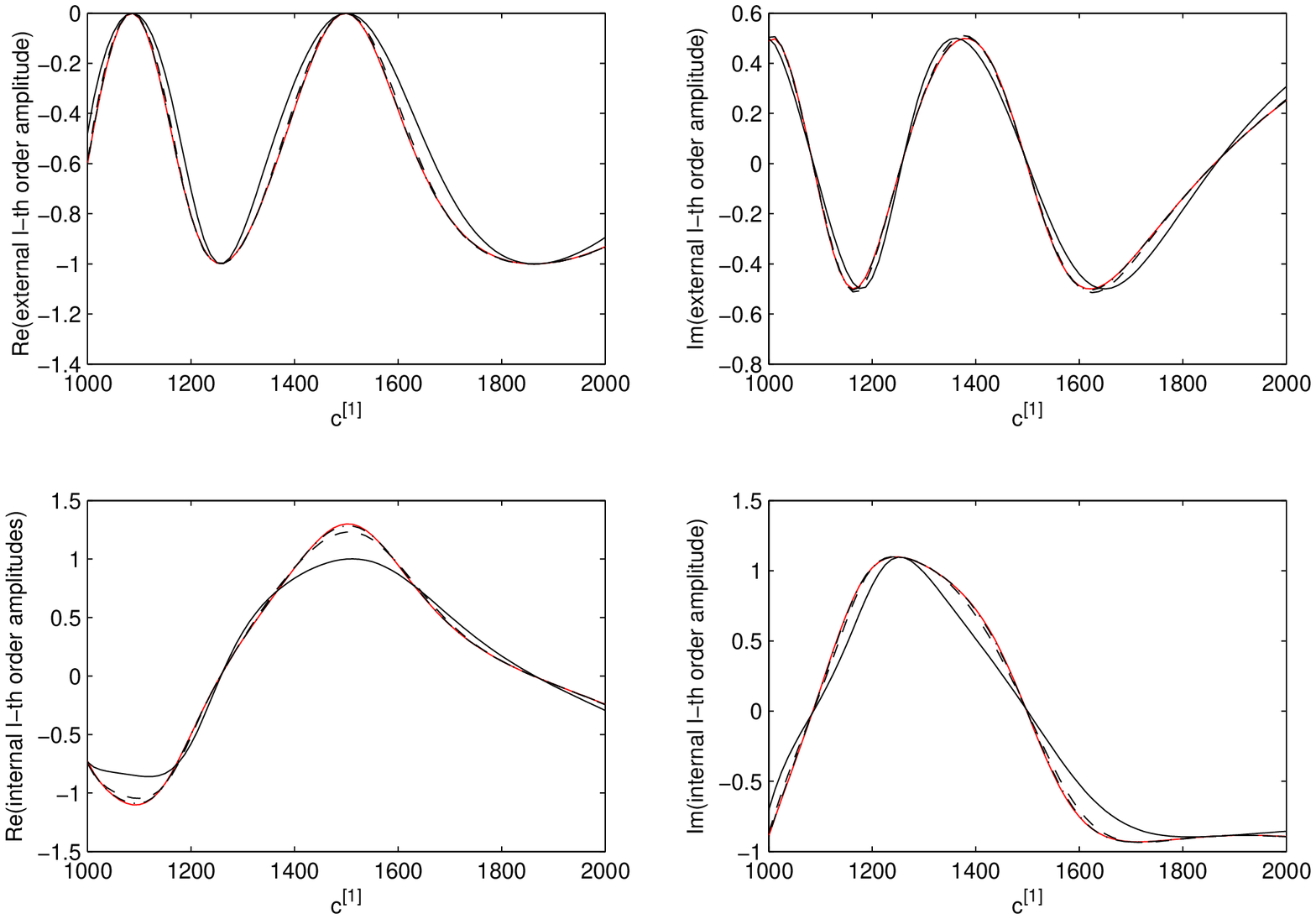}
\caption{Same as fig. \ref{fig1201} except that $l=2$.}
\label{fig1203}
\end{center}
\end{figure}
\begin{figure}[ptb]
\begin{center}
\includegraphics[width=0.75\textwidth]{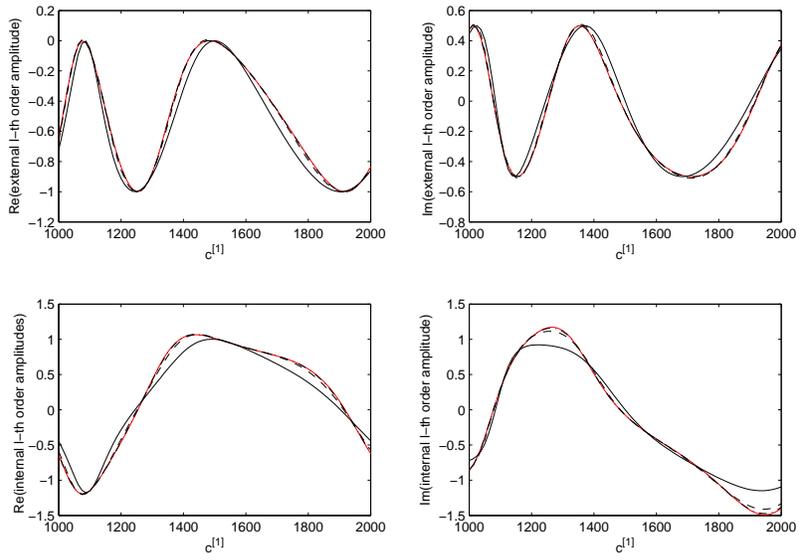}
\caption{Same as fig. \ref{fig1201} except that $l=3$.}
\label{fig1204}
\end{center}
\end{figure}
\clearpage
\newpage
We  observe fairly-substantial qualitative and quantitative differences between $\mathcal{A}_{l}^{[0]},~\mathcal{B}_{l}^{[1]}$ and $A_{l}^{[0]},~B_{l}^{[1]}$. These differences  seem to increase somewhat with the velocity contrast (but less than with mass density contrast observed previously) and, as previously, are reduced between the exact amplitudes and  $A_{l}^{[0](1)},~B_{l}^{[1](1)}$, and even more so, between the exact amplitudes and  $A_{l}^{[0](2)},~B_{l}^{[1](2)}$.
%%%%%%%%%%%%%%%%%%%%%%%%%%%%%%%%%%%%%%%%%%%%%%%%%%%%%%%%%%%%%
\subsection{Variation of $\Re(c^{[1]})$ for various $l$:  case of $\epsilon=-0.429$ density contrast at $f=20~KHz$}\label{fig13}
\begin{figure}[ht]
\begin{center}
\includegraphics[width=0.75\textwidth]{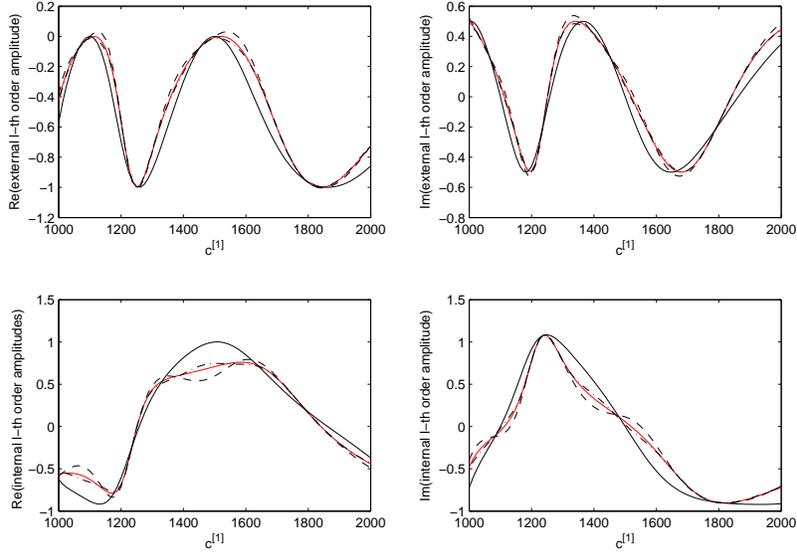}
\caption{Reflected and transmitted wavefield amplitudes as a function of $\Re(c^{[1]})$.  The upper left(right) panels depict the real(imaginary) parts of  $A_{l}^{[0]}$ (red),
$\mathcal{A}_{l}^{[0]}$ (black ------),
 $A_{l}^{[0](1)}$ (black - - - -), $A_{l}^{[0](2)}$ (black -.-.-.-).
The lower left(right) panels depict the real(imaginary) parts of  $B_{l}^{[1]}$ (red), $\mathcal{B}_{l}^{[1]}$ (black ------), $B_{l}^{[1](1)}$ (black - - - -), $B_{l}^{[1](2)}$ (black -.-.-.-). Case $\rho^{[1]}=700~Kgm^{-3}$, $\Im c^{[1]}=0~ms^{-1}$, $f=20000~Hz$, $l=0$.}
\label{fig1301}
\end{center}
\end{figure}
\begin{figure}[ptb]
\begin{center}
\includegraphics[width=0.75\textwidth]{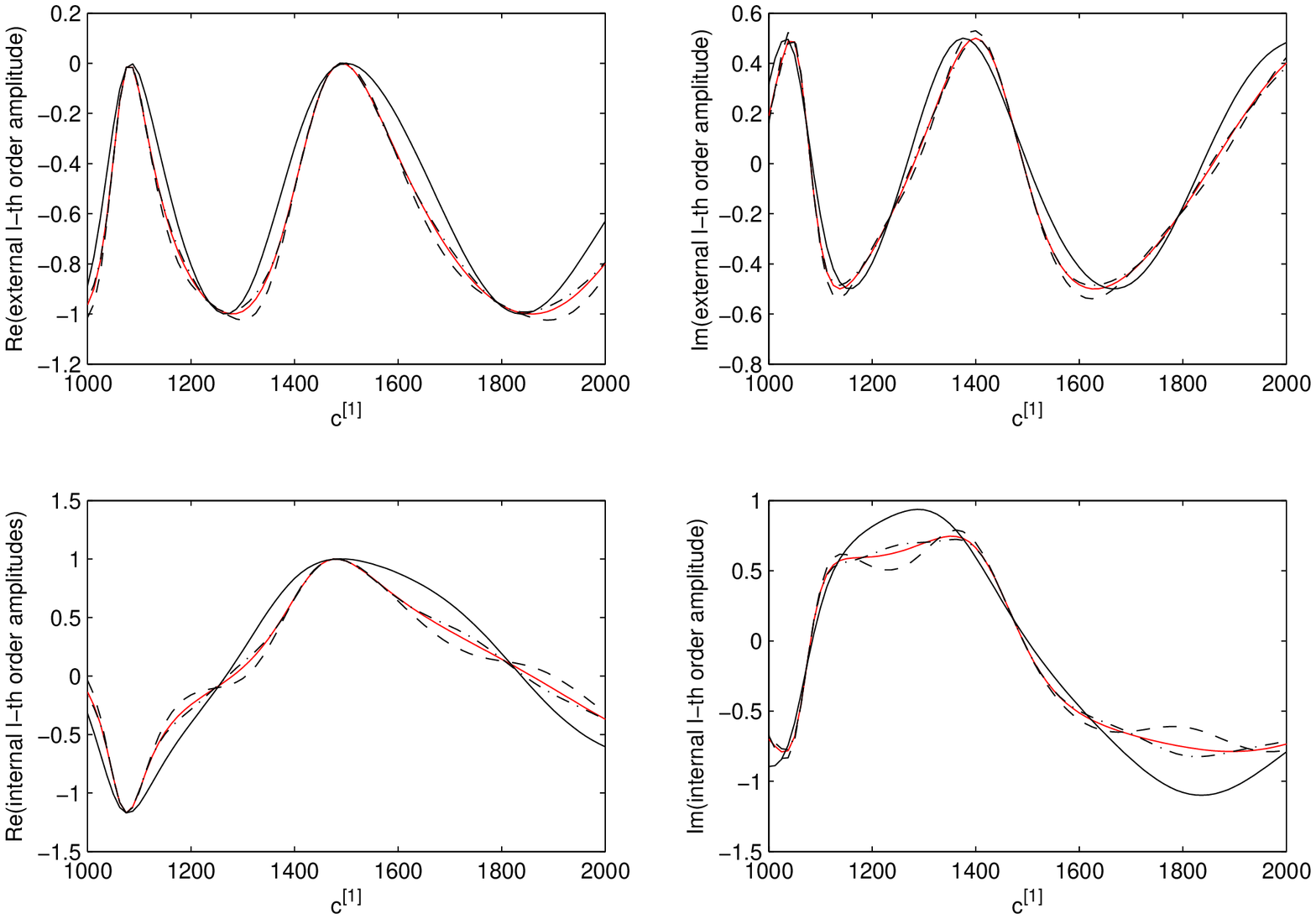}
\caption{Same as fig. \ref{fig1301} except that $l=1$.}
\label{fig1302}
\end{center}
\end{figure}
\begin{figure}[ptb]
\begin{center}
\includegraphics[width=0.75\textwidth]{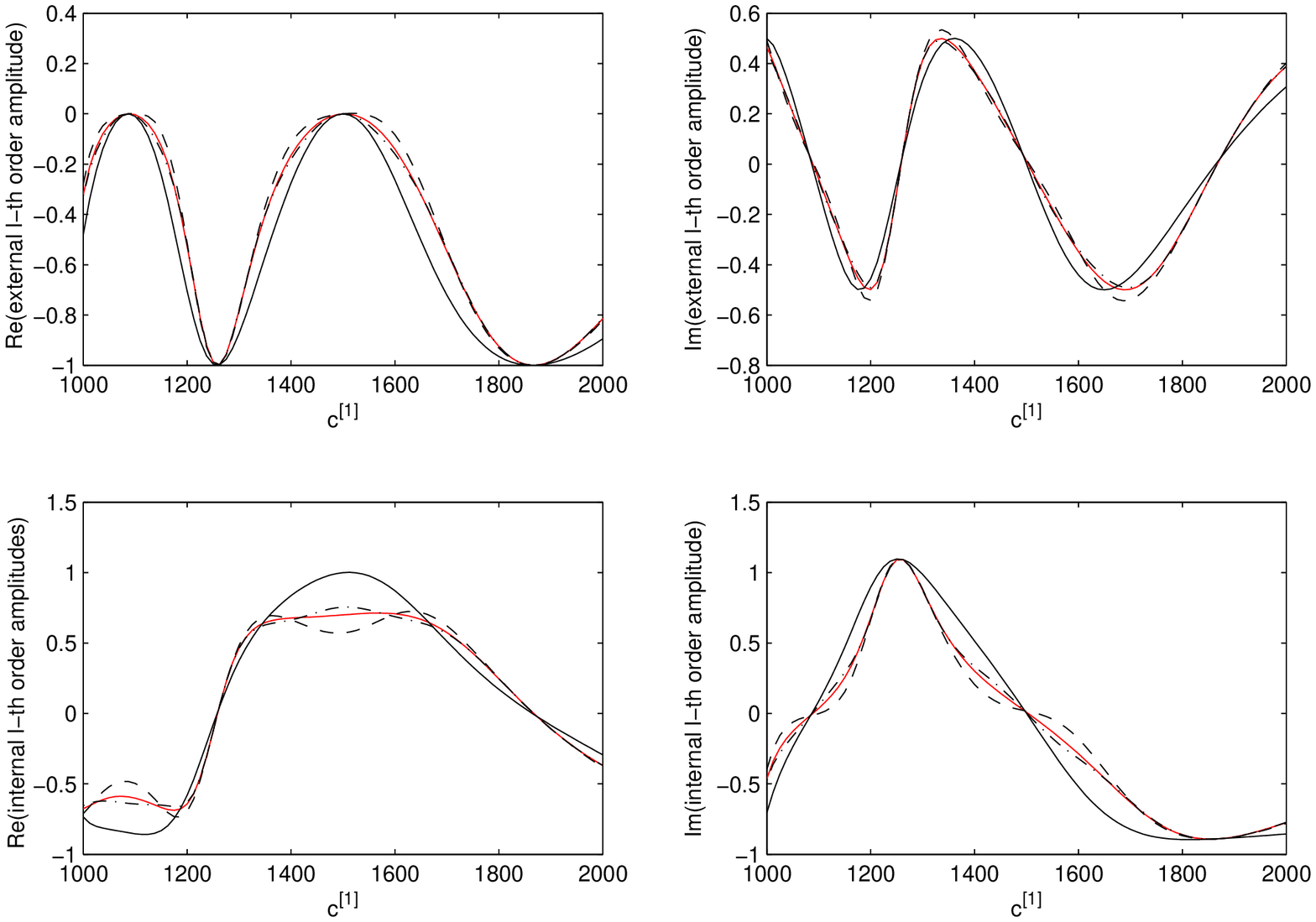}
\caption{Same as fig. \ref{fig1301} except that $l=2$.}
\label{fig1303}
\end{center}
\end{figure}
\begin{figure}[ptb]
\begin{center}
\includegraphics[width=0.75\textwidth]{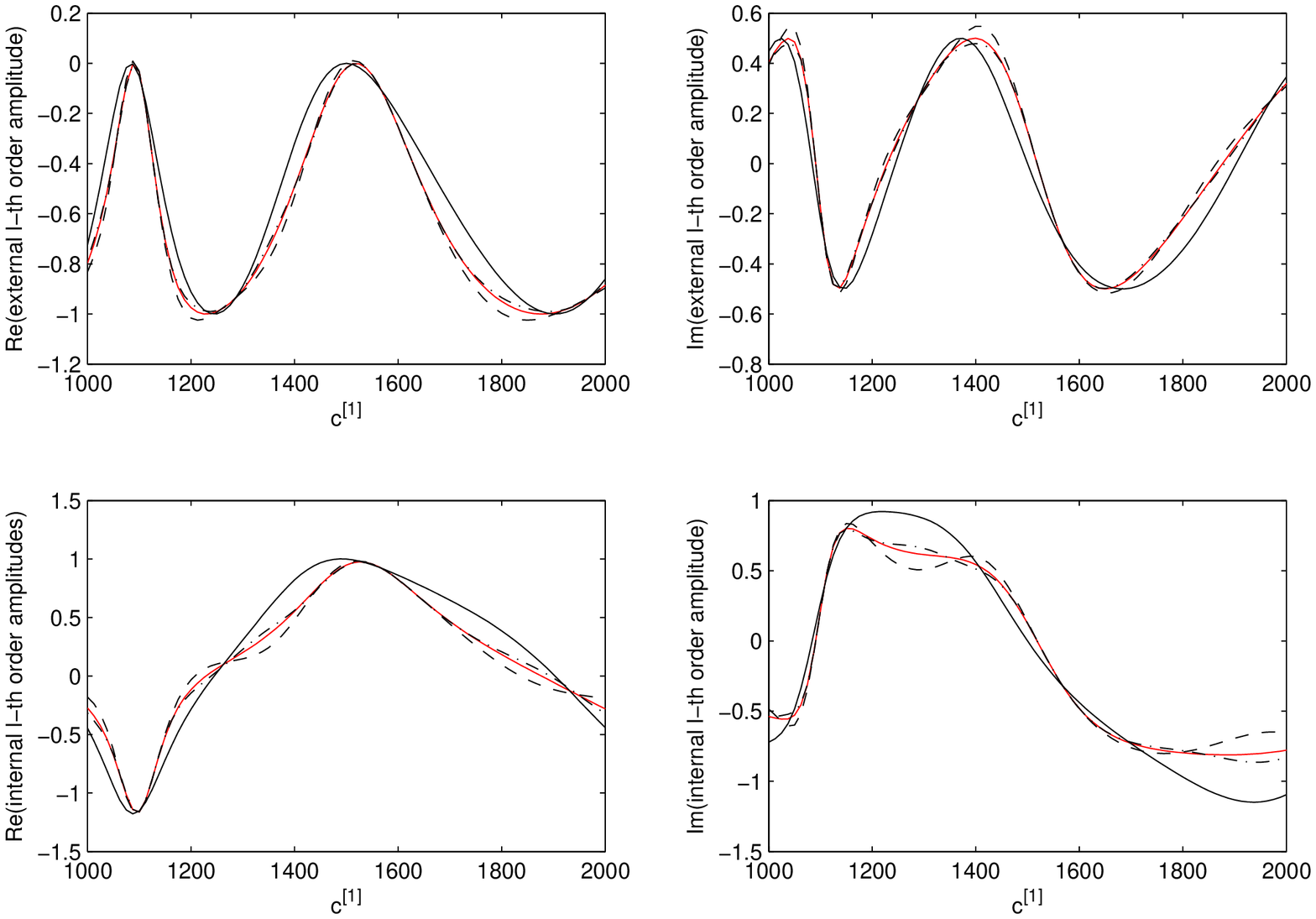}
\caption{Same as fig. \ref{fig1301} except that $l=3$.}
\label{fig1304}
\end{center}
\end{figure}
\clearpage
\newpage
We  observe very-substantial qualitative and quantitative differences between $\mathcal{A}_{l}^{[0]},~\mathcal{B}_{l}^{[1]}$ and $A_{l}^{[0]},~B_{l}^{[1]}$. These differences  seem to increase somewhat  with the velocity contrast (but less than with mass density contrast observed previously) and, as previously, are reduced between the exact amplitudes and  $A_{l}^{[0](1)},~B_{l}^{[1](1)}$, and even more so, between the exact amplitudes and  $A_{l}^{[0](2)},~B_{l}^{[1](2)}$.
%%%%%%%%%%%%%%%%%%%%%%%%%%%%%%%%%%%%%%%%%%%%%%%%%%%%%%%%%%%%%%%%%%%%%%%%%%%%%%
\section{Conclusions}
The approach, based on brute-force resolution of the partial differential equations (as they stand) of wave scattering, is certainly useful in contemporary studies and applications dealing with both forward and inverse scattering problems.  The principal drawback of this approach is that it is purely-numerical, and thus gives little insight as to what parameter (or group of parameters), contribute the most to the different phenomena that intervene in wave scattering. A convincing example of this is the phenomenon of normal mode resonances which may  be hard to detect in the numerical results if the frequency is not varied with very small steps, and certainly harder to explain from purely-numerical evidence. Another, practical, drawback of the numerical PDE approach is that it is computationally-intensive, even in 2D, scalar versions employed to get a first grasp on what are fundamentally 3D and/or vectorial problems, the latter requiring enormous computer resources out of reach of most research groups. This negative aspect of the brute-force PDE approach is particularly acute in inverse scattering problems  because the tendency is more and more to solve them by the nonlinear full-wave inversion (NLFWI) method which appeals to optimization algorithms that require  the resolution of (usually) many forward problems.

An alternative to the brute-force numerical resolution of the PDE is the integral equation (IE) approach. Actually, as we have shown herein, there exist many variants of this approach, the most appealing of which appears to be the one relying on domain integral (DI) representation of the scattered field because it is the easiest to employ when the obstacle is composed of several objects (thus is discontinuously-heterogeneous) or is continuously-heterogeneous. The other IE methods are interesting too because they involve less computations when the numerical stage is attained, but they require, at the outset, that the  objects composing the obstacle be homogeneous with respect to at least one of the constitutive parameters (usually the mass density in acoustic and elastodynamic problems, permeability in electromagnetic problems). In fact, when this assumption of homogeneity is made, it also greatly simplifies the DI formulation, a fact that was recognized very early by those concerned with biophysical, marine acoustical, and geophysical applications. The principal reason for  having adopted the DI approach in these applications was, and still is, that it leads, via the Born approximation and asymptotic forms of the Green's function, to what   is called diffraction tomography (DT), which constitutes an explicit,  functional rather than numerical  (although, somewhat distorted) form of solution of the inverse problem. Of course, the IE solutions (not necessarily of the DI variety) can also be incorporated in NLFWI schemes to speed up the numerical resolution of inverse problems.

Other than affording the advantages of simplicity and computational economy, one suspects that the constant mass density assumption (which is not usually invoked in the computational PDE approach)  must have some drawbacks when employed in conjuction with an IE technique. The most obvious one is that, in the inverse problem context, it is not possible, by this means, to obtain a quantitative, realistic retrieval of the mass density of an obstacle such as a female breast \cite{gd07,wb12} simply because this density is taken to be that of water. or of the surrounding tissue. The second drawback of the constant-density assumption is that it is based on blind faith, especially in the geophysical context.

Consequently, the question we addressed herein was: how reliable is the constant-density assumption, particularly in the 2D acoustical and geophysical contexts in which the governing PDE is the Bergman PDE (equivalent to the 2D SH Mal-Knopoff PDE) that one might want to solve by an IE method?. As shown herein, the IE method can take a variety of forms: BI-BI, DI-BI, BI-DI and DI-DI, the latter of which exists in several versions which we found to be equivalent. All these IE methods, which have been employed to numerically solve a variety of  both inverse and forward scattering problems, do not enable to answer the aforementioned question because they do not yield tractable mathematical (rather than purely-numerical)  solutions for 2D obstacles of general shape and composition. Moreover, all the IE techniques except the DI-DI one, rely on the initial assumption that the density of the obstacle be spatially-invariant, although different from that of the host medium (this being less-radical than assuming the density of both the obstacle and host to be spatially-invariant and equal, i.e., the constant density assumption). Thus, to compare the answers furnished by the four forms of the IE technique, we were obliged, at the outset, to make the assumption of a constant-density obstacle. But this was not sufficient, because none of the four IE techniques enable the obtention of a mathematically-tractable solution for a macroscopically-homogeneous (this term is used either for a medium containing no macroscopic objects or for a porous medium such as a solid network of connected microscopically-thin rods within a fluid for which a homogenization scheme enables its constitutive properties to be approximated \cite{of11,wi18} by those of an 'effective' homogeneous fluid or solid) obstacle of general shape. Consequently, we imposed the additional  assumption that the shape of the obstacle be canonical, this meaning that the governing PDE is separable in a given coordinate system and that the boundary of the obstacle can be represented by an equation $\mathfrak{k}=$const., where $\mathfrak{k}$ is one of the coordinates in the chosen coordinate system.

In fact, we chose the cylindrical coordinate system $r,\theta,z$ so that the obstacle is a cylinder \cite{of11} whose boundary is circular, i.e, $r=a=$const for all $\theta\in[0,2\pi$. For this type of macroscopically-homogeneous obstacle, the Bergman PDE can be solved \cite{of11,wi99} in closed form via domain decomposition and separation of variables (DD-SOV) so that, at worst, one thus disposes of a reference solution to which can be compared the solutions furnished by the various IE methods. We say 'at worst' because it is not evident that these IE methods can yield closed-form solutions even for such a homogeneous canonical-shaped obstacle so that the comparisons might have been no more than numerical in nature. It turns out (this was the result of a great deal of what is offered in this study) that all the IE schemes enable to re-generate the exact DD-SOV sollution which gives some legitimacy to these schemes, but still does not answer the question of the reliability of the constant-density supposition.

It could be thought that since we dispose of an exact solution, for the two constant-density media separated by a circular boundary, in convenient mathematical form, why not just mathematically compare this solution to its counterpart obtained by taking the two densities to be equal, or what is more interesting, nearly equal? Actually, there did not appear to exist an easy way to do this (particularly for the nearly-equal density case), but it was found that the DI-DI method, contrary to the other three IE methods, enables one to generate the solution in a form in which the density contrast $\epsilon$ emerges in a clearly-identifiable manner. From this type of solution, we were able to show that the  amplitudes of the external and internal cylindrical waves comprising the scattered field each take the form of a series in powers of $\epsilon$ and that the first term of each such  series is none other than the corresponding (external or internal wave) amplitude  resulting from the constant-density assumption.

This series form of the solution for the non-constant density problem thus constitutes the mathematical tool, supplemented by numerical examples extracted therefrom, by which we were able to answer the question of the reliability of the constant-density assumption, at least as concerns the model of a macroscopically-homogeneous obstacle having a circular boundary. Finally, we gave arguments that justify the employment of this model to yield useful information concerning the constant-density assumption even when the obstacle is neither macroscopically-homogeneous (but can be homogenized) nor enclosed within a physical circular boundary (but can be enclosed within a virtual circular boundary, a procedure that gives rise to no difficulties in the DI-DI method).
%%%%%%%%%%%%%%%%%%%%%%%%%%%%%%%%%%%%%%%%%%%%%%%%%%%%%%%%%%%%%%%%%%%%%%%%%%%%%%
%\clearpage
%\newpage
%

%%%%%%%%%%%%%%%%%%%%%%%%%%%%%%%%%%%%%%%%%%%%%%%%%%%%%%%%%%%%%%%%%%%%%%%%
\end{document}